\documentclass[a4paper,UKenglish,cleveref,autoref]{lipics-v2019}

\input{macros}
\boolfalse{compilePaper}

\nolinenumbers

\ispaper{
  \author{Delia Kesner}{IRIF, Universit\'{e} de Paris and CNRS}{}{}{}
  \author{Eduardo Bonelli}{Stevens Institute of Technology}{}{}{}
  \author{Andr\'{e}s Viso}{Universidad de Buenos Aires and Universidad Nacional de Quilmes}{}{}{}
}
{
  \author{Eduardo Bonelli}{Stevens Institute of Technology}{}{}{}
  \author{Delia Kesner}{IRIF, Universit\'{e} de Paris and CNRS}{}{}{}
  \author{Andr\'{e}s Viso}{Universidad de Buenos Aires and Universidad Nacional de Quilmes}{}{}{}
}

\ispaper{
  \authorrunning{D.~Kesner, E.~Bonelli, and A.~Viso}
  \Copyright{Delia Kesner, Eduardo Bonelli, and Andr\'{e}s Viso}
}{
  \authorrunning{E.~Bonelli, D.~Kesner, and A.~Viso}
  \Copyright{Eduardo Bonelli, Delia Kesner, and Andr\'{e}s Viso}
}

\begin{document}

\title{Strong Bisimulation for Control Operators}

% LIPIcs license is "CC-BY";  http://creativecommons.org/licenses/by/3.0/

\ccsdesc{
%F4 MATHEMATICAL LOGIC AND FORMAL LANGUAGES,
F4.1 Lambda calculus and related systems,
%F.3 LOGICS AND MEANINGS OF PROGRAMS,
F.3.2 Semantics of Programming Languages,
%F.3.3 Studies of Program Constructs,
%Type structure,
%D.3 PROGRAMMING LANGUAGES,
D.3.3 Language Constructs and Features.
%Data types and structures,
%Patterns.
}

\keywords{Lambda-mu calculus, proof-nets, strong bisimulation}

\ispaper{
  \category{Invited Talk}
  \EventEditors{Maribel Fern\'{a}ndez and Anca Muscholl}
  \EventNoEds{2}
  \EventLongTitle{28th EACSL Annual Conference on Computer Science Logic (CSL 2020)}
  \EventShortTitle{CSL 2020}
  \EventAcronym{CSL}
  \EventYear{2020}
  \EventDate{January 13--16, 2020}
  \EventLocation{Barcelona, Spain}
  \EventLogo{}
  \SeriesVolume{152}
  \ArticleNo{4}
}{
  \EventYear{2019}
  \ArticleNo{1}
}

\maketitle

  \begin{abstract}

    The purpose of this paper is to identify programs with
      control operators whose reduction semantics are in exact
      correspondence. This is achieved by
      introducing a 
    relation
    $\simeq$, defined over a revised presentation of Parigot's
      $\lmu$-calculus  we dub $\LMfull$.
     
      Our result builds on two fundamental ingredients: (1)
      factorization of $\lm$-reduction into multiplicative
      and exponential steps by means of explicit term operators
      of $\LMfull$, and
      (2) translation of $\LMfull$-terms into
      Laurent's polarized
      proof-nets (PPN) such that
      cut-elimination in PPN simulates our calculus.  Our proposed
      relation $\simeq$ is shown to characterize structural
      equivalence in PPN. Most notably, $\simeq$ is shown to be a
      strong bisimulation with respect to reduction in $\LMfull$, \ie
      two $\simeq$-equivalent terms have the exact same reduction
      semantics, a result which fails for Regnier's
      $\sigma$-equivalence in $\l$-calculus as well as for Laurent's
      $\sigma$-equivalence in $\lmu$.
  \end{abstract}

%%% Local Variables:
%%% mode: latex
%%% TeX-master: "main"
%%% End:

%%%%%%%%%%%%%%%%%%%%%%%%%%%%%%%%%%%%%%%%%%%%%%%%%%%%%%%%%%%%%%%%%%%%%%%%%%%%%%%
\section{Introduction}
\label{sec:introduction}
%%%%%%%%%%%%%%%%%%%%%%%%%%%%%%%%%%%%%%%%%%%%%%%%%%%%%%%%%%%%%%%%%%%%%%%%%%%%%%%

An important topic in the study of programming language theories is unveiling
structural similarities between expressions denoting programs. They are widely
known as \textit{structural equivalences}; equivalent expressions behaving
exactly in the same way. Process calculi are a rich source of examples. In CCS
expressions stand for processes in a concurrent system. For example, $P
\parallel Q$ denotes the parallel composition of processes $P$ and $Q$.
Structural equivalence includes equations such as the one stating that $P
\parallel Q$ and $Q \parallel P$ are equivalent. This minor reshuffling of
subexpressions has little impact on the behavior of the overall expression:
structural equivalence is a \emph{strong bisimulation} for process reduction.
\begin{wrapfigure}[6]{r}{0.20\textwidth}
\hspace{1em}
\begin{tikzcd}[ampersand replacement=\&]
o \arrow[rightsquigarrow]{d}[anchor=north,left]{}
  \&[-25pt] \simeq
  \&[-25pt] p \arrow[densely dashed,rightsquigarrow]{d}[anchor=north,left]{} \\ 
o'  
  \&[-25pt] \simeq
  \&[-25pt] p'
\end{tikzcd} 
\end{wrapfigure}
This paper is concerned with such notions of reshuffling of expressions in
\textit{$\lambda$-calculi with control operators}. The induced notion of
structural equivalence, in the sequel $\simeq$, should identify terms having
exactly the same reduction semantics too, that is, should be a strong
bisimulation with respect to reduction in these calculi. In other words,
$\simeq$ should be symmetric and moreover $o \simeq p$ and $o
\rightsquigarrow o'$ should imply the existence of $p'$ such that $p
\rightsquigarrow p'$ and $o' \simeq p'$, where $\rightsquigarrow$ denotes some
given notion of reduction for control operators (see
figure on the right).
\ignore{
\begin{equation}
\hspace{6cm}
% \scalebox{0.8}{
\begin{tikzcd}[ampersand replacement=\&]
o \arrow[rightsquigarrow]{d}[anchor=north,left]{}
  \&[-25pt] \simeq
  \&[-25pt] p \arrow[densely dashed,rightsquigarrow]{d}[anchor=north,left]{} \\ 
o'  
  \&[-25pt] \simeq
  \&[-25pt] p'
\end{tikzcd}
% }
\label{eq:example:bisimulation}
\end{equation}
}

Formulating such structural equivalences for  the $\lambda$-calculus is
hindered by the sequential (left-to-right) orientation in which expressions are
written. Consider for example the terms 
$\termapp{(\termabs{x}{\termapp{(\termabs{y}{t})}{u}})}{v}$
and $\termapp{\termapp{(\termabs{x}{\termabs{y}{t}})}{v}}{u}$.
% where $x\notin\fv{s}$.
They seem to have the same redexes, only permuted, similar to the situation
captured by the above mentioned CCS equation. A closer look, however, reveals
that this is not entirely correct. The former has two redexes (one indicated
below by underlining and another by overlining) and the latter has only one
(underlined):
\begin{equation}
\hfill
\underline{\termapp{(\termabs{x}{\overline{\termapp{(\termabs{y}{t})}{u}}})}{v}}
\mbox{ and }
\termapp{\underline{\termapp{(\termabs{x}{(\termabs{y}{t})})}{v}}}{u}
\hfill
\label{eq:example:permutation:lambda}
\end{equation}
The overlined redex on the left-hand side is not visible on the right-hand
side; it will only reappear, as a newly \emph{created} redex, once the
underlined redex is computed. Despite the fact that the syntax gets in the way,
Regnier~\cite{regnier94} proved that these terms behave in \emph{essentially}
the same way. More precisely, he introduced a structural equivalence for
$\lambda$-terms, known as \emph{$\sigma$-equivalence} and he proved that
$\sigma$-equivalent terms have head, leftmost, perpetual and, more generally,
maximal reductions of the same length. However, the mismatch between the terms
in (\ref{eq:example:permutation:lambda}) is unsatisfying since there clearly
seems to be an underlying strong bisimulation, which is not showing itself due
to a notational shortcoming. It turns out that through the graphical intuition 
provided by linear logic \textit{proof-nets}, one can define an enriched
$\lambda$-calculus that unveils a strong bisimulation for the intuitionistic
case~\cite{AccattoliBKL14}. Further details are described below.
In this paper, we resort to this same intuition
to explore whether it is possible to uncover a strong bisimulation behind a
notion of structural equivalence for the more challenging setting of classical
logic. Thus, we will not only capture structural equivalence on pure functions,
but also on \emph{programs with control operators}. In our case it is polarized
proof-nets (PPN) that will serve as semantic yardstick. We next briefly revisit
proof-nets and discuss how they help unveil structural equivalence as strong
bisimulation for $\lambda$-calculi. An explanation of the challenges that we
face in addressing the classical case will follow. 

\partitle{Proof-Nets.}
A \emph{proof-net} is a graph-like structure whose nodes denote logical
inferences and whose edges or wires denote the formula they operate on (\cf
Sec.~\ref{s:proofnets}). Proof-nets were introduced in the setting of linear
logic~\cite{Girard87}, which provides a mechanism to explicitly control the use
of resources by restricting the application of the \textit{structural} rules of
weakening and contraction. Proof-nets are equipped with an operational
semantics specified by graph transformation rules which captures cut
elimination in sequent calculus. The resulting cut elimination rules on
proof-nets are split into two different kinds: \emph{multiplicative}, that
essentially (linearly) reconfigure wires, and \emph{exponential}, which are the only ones
that are able to erase or duplicate (sub)proof-nets. The latter are considered
to introduce interesting or \emph{meaningful} computation. Most notably,
proof-nets abstract away the order in which certain rules occur in a sequent
derivation. As an example, assume three derivations of the judgements $\vdash
\Gam, A$, $\vdash \Delta, A^\bot, B$ and $\vdash \Lambda, B^\bot$, resp. The
order in which these derivations are composed via cuts into a single derivation
is abstracted away in the resulting proof-net: \[
\hfill
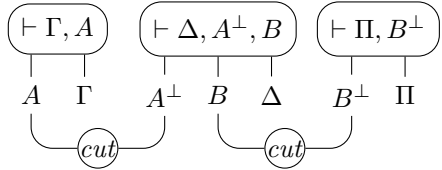
\begin{array}{ccc}
\input{proofnets/intro}
\end{array}
\hfill
\]

In other words, \emph{different} terms/derivations are represented by the
\emph{same} proof-net.  Hidden structural similarity between terms can thus be
studied by translating them to proof-nets. Moreover, following the Curry-Howard
isomorphism which relates computation and logic, this correspondence can be
extended not only to terms
themselves~\cite{DanosRegnier95,CosmoKP00,KesnerL05,AccattoliK12} but also to
their reduction behavior~\cite{Accattoli18}. In this paper, however, we
concentrate on identifying those different classical derivations which
translate to the same graph representation. As is standard in the literature,
the notion of proof-net identity we adopt includes simple equalities such as \emph{associativity}
of contraction nodes and other similar rewirings (\cf notion of
\emph{structural equivalence} of proof-nets\ispaper{}{ in Sec.~\ref{s:proofnets}}).

% \delia{Moreover,
% the fine-grained operational semantics of proof-nets cannot be
% expressed directly in lambda-calculi: simulation of
% $\beta$-reduction into proof-nets needs several intermediate states
% which cannot be expressed directly in the lambda-calculus
% itself. This mismatch between $\l$-terms  and proof-nets suggests
% the introduction of new term-languages being  able to capture equality of
% proof-nets. 
% }

\partitle{Intuitionistic $\sigma$-Equivalence.}
\label{p:inTuitionistic}
As mentioned before, Regnier introduced a notion of
\emph{$\sigma$-equivalence} on $\lambda$-terms (written here $\eqregnier$ and
depicted in Fig.~\ref{f:sigma-equivalence-lambda}), and proved that
$\sigma$-equivalent terms behave in essentially identical way. This
equivalence relation involves permuting certain redexes, and was unveiled
through the study of proof-nets. In particular, following Girard's encoding of
intuitionistic into linear logic~\cite{Girard87}, $\sigma$-equivalent terms
are mapped to the same proof-net (modulo multiplicative cuts and structural
equivalence).
\begin{figure}[h!]
\centering
$\begin{array}{rcll}
\termapp{(\termabs{x}{\termabs{y}{t}})}{u} & \simeq_{\rsigma_1} & \termabs{y}{\termapp{(\termabs{x}{t})}{u}} & \notatall{y}{u} \\
\termapp{(\termabs{x}{\termapp{t}{v}})}{u} & \simeq_{\rsigma_2} & \termapp{\termapp{(\termabs{x}{t})}{u}}{v} & \notatall{x}{v}
\end{array}$
\caption{Regnier's $\rsigma$-equivalence for $\l$-terms}
\label{f:sigma-equivalence-lambda}
\end{figure}

%L.~Regnier proved a series of results to justify what
%he meant by operationally indistinguishable: if $u \eqregnier v$, then
%$u$ and $v$ have leftmost, head, perpetual and, more generally,
%maximal reductions of the same length.
The reason why Regnier's result is not immediate is that redexes present on one
side of an equation may disappear on the other side of it, as illustrated in
the terms in (\ref{eq:example:permutation:lambda}).
% As an example, consider: 
% \begin{equation}
%        u=\underline{\termapp{(\termabs{x}{\overline{\termapp{(\termabs{y}{r})}{s}}})}{t}}  \simeq_{\rsigma_2} 
%        \termapp{\underline{\termapp{(\termabs{x}{(\termabs{y}{r})})}{t}}}{s}=v
%        \label{eq:regnier:sigma:not:strong:bisimulation}
% \end{equation}
%There are two redexes on the left which are marked via underlining and
%overlining.
One might rephrase this observation by stating that $\eqregnier$ is \emph{not a
strong bisimulation} over the set of $\lambda$-terms. If it were, then
establishing that $\sigma$-equivalent terms behave essentially in the same way
would be trivial.

Adopting a more refined view of $\lambda$-calculus, as suggested by linear
logic, which splits cut elimination on logical derivations into multiplicative
and exponential steps yields a decomposition of $\beta$-reduction into
multiplicative/exponential steps \emph{on terms}. The theory of \emph{explicit
substitutions} (a survey can be found in~\cite{KesnerES09}) provides a
convenient syntax to reflect these steps at the term level. Indeed,
$\beta$-reduction can be decomposed into two steps, namely $\rB$ (for
$\rB$eta), which acts \emph{at a distance}~\cite{AccattoliK12} in the
sense that the abstraction and the argument may be separated by an
arbitrary number of explicit substitutions,  and $\rS$ (for $\rS$ubstitution):
\begin{equation}
\hfill
\begin{array}{rcl}
\termapp{(\termabs{x}{t})\exsubs{x_1}{v_1} \ldots \exsubs{x_n}{v_n}}{u} & \rrule{\rB} & \termsubs{x}{u}{t}\exsubs{x_1}{v_1} \ldots \exsubs{x_n}{v_n} \\
\termsubs{x}{u}{t}                                                      & \rrule{\rS} & \substitute{x}{u}{t}
\end{array}
\hfill
\label{eq:split_of_beta}
\end{equation}
Firing the $\rB$-rule creates an \emph{explicit substitution} operator, written
$\termsubs{x}{u}{t}$, so that $\rB$ essentially reconfigures symbols, and
indeed reads as a multiplicative cut in proof-nets. The $\rS$-rule executes the
substitution by performing a replacement of  all free occurrences of $x$ in $t$
with $u$, written $\substitute{x}{u}{t}$, so that it is $\rS$ that performs
interesting or \emph{meaningful} computation and reads as an exponential cut in
proof-nets.  

A term without any occurrence of the left-hand side of rule $\rB$ is called a
$\rB$-normal form; we shall refer to these terms as \emph{canonical forms}.
Decomposition of $\beta$-reduction by means of the rules
in~(\ref{eq:split_of_beta}) prompts one to replace $\eqregnier$ 
(Fig.~\ref{f:sigma-equivalence-lambda}) with a new relation $\simeq_{\esigma}$
(Fig.~\ref{eq:sigma:es}). The latter is formed essentially by taking the
\emph{$\rB$-normal form} of each side of the $\simeq_{\rsigma}$  equations\footnote{Also included in $\simeq_{\esigma}$
is equation $\simeq_{\esigma_3}$ allowing commutation of orthogonal
(independent) substitutions.  Notice however that the $\rB$-expansion of
$\simeq_{\esigma_3}$-equivalent terms yields $\simeq_\rsigma$-equivalent terms
again. For example, the $\rB$-expansion of $\termsubs{x}{u}{\termsubs{y}{v}{t}}
\simeq_{\esigma_3} \termsubs{y}{v}{\termsubs{x}{u}{t}}$ yields
$\termapp{(\termabs{y}{\termapp{(\termabs{x}{t})}{u}})}{v}
\simeq_{\rsigma_1,\rsigma_2}
\termapp{(\termabs{x}{\termapp{(\termabs{y}{t})}{v}})}{u}$.}.
\begin{figure}[h!]
\centering
$
\begin{array}{rcll}
\termsubs{x}{u}{(\termabs{y}{t})}   & \simeq_{\esigma_1} & \termabs{y}{\termsubs{x}{u}{t}}     & \notatall{y}{u} \\
\termsubs{x}{u}{(\termapp{t}{v})}   & \simeq_{\esigma_2} & \termapp{\termsubs{x}{u}{t}}{v}     & \notatall{x}{v} \\
\termsubs{x}{u}{\termsubs{y}{v}{t}} & \simeq_{\esigma_3} & \termsubs{y}{v}{\termsubs{x}{u}{t}} & \notatall{y}{u},
  \notatall{x}{v}
\end{array}
$
\caption{Strong bisimulation for $\l$-terms with explicit substitutions}
\label{eq:sigma:es}
\end{figure}

Since $\rB$-reduction corresponds only to multiplicative
cuts in proof-nets, the translation of $\simeq_{\esigma}$-equivalent typed terms also yields
\emph{structurally equivalent} proof-nets. 
%\odelia{This may depicted as follows, where the arrow denotes the natural translation from
%terms with explicit substitutions~\cite{CosmoKP00,KesnerL05} to proof-nets:}
%% \begin{center}
%% % \scalebox{0.8}{
%% \begin{tikzcd}[ampersand replacement=\&]
%% t \arrow[->]{drrr}[left]{}
%%  \&[-25pt] \simeq_{\rsigma}
%%  \&[-25pt] u \arrow[->]{dr}[left]{} \& \&
%% \B{t} \arrow[->]{dl}[left]{}
%%  \&[-25pt] \simeq_{\esigma}
%%  \&[-25pt]
%% \B{u} \arrow[->]{dlll}[left]{} \\
%%  \& \& \& PN
%% \end{tikzcd}
%% % }
%% \end{center}
%\delia{figure erased}
In other words, $\simeq_{\esigma}$-equivalence classes of $\l$-terms with
explicit substitutions in $\rB$-normal form\ are in one-to-one correspondence with intuitionistic
linear logic proof-nets~\cite{AccattoliK12}.
% \odelia{This is however a \emph{static} correspondence, 
% that can be enlarged in order to capture
% a \emph{dynamic} one, 
% where reduction steps in the term-language are identified
% with those in proof-nets. One natural way to
% put such dynamic correspondence in evidence is 
% to enrich the term-language with new constructors (as weakening and contraction)
% inherited from linear logic~\cite{KesnerL05}, 
% another alternative approach is to modify the original
% Girard's cut elimination system~\cite{Accattoli18}.} 
Moreover, $\simeq_{\esigma}$ is a strong bisimulation with respect to
meaningful reduction (\ie $\rS$-reduction) over the extended set of terms that
includes explicit substitutions~\cite{AccattoliK12,AccattoliBKL14}. Indeed,
$\simeq_{\esigma}$ is symmetric, and moreover, $u \simeq_{\esigma} v$ and $u
\Rew{\rS} u'$ implies the existence of $v'$ such that $v \Rew{\rS} v'$ and $u'
\simeq_{\esigma} v'$. Note also that the $\rB$-normal form of
  both sides of~(\ref{eq:example:permutation:lambda}) are
  $\simeq_{\esigma}$-equivalent, thus repairing the mismatch.

\partitle{Classical $\sigma$-Equivalence.}
This work sets out to explore structural equivalence for
\emph{$\lambda$-calculi with control operators}. These calculi include
operations to manipulate the context in which a program is executed. We focus
here on Parigot's $\lm$-calculus~\cite{Parigot93}, which extends the
$\lambda$-calculus  with two new operations: $\termname{\alpha}{t}$
(\emph{named term}) and $\termcont{\alpha}{c}$ (\emph{$\mu$-abstraction}). The
former may be read as ``call continuation $\alpha$ with $t$ as
argument'' and the latter as ``record the current continuation as $\alpha$ and
continue as $c$''. Reduction in $\lm$ consists of the $\beta$-rule together
with: \[
\hfill
\begin{array}{rcl}
\termapp{(\termcont{\alpha}{c})}{u} & \rrule{\mu} & \termcont{\alpha}{\replace{}{\alpha}{u}{c}}
\end{array}
\hfill
\] where $\replace{}{\alpha}{u}{c}$, called here \emph{replacement}, replaces
all subexpressions of the form $\termname{\alpha}{t}$ in $c$ with
$\termname{\alpha}{(\termapp{t}{u})}$. Regnier's notion of $\sigma$-equivalence
for $\l$-terms was extended to  $\lm$ by Laurent~\cite{Laurent03} (\cf
Fig.~\ref{f:sigma-laurent} in Sec.~\ref{f:sigma-laurent}). Here is an example
of terms related by this extension, where the redexes are
underlined/overlined: \[
\hfill
\termapp{(\underline{\termapp{(\termabs{x}{\termcont{\alpha}{\termname{\gamma}{u}}})}{w}})}{v}
\eqlaurent
\overline{\termapp{(\termcont{\alpha}{\termname{\gamma}{\underline{\termapp{(\termabs{x}{u})}{w}}}})}{v}} 
\hfill
\] Once again, the fact that a harmless permutation of redexes has taken place
is not obvious. The term on the right has two redexes ($\mu$ and $\beta$) but
the one on the left only has one ($\beta$) redex. Another, more subtle, example
of terms related by Laurent's extension clearly suggests that operational
indistinguishability cannot rely on relating arbitrary $\mu$-redexes; the
underlined $\mu$-redex on the left does not appear at all on the right:
\begin{equation}
\hfill
\underline{\termapp{(\termcont{\al}{\termname{\al}{x}})}{y}}
\eqlaurent
\termapp{x}{y}
\hfill
\label{eq:thetaAndSBisim}
\end{equation}
Clearly, $\sigma$-equivalence on $\lmu$-terms \emph{fails to be a strong
bisimulation}. Nonetheless, Laurent proved properties for $\eqlaurent$ in
$\lmu$ similar to those of Regnier for $\eqregnier$ in $\lambda$. Again, one
has the feeling that there is a strong bisimulation hiding behind
$\sigma$-equivalence for $\lm$.
% \delia{esto queda medio descolgado, yo sacaria tambien la ultima frase}  \odelia{More precisely, he proved that $u
%   \eqlaurent v$ implies that $u$ is normalizable (resp. is head
%   normalizable, strongly normalizable) iff $v$ is normalizable
%   (resp. is head normalizable, strongly normalizable). } \delia{esto en todo caso se podria pasar a la seccion lmu}

\partitle{Towards a Strong Bisimulation for Control Operators.}
We seek to formulate a notion of equivalence for $\lambda\mu$ in the sense that it is concerned with harmless permutation of
redexes possibly involving control operators and
inducing a strong bisimulation.
As per the Curry-Howard isomorphism, proof normalization in classical logic
corresponds to computation in $\lambda$-calculi with control
operators~\cite{Griffin90,Parigot93}. Moreover, since classical logic can be
translated into \emph{polarized proof-nets} (PPN), as defined by
O.~Laurent~\cite{Laurent02,Laurent03}, we use PPNs to guide the development in
this work.
A first step towards our goal involves decomposing the $\mu$-rule as was done
for the $\beta$-rule with the rules in (\ref{eq:split_of_beta}): this produces a rule $\rM$
(for $\rM$u), to introduce an \emph{explicit replacement},
that also acts at a distance, and
another rule $\rRfull$ (for $\rRfull$eplacement), that executes replacements:
\begin{equation}
\hfill
\begin{array}{rcl}
\termapp{(\termcont{\alpha}{c}) \exsubs{x_1}{v_1} \ldots \exsubs{x_n}{v_n}}{u}
& \rrule{\rM}     & (\termcont{\alpha'}{\termrepl{\alpha'}{\alpha}{u}{c}})\exsubs{x_1}{v_1} \ldots \exsubs{x_n}{v_n} \\
\termrepl{\alpha'}{\alpha}{u}{c}
& \rrule{\rRfull} & \replace{\alpha'}{\alpha}{u}{c}
\end{array}
\hfill
\label{eq:split_of_mu}
\end{equation}
where $\replace{\alpha'}{\alpha}{u}{c}$ replaces each sub-expression of the
form $\termname{\al}{t}$ in $c$ by $\termname{\al'}{t u}$. Meaningful
computation is seen to be performed by $\rRfull$ rather than $\rM$. This
observation is further supported by the fact that both sides of the
$\rM$-rule translate into the same proof-net
\ispaper{(\cf Sec.~\ref{s:proofnets})}{(\cf Sec.~\ref{s:translation})}.

Therefore, we tentatively fix our notion of meaningful reduction to be $\rS
\cup \rRfull$ over the set of canonical forms, the latter now obtained by
taking \emph{both} $\rB$ and $\rM$-normal forms. However, in contrast to the
intuitionistic case where the decomposition of $\beta$ into a multiplicative
rule $\rB$ and an exponential rule $\rS$ suffices for unveiling the strong
bisimulation behind Regnier's $\sigma$-equivalence in $\lambda$-calculus, it
turns out that splitting the $\mu$-rule into $\rM$ and $\rRfull$ is not
fine-grained enough. There are various examples, that will be
developed in this paper, that illustrate that the
resulting relation is still not a strong bisimulation.
\begin{wrapfigure}[8]{r}{0.45\textwidth}
\hspace{2em}
\begin{tikzcd}[->,ampersand replacement=\&]
\termcont{\al'}{\termrepl{\al'}{\alpha}{y}{(\termname{\al}{x})}} \arrow[->]{d}[left]{\rRfull}
  \&[-25pt] \simeq
  \&[-25pt] \termapp{x}{y} \arrow[->]{d}{\rRfull} \arrow[phantom,negated]{d}{} \\ 
\termcont{\al'}{\termname{\al'}{\termapp{x}{y}}}
  \&[-25pt] \simeq
  \&[-25pt] \termapp{x}{y}
\end{tikzcd}
\caption{Failure of strong bisimulation}\label{eq:theta}
\end{wrapfigure}
One such example results from taking the $\rB\rM$ normal form of the terms
in equation~(\ref{eq:thetaAndSBisim}), as depicted in Fig.~\ref{eq:theta}.
This particular use of $\rRfull$ on the left seems innocuous. In fact we show
that in our proposed calculus and its corresponding translation to PPNs, both
terms $\termcont{\al'}{\termrepl{\al'}{\alpha}{y}{(\termname{\al}{x})}}$ and
$\termcont{\al'}{\termname{\al'}{\termapp{x}{y}}}$ denote structurally
equivalent PPNs (\cf \ispaper{Sec.~\ref{s:proofnets}}{Sec.~\ref{s:translation}}  for a detailed
discussion). In any case, this example prompts us to further inquire on the
fine structure of $\rRfull$. In particular, we will argue (Sec.~\ref{sec:refining_replacement})
 that rule $\rRfull$ should be further
decomposed into \emph{several} independent notions, each one behaving
differently with respect to PPNs, and thus with respect to our strong
bisimulation $\simeq$. Identifying these notions and their interplay in order
to expose the strong bisimulation hidden behind Laurent's $\sigma$-equivalence
is the challenge we address in this work.

\partitle{Contributions.}
The multiplicative/exponential splitting of the intuitionistic case applied to
the classical case, falls noticeably short in identifying programs with control
operators whose reduction semantics are in exact correspondence. The need to
further decompose rule $\rRfull$ is rather unexpected,
and our proposed
decomposition turns out to be
subtle yet admits a natural translation to PPN. Moreover, it allows us to
obtain a novel and far from obvious strong bisimulation result, highlighting
the deep correspondence between PPNs and classical term calculi. Our
contributions may be summarized as follows:

\begin{enumerate}
  \item A refinement of $\lm$, called $\LMfull$-calculus, including explicit
  substitutions for variables (resp. explicit replacement for names), \ispaper{and being confluent (Thm.~\ref{t:confluence_of_LM})}{.} \ispaper{}{$\LMfull$ is proved to be confluent
  (Thm.~\ref{t:confluence_of_LM}).}
  
\item A natural interpretation of $\LMfull$ into
  PPN. More precisely, $\LMfull$-reduction  can be implemented by PPN cut
  elimination (Thm.~\ref{t:simulation-into-ppn})\ispaper{}{, in the sense that one-step
  reduction in $\LMfull$ translates to a reduction sequence in PPN}.
  
  \item A notion of structural equivalence $\simeq$ for $\LMfull$
  which:
  \begin{enumerate}
    \item characterizes PPN modulo structural
    equivalence (Thm.~\ref{t:completeness}); 
    
    \item is conservative over Laurent's original equivalence
    $\eqlaurent$ (Thm.~\ref{t:equivalencia-sigma-Sigma}); 
    
  \item is a strong bisimulation with respect to
    meaningful steps
    (Thm.~\ref{t:bisimulation}).
  \end{enumerate}
\end{enumerate}

\ignore{ 
Here is a brief summary explaining how $\simeq$ relates to the other relations
mentioned in this Introduction:
\begin{center}
\begin{tabular}{|c|c|c|c|c|}
\hline
\textbf{Theory} & \textbf{Calculus} & \textbf{Ref.} & \textbf{Logic} & \textbf{S. Bisim}\\
%    \textbf{Theory} & & & & \textbf{Bisim.}\\
\hline
$\eqregnier$ & $\lambda$ &  \cite{regnier94} & Intuit. & No
\\\hline
$\simeq_\esigma$ & $\Lambda_{\hat{P}},\lambda_{\mathsf{sub}}^{\sim}$ & \cite{AccattoliK12,AccattoliBKL14} & Intuit. & Yes \\\hline
$\eqlaurent$ & $\lmu$ & \cite{Laurent03} & Classical & No \\\hline
$\simeq$ & $\LMfull$ & This paper  & Classical & Yes \\\hline
\end{tabular}
\end{center}
}
% \delia{tambien hablar del resultado de confluence y del resultado de
% PSN, es el lugar donde uno va a mirar para entender bien cual es el
% aporte del paper, esta seccion deberia ser mas extensa.}

\partitle{Structure of the Paper.}  
Sec.~\ref{s:lmu-calculus} and~\ref{s:LM-calculus} present $\lmu$ and $\LMfull$,
resp. Sec.~\ref{sec:refining_replacement} presents a further refinement of
$\LMfull$. Sec.~\ref{s:types} defines typed $\LMfull$-objects,
Sec.~\ref{s:proofnets} defines polarized proof-nets, and
\ispaper{}{Sec.~\ref{s:translation}} presents the translation from the
former to the latter. Sec.~\ref{s:Sigma-equivalence} presents our
  equivalence 
  $\simeq$. Its properties are discussed and proved in Sec.~\ref{s:completeness}
and Sec.~\ref{s:bisimulation}. Finally, Sec.~\ref{s:conclusion} concludes and
describes related work. Most proofs \ispaper{can be found in~\cite{TR}
including extended details on PPNs, whose presentation has been
  abridged in this paper.}{are
  relegated to the Appendix.}

%%% Local Variables:
%%% mode: latex
%%% TeX-master: "main"
%%% End:

%%%%%%%%%%%%%%%%%%%%%%%%%%%%%%%%%%%%%%%%%%%%%%%%%%%%%%%%%%%%%%%%%%%%%%%%%%%%%%%
\section{\texorpdfstring{The $\lmu$-calculus}{The lm-calculus}}
\label{s:lmu-calculus}
%%%%%%%%%%%%%%%%%%%%%%%%%%%%%%%%%%%%%%%%%%%%%%%%%%%%%%%%%%%%%%%%%%%%%%%%%%%%%%%

\partitle{Preliminary Concepts.}
\label{s:preliminary}
A rewrite system $\R$ is a set of objects and a binary (one-step)
\emph{reduction} relation $\Rew{\R}$ over those objects. We write $\Rewn{\R}$
(resp. $\Rewplus{\R}$) for the reflexive-transitive (resp. transitive) closure
of $\Rew{\R}$. \ignore{, and $\Rew{\R}^n$ for the composition of $n$-steps of
$\Rew{\R}$, thus $t \Rew{\R}^n u$ denotes a finite $\R$-reduction sequence of
length $n$ from $t$ to $u$.} A term $t$ is in $\R$-normal form, written $t \in
\R$-nf or simply $t \in \R$, if there is no $t'$ s.t. $t \Rew{\R} t'$.

\partitle{Syntax.}
We fix a countable infinite set of \deft{variables} $x, y, z, \ldots$ and
\deft{continuation names} $\al, \beta, \ga, \ldots$. The set of \deft{objects}
$\objects{\lmu}$, \deft{terms} $\terms{\lmu}$, \deft{commands}
$\commands{\lmu}$ and \deft{contexts} of the $\lmu$-calculus are given by the
following grammar: \[
\hfill
\begin{array}{l@{\enspace}l@{\enspace}l@{\enspace}l}
\textbf{Objects}          & o         & \Coloneq  & t \mid c \\
\textbf{Terms}            & t         & \Coloneq  & x \mid \termapp{t}{t} \mid \termabs{x}{t} \mid \termcont{\al}{c} \\
\textbf{Commands}         & c         & \Coloneq  & \termname{\al}{t} \\
\textbf{Contexts}         & \ctxt{O}  & \Coloneq  & \ctxt{T} \mid \ctxt{C} \\
\textbf{Term Context}     & \ctxt{T}  & \Coloneq  & \Box \mid \termapp{\ctxt{T}}{t} \mid \termapp{t}{\ctxt{T}} \mid \termabs{x}{\ctxt{T}} \mid \termcont{\alpha}{\ctxt{C}} \\
\textbf{Command Context}  & \ctxt{C}  & \Coloneq  & \boxdot \mid \termname{\alpha}{\ctxt{T}}
\end{array}
\hfill
\] The term  $\termapp{(\ldots(\termapp{(\termapp{t}{u_1})}{u_2})\ldots)}{u_n}$
abbreviates as $\termapp{\termapp{\termapp{t}{u_1}}{u_2}\ldots}{u_n}$\ignore{or
  $\termapp{t}{\vec{u}}$ when $n$ is clear from the context}.
The grammar extends $\l$-terms with two new constructors:
commands $\termname{\al}{t}$ and $\mu$-abstractions $\termcont{\al}{c}$.
Regarding contexts, there are two holes $\Box$ and $\boxdot$ of sort term
($\mathsf{t}$) and command ($\mathsf{c}$) respectively. We write
$\ctxtapp{\ctxt{O}}{o}$ to denote the replacement of the hole $\Box$ (resp.
$\boxdot$) by a term (resp. by a command). We often decorate contexts or
functions over expressions with sorts $\mathsf{t}$ and $\mathsf{c}$.
For example, $\ctxtSort{O}{t}$ is a context $\ctxt{O}$ with a hole of sort
\emph{term}. The subscript is omitted if it is clear from the context.

\deft{Free} and \deft{bound variables} of objects are defined as expected, in
particular $\fv{\termcont{\al}{c}} \eqdef \fv{c}$ and $\fv{\termname{\al}{t}}
\eqdef \fv{t}$. \deft{Free names} of objects are defined as follows: $\fn{x}
\eqdef \emptyset$, $\fn{\termabs{x}{t}} \eqdef \fn{t}$, $\fn{\termapp{t}{u}}
\eqdef \fn{t} \cup \fn{u}$, $\fn{\termcont{\al}{c}} \eqdef \fn{c} \setminus
\{\al\}$, and $\fn{\termname{\al}{t}} \eqdef \fn{t} \cup \{\al\}$. \deft{Bound
names} are defined accordingly. We use $\fvp{x}{o}$ (resp. $\fnp{\alpha}{o}$)
to denote the number of free occurrences of the variable $x$ (resp. name
$\alpha$) in $o$. 
We write
$\notatall{x}{o}$ (resp. $\notatall{\alpha}{o}$) if $x \notin \fv{o}$ and
$x \notin \bv{o}$ (resp. $\alpha \notin \fn{o}$ and $\alpha \notin \bn{o}$).
  This notion is extended to contexts as expected.

\ignore{$$
% \scalebox{.90}{$
\begin{array}{cc}
\begin{array}{r@{\enspace}c@{\enspace}l}
\fn{x}                 & \eqdef & \emptyset \\
\fn{\termabs{x}{t}}    & \eqdef & \fn{t} \\  
\fn{\termapp{t}{u}}    & \eqdef & \fn{t} \cup \fn{u} 
\end{array}
&
\begin{array}{r@{\enspace}c@{\enspace}l}
\fn{\termcont{\al}{c}} & \eqdef & \fn{c} \setminus \{\al\} \\
\fn{\termname{\al}{t}} & \eqdef & \fn{t} \cup \{\al\}
\end{array}
\end{array}
% $}
$$}

We work with the standard notion of \deft{$\alpha$-conversion} \ie renaming of
bound variables and names, thus for example
$\termname{\delta}{\termapp{(\termcont{\al}{\termname{\al}{(\termabs{x}{x})}})}{z}}
\equiv_{\alpha}
\termname{\delta}{\termapp{(\termcont{\beta}{\termname{\beta}{(\termabs{y}{y})}})}{z}}$. In particular, when using two different
  symbols to denote bound
  variables (resp. names), we assume that they are distinct,  without explicitly
  mentioning it.

\partitle{Semantics.}
\deft{Application} of the \deft{substitution} $\isubs{x}{u}$ to the object $o$,
written $o\isubs{x}{u}$, may require $\alpha$-conversion in order to avoid
capture  of free variables/names, and it is defined as expected.
\ignore{$$
\begin{array}{r@{\enspace}c@{\enspace}l}
(x)\isubs{x}{u}                    & \eqdef & u \\
(y)\isubs{x}{u}                    & \eqdef & y \\
(\termapp{t}{v})\isubs{x}{u}       & \eqdef & \termapp{t\isubs{x}{u}}{v\isubs{x}{u}} \\
(\termabs{z}{t})\isubs{x}{u}       & \eqdef & \termabs{z}{t\isubs{x}{u}} \\
(\termcont{\ga}{\Com})\isubs{x}{u} & \eqdef & \termcont{\ga}{\Com\isubs{x}{u}} \\
(\termname{\al}{t})\isubs{x}{u}    & \eqdef & \termname{\al}{t\isubs{x}{u}}
\end{array} $$
}
\deft{Application} of the \deft{replacement} $\ire{\al}{\al'}{u}$ to an object
$o$, where $\al \neq \al'$, written $o\ire{\al}{\al'}{u}$, passes the term $u$
as an argument to any sub-command of $o$ of the form $\termname{\al}{t}$ and
changes the name of $\alpha$ to $\alpha'$. This operation is also defined
modulo $\alpha$-conversion in order to avoid the capture of free
variables/names. Formally: \label{page-replacement} \[
\hfill
\begin{array}{r@{\enspace}c@{\enspace}llll}
x\ire{\al}{\al'}{u}                     & \eqdef & x \\
(\termapp{t}{v})\ire{\al}{\al'}{u}      & \eqdef & \termapp{t\ire{\al}{\al'}{u}}{v\ire{\al}{\al'}{u}} \\
(\termabs{x}{t})\ire{\al}{\al'}{u}      & \eqdef & \termabs{x}{t\ire{\al}{\al'}{u}} & \notatall{x}{u} \\
(\termcont{\beta}{c})\ire{\al}{\al'}{u} & \eqdef & \termcont{\beta}{c\ire{\al}{\al'}{u}} & \notatall{\beta}{(u,\al,\al')} \\
(\termname{\al}{c})\ire{\al}{\al'}{u}   & \eqdef & \termname{\al'}{(\termapp{c\ire{\al}{\al'}{u}}{u})} \\
(\termname{\beta}{c})\ire{\al}{\al'}{u} & \eqdef & \termname{\beta}{c\ire{\al}{\al'}{u}} & \beta \neq \al
\end{array}
\hfill
\] For example, if $\Id = \termabs{z}{z}$, then 
$(\termapp{{(\termcont{\al}{\termname{\al}{x}})}}{(\termabs{z}{\termapp{z}{x}})})\isubs{x}{\Id}$
is equal to
$\termapp{{(\termcont{\al}{\termname{\al}{\Id}})}}{(\termabs{z}{\termapp{z}{\Id}})}$,
and
$(\termname{\al}{\termapp{x}{(\termcont{\beta}{\termname{\al}{y}})}})\ire{\al}{\al'}{\Id}
=
\termname{\al'}{\termapp{\termapp{x}{(\mu\beta.\termname{\al'}{\termapp{y}{\Id}})}}{\Id}}$.

\begin{definition}
The $\l\mu$-calculus is given by the set $\objects{\lmu}$ and the
\deft{$\l\mu$-reduction relation} $\Rew{\lmu}$, defined as the closure by
\textit{all} contexts of the following rewriting
rules\footnote{Parigot~\cite{Parigot93}'s $\mu$-rule
$\termapp{(\termcont{\al}{c})}{u} \rrule{\mu}
\termcont{\al}{c\iredos{\al}{u}}$, relies on a binary replacement operation
$\iredos{\al}{u}$ assigning $\termname{\al}{\termapp{(t\iredos{\al}{u})}{u}}$
to $\termname{\al}{t}$ (thus not changing the name of the command). We remark
that $\termcont{\al}{c\iredos{\al}{u}} \equiv_{\alpha}
\termcont{\al'}{c\ire{\al}{\al'}{u}}$. \ignore{; thus \eg
$\termcont{\al}{(\termname{\al}{x})\iredos{\al}{u}} =
\termcont{\al}{\termname{\al}{\termapp{x}{u}}} \equiv_{\alpha}
\termcont{\gamma}{\termname{\gamma}{\termapp{x}{u}}} =
\termcont{\gamma}{(\termname{\al}{x})\ire{\al}{\gamma}{u}}$.} We adopt here the
ternary presentation of the replacement operator~\cite{KesnerV19}, because it naturally extends
to that of the $\LMfull$-calculus in Sec.~\ref{s:LM-calculus}.} (equivalently,
$\mathbin{\Rew{\lmu}} \eqdef \ctxtapp{\ctxtSort{O}{t}}{\rrule{\beta} \cup
\rrule{\mu}}$: \[
\hfill
\begin{array}{lll}
\termapp{(\termabs{x}{t})}{u}    & \rrule{\beta} & t\isubs{x}{u} \\
\termapp{(\termcont{\al}{c})}{u} & \rrule{\mu}   & \termcont{\al'}{c\ire{\al}{\al'}{u}}
\end{array}
\hfill
\]
\end{definition}

Various control operators can be expressed in the
$\lmu$-calculus~\cite{DBLP:conf/lpar/Groote94,Laurent:2003}.
% \cdelia{
% A typical example is
% $\textbf{call-cc} \eqdef
% \termabs{y}{\termcont{\al}{\termname{\al}{\termapp{y}{(\termabs{x}{\termcont{\beta}{\termname{\al}{x}}})}}}}$~\cite{Griffin90} 
% which gives rise to the following reduction sequence: $$
% % \begin{center}
% % \scalebox{0.8}{$
% \begin{array}{cl}
%             & \termapp{\termapp{\termapp{\textbf{call-cc}}{t}}{u_1}}{\dots u_n} \\
% \Rew{\beta} & \termapp{\termapp{(\termcont{\al}{\termname{\al}{\termapp{t}{(\termabs{x}{\termcont{\beta}{\termname{\al}{x}}})}}})}{u_1}}{\dots u_n} \\
% \Rew{\mu}   & \termapp{\termapp{(\termcont{\al}{\termname{\al}{\termapp{\termapp{t}{(\termabs{x}{\termcont{\beta}{\termname{\al}{\termapp{x}{u_1}}}})}}{u_1}}})}{u_2}}{\dots u_n} \\
% \Rewn{\mu}  & \termcont{\al}{\termname{\al}{\termapp{\termapp{\termapp{t}{(\termabs{x}{\termcont{\beta}{\termname{\al}{\termapp{\termapp{x}{u_1}}{\dots u_n}}}})}}{u_1}}{\dots u_n}}}
% \end{array} $$
% % $}
% % \end{center}
% }{
A typical example of expressiveness of the $\lmu$-calculus is the control
operator \textbf{call-cc}~\cite{Griffin90}, specified by the term
$\termabs{x}{\termcont{\al}{\termname{\al}{\termapp{x}{(\termabs{y}{\termcont{\delta}{\termname{\al}{y}}})}}}}$.
The term \textbf{call-cc} is assigned the type
$\typefunc{(\typefunc{(\typefunc{A}{B})}{A})}{A}$ (Peirce's Law) in the simply
typed $\lmu$-calculus, thus capturing classical logic.

\partitle{The Notion of $\sigma$-Equivalence for $\lmu$-Terms.}
As in $\lambda$-calculus, structural equivalence for $\lm$ captures inessential
permutation of redexes, but this time also involving the control constructs.
Laurent's notion of $\sigma$-equivalence for $\lm$-terms~\cite{Laurent03}
(written here also $\eqlaurent$) is depicted in Fig.~\ref{f:sigma-laurent}.
\begin{figure}[h!]
\[
\hfill
\begin{array}{r@{\hspace{3.8pt}}c@{\hspace{3.8pt}}ll}
\termapp{(\termabs{y}{\termabs{x}{t}})}{v}                                                                  & \simeq_{\lsigma_1} & \termabs{x}{\termapp{(\termabs{y}{t})}{v}} & \notatall{x}{v} \\
\termapp{(\termabs{x}{\termapp{t}{v}})}{u}                                                                  & \simeq_{\lsigma_2} & \termapp{\termapp{(\termabs{x}{t})}{u}}{v} & \notatall{x}{v} \\
\termapp{(\termabs{x}{\termcont{\alpha}{\termname{\beta}{u}}})}{w}                                          & \simeq_{\lsigma_3} & \termcont{\alpha}{\termname{\beta}{\termapp{(\termabs{x}{u})}{w}}} & \notatall{\al}{w} \\
\termname{\alpha'}{\termapp{(\termcont{\alpha}{\termname{\beta'}{\termapp{(\termcont{\beta}{c})}{w}}})}{v}} & \simeq_{\lsigma_4} & 
\termname{\beta'}{\termapp{(\termcont{\beta}{\termname{\alpha'}{\termapp{(\termcont{\alpha}{c})}{v}}})}{w}} & \notatall{\al}{w},
\notatall{\beta}{v}, \beta \neq \al', \al \neq \beta' \\
\termname{\alpha'}{\termapp{(\termcont{\alpha}{\termname{\beta'}{\termabs{x}{\termcont{\beta}{c}}}})}{v}}   & \simeq_{\lsigma_5} & 
\termname{\beta'}{\termabs{x}{\termcont{\beta}{\termname{\alpha'}{\termapp{(\termcont{\alpha}{c})}{v}}}}} & \notatall{x}{v},
\notatall{\beta}{v}, \beta \neq \al', \al \neq \beta' \\
\termname{\alpha'}{\termabs{x}{\termcont{\alpha}{\termname{\beta'}{\termabs{y}{\termcont{\beta}{c}}}}}}     & \simeq_{\lsigma_6} & 
\termname{\beta'}{\termabs{y}{\termcont{\beta}{\termname{\alpha'}{\termabs{x}{\termcont{\alpha}{c}}}}}} & \beta \neq \al', \al \neq \beta' \\
\termname{\alpha}{\termcont{\beta}{c}}                                                                      & \simeq_{\lsigma_7} & c\iren{\beta}{\alpha} \\
\termcont{\alpha}{\termname{\alpha}{v}}                                                                     & \simeq_{\lsigma_8} & v & \notatall{\al}{v}
\end{array}
\hfill
\]
\caption{Laurent's $\lsigma$-equivalence for $\lmu$-terms}
\label{f:sigma-laurent}
\end{figure} 
The first two equations are exactly those of Regnier (hence $\eqlaurent$ on
$\lmu$-terms strictly extends $\eqregnier$ on $\l$-terms); the remaining ones
involve interactions between control operators themselves or control operators
and application and abstraction.

Laurent proved properties for $\eqlaurent$ similar to those of Regnier for
$\eqregnier$. More precisely, $u \eqlaurent v$ implies that $u$ is normalizable
(resp. is head normalizable, strongly normalizable) iff $v$ is normalizable
(resp. is head normalizable, strongly normalizable)~\cite[Prop.~35]{Laurent03}.
Based on Girard's encoding of classical into linear logic~\cite{Girard87}, he
also proved that the translation of the left and right-hand sides of the
equations of $\eqlaurent$, in a typed setting, yield structurally equivalent
PPNs~\cite[Thm.~41]{Laurent03}. These results are non-trivial because the left and right-hand side
\begin{wrapfigure}[8]{r}{0.35\textwidth}
\hspace{1em}
\begin{tikzcd}[->,ampersand replacement=\&]
\termapp{(\termcont{\al}{\termname{\al}{x}})}{y} \arrow{d}[left]{\mu}
  \&[-25pt] \simeq_{\lsigma_8}
  \&[-25pt] \termapp{x}{y} \arrow{d}{\mu} \arrow[phantom,negated]{d}{} \\ 
\termcont{\al}{\termname{\al}{\termapp{x}{y}}}  
\&[-25pt] \simeq_{\lsigma_8}
  \&[-25pt] \termapp{x}{y}
\end{tikzcd}
\caption{Laurent's $\simeq_\sigma$ equivalence not a strong bisimulation}
\label{eq:intro:theta}
\end{wrapfigure}
of the equations in Fig.~\ref{f:sigma-laurent} do not have the same $\beta$ and
$\mu$ redexes. For example, $\termapp{(\termcont{\al}{\termname{\al}{x}})}{y}$
and $\termapp{x}{y}$ are related by equation $\lsigma_8$, however the former
has a $\mu$-redex (more precisely it has a \textit{linear} $\mu$-redex) and the
latter has none. Indeed, as mentioned in Sec.~\ref{sec:introduction} (\cf the
terms in~(\ref{eq:thetaAndSBisim})), $\eqlaurent$ is not a strong bisimulation
with respect to $\lm$-reduction (\cf Fig.~\ref{eq:intro:theta}). There are
other examples illustrating that $\eqlaurent$ is not a strong bisimulation (\cf
Sec.~\ref{s:Sigma-equivalence}). It seems natural to wonder whether, just like
in the intuitionistic case, a more refined notion of $\lmu$-reduction could
change this state of affairs; that is a challenge we take up in this paper.

%%%%%%%%%%%%%%%%%%%%%%%%%%%%%%%%%%%%%%%%%%%%%%%%%%%%%%%%%%%%%%%%%%%%%%%%%%%%%%%
\section{\texorpdfstring{The $\LMfull$-calculus}{The LM-calculus}}
\label{s:LM-calculus}
%%%%%%%%%%%%%%%%%%%%%%%%%%%%%%%%%%%%%%%%%%%%%%%%%%%%%%%%%%%%%%%%%%%%%%%%%%%%%%%

We now extend the syntax of $\lmu$ to that of $\LMfull$. We
again fix a countable infinite set of \deft{variables} $x, y, z, \ldots$ and
\deft{continuation names} $\al, \beta, \ga, \ldots$.  The set of \deft{objects}
$\objects{\LMfull}$, \deft{terms} $\terms{\LMfull}$, \deft{commands}
$\commands{\LMfull}$, \deft{stacks} and \deft{contexts} are given by the
following grammar: \[
\hfill
\begin{array}{llll}
\textbf{Objects}               & o        & \Coloneq & t \mid c \mid s \\
\textbf{Terms}                 & t        & \Coloneq & x \mid \termapp{t}{t} \mid \termabs{x}{t} \mid \termcont{\alpha}{c} \mid \termsubs{x}{t}{t} \\
\textbf{Commands}              & c        & \Coloneq & \termname{\alpha}{t} \mid \termrepl{\alpha'}{\alpha}{s}{c} \\
\textbf{Stacks}                & s        & \Coloneq & \termemst \mid \termpush{t}{s} \\
\textbf{Contexts}              & \ctxt{O} & \Coloneq & \ctxt{T} \mid \ctxt{C} \mid \ctxt{S}\\
\textbf{Term Contexts}         & \ctxt{T} & \Coloneq & \Box \mid \termapp{\ctxt{T}}{t} \mid \termapp{t}{\ctxt{T}} \mid \termabs{x}{\ctxt{T}} \mid \termcont{\alpha}{\ctxt{C}} \mid \termsubs{x}{t}{\ctxt{T}} \mid \termsubs{x}{\ctxt{T}}{t} \\
\textbf{Command Contexts}      & \ctxt{C} & \Coloneq & \boxdot \mid \termname{\alpha}{\ctxt{T}} \mid \termrepl{\alpha'}{\alpha}{s}{\ctxt{C}} \mid \termrepl{\alpha'}{\alpha}{\ctxt{S}}{c} \\
\textbf{Stack Contexts}        & \ctxt{S} & \Coloneq & \termpush{\ctxt{T}}{s} \mid \termpush{t}{\ctxt{S}} \\
\textbf{Substitution Contexts} & \ctxt{L} & \Coloneq & \Box \mid \termsubs{x}{t}{\ctxt{L}} \\
\textbf{Replacement Contexts}  & \ctxt{R} & \Coloneq & \boxdot \mid  \termrepl{\alpha'}{\alpha}{s}{\ctxt{R}}
\end{array}
\hfill
\]

Terms  of  $\lmu$ are enriched with \deft{explicit substitutions} of the form
$\termsubs{x}{u}{t}$. Commands of $\lmu$ are enriched with \deft{explicit
replacements} of the form $\termrepl{\al'}{\al}{s}{c}$, where $\alpha \neq
\alpha'$, and $\alpha'$ is called a \deft{replacement name}. Stacks are empty
($\termemst$) or non-empty ($\termpush{t}{s}$). Explicit replacements with empty stacks (\ie
$\exrepl{\alpha}{\beta}{\termemst}$) are called \deft{renaming replacements},
otherwise \deft{stack replacements}. 

Stack concatenation, denoted $\termpush{s}{s'}$, is defined as
expected\ispaper{}{: if $s
= \termpush{t_1}{\termpush{\ldots}{\termpush{t_n}{\termemst}}}$, then
$\termpush{s}{s'} = \termpush{t_1}{\termpush{\ldots}{\termpush{t_n}{s'}}}$},
where $\termpush{\_}{\_}$ is associative and $\termemst$ is the neutral
element. We often write
$\termpush{t_1}{\termpush{\ldots}{\termpush{t_n}{\termemst}}}$ simply as
$\termpush{t_1}{\termpush{\ldots}{t_n}}$ (thus, in particular,
$\termpush{u}{\termemst}$ is abbreviated as $u$).  Moreover, given a term $u$,
we use the abbreviation $\termconc{u}{s}$ for the term $u$ if $s = \termemst$
and $(\termapp{(\termapp{u}{t_1})}{\ldots) t_n}$ if $s =
\termpush{t_1}{\termpush{\ldots}{t_n}}$. This operation is left-associative,
hence $\termconc{\termconc{u}{s_1}}{s_2}$ means
$\termconc{(\termconc{u}{s_1})}{s_2}$. 

\deft{Free} and \deft{bound variables} of $\LMfull$-objects are extended as
expected. In particular,
%\delia{dejo solamente los casos
%interesantes nuevos con respecto a $\lmu$ y saco el sort para hacer mas light
%la notacion, ya que no los usamos luego con sort}
\ignore{ $$
\begin{array}{ll}
\begin{array}{llll}
\fvSort{t}{x} & \eqdef & \{x\}\\
\fvSort{t}{\termapp{t}{u}} & \eqdef & \fvSort{t}{t}\cup \fvSort{t}{u}\\
\fvSort{t}{\termabs{x}{t}} & \eqdef & \fvSort{t}{t}\setminus \{x\}\\
\fvSort{t}{\termcont{\beta}{t}} & \eqdef & \fvSort{t}{t}\\
\fvSort{t}{\termsubs{x}{u}{t}} & \eqdef & \fvSort{t}{t}\setminus\{x\}\cup\fvSort{t}{u}\\
\fvSort{c}{\termname{\alpha}{t}} & \eqdef & \fvSort{t}{t}\\ 
\fvSort{c}{\termrepl{\gamma'}{\gamma}{s}{c}} & \eqdef & \fvSort{c}{c}\cup\fvSort{s}{s} \\
\fvSort{s}{t} & \eqdef & \fvSort{t}{t} \\
\fvSort{s}{t\cdot s}  & \eqdef & \fvSort{t}{t}\cup\fvSort{s}{s}
\end{array}
&
\begin{array}{llll}
\fnSort{t}{x} & \eqdef & \{\}\\
\fnSort{t}{\termapp{t}{u}} & \eqdef & \fnSort{t}{t}\cup \fnSort{t}{u}\\
\fnSort{t}{\termabs{x}{t}} & \eqdef & \fnSort{t}{t}\\
\fnSort{t}{\termcont{\beta}{t}} & \eqdef & \fnSort{t}{t}\setminus\{\beta\}\\
\fnSort{t}{\termsubs{x}{u}{t}} & \eqdef & \fnSort{t}{t}\cup\fnSort{t}{u}\\
\fnSort{c}{\termname{\alpha}{t}} & \eqdef & \fnSort{t}{t}\cup\{\alpha\}\\ 
\fnSort{c}{\termrepl{\gamma'}{\gamma}{s}{c}} & \eqdef & \fnSort{c}{c}\setminus \{\gamma\}\cup\{\gamma'\}\cup\fnSort{s}{s} \\
\fnSort{s}{t} & \eqdef & \fnSort{t}{t} \\
\fnSort{s}{t\cdot s}  & \eqdef & \fnSort{t}{t}\cup\fnSort{s}{s}
\end{array}
\end{array} $$
}
$\fv{\termsubs{x}{u}{t}} \eqdef \fv{t} \setminus \{x\} \cup \fv{u}$, and
$\fv{\termrepl{\gamma'}{\gamma}{s}{c}} \eqdef \fv{c} \cup \fv{s}$, while
$\fn{\termsubs{x}{u}{t}} \eqdef \fv{t} \cup \fv{u}$, and
$\fn{\termrepl{\gamma'}{\gamma}{s}{c}} \eqdef \fn{c} \setminus \{\gamma\}
\cup \{\gamma'\} \cup \fn{s}$. 
 We work, as usual, modulo $\alpha$-conversion so that bound variables and names
can be renamed. Thus \eg $\termsubs{x}{u}{x} \equiv_{\alpha}
\termsubs{y}{u}{y}$ and $\termcont{\gamma}{\termname{\gamma}{x}}
\equiv_{\alpha} \termcont{\beta}{\termname{\beta}{x}}$. In particular, we will
always assume by $\alpha$-conversion, that $x \notin \fv{u}$ in the term
$\termsubs{x}{u}{t}$ and $\alpha \notin \fn{s}$ in the command
$\termrepl{\alpha'}{\alpha}{s}{c}$. 

The notions of free and bound variables and names are extended to contexts by
defining $\fv{\Box} = \fv{\boxdot} = \fn{\Box} = \fn{\boxdot} = \emptyset$. A
variable $x$ \deft{occurs bound} in $\ctxt{O}$ if, for any fresh variable $y
\neq x$, it occurs in $\ctxtapp{\ctxt{O}}{y}$ but not free.
Similarly for names. Thus for example $x$ is bound in $\termabs{x}{\Box}$ and
$\termapp{(\termabs{x}{x})}{\Box}$ and $\alpha$ is bound in
$\termrepl{\alpha'}{\alpha}{s}{\boxdot}$. We use $\fv{o_1,o_2}$ (resp.
$\fn{o_1,o_2}$) to abbreviate $\fv{o_1} \cup \fv{o_2}$. (resp. $\fn{o_1} \cup
\fn{o_2}$) and also $\fn{o,\alpha}$ to abbreviate $\fn{o} \cup \{\alpha\}$. An
object $o$ is \deft{free for a context} $\ctxt{O}$, written
$\freeFor{o}{\ctxt{O}}$, if the bound variables and bound names of $\ctxt{O}$
do not occur free in $o$. Thus for example
$\freeFor{zy}{\termabs{x}{(\termsubs{x'}{w}{\Box})}}$ holds but
$\freeFor{xy}{\termabs{x}{\Box}}$ does not hold. This notation is naturally
extended to sets, \ie $\freeFor{S}{\ctxt{O}}$ means that the bound variables
and bound names of $\ctxt{O}$ do not occur free in any element of $S$. We write
$\notatall{x}{o}$ (resp. $\notatall{\alpha}{o}$) if $x \notin \fv{o}$ and
$x \notin \bv{o}$ (resp. $\alpha \notin \fn{o}$ and $\alpha \notin \bn{o}$).
This notion is extended to contexts as expected.

%%%%%%%%%%%%%%%%%%%%%%%%%%%%%%%%%%%%%%%%
%%\subsection{Substitution}
%%%%%%%%%%%%%%%%%%%%%%%%%%%%%%%%%%%%%%%%

As in Sec.~\ref{s:lmu-calculus}, we use $\substitute{x}{u}{o}$ and
$\replace{\alpha'}{\alpha}{s}{o}$ to denote, respectively, the natural
extensions of the \deft{substitution} and \deft{replacement} operations to
$\LMfull$-objects. Both are defined modulo $\alpha$-conversion to avoid capture
of free variables/names. While the first notion is standard, we formalise the
second one.

\begin{definition}
\label{def:replacement}
Given $\alpha\notin \fn{s}$, the replacement
$\replace{\alpha'}{\alpha}{s}{o}$ is defined as follows: \[
\hfill
\begin{array}{rcll}
\replace{\alpha'}{\alpha}{s}{x}                                               & \eqdef & x \\
\replace{\alpha'}{\alpha}{s}{(\termapp{t}{u})}                                & \eqdef & \termapp{\replace{\alpha'}{\alpha}{s}{t}}{\replace{\alpha'}{\alpha}{s}{u}} \\
\replace{\alpha'}{\alpha}{s}{(\termabs{x}{t})}                                & \eqdef & \termabs{x}{\replace{\alpha'}{\alpha}{s}{t}} & \notatall{x}{s} \\
\replace{\alpha'}{\alpha}{s}{(\termcont{\beta}{c})}                           & \eqdef & \termcont{\beta}{\replace{\alpha'}{\alpha}{s}{c}} &
\notatall{\beta}{(s, \alpha, \al')} \\
\replace{\alpha'}{\alpha}{s}{\termsubs{x}{u}{t}}                              & \eqdef & \termsubs{x}{\replace{\alpha'}{\alpha}{s}{u}}{\replace{\alpha'}{\alpha}{s}{t}} & \notatall{x}{s} \\
\replace{\alpha'}{\alpha}{s}{(\termname{\alpha}{t})}                          & \eqdef & \termname{\alpha'}{(\termconc{\replace{\alpha'}{\alpha}{s}{t}}{s})} \\
\replace{\alpha'}{\alpha}{s}{(\termname{\beta}{t})}                           & \eqdef & \termname{\beta}{\replace{\alpha'}{\alpha}{s}{t}} & \alpha \neq \beta\\
\replace{\alpha'}{\alpha}{s}{\termrepl{\beta}{\gamma}{s'}{c}}                 & \eqdef & \termrepl{\beta}{\gamma}{\replace{\alpha'}{\alpha}{s}{s'}}{\replace{\alpha'}{\alpha}{s}{c}} & \alpha \neq \beta \\
\replace{\alpha'}{\alpha}{s}{\termrepl{\alpha}{\gamma}{\termemst}{c}}         & \eqdef & \termrepl{\alpha'}{\beta}{\termemst}{\termrepl{\beta}{\gamma}{s}{\replace{\alpha'}{\alpha}{s}{c}}} & s \neq \termemst, \beta \mbox{ fresh} \\
\replace{\alpha'}{\alpha}{\termemst}{\termrepl{\alpha}{\gamma}{\termemst}{c}} & \eqdef & \termrepl{\alpha'}{\gamma}{\termemst}{\replace{\alpha'}{\alpha}{\termemst}{c}} &  \\
\replace{\alpha'}{\alpha}{s}{\termrepl{\alpha}{\gamma}{s'}{c}}                & \eqdef & \termrepl{\alpha'}{\gamma}{\replace{\alpha'}{\alpha}{s}{s'} \cdot s}{\replace{\alpha'}{\alpha}{s}{c}} & s' \neq \termemst \\
\replace{\alpha'}{\alpha}{s}{\termemst}                                       & \eqdef & \termemst & \\
\replace{\alpha'}{\alpha}{s}{(\termpush{t}{s'})}                              & \eqdef & \termpush{\replace{\alpha'}{\alpha}{s}{t }}{\replace{\alpha'}{\alpha}{s}{s'}}   & \\
 \end{array}
\hfill
\]
\end{definition}

\Eg $\replace{\gamma}{\al}{\termpush{y_1}{y_2}}{(\termname{\al}{x})} =
\termname{\gamma}{\termapp{\termapp{x}{y_1}}{y_2}}$, and
$\replace{\gamma}{\al}{y_1}{\termrepl{\al}{\beta}{z_1}{(\termname{\al}{x})}}
=
\termrepl{\gamma}{\beta}{\termpush{z_1}{y_1}}{(\termname{\gamma}{\termapp{x}{y_1}})}$,
while
$\replace{\gamma}{\al}{\termpush{y_1}{y_2}}{\termrepl{\al}{\beta}{\termemst}{(\termname{\al}{x})}}
=
\termrepl{\gamma}{\gamma'}{\termemst}{\termrepl{\gamma'}{\beta}{\termpush{y_1}{y_2}}{(\termname{\gamma}{\termapp{\termapp{x}{y_1}}{y_2}})}}$.

When $s = \termemst$,  the replacement operation $\replace{\alpha'}{\alpha}{s}{\_}$ is called a \deft{renaming}. 
Most of the cases in the definition above are straightforward, we only comment
on the interesting ones. When the (meta-level) replacement operator affects a
renaming replacement, \ie in the case
$\replace{\alpha'}{\alpha}{s}{c\exrepl{\alpha}{\gamma}{\emst}}$, the renaming
$\exrepl{\al}{\gamma}{\termemst}$ is \textit{blocking} the replacement, so that
an explicit replacement $\exrepl{\beta}{\gamma}{s}$ with a fresh name $\beta$
is created, and $\beta$ is then renamed to $\al'$. Regarding the last clause of
the definition for commands, since the explicit replacement
$\exrepl{\al}{\gamma}{s'}$ is blocking, it accumulates any additional arguments
for $\gamma$, hence why \textit{stacks} (\ie sequences of terms) are used,
instead of terms, as target for names.

The two operations $\substitute{x}{u}{o}$, and
$\replace{\alpha'}{\alpha}{s}{o}$ are extended to contexts as expected.

%%%%%%%%%%%%%%%%%%%%%%%%%%%%%%%%%%%%%%%%
%{\bf $\LMfull$-Reduction.}
%%%%%%%%%%%%%%%%%%%%%%%%%%%%%%%%%%%%%%%%
\begin{definition}
The \deft{$\LMfull$-calculus} is given by the set of objects
$\objects{\LMfull}$ and the \deft{$\LMfull$-reduction relation}
$\Rew{\LMfull}$, defined as the closure by {\it all} contexts of the rewriting
rules:
\[
\hfill
\begin{array}{c@{\qquad}c}
\begin{array}{rcl}
\termapp{\ctxtapp{\ctxt{L}}{\termabs{x}{t}}}{u}
  & \rrule{\rB}     & \ctxtapp{\ctxt{L}}{\termsubs{x}{u}{t}} \\
\termsubs{x}{u}{t}
  & \rrule{\rS}     & \substitute{x}{u}{t} 
\end{array}
&
\begin{array}{rcl}
\termapp{\ctxtapp{\ctxt{L}}{\termcont{\alpha}{c}}}{u}
  & \rrule{\rM}     & \ctxtapp{\ctxt{L}}{\termcont{\alpha'}{\termrepl{\alpha'}{\alpha}{\termpush{u}{\termemst}}{c}}} \\
\termrepl{\alpha'}{\alpha}{s}{c}
  & \rrule{\rRfull} & \replace{\alpha'}{\alpha}{s}{c}
\end{array}
\end{array}
\hfill
\] where $\rrule{\rB}$ and $\rrule{\rM}$ are both constrained by the condition
$\freeFor{u}{\ctxt{L}}$, and $\rrule{\rM}$ also requires
$\notatall{\alpha'}{(c,u,\alpha,\ctxt{L})}$, thus both rules pull the list
context $\ctxt{L}$ out by avoiding the capture of free variables/names of $u$.
Equivalently, $\mathbin{\Rew{\LMfull} \eqdef
\ctxtapp{\ctxtSort{O}{t}}{\rrule{\rB} \cup \rrule{\rS} \cup \rrule{\rM}} \cup
\ctxtapp{\ctxtSort{O}{c}}{\rrule{\rRfull}}}$. Given $X \in \{\rB, \rS, \rM,
\rRfull\}$, we write $\Rew{X}$  for the closure by all contexts of $\rrule{X}$.
\end{definition}

Note that $\rB$ and $\rM$ above, also
presented in~(\ref{eq:split_of_beta}) and~(\ref{eq:split_of_mu}) of the introduction,  operate
\emph{at a distance}~\cite{AccattoliK12}, a
  characteristic in line with our semantical development
being guided by Proof-Nets.  Also, following Parigot~\cite{Parigot93}, one might
be tempted to rephrase the reduct of $\rM$ with a binary constructor, writing
$\ctxtapp{\ctxt{L}}{\termcont{\alpha}{\termrepl{}{\alpha}{u}{c}}}$. But this is
imprecise since free occurrences of $\alpha$ in $c$ cannot be bound to both
$\mu \alpha$ and $\termrepl{}{\alpha}{u}{}$. The subscript $\alpha'$ in
$\termrepl{\alpha'}{\alpha}{u}{c}$ shall replace $\alpha$ in $c$ as described
above. The $\LMfull$-calculus implements the $\lmu$-calculus by means of more
atomic steps, \ie

\begin{lemma}
Let $o \in \objects{\lmu}$.
If $o \Rew{\lmu} o'$, then $o \Rewn{\LMfull} o'$.
\end{lemma}

\ispaper{}{\partitle{Confluence of $\LMfull$-Reduction.}}
Just like $\lmu$, the $\LMfull$-calculus is confluent too. This is proved
by using the interpretation method~\cite{Hardin1989}, where $\LMfull$ is
interpreted into $\lmu$ by means of
\ispaper{a suitable projection function. }{the following projection function from
$\lmu$-objects to $\LMfull$-objects:
\[
\hfill
\begin{array}{cc}
\begin{array}{r@{\enspace}c@{\enspace}l}
\project{(x)}                    & \eqdef & x \\
\project{(\termapp{t}{u})}       & \eqdef & \termapp{\project{t}}{\project{u}} \\
\project{(\termabs{x}{t})}       & \eqdef & \termabs{x}{\project{t}} \\
\project{(\termsubs{x}{u}{t})}   & \eqdef & \substitute{x}{\project{u}}{\project{t}} \\
\project{(\termcont{\alpha}{c})} & \eqdef & \termcont{\alpha}{\project{c}}       
\end{array}
&
\begin{array}{r@{\enspace}c@{\enspace}l}
\project{(\termname{\alpha}{t})}             & \eqdef & \termname{\alpha}{\project{t}} \\
\project{(\termrepl{\alpha'}{\alpha}{s}{c})} & \eqdef & \replace{\al'}{\al}{\project{s}}{\project{c}} \\ \\
\project{(\termemst)}                        & \eqdef & \termemst \\
\project{(\termpush{t}{s})}                  & \eqdef & \termpush{\project{t}}{\project{s}}
\end{array}
\end{array}
\hfill
\] The use of $\replace{\al'}{\al}{\project{s}}{\project{c}}$ above refers to
replacement of $\LMfull$ but acting on $\lmu$-terms.
}
\begin{toappendix}
\begin{theorem}
\label{t:confluence_of_LM}
The $\Rew{\LMfull}$ relation is confluent.
\end{theorem}
\end{toappendix}

\begin{proof}
By the interpretation method~\cite{Hardin1989}, using confluence of
$\Rew{\lm}$~\cite{Parigot93}. Details
\ispaper{in~\cite{TR}.}{in App.~\ref{a:confluence}.}
\end{proof}

%%%Local Variables:
%%% mode: latex
%%% TeX-master: "main"
%%% End:

%%%%%%%%%%%%%%%%%%%%%%%%%%%%%%%%%%%%%%%%%%%%%%%%%%%%%%%%%%%%%%%%%%%%%%%%%%%%%%%
\section{Refining Replacement}
\label{sec:refining_replacement}
%%%%%%%%%%%%%%%%%%%%%%%%%%%%%%%%%%%%%%%%%%%%%%%%%%%%%%%%%%%%%%%%%%%%%%%%%%%%%%%

Now that we have introduced the relation $\LMfull$ and hence the reader has a
clearer picture of the presentation/implementation of $\lambda\mu$ we will be
working with, we briefly revisit our objective. We seek to identify a relation
$\simeq$ on a \emph{subset} of $\LMfull$-terms (which we call \emph{canonical})
that is a strong bisimulation for a \emph{subset} of $\LMfull$-reduction (which
we call \emph{meaningful}). The proof-net translation of the intuitionistic
case suggests that $\simeq$ be defined on the subset of $\LMfull$ terms that
are in $\rB\rM$-normal form, since a term and its $\rB\rM$-reduct are
essentially different syntactic presentations of the same thing. Consequently,
we can in principle declare $\rS\cup\rRfull$ as the subset of
$\LMfull$-reduction that is \emph{meaningful}. However the proof-net
translation of $\LMfull$ suggests that meaningful reduction for $\LMfull$
is the exponential one, manipulating \emph{boxes} (\cf Sec.~\ref{s:proofnets}), which allow
in particular their erasure and duplication, and unfortunately $\rRfull$
also includes cases without any box manipulation. Moreover, as hinted at
in the introduction, when incorporating the $\rB\rM$-normal form of Laurent's
$\sigma$-equivalence equations into $\simeq$, one immediately realizes that
strong bisimulation fails. The heart of the matter is that $\rRfull$ is too
course-grained and that we should break it down, weeding out those instances
that present an obstacle to strong bisimulation. Indeed, one can
distinguish between linear and non-linear instances of $\rRfull$, the
translation of the latter involving boxes while the former's not. This section
presents a refinement of $\rRfull$, in four stages,
identifying a subset of
  replacement $\rRfull$ we dub \emph{meaningful
    replacement reduction}. The latter, together with 
  $\rS$, will conform
  the whole notion of  \emph{meaningful reduction} (Def~\ref{def:relation-reduce}).

\partitle{Stage 1: Renaming vs Stack Replacement.}
In this first stage we split $\rRfull$ according to
the nature of the explicit replacement, renaming or stack: \[
\hfill
\begin{array}{rlll}
  \termrepl{\alpha'}{\alpha}{\termemst}{c}
  & \rrule{\rRstackvacio} & \replace{\alpha'}{\alpha}{\termemst}{c} & \\
  \termrepl{\alpha'}{\alpha}{s}{c}
  & \rrule{\rRstacknovacio} & \replace{\alpha'}{\alpha}{s}{c}&
  \text{ if $s \neq  \termemst$} \\
\end{array}
\hfill
\] 
Accordingly, we call $\rRstackvacio$ the renaming replacement rule and
$\rRstacknovacio$ the stack replacement rule.
%\ispaper{}{
\begin{wrapfigure}[\ispaper{9}{7}]{r}{0.58\textwidth}
%\begin{equation}
%\hfill
\scalebox{0.85}{
\begin{tikzcd}[ampersand replacement=\&]
\termrepl{\alpha'}{\alpha}{s}{(\termname{\alpha}{\termcont{\beta}{c}})} \arrow{d}[left]{\rRfull}
  \&[-30pt] \simeq_{\lsigma_7}
  \&[-30pt] \termrepl{\alpha'}{\alpha}{s}{c \iren{\beta}{\alpha}} \arrow{dd}[left]{\rRfull} \\
\termname{\alpha'}{\termapp{(\termcont{\beta}{\replace{\alpha'}{\alpha}{s}{c}})}{s}}
  \arrow[]{d}[left]{\BMName}
  \&[-30pt] 
  \&[-30pt] \\
\termname{\alpha'}{\termcont{\beta'}{\BMName{(\termrepl{\beta'}{\beta}{s}{\replace{\alpha'}{\alpha}{s}{c}})}}}
  \&[-30pt] ???
  \&[-30pt] \BMName{(\replace{\alpha'}{\alpha}{s}{c \iren{\beta}{\alpha}})}
\end{tikzcd}
}
\hfill
\caption{Implicit renaming and strong bisimulation}\label{e:rho-breaks-strong-bisimulation:pre}
%\end{equation}
\end{wrapfigure}
%}
The renaming replacement rule unfortunately throws away important information
that is required for our strong-bisimulation. As an example, consider the
equation $\lsigma_7$ of 
%\ispaper{
%\begin{wrapfigure}[9]{r}{0.58\textwidth}
%%\begin{equation}
%%\hfill
%\scalebox{0.85}{
%\begin{tikzcd}[ampersand replacement=\&]
%\termrepl{\alpha'}{\alpha}{s}{(\termname{\alpha}{\termcont{\beta}{c}})} \arrow{d}[left]{\rRfull}
  %\&[-30pt] \simeq_{\lsigma_7}
  %\&[-30pt] \termrepl{\alpha'}{\alpha}{s}{c \iren{\beta}{\alpha}} \arrow{dd}[left]{\rRfull} \\
%\termname{\alpha'}{\termapp{(\termcont{\beta}{\replace{\alpha'}{\alpha}{s}{c}})}{s}}
  %\arrow[]{d}[left]{\BMName}
  %\&[-30pt] 
  %\&[-30pt] \\
%\termname{\alpha'}{\termcont{\beta'}{\BMName{(\termrepl{\beta'}{\beta}{s}{\replace{\alpha'}{\alpha}{s}{c}})}}}
  %\&[-30pt] ???
  %\&[-30pt] \BMName{(\replace{\alpha'}{\alpha}{s}{c \iren{\beta}{\alpha}})}
%\end{tikzcd}
%}
%\hfill
%\caption{Implicit renaming and strong bisimulation}\label{e:rho-breaks-strong-bisimulation:pre}
%%\end{equation}
%\end{wrapfigure}
%}{}
Fig.~\ref{f:sigma-laurent}, but under a $\LMfull$ context containing an explicit replacement $\exrepl{\alpha}{\alpha'}{s}$, as depicted in
Fig.~\ref{e:rho-breaks-strong-bisimulation:pre}.
Firing $\exrepl{\alpha}{\alpha'}{s}$ on the left leads to the
creation of another stack replacement redex on the left, with no
such redex mimicking it on the right. Indeed, the term on the lower left hand
corner has a pending explicit replacement and hence cannot be equated with the
term on the lower right hand corner.

As for our notion of \emph{meaningful replacement} reduction, this leads us to
disregard the renaming replacement rule, leaving renaming replacements in terms
as is, that is, \emph{without executing them}. An adaptation of $\lsigma_7$ to $\LMfull$ will later be adopted in our
$\simeq$ relation (\cf Sec.~\ref{s:Sigma-equivalence}) where 
(implicit) renaming is left pending as an explicit renaming replacement. In
summary, at this stage, our notion of meaningful replacement reduction is taken to be just
$\rRstacknovacio$.

\partitle{Stage 2: A First Refinement of Stack Replacement.}
The next stage in the refinement process is to further split the stack
replacement rule $\rRstacknovacio$ based on the number of free occurrences of
the name to be replaced, as indicated by $\fnp{\alpha}{c}$ below. The
motivation behind this split is that some instances of the replacement rule
that replace exactly one name, actually relate terms that are structurally
equivalent when translated to PPNs, hence perform no meaningful computation.
Thus we consider the following rules: \[
\hfill
\begin{array}{rcll}
  \termrepl{\alpha'}{\alpha}{s}{c}
  & \rrule{\rRdiffu} & \replace{\alpha'}{\alpha}{s}{c}&
  \text{ if $\fnp{\alpha}{c}\neq 1$ and $s \neq \termemst$} \\
\termrepl{\alpha'}{\alpha}{s}{c}
  & \rrule{\rRequ} & \replace{\alpha'}{\alpha}{s}{c} &
  \text{ if $\fnp{\alpha}{c}=1$ and $s \neq \termemst$ } \\
\end{array}
\hfill
\] In rule $\rRdiffu$ we immediately recognize as involving
substantial work due to duplication or erasure of the stack $s$. Hence, we
update our current notion of meaningful replacement computation by replacing our former
selection of  $\rRstacknovacio$ with the more restricted $\rRdiffu$.

We next have to determine whether rule $\rRequ$,
or refinements thereof, should too be judged as meaningful. For that we must
take a closer look at its behavior for specific instances of the command $c$
based on the possible (unique) occurrence of the name $\alpha$:
on a named term (Stage 3) or on an explicit replacement (Stage 4).

\partitle{Stage 3: Stack Replacement on Named Term.}
In the right-hand side of rule $\rRequ$ the command $c$ is traversed
by the meta-level replacement $\replaceo{\alpha'}{\alpha}{s}$ until the unique
free occurrence $\alpha$ is reached. Since free names only occur in commands,
the left hand-side of $\rRequ$ has necessarily one of the following forms: \[
\hfill
\termrepl{\alpha'}{\alpha}{s}{\ctxtapp{\ctxt{C}}{\termname{\alpha}{t}}}
\qquad \qquad
\termrepl{\alpha'}{\alpha}{s}{\ctxtapp{\ctxt{C}}{\termrepl{\alpha}{\beta}{s'}{c'}}}
\hfill
\] for some context $\ctxt{C}$ and where $\alpha$ does not occur free in
$\ctxt{C}, t, c', s'$. We next focus on the first case leaving the second to
Stage 4. The first case gives rise to the rule
$\prelineal$: \[
\hfill
\begin{array}{rcll}
\termrepl{\alpha'}{\alpha}{s}{\ctxtapp{\ctxt{C}}{\termname{\alpha}{t}}}
      & \rrule{\prelineal} & 
\ctxtapp{\ctxt{C}}{\termname{\alpha'}{\termconc{t}{s}}} &  \text{if $\notatall{\al}{(t, \ctxt{C})}, s \neq \termemst$}
\end{array}
\hfill
\] At this point in our development, we pause and briefly discuss
\emph{linear $\mu$-redexes} before getting back to $\prelineal$. From the very
beginning, there was never any hope for $\sigma$-equivalence
(Fig.~\ref{f:sigma-laurent}) to be a strong bisimulation since, as mentioned by
Laurent~\cite{Laurent03}, it does not distinguish between terms with
\emph{linear} $\mu$-redexes. A $\mu$-redex in $\lmu$ is \emph{linear} if it has
the form
$\termapp{(\termcont{\alpha}{\ctxtapp{\ctxt{Q}}{\termname{\alpha}{u}}})}{v}$
with $\notatall{\al}{(u,\ctxt{Q})}$ and $\ctxt{Q}$ defined as follows:
\[
\hfill
\begin{array}{c@{\qquad}c}
\begin{array}{r@{\enspace}c@{\enspace}l}
\ctxt{P} & \Coloneq & \Box \mid \termapp{\ctxt{P}}{t} \mid \termabs{x}{\ctxt{P}} \mid \termcont{\alpha}{\termname{\beta}{\ctxt{P}}} \\
\end{array}
&
\begin{array}{r@{\enspace}c@{\enspace}l}
\ctxt{Q} & \Coloneq & \boxdot \mid \termname{\beta}{\ctxtapp{\ctxt{P}}{\termcont{\gamma}{\boxdot}}}
\end{array}
\end{array}
\hfill
\] An example is the term $\termapp{(\termcont{\al}{\termname{\al}{x}})}{y}$,
one of the two terms of (\ref{eq:thetaAndSBisim}) mentioned in 
Sec.~\ref{sec:introduction}. Such $\mu$-reduction steps reducing linear
$\mu$-redexes hold no operational meaning: if $o$ linearly $\mu$-reduces to
$o'$, then their graphical interpretation yield PPNs whose multiplicative
normal form are structurally equivalent~\cite[Thm.~41]{Laurent03} (revisited in
Sec.~\ref{s:completeness} as Thm.~\ref{t:laurent}).

Returning to our development, we next need to identify what a linear/non-linear
split of rule $\prelineal$ looks like in our setting of $\LMfull$, the
intuition arising, again, from PPN. For that we introduce \emph{linear contexts}.

\begin{definition}
\label{def:linear_contexts}
There are four sets of linear contexts, each denoted using the expressions
$\ctxt{XY}$, with $\ctxt{X}, \ctxt{Y} \in \{\ctxt{T}, \ctxt{C}\}$. The letters
$\ctxt{X}$ and $\ctxt{Y}$ in the expression $\ctxt{XY}$ denote the sort of the
object with which the hole will be filled and the sort of the resulting term,
resp.: \eg $\ctxtLTC$ denotes a context that takes a command and outputs a
term. \[
\hfill
\begin{array}{r@{\enspace}r@{\enspace}c@{\enspace}l}
\textbf{(Linear TT Contexts)} & 
\ctxtLTT & \Coloneq & \Box \mid \termapp{\ctxtLTT}{t} \mid \termabs{x}{\ctxtLTT} \mid \termcont{\alpha}{\ctxtLCT} \mid \termsubs{x}{t}{\ctxtLTT} \\
\textbf{(Linear TC Contexts)} & 
\ctxtLTC & \Coloneq & \termapp{\ctxtLTC}{t} \mid \termabs{x}{\ctxtLTC} \mid \termcont{\alpha}{\ctxtLCC} \mid \termsubs{x}{t}{\ctxtLTC} \\
\textbf{(Linear CC Contexts)} & 
\ctxtLCC & \Coloneq & \boxdot \mid
\termname{\alpha}{\ctxtLTC} \mid \termrepl{\alpha'}{\alpha}{s}{\ctxtLCC} \\
\textbf{(Linear CT Contexts)} & 
\ctxtLCT & \Coloneq &
\termname{\alpha}{\ctxtLTT} \mid \termrepl{\alpha'}{\alpha}{s}{\ctxtLCT}
\end{array}
\hfill
\]
\end{definition}

\noindent For example, $\termname{\alpha}{\boxdot}$ is a $\ctxtLTC$
context and
$\termrepl{\alpha'}{\alpha}{u}{\termname{\beta}{(\termapp{\Box}{v})}}$
is a $\ctxtLTT$ context. \ispaper{}{An alternative definition of
  linear context, used in some technical proofs, can be found in the
  Appendix}.

Given the above definition of linear contexts we can now split $\prelineal$
into its linear and non-linear versions: 
\[
\hfill
\begin{array}{rlll}
\termrepl{\alpha'}{\alpha}{s}{\ctxtapp{\ctxt{C}}{\termname{\alpha}{t}}}   & \rrule{\nonprelineal} & \ctxtapp{\ctxt{C}}{\termname{\alpha'}{\termconc{t}{s}}} &  \text{if $\ctxt{C}$ not linear, $\notatall{\al}{(t,\ctxt{C})}, s \neq \termemst$} \\
\termrepl{\alpha'}{\alpha}{s}{\ctxtapp{\ctxt{LCC}}{\termname{\alpha}{t}}} & \rrule{\linealrule}   & 
\ctxtapp{\ctxt{LCC}}{\termname{\alpha'}{\termconc{t}{s}}} & \text{if $\notatall{\al}{(t,\ctxt{LCC})}$, $s \neq \termemst$}
\end{array}
\hfill
\] The named term $\termname{\alpha}{t}$ in the non-linear rule $\nonprelineal$
could be duplicated or erased and thus this rule joins $\rRdiffu$ as part of
meaningful replacement reduction. However, as was the case above for linear $\mu$-redexes,
rule $\linealrule$ has no meaningful computational content and hence will be
incorporated into $\simeq$ by taking its canonical normal form  ($\ctxt{LCC}$
and $t$ below are assumed in canonical normal form). The new equation is called
$\lineal$:
\[
\hfill
\termrepl{\alpha'}{\alpha}{s}{\ctxtapp{\ctxt{LCC}}{\termname{\alpha}{t}}}
\simeq_{\lineal}  
\ctxtapp{\ctxt{LCC}}{\termname{\alpha'}{\BMC{\termconc{t}{s}}}}
\quad\text{if $\notatall{\al}{(t,\ctxt{LCC})}$, $s \neq \termemst$}
\hfill
\] The notation $\BMC{\_}$
denotes the canonical form of an object.
A precise definition will be presented in Stage 4.

Summarizing our results of Stage 3, meaningful replacement reduction consists for
the moment of $\rRdiffu$ and $\nonprelineal$. In the next and final stage, we
analyze the case where  the left hand-side of $\rRequ$ has the form
$\termrepl{\alpha'}{\alpha}{s}{\ctxtapp{\ctxt{C}}{\termrepl{\alpha}{\beta}{s'}{c'}}}$.

\partitle{Stage 4: Stack Replacement on Explicit Replacements.}
Suppose the left hand-side of $\rRequ$ has the form
$\termrepl{\alpha'}{\alpha}{s}{\ctxtapp{\ctxt{C}}{\termrepl{\alpha}{\beta}{s'}{c'}}}$.
This gives rise to the two
instances $\rRswap$ and
$\composition$: \[
\hfill
\begin{array}{rcll}
\termrepl{\alpha'}{\alpha}{s}{\ctxtapp{\ctxt{C}}{\termrepl{\alpha}{\beta}{\termemst}{c'}}}  & \rrule{\rRswap}       & \ctxtapp{\ctxt{C}}{\termrepl{\alpha'}{\alpha}{\termemst}{\termrepl{\alpha}{\beta}{s}{c'}}} & \text{if $\notatall{\al}{(c',\ctxt{C})}, s \neq \termemst$} \\
\termrepl{\alpha'}{\alpha}{s}{\ctxtapp{\ctxt{C}}{\termrepl{\alpha}{\beta}{s'}{c'}}}         & \rrule{\composition}  & \ctxtapp{\ctxt{C}}{\termrepl{\alpha'}{\beta}{\termpush{s'}{s}}{c'}} & \text{if $\notatall{\al}{(c',\ctxt{C},s')}, s',s \neq \termemst$}
\end{array}
\hfill
\] If we now consider linear/non-linear variants of the above two rules,
depending on whether the context $\ctxt{C}$ in their LHSs is a linear context
or not (in the sense of Def.~\ref{def:linear_contexts}), we end up with the
following four rules, where we use letters $r$ and $r'$ to denote
  non-empty stacks. 
\[\kern-2em
\hfill
 \begin{array}{rlll}
 \termrepl{\alpha'}{\alpha}{r}{\ctxtapp{\ctxt{C}}{\termrepl{\alpha}{\beta}{\termemst}{c'}}}    & \rrule{\nonlinrRswap}       & \ctxtapp{\ctxt{C}}{\termrepl{\alpha'}{\alpha}{\termemst}{\termrepl{\alpha}{\beta}{r}{c'}}} & \text{$\ctxt{C}$ not linear, $\notatall{\al}{(c',\ctxt{C})}$} \\
 \termrepl{\alpha'}{\alpha}{r}{\ctxtapp{\ctxt{C}}{\termrepl{\alpha}{\beta}{r'}{c'}}}           & \rrule{\nonlincomposition}  & \ctxtapp{\ctxt{C}}{\termrepl{\alpha'}{\beta}{\termpush{r'}{r}}{c'}} & \text{$\ctxt{C}$ not linear, $\notatall{\al}{(c',\ctxt{C},s')}$} \\
 \termrepl{\alpha'}{\alpha}{r}{\ctxtapp{\ctxt{LCC}}{\termrepl{\alpha}{\beta}{\termemst}{c'}}}  & \rrule{\rW}                 & \ctxtapp{\ctxt{LCC}}{\termrepl{\alpha'}{\alpha}{\termemst}{\termrepl{\alpha}{\beta}{r}{c'}}} & \text{$\notatall{\al}{(c',\ctxt{LCC})}, \freeFor{\{s, \al'\}}{\ctxtLCC}$} \\
 \termrepl{\alpha'}{\alpha}{r}{\ctxtapp{\ctxt{LCC}}{\termrepl{\alpha}{\beta}{r'}{c'}}}         & \rrule{\rC}                 & \ctxtapp{\ctxt{LCC}}{\termrepl{\alpha'}{\beta}{\termpush{r'}{r}}{c'}} & \text{$\notatall{\al}{(c',\ctxt{LCC},s')}, \freeFor{\{s,\al'\}}{\ctxtLCC}$}
       \end{array} 
\]
The rules involving non-linear contexts, namely $\nonlinrRswap$ and
$\nonlincomposition$ join $\rRdiffu$ and $\nonprelineal$ in conforming
meaningful replacement computation. Indeed, although
there is a unique occurrence of
$\alpha$ which is target of the explicit replacement on the LHS of these rules,
the stack $s$ could be duplicated or erased. This concludes our
deconstruction of rule $\rRfull$.

\begin{definition}
\label{def:meaningful_reduction}
\deft{Meaningful replacement reduction}, written $\Rew{\rR}$, is defined by the
contextual closure of the reduction rules
$\{\rRdiffu,\nonprelineal,\nonlinrRswap,\nonlincomposition\}$.
\end{definition}

Rules $\rM$, which together with $\rB$ compute canonical forms, create new
explicit replacements. These explicit replacements may be rearranged using
rules $\rW$ and $\rC$ leading to the following notion of canonical form
computation:

\begin{definition}
\label{def:canonical}
\deft{Canonical forms} are terms in $\rB$, $\rM$, $\rC$ and $\rW$-normal form.
The reduction relation $\Rew{\rB\rM\rC\rW}$ is easily seen to be confluent and
terminating, thus, from now on, the notation $\BMC{o}$ stands for the (unique)
$\rB\rM\rC\rW$-normal form of an object $o$. It will be shown later that
$\Can$-reduction on $\LMfull$-objects corresponds to multiplicative cuts in
PPNs (\cf \ispaper{Thm.~\ref{t:simulation-into-ppn}}{Lem.~\ref{l:bmc-to-proofnets}}).
\end{definition}

 In summary, we shall define a strong bisimulation
$\simeq$ (Sec.~\ref{s:Sigma-equivalence}) defined
exclusively on canonical forms and where reduction is taken to be
meaningful:
\begin{definition}
  \label{def:relation-reduce}
\deft{Meaningful reduction} is a relation of canonical forms defined
 as follows:
$$t \reduce t'\text{ iff }t \rewrite{\rS\rR} u \text{ and }
t' = \BMC{u}$$ where $\Rew{\rS\rR}$ is $\Rew{\rS} \cup
  \Rew{\rR}$. We may write $\reduce_\rS$ or $\reduce_\rR$ to
emphasize a meaningful step corresponding to rule $\rS$ or $\rR$
respectively.
\end{definition}

The following diagram summarizes this section's findings. \[
%\kern-1em
\hfill
%\scalebox{0.8}{\input{diagrama}}
\scalebox{0.75}
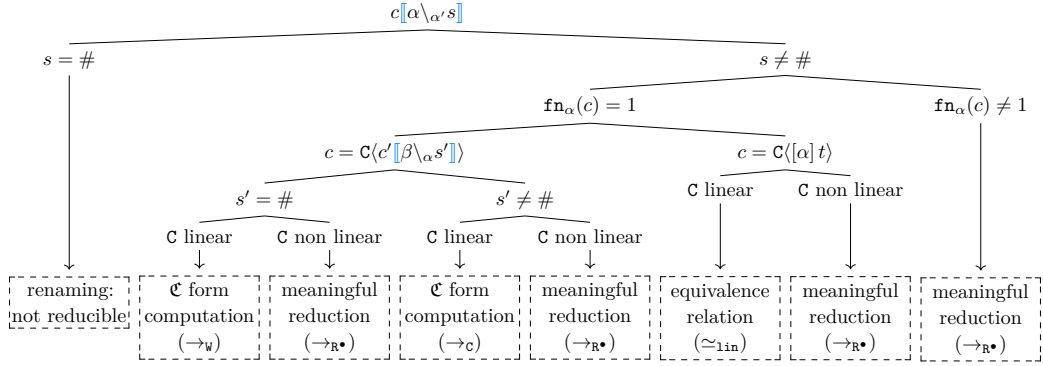
{\begin{tikzpicture}
\node (root)    [] {$\termrepl{\alpha'}{\alpha}{s}{c}$};

\node (seq)     [below of=root, xshift=-17.875em, yshift=0.5em] {$s = \termemst$};
\node (snq)     [below of=root, xshift= 17.875em, yshift=0.5em] {$s \neq \termemst$};

\node (aeq)     [below of=snq, xshift=-9.75em, yshift=0.5em]    {$\fnp{\alpha}{c} = 1$};
\node (anq)     [below of=snq, xshift= 9.75em, yshift=0.5em]    {$\fnp{\alpha}{c} \neq 1$};

\node (repl)    [below of=aeq, xshift=-9.75em, yshift=0.5em]    {$c = \ctxtapp{\ctxt{C}}{\termrepl{\alpha}{\beta}{s'}{c'}}$};
\node (name)    [below of=aeq, xshift= 9.75em, yshift=0.5em]    {$c = \ctxtapp{\ctxt{C}}{\termname{\alpha}{t}}$};

\node (seq')    [below of=repl, xshift=-6.5em, yshift=0.5em]    {$s' = \termemst$};
\node (snq')    [below of=repl, xshift= 6.5em, yshift=0.5em]    {$s' \neq \termemst$};

\node (seql)    [below of=seq', xshift=-3.25em, yshift=1em]     {\parbox{5pc}{\centering $\ctxt{C}$ linear}};
\node (seqnl)   [below of=seq', xshift= 3.25em, yshift=1em]     {\parbox{5pc}{\centering $\ctxt{C}$ non linear}};
\node (snql)    [below of=snq', xshift=-3.25em, yshift=1em]     {\parbox{5pc}{\centering $\ctxt{C}$ linear}};
\node (snqnl)   [below of=snq', xshift= 3.25em, yshift=1em]     {\parbox{5pc}{\centering $\ctxt{C}$ non linear}};
\node (namel)   [below of=name, xshift=-3.25em, yshift=1em]     {\parbox{5pc}{\centering $\ctxt{C}$ linear}};
\node (namenl)  [below of=name, xshift= 3.25em, yshift=1em]     {\parbox{5pc}{\centering $\ctxt{C}$ non linear}};

\node (rename)  [below of=seq,    yshift=-9.375em]  {\dashnode{5pc}{renaming: not reducible}};
\node (swap)    [below of=seql,   yshift=-1.25em]   {\dashnode{5pc}{$\Can$ form computation ($\Rew{\rW}$)}};
\node (rb1)     [below of=seqnl,  yshift=-1.25em]   {\dashnode{5pc}{meaningful reduction ($\Rew{\rR}$)}};
\node (comp)    [below of=snql,   yshift=-1.25em]   {\dashnode{5pc}{$\Can$ form computation ($\Rew{\rC}$)}};
\node (rb2)     [below of=snqnl,  yshift=-1.25em]   {\dashnode{5pc}{meaningful reduction ($\Rew{\rR}$)}};
\node (eqlin)   [below of=namel,  yshift=-3.625em]  {\dashnode{5pc}{equivalence relation ($\simeq_{\lineal}$)}};
\node (rb3)     [below of=namenl, yshift=-3.625em]  {\dashnode{5pc}{meaningful reduction ($\Rew{\rR}$)}};
\node (rb4)     [below of=anq,    yshift=-7.875em]  {\dashnode{5pc}{meaningful reduction ($\Rew{\rR}$)}};

\draw (root.south) -- node{} (seq.north)
      (root.south) -- node{} (snq.north)
      (snq.south)  -- node{} (aeq.north)
      (snq.south)  -- node{} (anq.north)
      (aeq.south)  -- node{} (repl.north)
      (aeq.south)  -- node{} (name.north)
      (repl.south) -- node{} (seq'.north)
      (repl.south) -- node{} (snq'.north)
      (seq'.south) -- node{} (seql.north)
      (seq'.south) -- node{} (seqnl.north)
      (snq'.south) -- node{} (snql.north)
      (snq'.south) -- node{} (snqnl.north)
      (name.south) -- node{} (namel.north)
      (name.south) -- node{} (namenl.north);
\draw[->] (seq.south)    -- node{} (rename.north);
\draw[->] (seql.south)   -- node{} (swap.north);
\draw[->] (seqnl.south)  -- node{} (rb1.north);
\draw[->] (snql.south)   -- node{} (comp.north);
\draw[->] (snqnl.south)  -- node{} (rb2.north);
\draw[->] (namel.south)  -- node{} (eqlin.north);
\draw[->] (namenl.south) -- node{} (rb3.north);
\draw[->] (anq.south)    -- node{} (rb4.north);
\end{tikzpicture}
}
\hfill
\]

%%%Local Variables:
%%% mode: latex
%%% TeX-master: "main"
%%% End:

%%%%%%%%%%%%%%%%%%%%%%%%%%%%%%%%%%%%%%%%%%%%%%%%%%%%%%%%%%%%%%%%%%%%%%%%%%%%%%%
\section{Types}
\label{s:types}
%%%%%%%%%%%%%%%%%%%%%%%%%%%%%%%%%%%%%%%%%%%%%%%%%%%%%%%%%%%%%%%%%%%%%%%%%%%%%%%

In this section we introduce simple types for $\LMfull$, which extends the
type system in~\cite{Parigot93} to our syntax. 
\deft{Types} are generated by the following grammar: $$
% \begin{center}
% \scalebox{0.8}{$
\begin{array}{llll}
\textbf{Term Types}  & A & \Coloneq & \iota \mid A \tarrow B \\
\textbf{Stack Types} & S & \Coloneq & \typeemst \mid \push{A}{S}
\end{array} $$
% $}
% \end{center}
where $\iota$ is a base type. 
The type constructor $\typepush{\_}{\_}$ should be understood as a
non-commutative conjunction, which translates to a tensor in linear logic (see
\ispaper{~\cite{TR}}{Sec.~\ref{s:translation}}). The arrow is right associative. We use the
abbreviation $A_1 \cdot A_2 \cdot \ldots \cdot A_n \cdot \temst \tarrow B$ for
the type $A_1 \tarrow A_2 \ldots \tarrow A_n \tarrow B$ (in particular,
$\typefunc{\typeemst}{B}$ is equal to $B$ so $\typeemst$ is the left neutral
element for the functional type). \deft{Variable assignments} ($\Gam$), are functions from
variables to types; we write $\emptyset$ for the empty variable assignment.  Similarly, \deft{name assignments}
($\Del$), are functions from names to types. We write $\Gam \cup
\Gam'$ and $\Del \cup \Del'$ for the \deft{compatible union}
between assignments meaning that if
  $x\in\dom{\Gam\cap\Gam'}$ then $\Gam(x)=\Gam'(x)$, and similarly for
  $\Del$ and $\Del'$.
  When $\dom{\Gam}$ and $\dom{\Gam'}$ are disjoint
we may write $\Gam, \Gam'$. The same for name
assignments.

The \deft{typing rules} are presented in Fig.~\ref{f:typing-rules}. There are
three kinds of \deft{typing judgements}: $\Gam \vdash t:A \mid \Del$ for terms,
$\Gam \vdash c \mid \Del$ for commands and $\Gam \vdash s:S \mid \Del$ for
stacks. The notation $\Gam, (x:A)^{\esta}$ (resp. $\Del, (\alpha:A)^{\esta}$)
is used to denoted either $\Gam, x:A$ or $\Gam$ (resp. either $\Del, \alpha:A$
or $\Del$), \ie the assumption $x:A$ occurs at most once in $\Gam,
(x:A)^{\esta}$. Commands have no type, \cf rules $(\nametr)$ and $(\trepl)$,
and stacks are typed with stack types, which are heterogeneous lists, \ie each
component of the list can be  typed with a different type. The interesting rule
is $(\trepl)$, which is a logical modus ponens rule, where the fresh variable
$\al'$ may be already present in the name assignment of the command $c$ or the
stack $s$, thus the notation $\Del \cup \Del', \al':B$ means in particular that
$\al'$ is neither in $\Delta$ nor in $\Delta'$.

\begin{figure*}[t]
\centering
\scalebox{0.95}{$
\begin{array}{c}
\Rule{\vphantom{\Gamma}}
     {x:A \vdash x:A \mid \emptyset}
     {(\tax)}
\qquad
\Rule{\Gam \vdash t:A \tarrow B \mid \Del
      \quad
      \Gam' \vdash u:A  \mid \Del'
     }
     {\Gam \cup \Gam' \vdash tu:B \mid \Del \cup \Del'}
     {(\app)}
\\
\\
\Rule{\Gam, (x:A)^{\esta} \vdash t:B \mid \Del}
     {\Gam \vdash \l x. t: A \tarrow B \mid \Del}
     {(\abs)}
\qquad
\Rule{\Gam \vdash c \mid \Del, (\alpha:A)^{\esta}}
     {\Gam \vdash \termcont{\alpha}{c}: A \mid \Del}
     {(\mutr)}
\qquad
\Rule{\Gam \vdash t:A \mid \Del, (\alpha: A)^{\esta}}
     {\Gam \vdash \termname{\alpha}{t} \mid \Del, \alpha: A}
     {(\nametr)}
\\
\\
\Rule{\Gam, (x:B)^{\esta} \vdash t:A \mid \Del
      \quad
      \Gam' \vdash u:B \mid \Del'
     }
     {\Gam \cup \Gam' \vdash t\exsubs{x}{u}:A \mid \Del \cup  \Del'}
     {(\subtr)}
\\
\\
\Rule{\Gam \vdash c \mid  \Del, (\alpha: S \tarrow B)^{\esta}, (\alpha':B)^{\esta}
      \quad
      \Gam' \vdash s: S  \mid \Del', (\alpha':B)^{\esta}
     }
     {\Gam \cup \Gam' \vdash c\exrepl{\alpha'}{\alpha}{s} \mid  \Del \cup \Del', \alpha': B}
     {(\trepl)} 
\\
\\
\Rule{\phantom{\Gamma}}
     {\emptyset  \vdash \termemst: \typeemst \mid \emptyset}
     {(\stacku)}
\qquad
\Rule{\Gam \vdash t: A \mid  \Del
      \quad
      \Gam' \vdash s: S \mid \Del'
     }
     {\Gam \cup \Gam' \vdash \termpush{t}{s}: \typepush{A}{S} \mid \Del \cup \Del'}
     {(\stackd)}
\end{array}
$}
\caption{Typing Rules for the $\LMfull$-calculus}
\label{f:typing-rules}
\end{figure*}

We use the abbreviation $\Gam \vdash o:\type \mid \Del$
  if $o=t$ and $\type = A$, or
  $o=c$ and there is no type, or
  $o=s$ and $\type=S$. 
  We write $\pi \dem \Gam \vdash o:\type \mid \Del$  if $\pi$ is a type derivation concluding with $\Gam \vdash o:\type \mid \Del$.
The typing system enjoys the following properties:

\begin{toappendix}
\begin{lemma}[Relevance]
\label{l:relevance}
Let $o\in\objects{\LMfull}$. If $\pi \dem \Gam \vdash o:\type  \mid \Del$, then $\dom{\Gam} = \fv{o}$ and
$\dom{\Del} = \fn{o}$.
\end{lemma}
\end{toappendix}

\begin{toappendix}
\begin{lemma}[Preservation of Types for $\eqlaurent$]
  \label{l:PTLaurent}
 Let $o \in \objects{\lmu}$.
If $\pi \dem \Gam \vdash o:\type \mid \Del$ and $o \eqlaurent o'$, then there
exist $\pi' \dem \Gam \vdash o':\type \mid \Del'$. 
\end{lemma}
\end{toappendix}

\begin{toappendix}
\begin{lemma}[Subject Reduction]
\label{l:SR}
Let $o \in \objects{\LMfull}$ s.t. $\pi_o \dem \Gam \vdash o:\type \mid \Del$.
If $o \Rew{\LMfull} o'$, then there exist $\Gam' \subseteq \Gam$ and $\Del'
\subseteq \Del$ and $\pi_{o'}$ s.t. $\pi_{o'} \dem \Gam' \vdash o' : \type \mid
\Del'$.
\end{lemma}
\end{toappendix}

Remark that free variables and names of objects decrease in
  the case of erasing reduction steps, as for example $(\l x. y) z \Rew{} y$
  or $(\mu \al. \termname{\gamma} x)z \Rew{}  \mu \al. \termname{\gamma} x$.

  From now on, when $o \eqlaurent o'$ (resp. 
  $o \Rew{\LMfull} o'$), we will refer to $\pi_o$ and $\pi_{o'}$ as two
  \deft{related} typing derivations, \ie $\pi_{o'}$ is
  obtained from $\pi_o$ by the proof of Lem.~\ref{l:PTLaurent} (resp. 
    Lem.~\ref{l:SR}).

%%% Local Variables:
%%% mode: latex
%%% TeX-master: "main"
%%% End:

\ispaper{
  %%%%%%%%%%%%%%%%%%%%%%%%%%%%%%%%%%%%%%%%%%%%%%%%%%%%%%%%%%%%%%%%%%%%%%%%%%%%%%%
\section{Polarized Proof Nets}
\label{s:proofnets}
%%%%%%%%%%%%%%%%%%%%%%%%%%%%%%%%%%%%%%%%%%%%%%%%%%%%%%%%%%%%%%%%%%%%%%%%%%%%%%%

Laurent~\cite{Laurent02} introduced Polarized Linear Logic (LLP), a proof
system based on polarities on linear logic formulae. It is equipped with a
corresponding notion of Polarized Proof-Nets (PPN) which allows for a simpler
correctness criterion.

One particularly interesting feature is the translation of classical logic into
LLP obtained by interpreting $A\rightarrow B$ as $\formbang{A} \multimap B$,
which is a straightforward extension of that from intuitionistic logic to LLP,
thus capturing the translation from $\lambda$-calculus to LLP.

As mentioned in Sec.~\ref{sec:introduction}, it is possible to translate
$\LMfull$-objects to PPNs. More precisely, we translate typing derivations
$\pi \dem\Gam \vdash o : \type \mid \Del$. The translations of $\type$ and
(formulae in) $\Del$ are called \emph{output} formulae whereas, the
translations of formulae in $\Gamma$ are called \emph{input} formulae, given
that they correspond to the $\lambda$-variables. Following~\cite{Laurent03},
this gives the following formulae categories: \[
\hfill
\begin{array}{lccll}
\textbf{Formulae}             & F & \Coloneq & N \mid P \\
\textbf{Negative formulae}    & N & \Coloneq & O \mid \formder{Q} \\
\textbf{Positive formulae}    & P & \Coloneq & Q \mid \formbang{O} \\
\textbf{Output formulae}      & O & \Coloneq & \formbase \mid \formpar{\formder{Q}}{O} \\  
\textbf{Anti-output formulae} & Q & \Coloneq & \formneg{\formbase} \mid \formten{\formbang{O}}{Q}
\end{array}
\hfill
\] Negation is involutive $({\formbase}^{\bot \bot} = \formbase)$ with
$\formneg{(\formpar{\formder{Q}}{O})} =
\formten{\formbang{\formneg{Q}}}{\formneg{O}}$ and
$\formneg{(\formder{Q})} = \formbang{\formneg{Q}}$.

\begin{definition}
\label{def:proof_structure}
A \deft{proof-structure} is a finite acyclic oriented graph built over the
alphabet of nodes represented below (where the orientation is the top-bottom
one): \[
\hfill
\begin{array}{cccc}
\textbf{Axiom} & \textbf{Cut} & \textbf{Weakening} & \textbf{Contraction} \\ 
\scalebox{0.8}{\begin{tikzpicture}[node distance=\PNdist]
\PNnode{A1}{\textit{ax}};
\node (N1) [below of=A1, xshift=-\PNdist] {$\formneg{O}$};
\node (N2) [below of=A1, xshift= \PNdist] {$O$};

\draw[PNarrow] (A1.west) -| node{} (N1.north)
  	  	       (A1.east) -| node{} (N2.north);
\end{tikzpicture}
}
&
\scalebox{0.8}{\begin{tikzpicture}[node distance=\PNdist]
\PNnode{C1}{\textit{cut}};
\node (N1) [above of=C1, xshift=-\PNdist] {$N$};
\node (N2) [above of=C1, xshift= \PNdist] {$\formneg{N}$};

\draw[PNarrow] (N1.south) |- node{} (C1.west)
  	  	       (N2.south) |- node{} (C1.east);
\end{tikzpicture}
}
&
\scalebox{0.8}{\begin{tikzpicture}[node distance=\PNdist]
\PNnode{W1}{\textit{w}};
\node (N1) [below of=W1] {$N$};

\draw[PNarrow] (W1.south) -- node{} (N1.north);
\end{tikzpicture}
}
&
\scalebox{0.8}{\begin{tikzpicture}[node distance=\PNdist]
\PNnode{C1}{\textit{c}};
\node (N1) [above of=C1, xshift=-\PNdist] {$N$};
\node (N2) [above of=C1, xshift= \PNdist] {$N$};
\node (N3) [below of=C1] {$N$};

\draw[PNarrow] (N1.south) -- node{} (C1.north west)
  	  	       (N2.south) -- node{} (C1.north east)
  	  	       (C1.south) -- node{} (N3.north);
\end{tikzpicture}
}
\\
\\
\textbf{Tensor} & \textbf{Par} & \textbf{Dereliction} & \textbf{Box} \\ 
\scalebox{0.8}{\input{proofnets/tensor.tex}}
&
\scalebox{0.8}{\input{proofnets/par.tex}}
&
\scalebox{0.8}{\begin{tikzpicture}[node distance=\PNdist]
\PNnode{D1}{\textit{d}};
\node (N1) [above of=D1] {$Q$};
\node (N2) [below of=D1] {$\formder{Q}$};

\draw[PNarrow] (N1.south) -- node{} (D1.north)
               (D1.south) -- node{} (N2.north);
\end{tikzpicture}
}
&
\scalebox{0.8}{\input{proofnets/bang.tex}}
\end{array}
\hfill
\]
\end{definition}
Remark that each wire is labelled with a formula. In particular, conclusions in
boxes have only one bang formula $\formbang{O}$, all others being negative
formulae.

A \deft{polarized proof-net} (PPN) is a proof-structure satisfying a simple
correctness criterion~\cite{Laurent03}. We refer the reader to
op.cit. and simply mention that since our proof-structures are obtained from
translating \emph{typed terms}, this criterion is always met. \deft{Structural
equivalence} for PPNs, is based on a set of axioms that allow a reordering of
weakening and contraction nodes. By lack of space we don't provide all the
technical details here, but we refer the interested to~\cite{TR} for further
details on this standard relation. In the sequel, \deft{polarized proof-net
equality}, written $\equiv$, is always taken modulo structural equivalence.

The reduction relation for PPNs, denote by $\Rew{\redpn}$, is given by a set
of {\em cut elimination rules}, split into \emph{multiplicative} and
\emph{exponential} rules. By lack of space we don't include the rules here, but
we refer the interested to~\cite{TR} for further technical details. The major
point is that only exponential cuts deal with erasure and  duplication of boxes
and $\nomformten$-trees, these last ones used to interpret stacks in the term
language. Exponential cuts are considered as the meaningful rules of PPNs,
because of their erasure/duplication power. Multiplicative cuts are confluent
and terminating, and we thus use \emphdef{multiplicative normal-forms} as a
technical tool to define the translation of typed $\LMfull$-objects to PPNs.

More precisely, the translation from $\LMfull$ to PPNs guiding the semantical
development of our work is an extension of that introduced in~\cite{Laurent03}
for $\lmu$-terms, based in turn on Girard's translation of classical formulae
to linear logic. We do not give any technical detail in this abstract, formal
definitions are fully developed in~\cite{TR}. For the sequel, it is sufficient
to keep in mind that formulae are translated to polarized linear logic, and
type derivations to PPNs. We use the notation $\topn{\pi}$ to denote the
translation of the typing derivation $\pi$.

Without entering into details, it is worth mentioning that $\topn{\pi}$ does
not preserve $\LMfull$-reduction, so that we extend it  to  a new one, written
$\topnd{\_}$, in such a way that $\topnd{\pi}$ is the multiplicative
normal-form of $\topn{\pi}$. Then, the following property holds.

\begin{theorem}
\label{t:simulation-into-ppn}
Let $o, p$ be typed $\LMfull$-objects. If $o \Rew{\LMfull} p$, and $\pi_o,
\pi_p$ are two related corresponding type derivations for $o$ and $p$ resp.,
then $\topnd{\pi_o} \Rewn{\redpn} \topnd{\pi_{p}}$.
\end{theorem}

\begin{proof}
By induction on $\Rew{\LMfull}$ by adapting Lem. 18 and 19 in~\cite{Laurent02}.
More precisely, the only $\LMfull$-reduction steps involving \emph{exponential}
rules on the PPNs side are $\Rew{\rS}$ and the non-linear instances of
$\Rew{\rRfull}$ (called $\Rew{\rR}$ in Sec.~\ref{sec:refining_replacement}),
while $\Rew{\rB}$, $\Rew{\rC}$ and $\Rew{\rW}$ are translated to
\emph{multiplicative} cuts, and both $\Rew{\rM}$ and $\Rew{\linealrule}$ give
the identity.
\end{proof}

%%%Local Variables:
%%% mode: latex
%%% TeX-master: "main"
%%% End:         

}{
  %%%%%%%%%%%%%%%%%%%%%%%%%%%%%%%%%%%%%%%%%%%%%%%%%%%%%%%%%%%%%%%%%%%%%%%%%%%%%%%
\section{Proof-nets}
\label{s:proofnets}
%%%%%%%%%%%%%%%%%%%%%%%%%%%%%%%%%%%%%%%%%%%%%%%%%%%%%%%%%%%%%%%%%%%%%%%%%%%%%%%

%\oedu{\delia{This section introduces polarized proof-nets, a graphical formalism
%defined in~\cite{Laurent03} to interpret polarized formulae. }}

% \eduso{Laurent~\cite{Laurent02} introduced Polarized Linear Logic (LLP), a
% proof system on polarized formulae which \cdelia{relaxes}{extends} the use of structural
% rules, \delia{usually defined on ? formulae},  on negative formulae. It is accompanied by a corresponding
% notion of Polarized Proof-Nets (PPN). There are several reasons why
% LLP/PPN is convenient for encoding $\lmu$, the salient being
%   that  it allows a translation from classical logic into LLP
% that translates $A\rightarrow B$ as $\formbang{A} \multimap B$, thus
% allowing the translation to be a straightforward extension of that of
% the $\lambda$-calculus. In fact, we use a fragment of the
% multiplicative exponential PPNs of~\cite{Laurent02} presented
% in~\cite{Laurent03}.}{}

Laurent~\cite{Laurent02} introduced Polarized Linear Logic (LLP), a proof
system based on polarities on linear logic formulae: 
\begin{equation}
\hfill
\begin{array}{llccll}
\textbf{Negative formulae} & N & \Coloneq & \formbase \mid \formpar{N}{N} \mid \formder{P} \\
\textbf{Positive formulae} & P & \Coloneq & \formneg{\formbase} \mid \formten{P}{P} \mid \formbang{N} \\
\end{array}
\label{eq:LLP_polarities}
\hfill
\end{equation}
where $\formbase$ is assumed to be any atomic formula. LLP extends the use of
structural rules, usually defined only on ? formulae, to negative formulae. It
is equipped with a corresponding notion of Polarized Proof-Nets (PPN) which
allows for a simpler correctness criterion.

One feature of particular interest is that it is possible to translate
classical logic into LLP by interpreting $A\rightarrow B$ as $\formbang{A}
\multimap B$, obtaining a translation which is a straightforward extension of
that from intuitionistic logic to LLP, thus capturing the translation from
$\lambda$-calculus to LLP. In fact, this translation extends the one from
intuitionistic logic to standard, \ie non-polarized, linear logic. Indeed,
following Girard's translation of classical formulae, one may interpret the
simple types from Sec.~\ref{s:types} as LLP types:  \[
\hfill
\begin{array}{rcl}
\toform{\typebase}         & \eqdef & \formbase \\
\toform{(\typefunc{A}{B})} & \eqdef & \formbang{(\toform{A})} \multimap \toform{B} \equiv  \formpar{\formder{(\formneg{\toform{A}})}}{\toform{B}}
\end{array}
\hfill
\] The image of this translation is a strict subset of the set of polarized
formulae, namely the following \deft{output formulae} (presented together with
their duals \deft{anti-output formulae}):  \[
\hfill
\begin{array}{lccll}
\textbf{Output formulae}      & O & \Coloneq & \formbase \mid \formpar{\formder{Q}}{O} \\  
\textbf{Anti-output formulae} & Q & \Coloneq & \formneg{\formbase} \mid \formten{\formbang{O}}{Q}
\end{array}
\hfill
\] As described in the upcoming Sec.~\ref{s:translation},
following~\cite{Laurent02},  we too will translate typing derivations of
judgements of the form $\pi \dem\Gam \vdash o : \type \mid \Del$ into PPNs.
The translations of $\type$ and (formulae in) $\Del$ turn out to be output
formulae whereas, the translations of formulae in $\Gamma$ will be of the form
$\formder{Q}$, so that it seems reasonable  to dub them \deft{input formulae},
given that they correspond to the $\lambda$-variables. Since output formulae
are a subset of the negative formulae and anti-output formale are a subset of
the positive formulae one arrives at the following formulae
categories~\cite{Laurent03}: \[
\hfill
\begin{array}{lccll}
\textbf{Formulae}             & F & \Coloneq & N \mid P \\
\textbf{Negative formulae}    & N & \Coloneq & O \mid \formder{Q} \\
\textbf{Positive formulae}    & P & \Coloneq & Q \mid \formbang{O} \\
\textbf{Output formulae}      & O & \Coloneq & \formbase \mid \formpar{\formder{Q}}{O} \\  
\textbf{Anti-output formulae} & Q & \Coloneq & \formneg{\formbase} \mid \formten{\formbang{O}}{Q}
\end{array}
\hfill
\] Negation is involutive $({\formbase}^{\bot \bot} = \formbase)$ with
$\formneg{(\formpar{\formder{Q}}{O})} =
\formten{\formbang{\formneg{Q}}}{\formneg{O}}$ and
$\formneg{(\formder{Q})} = \formbang{\formneg{Q}}$.

\begin{definition}
\label{def:proof_structure}
A \deft{proof-structure} is a finite acyclic oriented graph built over the
alphabet of nodes represented below (where the orientation is the top-bottom
one): \[
\hfill
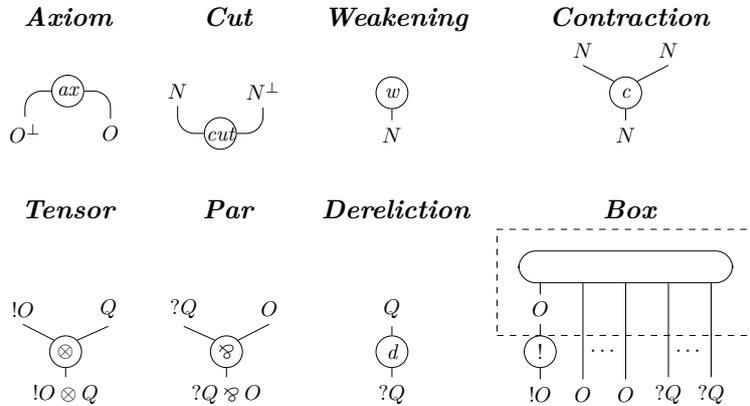
\begin{array}{cccc}
\textbf{Axiom} & \textbf{Cut} & \textbf{Weakening} & \textbf{Contraction} \\ 
\scalebox{0.8}{\begin{tikzpicture}[node distance=\PNdist]
\PNnode{A1}{\textit{ax}};
\node (N1) [below of=A1, xshift=-\PNdist] {$\formneg{O}$};
\node (N2) [below of=A1, xshift= \PNdist] {$O$};

\draw[PNarrow] (A1.west) -| node{} (N1.north)
  	  	       (A1.east) -| node{} (N2.north);
\end{tikzpicture}
}
&
\scalebox{0.8}{\begin{tikzpicture}[node distance=\PNdist]
\PNnode{C1}{\textit{cut}};
\node (N1) [above of=C1, xshift=-\PNdist] {$N$};
\node (N2) [above of=C1, xshift= \PNdist] {$\formneg{N}$};

\draw[PNarrow] (N1.south) |- node{} (C1.west)
  	  	       (N2.south) |- node{} (C1.east);
\end{tikzpicture}
}
&
\scalebox{0.8}{\begin{tikzpicture}[node distance=\PNdist]
\PNnode{W1}{\textit{w}};
\node (N1) [below of=W1] {$N$};

\draw[PNarrow] (W1.south) -- node{} (N1.north);
\end{tikzpicture}
}
&
\scalebox{0.8}{\begin{tikzpicture}[node distance=\PNdist]
\PNnode{C1}{\textit{c}};
\node (N1) [above of=C1, xshift=-\PNdist] {$N$};
\node (N2) [above of=C1, xshift= \PNdist] {$N$};
\node (N3) [below of=C1] {$N$};

\draw[PNarrow] (N1.south) -- node{} (C1.north west)
  	  	       (N2.south) -- node{} (C1.north east)
  	  	       (C1.south) -- node{} (N3.north);
\end{tikzpicture}
}
\\
\\
\textbf{Tensor} & \textbf{Par} & \textbf{Dereliction} & \textbf{Box} \\ 
\scalebox{0.8}{\input{proofnets/tensor.tex}}
&
\scalebox{0.8}{\input{proofnets/par.tex}}
&
\scalebox{0.8}{\begin{tikzpicture}[node distance=\PNdist]
\PNnode{D1}{\textit{d}};
\node (N1) [above of=D1] {$Q$};
\node (N2) [below of=D1] {$\formder{Q}$};

\draw[PNarrow] (N1.south) -- node{} (D1.north)
               (D1.south) -- node{} (N2.north);
\end{tikzpicture}
}
&
\scalebox{0.8}{\input{proofnets/bang.tex}}
\end{array}
\hfill
\]
\end{definition}
Remark that each wire is labelled with a formula. In particular, conclusions in
boxes have only one bang formula $\formbang{O}$, all others being negative
formulae. When using an uppercase greek letter to label wire, \eg $\Gamma$, we
mean that there are actually $|\Gamma|$ wires, each one labelled with a formula
in $\Gamma$. This abbreviated notation will be particularly useful to define
proof-net reduction and term translations.

A \deft{polarized proof-net} (PPN)~\footnote{The
notion of negative and positive formulae of~\cite{Laurent03}, that we also use
here, is slightly more restrictive than that of the
grammars~(\ref{eq:LLP_polarities}), which follows~\cite{Laurent02}.
%\odelia{Accordingly, the set of PPN of~\cite{Laurent02} are more general. For example,
%the tensor node in Def.~\ref{def:proof_structure} requires the ``left''
%argument to be of the form $!O$ rather than an arbitrary positive formula, the
%latter being allowed in~\cite{Laurent02}. }
This paper follows the presentation
in~\cite{Laurent03}.} is a proof-structure satisfying a simple correctness
criterion described in~\cite{Laurent03}. We refer the reader to op.cit. and
simply mention that since our proof-structures are obtained from translating
typed terms, this criterion is always met. Structural equivalence for
PPNs, is based on the five identities in
Fig.~\ref{fig:structural}. The first equation specifies associativity of
contraction nodes, the second one axiomatizes permeability of contraction
w.r.t. boxes, the third one specifies neutrality of weakening w.r.t. the
contraction operation, the fourth one pushes final weakening nodes to the top
level, and the fifth one removes final weakening nodes. Then, the
\deft{structural equivalence} is defined as the closure of the above
identities, where the first three identities are closed by any context, and the
last two are only closed by contexts not binding the weakening wire (hence our
referring to them as \emph{final} weakening nodes). In the sequel,
\deft{polarized proof-net equality}, written $\equiv$, is always taken modulo
structural equivalence, as done in~\cite{Laurent03}.

\begin{figure}
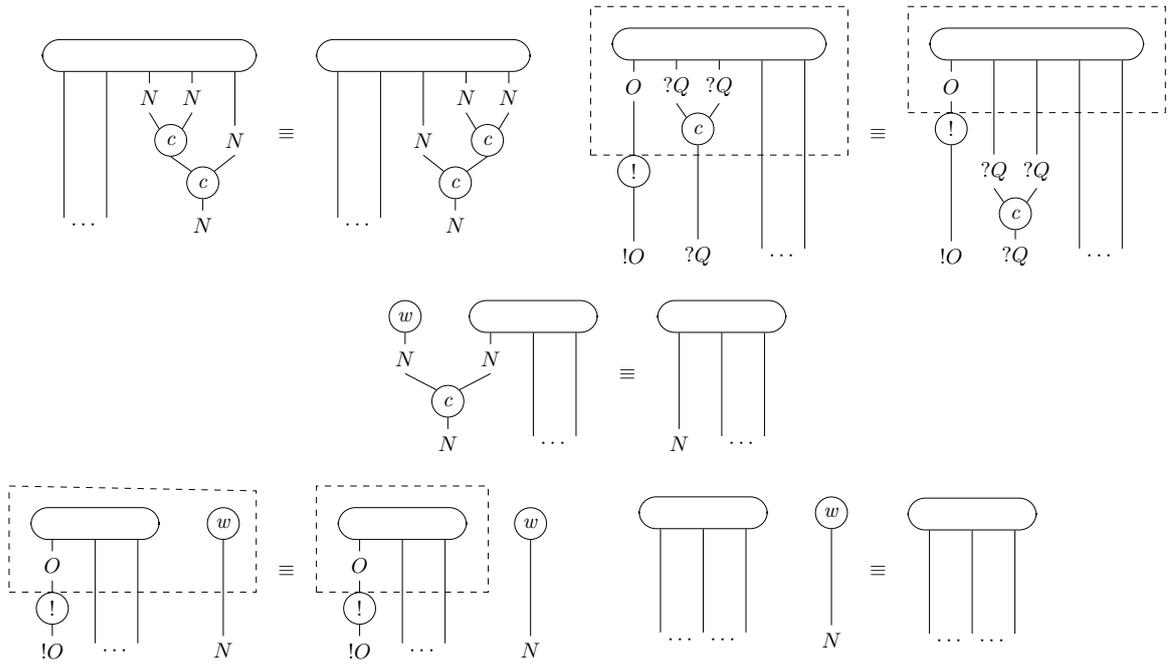
%
\begin{center}
\scalebox{0.8}{$
\kern-2em
\begin{array}{rcl@{\qquad}rcl}
\input{proofnets/Aleft} & \equiv & \input{proofnets/Aright} &
\input{proofnets/Bleft} & \equiv & \input{proofnets/Bright} \\
\\
\multicolumn{6}{c}{
\begin{array}{rcl}
\input{proofnets/Cleft} & \equiv & \input{proofnets/Cright}
\end{array}} \\
\\
\input{proofnets/Dleft} & \equiv & \input{proofnets/Dright} &
\input{proofnets/Eleft} & \equiv & \input{proofnets/Eright}
\end{array}
$}
\end{center}
\caption{Identities for structural equivalence of polarized proof-nets}%
\label{fig:structural}%
\end{figure}

The reduction relation for PPNs, denote by $\Rew{\redpn}$, is given by the set
of {\em cut elimination rules} appearing in
Figures~\ref{f:cut-elim-MELL-mult},~\ref{f:cut-elim-MELL-box}
and~\ref{f:cut-elim-MELL-tensor}. While the first two figures contain the
standard cut elimination rules for MELL~\cite{Girard87}, split into
multiplicative and exponential rules respectively, the third one contains new
rules specific to PPNs~\cite{Laurent03} which are also exponential. Rules in
Figure.~\ref{f:cut-elim-MELL-tensor} deal with cuts between
weakening/contraction/box and $\nomformten$-trees. A \deft{$\nomformten$-tree},
used to interpret a stack in the term language, is just a box or an axiom, or
it is formed by a $\nomformten$-node whose premises are a box and a
$\nomformten$-tree. Remark that $\weaktencut$ (resp. $\conttencut$ and
$\boxtencut$) is similar to $\weakboxcut$ (resp. $\contboxcut$ and
$\boxboxcut$) if one considers $\nomformten$-trees as boxes. The reduction
relation $\Rew{\redpn}$ is confluent and strongly normalising (Cor. 14 and Cor.
15 in~\cite{Laurent03}). Moreover, also $\{\axcut,\multcut\}$ is confluent and
strongly normalising, its normal forms are called \emphdef{multiplicative
normal-forms}, a technical tool used later on to define the translation of
typed $\LMfull$-objects to PPNs.

\begin{figure}[h]
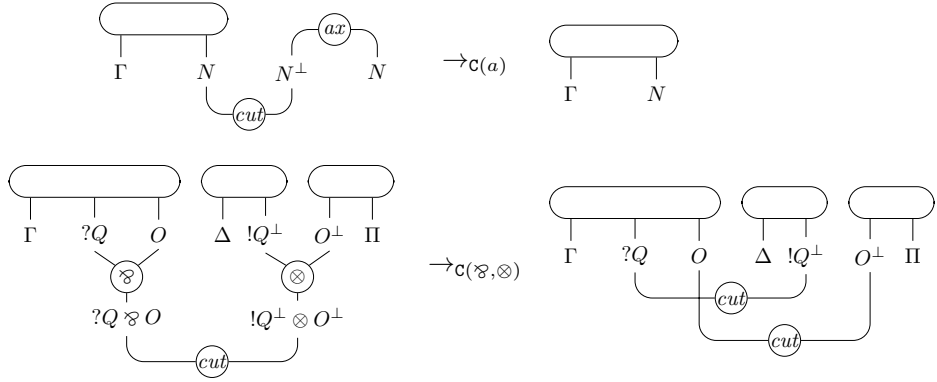

\[
\hfill
\begin{array}{rcl}
\scalebox{.8}{\input{figures/axcutleft}}   & \Rew{\axcut}   & \scalebox{.8}{\input{figures/axcutright}}
\\
\\
\scalebox{.8}{\input{figures/multcutleft}} & \Rew{\multcut} & \scalebox{.8}{\input{figures/multcutright}}
\end{array}
\hfill
\]
\caption{Cut elimination for PPN: multiplicative cuts (1/3)}
\label{f:cut-elim-MELL-mult}
\end{figure}

\begin{figure}
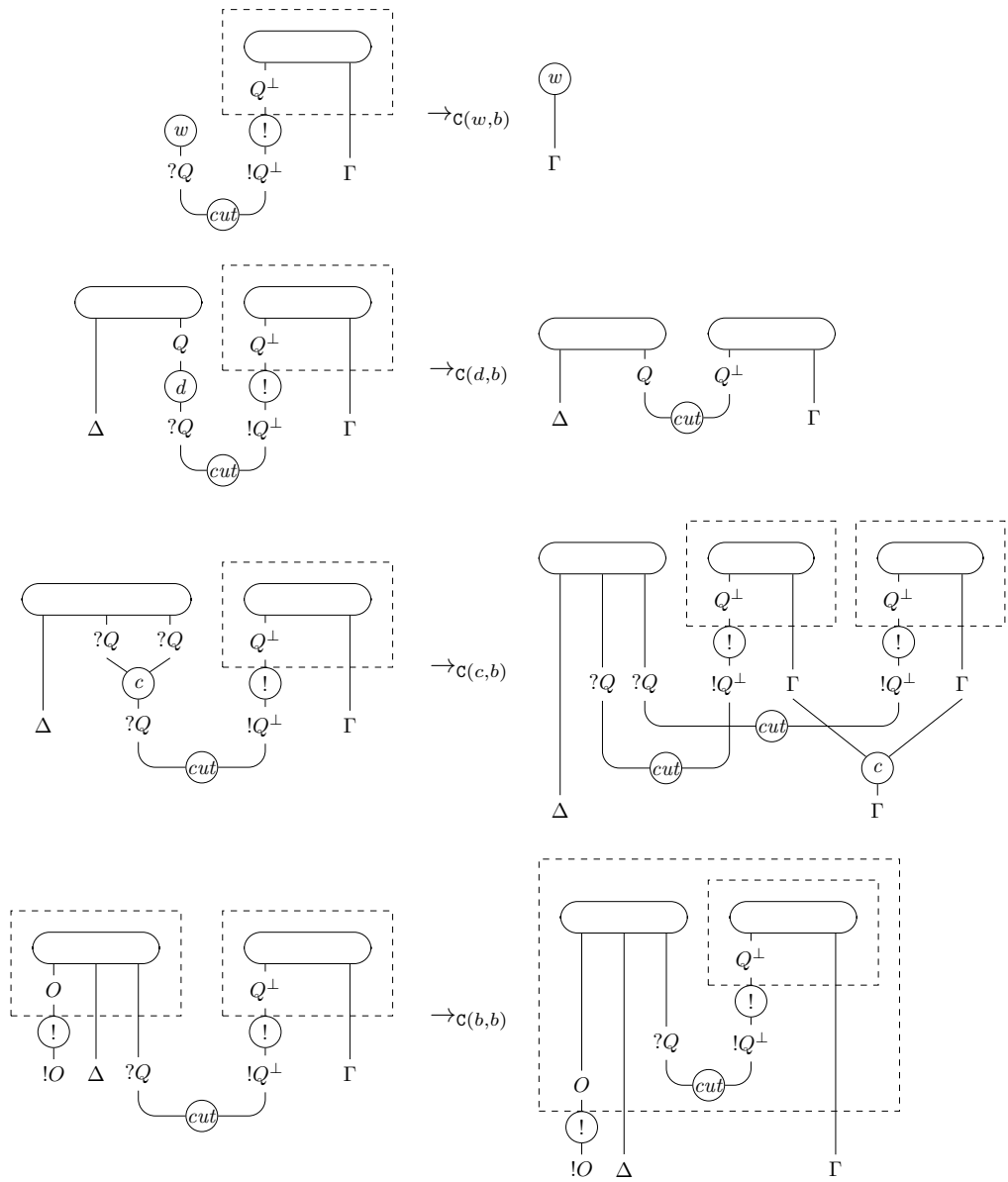

\[
\hfill
\begin{array}{rcl}
\scalebox{.8}{\input{figures/weakboxcutleft}} & \Rew{\weakboxcut} & \scalebox{.8}{\input{figures/weakboxcutright}}
\\
\\
\scalebox{.8}{\input{figures/derboxcutleft}}  & \Rew{\derboxcut}  & \scalebox{.8}{\input{figures/derboxcutright}}
\\
\\
\scalebox{.8}{\input{figures/conboxcutleft}}  & \Rew{\contboxcut} & \scalebox{.8}{\input{figures/conboxcutright}}
\\
\\
\scalebox{.8}{\input{figures/boxboxcutleft}}  & \Rew{\boxboxcut}  & \scalebox{.8}{\input{figures/boxboxcutright}}
\end{array}
\hfill
\]
\caption{Cut elimination for PPN: exponential cuts (2/3)}
\label{f:cut-elim-MELL-box}
\end{figure}

\begin{figure}
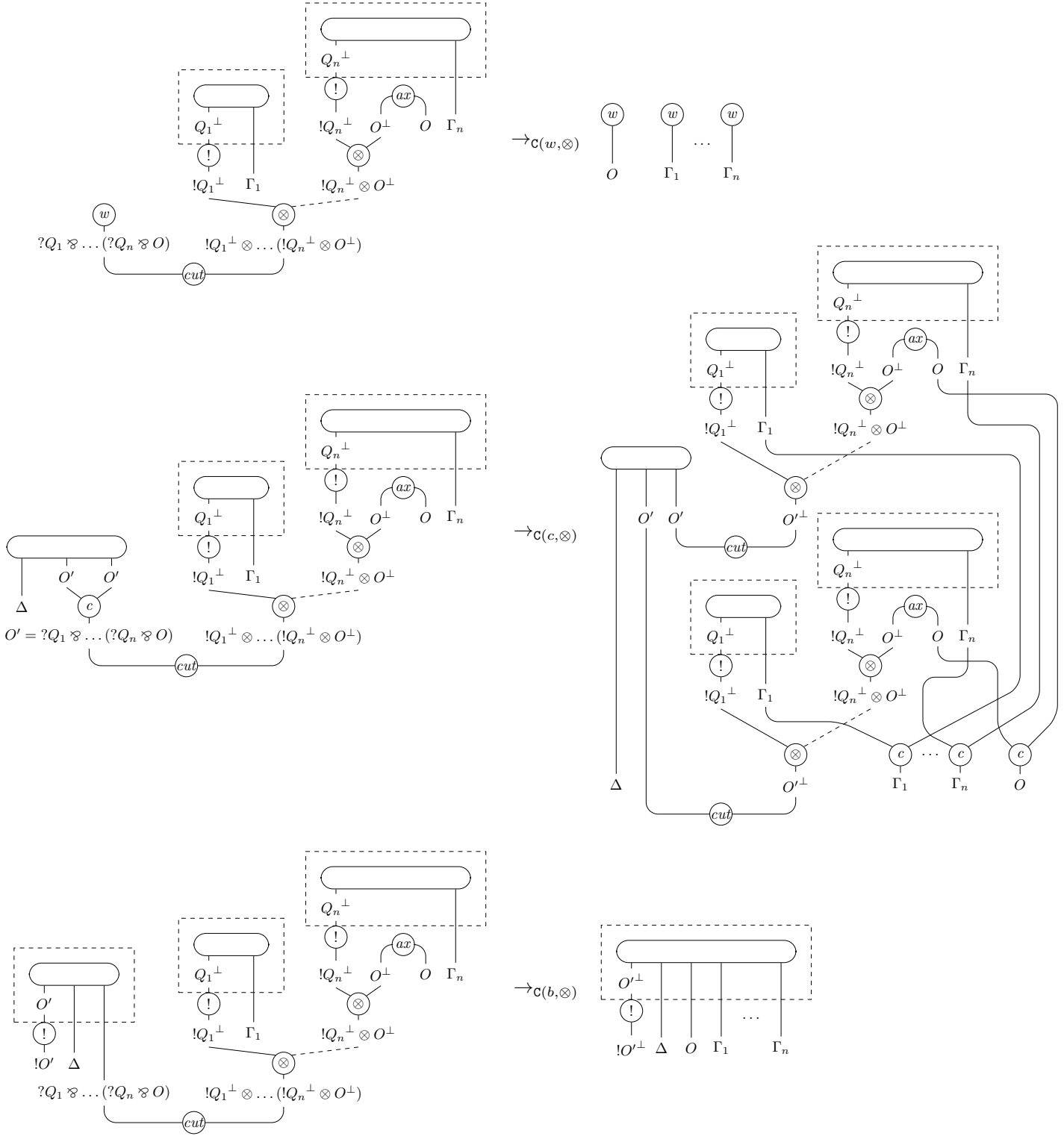

\[
\hfill
\kern-8em
\begin{array}{rcl}
\scalebox{.75}{\input{figures/weaktensorcutleft}} & \Rew{\weaktencut} & \scalebox{.75}{\input{figures/weaktensorcutright}}
%\\
\\[-2em]
\scalebox{.75}{\input{figures/contensorcutleft}}  & \Rew{\conttencut} & \scalebox{.75}{\input{figures/contensorcutright}}
\\
\\
\scalebox{.75}{\input{figures/boxtensorcutleft}}  & \Rew{\boxtencut}  & \scalebox{.75}{\input{figures/boxtensorcutright}}
\end{array}
\hfill
\]
\caption{Cut elimination for MELL: exponential cuts (3/3)}
\label{f:cut-elim-MELL-tensor}
\end{figure}

%%%Local Variables:
%%% mode: latex
%%% TeX-master: "main"
%%% End:         

}
\ispaper{
  \input{translation-paper}
}{
  %%%%%%%%%%%%%%%%%%%%%%%%%%%%%%%%%%%%%%%%%%%%%%%%%%%%%%%%%%%%%%%%%%%%%%%%%%%%%%%
\section{Translation to proof-nets}
\label{s:translation}
%%%%%%%%%%%%%%%%%%%%%%%%%%%%%%%%%%%%%%%%%%%%%%%%%%%%%%%%%%%%%%%%%%%%%%%%%%%%%%%

This section  presents a translation of $\LMfull$-objects to
PPNs, which is an extension of that introduced in~\cite{Laurent03} for
$\lmu$-terms.

We start by translating the set of types defined in Sec.~\ref{s:types} into
polarized formulae, and more precisely, into output formulae. The
\deft{translation of term types to polarized output formulae},
written
$\toform{\_}$,  follows Girard's translation of classical formulae, and
is inductively defined as follows: $$
\begin{array}{rcl}
\toform{\typebase}         & \eqdef & \formbase \\
\toform{(\typefunc{A}{B})} & \eqdef & \formpar{\formder{(\formneg{\toform{A}})}}{\toform{B}}
\end{array} $$
%For each stack type $S$ and each term type $B$, we write $\toform{S}_{B}$
%for the
%formula $\toform{(\typefunc{S}{B})}$.
The \deft{translation of stack types to polarized output formulae}  is
parameterized over a type $B$ (the type of the empty stack), is written
$\toform{\_}_B$, and defined as follows:
$$
\begin{array}{rcl}
\toform{\typeemst}_B   & \eqdef & \toform{B} \\
\toform{(\push{A}{S})} & \eqdef & \formpar{\formder{(\formneg{\toform{A}})}}{\toform{S}_B}
\end{array} $$

The translation $\toform{\_}$ is also extended to assignments:
If $\Gam = x_1:A_1, \ldots, x_n:A_n$ and $\Del = \al_1:B_1, \ldots, \al_m:B_m$,
then $\toform{\Gam} = \toform{A_1}, \ldots, \toform{A_n}$ and $\toform{\Del} =
\toform{B_1}, \ldots, \toform{B_m}$. We use the
abbreviated notation $\formder{(\formneg{\toform{\Gam}})}$ for
$\formder{(\formneg{\toform{A_1}})}, \ldots, \formder{(\formneg{\toform{A_n}})}$. 

The \deft{translation of (typed) $\LMfull$-objects to polarized proof-nets} is
defined by induction on typing derivations. Given $\derivable{\pi}{\Gam \vdash
o: T \mid \Del}{}$ we generally write $\topn{\pi}$ to denote its translation.
Sometimes, for convenience, we also use the notation $\topn{(\Gam \vdash o: T \mid
\Del)}$. In all the cases, every conclusion in
$\formder{\formneg{\toform{\Gam}}}$ (resp. in $\toform{\Del}$) is labelled with
the name of the corresponding variable (resp. name). The translation of terms
and stacks has a \deft{distinguished} conclusion, while the translation of
commands does not. The distinguished conclusion of a term is an output formula
and that of a stack is an anti-output formula.

\begin{itemize}
\item ({\bf Terms}) A typing derivation $\pi \dem \Gam \vdash t:A
\mid \Del$ is translated to a proof-net $\topn{\pi}$ with conclusions
$\formder{\formneg{\toform{\Gam}}}, \toform{A}, \toform{\Del}$, where
$\toform{A}$ (an output formula) is the distinguished one. 

\item ({\bf Commands}) A typing derivation $\pi \dem \Gam \vdash c
\mid \Del$ is translated to a proof-net $\topn{\pi}$ with conclusions
$\formder{\formneg{\toform{\Gam}}}, \toform{\Del}$.

\item ({\bf Stacks}) A typing derivation $\pi \dem
  \Gam \vdash s:S \mid \Del$ is translated, for every type $B$, into 
  a polarized proof-net $\topn{\pi}_B$ --which is a $\otimes$-tree--
  with conclusions $\formder{\formneg{\toform{\Gam}}},
  \formneg{\toform{(\typefunc{S}{B})}}, \toform{B}, \toform{\Del}$, where
  $\formneg{\toform{(\typefunc{S}{B})}}$  (anti-output formula) is the
  distinguished conclusion. The parameter $B$ in the translation of
  stacks corresponds to the codomain of the functional type that
  consumes the stack (\cf typing rule $\trepl$).
\end{itemize}

The translation from typed $\LMfull$-terms to PPNs
extends the one in~\cite{Laurent03} to the new  contructors of
$\LMfull$. It is
    defined over typing derivations, each defining clause below
    depicting a possible typing rule from Fig.~\ref{f:typing-rules} with
    which the typing derivation may end. In the following pictures we
  use \tikz{\draw[PNarout] (0,0) -- node{} (0.5,0);} for the
  distinguished conclusion and \tikz{\draw[PNarold] (0,0) -- node{}
    (0.5,0);} for the old distinguished one in the inductive step. We
  also use $(\alpha)$ to denote the old name of the distinguished
  conclusion and $(x)$ to denote the old name of a wire representing a
  free variable.  
\begin{remark}
  In all the type derivations based on two different
  premises, the translation adds contractions nodes for sharing common
  variables and names, but we omit these contraction nodes in the
  figures to make them more readable. Moreover, we only draw the wires
  that actively participate in the construction, thus omitting those
  conclusions of the obtained proof-nets which are not relevant for
  the translation itself.
\end{remark}

\begin{itemize}
  \item $\topn{(x:A \vdash x:A \mid \emptyset)}$
  $$\scalebox{.75}{\input{proofnets/tax}}$$

  \item $\topn{(\Gam \cup \Gam' \vdash tu:B \mid \Del \cup \Del')}$
  $$\scalebox{.75}{\input{proofnets/app}}$$
  We add contraction nodes for all the conclusions corresponding to common
  variables/names of $t$ and $u$.

  \item $\topn{(\Gam \vdash \l x. t: A \tarrow B \mid \Del)}$ \\
\setlength{\itemwidth}{.5\hsize-\leftmargin-\itemindent-4\tabcolsep}
  \begin{tabular}{clcl}
\labelitemii & \parbox[t]{\itemwidth}{$x \in \fv{t}$} &
\labelitemii & \parbox[t]{\itemwidth}{$x \notin \fv{t}$, thus $x \notin \dom{\Gam}$} \\[1em]
\multicolumn{2}{c}{\scalebox{.8}{\input{proofnets/absin}}} &
\multicolumn{2}{c}{\scalebox{.8}{\input{proofnets/absnotin}}}
  \end{tabular}
  
  \item $\toform{(\Gam \vdash \termcont{\alpha}{c}: A \mid \Del)}$ \\
  \begin{tabular}{clcl}
\labelitemii & \parbox[t]{\itemwidth}{$\alpha \in \fn{c}$} &
\labelitemii & \parbox[t]{\itemwidth}{$\alpha \notin \fn{c}$, thus $\alpha \notin \dom{\Del}$} \\[1em]
\multicolumn{2}{c}{\scalebox{.8}{\begin{tikzpicture}[node distance=\PNdist]
\node[PNsub] (TC) [minimum width=\PNdist*4] {$\topn{(\Gam \vdash c \mid \Del, \alpha:A)}$};
\node (N1) [below of=TC, yshift=-\PNdist] {$\toform{A}$};

\draw[PNarout] (TC.south) -- node[left]{$(\alpha)$} (N1.north);
\end{tikzpicture}
}} &
\multicolumn{2}{c}{\scalebox{.8}{\begin{tikzpicture}[node distance=\PNdist]
\PNnode{W1}{\textit{w}};
\node[PNsub] (TC) [right of=W1, xshift=\PNdist*2, minimum width=\PNdist*4] {$\topn{(\Gam \vdash c \mid \Del)}$};
\node (N1) [below of=W1, yshift=-\PNdist] {$\toform{A}$};

\draw[PNarout] (W1.south) -- node{} (N1.north);
\end{tikzpicture}
}}
  \end{tabular}
  
  \item $\topn{(\Gam \cup \Gam' \vdash t\exsubs{x}{u}:A \mid \Del \cup \Del')}$ \\
  \begin{tabular}{clcl}
\labelitemii & \parbox[t]{\itemwidth}{$x \in \fv{t}$} &
\labelitemii & \parbox[t]{\itemwidth}{$x \notin \fv{t}$, thus $x \notin \dom{\Gam}$} \\[1em]
\multicolumn{2}{c}{\scalebox{.8}{\input{proofnets/subin}}} &
\multicolumn{2}{c}{\scalebox{.8}{\input{proofnets/subnotin}}}
  \end{tabular} \\
  We add contraction nodes for all the conclusions corresponding to common
  variables/names of $t$ and $u$.
  
  \item $\topn{(\Gam \vdash \termname{\alpha}{t} \mid \Del, \alpha: A)}$ \\
  \begin{tabular}{clcl}
\labelitemii & \parbox[t]{\itemwidth}{$\alpha \in \fn{t}$} &
\labelitemii & \parbox[t]{\itemwidth}{$\alpha \notin \fn{t}$, thus $\alpha \notin \dom{\Del}$} \\[1em]
\multicolumn{2}{c}{\scalebox{.8}{\input{proofnets/namein}}} &
\multicolumn{2}{c}{\scalebox{.8}{\begin{tikzpicture}[node distance=\PNdist]
\node[PNsub] (TT) [minimum width=\PNdist*4] {$\topn{(\Gam \vdash t:A \mid \Del)}$};
\node (N1) [below of=TT, yshift=-\PNdist] {$\toform{A}$};
\node (DM) [below of=N1, yshift=-\PNdist*1.5] {\vphantom{$\toform{A}$}};

\draw[PNarold] (TT.south) -- node[left]{$\alpha$} (N1.north);
\end{tikzpicture}
}}
  \end{tabular}
  
  \item $\topn{(\Gam \cup \Gam' \vdash c\exrepl{\alpha'}{\alpha}{s} \mid  \Del \cup \Del', \alpha': B)}$ \\
  We illustrate below the cases where $\alpha' \notin \fn{s}$ \\
  \begin{tabular}{clcl}
\labelitemii & \parbox[t]{\itemwidth}{$\alpha, \alpha' \in \fn{c}$} &
\labelitemii & \parbox[t]{\itemwidth}{$\alpha \in \fn{c}$ and $\alpha' \notin \fn{c}$, thus $\alpha' \notin \dom{\Del}$} \\[1.5em]
\multicolumn{2}{c}{\scalebox{.8}{\input{proofnets/treplinin}}}\kern-1em &
\multicolumn{2}{c}{\scalebox{.8}{\input{proofnets/treplinnotin}}}
\\
\labelitemii & \parbox[t]{\itemwidth}{$\alpha \notin \fn{c}$ and $\alpha' \in \fn{c}$, thus $\alpha \notin \dom{\Del}$} &
\labelitemii & \parbox[t]{\itemwidth}{$\alpha, \alpha' \notin \fn{c}$, thus $\alpha, \alpha' \notin \dom{\Del}$} \\[1.5em]
\multicolumn{2}{c}{\scalebox{.8}{\input{proofnets/treplnotinin}}} &
\multicolumn{2}{c}{\scalebox{.8}{\input{proofnets/treplnotinnotin}}}
  \end{tabular} \\
  We add contraction nodes for all the conclusions corresponding to common
  variables/names of $c$ and $s$. For the cases where $\alpha'
  \in \fn{s}$ we just add an extra contraction node with the conclusion
  $\alpha'$.
  
  \item $\topn{(\emptyset \vdash \termemst: \temst \mid \emptyset)}_{B}$
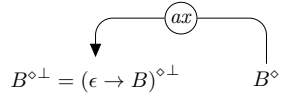
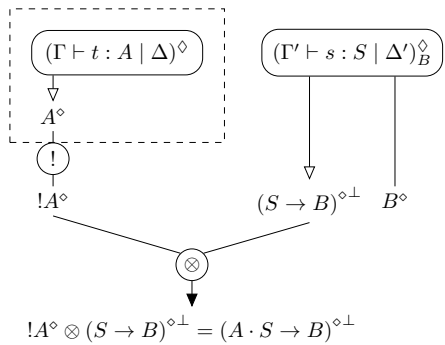
  $$\scalebox{.8}{\begin{tikzpicture}
\PNnode{A1}{\textit{ax}};
\node (N1) [below of=A1, xshift=\PNdist*2] {$\toform{B}$};
\node (N2) [below of=A1, xshift=-\PNdist*2] {$\formneg{\toform{B}} = \formneg{\toform{(\typefunc{\temst}{B})}}$};

\draw[PNarrow] (A1.east) -| node{} (N1.north);
\draw[PNarout] (A1.west) -| node{} (N2.north);
\end{tikzpicture}
}$$
  
  \item $\topn{(\Gam \cup \Gam' \vdash t \cdot s: \push{A}{S} \mid \Del \cup \Del')_{B}}$
  $$\scalebox{.8}{\input{proofnets/stackd}}$$
  We add contraction nodes for all the conclusions corresponding to
  common variables/names of $t$ and $s$. 
\end{itemize}

%%% Local Variables:
%%% mode: latex
%%% TeX-master: "main"
%%% End:

The translation $\topn{\_}$ does not preserve reduction, typically for rule
$\rRfull$. Here is an example: \[\hfill
\termrepl{\alpha'}{\alpha}{s}{\termrepl{\alpha}{\gamma}{\termemst}{c}}
\Rew{\rRfull}
\replace{\alpha'}{\alpha}{s}{\termrepl{\alpha}{\gamma}{\termemst}{c}} =
\termrepl{\alpha'}{\beta}{\termemst}{\termrepl{\beta}{\gamma}{s}{\replace{\alpha'}{\alpha}{s}{c}}}
\hfill \]
The translation of the outermost explicit renaming on the right-hand side
introduces a multiplicative cut on the translated PPN that cannot be reached
from any sub-PPN resulting from the translation of the left-hand side. Thus,
we extend the translation $\topn{\_}$ to a new one, written $\topnd{\_}$: $\topnd{\pi}$ is defined as the multiplicative normal-form of
$\topn{\pi}$. The following property holds.

\begin{theorem}
\label{t:simulation-into-ppn}
Let $o, p$ be typed $\LMfull$-objects. If $o \Rew{\LMfull} p$, and $\pi_o,
\pi_p$ are two related corresponding type derivations for $o$ and $p$ resp.,
then $\topnd{\pi_o} \Rewn{\redpn} \topnd{\pi_{p}}$.
\end{theorem}

\begin{proof}
By induction on $\Rew{\LMfull}$ by adapting Lem. 18 and 19 in~\cite{Laurent02}.
%\odelia{In particular, since stacks are translated to $\nomformten$-trees (\cf
%the two last cases of the translation of typed $\LMfull$-objects to PPNs), the
%translation of rule $\rRfull$ uses the \andres{(exponential)} rules in
%Figure~\ref{f:cut-elim-MELL-tensor}. }
The only 
$\LMfull$-reduction steps on terms involving exponential rules
on their translations to PPNs
(Fig.~\ref{f:cut-elim-MELL-box} and~\ref{f:cut-elim-MELL-tensor}) are
$\Rew{\rS}$ and the non-linear instances of $\Rew{\rRfull}$ (called $\Rew{\rR}$
in Sec.~\ref{sec:refining_replacement}), while $\Rew{\rB}$, $\Rew{\rC}$ and
$\Rew{\rW}$ are translated to multiplicative cuts
(Fig.~\ref{f:cut-elim-MELL-mult}), and both  $\Rew{\rM}$ and
$\Rew{\linealrule}$ give the identity.
\end{proof}

%%%Local Variables:
%%% mode: latex
%%% TeX-master: "main"
%%% End:

}
%\input{decomposing-replacement}  % no se usa mas
%%%%%%%%%%%%%%%%%%%%%%%%%%%%%%%%%%%%%%%%%%%%%%%%%%%%%%%%%%%%%%%%%%%%%%%%%%%%%%%
\section{\texorpdfstring{Structural Equivalence for $\LMfull$}{Structural Equivalence for LM}}
\label{s:Sigma-equivalence}
%%%%%%%%%%%%%%%%%%%%%%%%%%%%%%%%%%%%%%%%%%%%%%%%%%%%%%%%%%%%%%%%%%%%%%%%%%%%%%%

\ignore{agregar el ejemplo de no strong bisimulation $u :=
(\termcont{\al}{\termname{\al}{\termcont{\beta}{\termname{\beta}{x}}}}) y
\simeq_{\lsigma_7} (\termcont{\al}{\termname{\al}{x}})y =: v$.}

We introduce our notion of structural equivalence for $\LMfull$,
written $\simeq$, breaking down the presentation into the three key
tools on which we have based our development: canonical forms, linear
contexts and renaming replacements. Finally, we introduce $\simeq$
itself.

\ignore{This section introduces a structural equivalence $\simeq$ for
$\LMfull$. As explained in the introduction, our equivalence relation $\simeq$
on $\LMfull$-terms is inspired from that of O. Laurent $\eqlaurent$ on
$\lmu$-terms, in particular $\simeq$-terms have the same PN representation
(Lem.~\ref{l:ida}). However, while $\eqlaurent$ $\lmu$-terms do not behave
\textit{exactly} the same, because they do not induce the same graph reduction,
our $\simeq$ relation turns out to be a strong bisimulation
(Thm.~\ref{t:bisimulation}).}

\partitle{Canonical Forms.}
As discussed in Sec.~\ref{sec:introduction}, the
initial intuition in defining a strong bisimulation for $\LMfull$ arises from
the intuitionistic case: Regnier's equivalence $\simeq_{\rsigma}$ is not a
strong bisimulation, but taking the $\rB$-normal form of the left and right
hand sides of these equations, results in a strong bisimulation
$\simeq_{\esigma}$ on $\l$-terms with explicit substitutions. In the classical
case, we similarly begin from Laurent's $\eqlaurent$ relation on $\lmu$-terms
and consider the canonical forms, realised by the $\BMCform$
(Def.~\ref{def:canonical}), resulting in the relation $\simeq_{\elsigma}$ on
$\LMfull$-terms (Fig.~\ref{fig:eqlaurent:preliminar}).
This equational theory would be the natural candidate for our strong
bisimulation, but unfortunately it is not the case as we explain below.
\ignore{
\begin{figure*}[h!]
\centering
\[\kern-3em
\begin{array}{rcll}
\termsubs{x}{u}{(\termabs{y}{t})}                         & \simeq_{\elsigma_1} & \termabs{y}{\termsubs{x}{u}{t}}     & \cdelia{y \notin \fv{u}}{\notatall{y}{u}} \\
\termsubs{x}{u}{(\termapp{t}{v})}                         & \simeq_{\elsigma_2} & \termapp{\termsubs{x}{u}{t}}{v}     & \cdelia{x \notin \fv{v}}{\notatall{x}{v}} \\
\termsubs{x}{v}{(\termcont{\alpha}{\termname{\beta}{u}})} & \simeq_{\elsigma_3} & \termcont{\alpha}{\termname{\beta}{\termsubs{x}{v}{u}}} & \notatall{\alpha}{v} \\
\termname{\alpha'}{(\termcont{\alpha''}{\termrepl{\alpha''}{\alpha}{u}{\termname{\beta'}{(\termcont{\beta''}{\termrepl{\beta''}{\beta}{v}{c}})}}})} & \simeq_{\elsigma_4} & \termname{\beta'}{(\termcont{\beta''}{\termrepl{\beta''}{\beta}{v}{\termname{\alpha'}{(\termcont{\alpha''}{\termrepl{\alpha''}{\alpha}{u}{c}})}}})} & \notatall{\alpha}{v}, \notatall{\beta}{w}, \delia{\beta'' \neq \al'}, \delia{\al'' \neq \beta'}\\
\termname{\alpha'}{(\termcont{\alpha''}{\termrepl{\alpha''}{\alpha}{v}{(\termname{\beta'}{\termabs{x}{\termcont{\beta}{c}}})}})} & \simeq_{\elsigma_5} & \termname{\beta'}{\termabs{x}{\termcont{\beta}{\termname{\alpha'}{(\termcont{\alpha''}{\termrepl{\alpha''}{\alpha}{v}{c}})}}}} & \notatall{x}{v}, \notatall{\beta}{v}, \delia{\beta'' \neq \al'}, \delia{\al'' \neq \beta'} \\
\termname{\alpha'}{\termabs{x}{\termcont{\alpha}{\termname{\beta'}{\termabs{y}{\termcont{\beta}{c}}}}}}                          & \simeq_{\elsigma_6} & \termname{\beta'}{\termabs{y}{\termcont{\beta}{\termname{\alpha'}{\termabs{x}{\termcont{\alpha}{c}}}}}}  \\
\termname{\alpha}{\termcont{\beta}{c}}  & \simeq_{\elsigma_7}    & c\iren{\beta}{\alpha} \\
\termcont{\alpha}{\termname{\alpha}{t}} & \simeq_{\elsigma_{8}} & t & \notatall{\alpha}{t} \\
\termsubs{x}{u}{\termsubs{y}{v}{t}}                       & \simeq_{\elsigma_9} & \termsubs{y}{v}{\termsubs{x}{u}{t}} & \cdelia{y \notin \fv{u}, x \notin \fv{v}}{\notatall{y}{u}, \notatall{x}{v}} \\
\end{array}
\]
\caption{A first reformulation of Laurent's $\eqlaurent$ on $\LMfull$-terms}
\label{fig:eqlaurent:preliminar}
\end{figure*}
}

\begin{figure*}[h!]
\centering
\[
\begin{array}{rcl}
\termsubs{x}{u}{(\termabs{y}{t})}                         & \simeq_{\elsigma_1} & \termabs{y}{\termsubs{x}{u}{t}}       \\
\termsubs{x}{u}{(\termapp{t}{v})}                         & \simeq_{\elsigma_2} & \termapp{\termsubs{x}{u}{t}}{v}       \\
\termsubs{x}{v}{(\termcont{\alpha}{\termname{\beta}{u}})} & \simeq_{\elsigma_3} & \termcont{\alpha}{\termname{\beta}{\termsubs{x}{v}{u}}}   \\
\termname{\alpha'}{(\termcont{\alpha''}{\termrepl{\alpha''}{\alpha}{u}{\termname{\beta'}{(\termcont{\beta''}{\termrepl{\beta''}{\beta}{v}{c}})}}})} & \simeq_{\elsigma_4} & \termname{\beta'}{(\termcont{\beta''}{\termrepl{\beta''}{\beta}{v}{\termname{\alpha'}{(\termcont{\alpha''}{\termrepl{\alpha''}{\alpha}{u}{c}})}}})}  \\
\termname{\alpha'}{(\termcont{\alpha''}{\termrepl{\alpha''}{\alpha}{v}{(\termname{\beta'}{\termabs{x}{\termcont{\beta}{c}}})}})} & \simeq_{\elsigma_5} & \termname{\beta'}{\termabs{x}{\termcont{\beta}{\termname{\alpha'}{(\termcont{\alpha''}{\termrepl{\alpha''}{\alpha}{v}{c}})}}}}   \\
\termname{\alpha'}{\termabs{x}{\termcont{\alpha}{\termname{\beta'}{\termabs{y}{\termcont{\beta}{c}}}}}}                          & \simeq_{\elsigma_6} & \termname{\beta'}{\termabs{y}{\termcont{\beta}{\termname{\alpha'}{\termabs{x}{\termcont{\alpha}{c}}}}}}  \\
\termname{\alpha}{\termcont{\beta}{c}}  & \simeq_{\elsigma_7}    & c\iren{\beta}{\alpha} \\
\termcont{\alpha}{\termname{\alpha}{t}} & \simeq_{\elsigma_{8}} & t   \\
\termsubs{x}{u}{\termsubs{y}{v}{t}}                       & \simeq_{\elsigma_9} & \termsubs{y}{v}{\termsubs{x}{u}{t}}   \\
\end{array}
\]
%\begin{tabular}{l}
\parbox{\textwidth}{
Conditions for the equations:  
$\elsigma_1: \notatall{y}{u}$,
$\elsigma_2: \notatall{x}{v}$,
$\elsigma_3: \notatall{\alpha}{v}$, 
$\elsigma_4: \notatall{\alpha}{v}, \notatall{\beta}{w}, \beta'' \neq \al', \al'' \neq \beta'$,
$\elsigma_5: \notatall{x}{v}, \notatall{\beta}{v}, \beta'' \neq \al', \al'' \neq \beta'$,
$\elsigma_8: \notatall{\alpha}{t}$,
$\elsigma_9: \notatall{y}{u}, \notatall{x}{v}$.
}
%\end{tabular}
\caption{A first reformulation of Laurent's $\eqlaurent$ on $\LMfull$-terms}
\label{fig:eqlaurent:preliminar}
\end{figure*}

\partitle{Linear Contexts.}
The first three equations in Fig.~\ref{fig:eqlaurent:preliminar} together with
equation ${\elsigma_9}$ can be generalized by noting that explicit substitution
commutes with {\it linear} contexts (\cf Def.~\ref{def:linear_contexts}), the
latter being the contexts that cannot be erased, nor duplicated, \ie in
proof-net parlance, they do not lay inside a box. The same situation arises
between linear contexts and explicit replacements (\cf equations
${\elsigma_4}\mbox{-}{\elsigma_5}$). Thus, linear contexts can be traversed by
any {\it independent} explicit operator (substitution/replacement). Thanks to
linear contexts, equations
${\elsigma_1}\mbox{-}{\elsigma_2}\mbox{-}{\elsigma_3}\mbox{-}{\elsigma_9}$ and
also ${\elsigma_4}\mbox{-}{\elsigma_5}$ from
Fig.~\ref{fig:eqlaurent:preliminar} can be subsumed by (and hence replaced
with) the commutation between linear contexts and explicit operators as
specified by the following equations, which will be part of our equivalence
$\simeq$. \[ \hfill
\begin{array}{rcl}
\termsubs{x}{u}{\ctxtapp{\ctxtLTT}{v}}               & \simeq_{\eqexsubs} & \ctxtapp{\ctxtLTT}{\termsubs{x}{u}{v}} \\
\termrepl{\alpha'}{\alpha}{s}{\ctxtapp{\ctxtLCC}{c}} & \simeq_{\eqexrepl} & \ctxtapp{\ctxtLCC}{\termrepl{\alpha'}{\alpha}{s}{c}}
\end{array} \hfill \]
The first equation is constrained by the condition $\notatall{x}{\ctxtLTT}$ and
$\freeFor{u}{\ctxtLTT}$ while the second one by $\notatall{\alpha}{\ctxtLCC}$
and $\freeFor{\{s,\alpha'\}}{\ctxtLCC}$, which essentially prevent any capture
of free variables/names. 

\partitle{Renaming Replacements.}
As mentioned in Sec.~\ref{sec:refining_replacement}\ispaper{}{ (following
Fig.~\ref{e:rho-breaks-strong-bisimulation:pre})}, we adapt
$\simeq_{\elsigma_7}$ by transforming implicit
(meta-level) renaming into explicit replacement.
The new equation becomes: \[
\hfill
\termname{\alpha}{\termcont{\beta}{c}} \simeq_{\eqrho} c\neren{\beta}{\alpha}
\hfill
\] and the situation of the diagram in Fig.~\ref{e:rho-breaks-strong-bisimulation:pre} is
modified as follows: \[
\hfill
\kern-1em
\scalebox{0.85}{
\begin{tikzcd}[ampersand replacement=\&]
\termrepl{\alpha'}{\alpha}{s}{(\termname{\alpha}{\termcont{\beta}{c}})}
  \arrow{d}[left]{\rR}
  \&[-30pt] \simeq_{\eqrho}
  \&[-30pt] \termrepl{\alpha'}{\alpha}{s}{c \neren{\beta}{\alpha}}
  \arrow[rightsquigarrow]{d}[left]{\rR} \\
\termname{\alpha'}{(\termcont{\beta}{\replace{\alpha'}{\alpha}{s}{c}})\tconc s}
  \arrow[]{d}[left]{\Can}
  \&[-30pt] 
  \&[-30pt] \BMC{\replace{\alpha'}{\alpha}{s}{c \neren{\beta}{\alpha}}}
  \arrow[equal]{d}[left]{def} \\
\termname{\alpha'}{\termcont{\beta'}{\BMC{\termrepl{\beta'}{\beta}{s}{\replace{\alpha'}{\alpha}{s}{c}}}}}
  \&[-30pt] \simeq_{\eqrho}
  \&[-30pt]
  \termrepl{\beta'}{\beta}{s}{\BMC{\replace{\alpha'}{\alpha}{s}{c}}}\neren{\beta'}{\alpha'}
\end{tikzcd}
}
\hfill 
\] Note how the behaviour of replacement over renaming replacements (\cf
Def.~\ref{def:replacement}) plays a key role in the bottom right corner of the
previous diagram.
\ignore{: $$\replace{\alpha'}{\alpha}{s}{c \neren{\beta}{\alpha}} \eqdef
\termrepl{\beta'}{\beta}{s}{\replace{\alpha'}{\alpha}{s}{c}}\neren{\beta'}{\alpha'}$$}

\partitle{The Relation $\simeq$ and Admissible Equalities.}
Given all these considerations, we define:

\begin{definition}
\label{def:Sigma_equivalence}
\deft{\regnier-equivalence}, written $\simeq$, is a relation over terms in
canonical normal form. It is defined as the smallest reflexive, symmetric and
transitive relation over terms in canonical normal form, that is closed under
the following axioms: 
$$
\begin{array}{rcll}
\termsubs{x}{u}{\ctxtapp{\ctxtLTT}{v}}                                                                  & \simeq_{\eqexsubs} & \ctxtapp{\ctxtLTT}{\termsubs{x}{u}{v}} & \notatall{x}{\ctxtLTT}, \freeFor{u}{\ctxtLTT} \\
\termrepl{\alpha'}{\alpha}{s}{\ctxtapp{\ctxtLCC}{c}}                                                    & \simeq_{\eqexrepl} & \ctxtapp{\ctxtLCC}{\termrepl{\alpha'}{\alpha}{s}{c}} & \notatall{\alpha}{\ctxtLCC}, \freeFor{\{s, \alpha'\}}{\ctxtLCC} \\
\termrepl{\alpha'}{\alpha}{s}{(\termname{\alpha}{u})}                                                   & \simeq_{\eqlinear} & \termname{\alpha'}{\BMC{\termconc{u}{s}}} & \notatall{\al}{u},  s\neq \termemst  \\
\termname{\alpha'}{\termabs{x}{\termcont{\alpha}{\termname{\beta'}{\termabs{y}{\termcont{\beta}{u}}}}}} & \simeq_{\eqpoppop} & \termname{\beta'}{\termabs{y}{\termcont{\beta}{\termname{\alpha'}{\termabs{x}{\termcont{\alpha}{u}}}}}} & \alpha\neq\beta', \alpha'\neq\beta \\
\termname{\alpha}{\termcont{\beta}{c}}                                                                  & \simeq_{\eqrho}    & c \exrepl{\alpha}{\beta}{\termemst} & \\
\termcont{\alpha}{\termname{\alpha}{t}}                                                                 & \simeq_{\eqtheta}  & t & \notatall{\alpha}{t}
\end{array}
$$
\end{definition}

\ignore{
This notion of equivalence allows to deduce a natural form of permutation of
independent renamings. More precisely, 
\begin{lemma}
Let $\beta \neq \alpha'$ and $\alpha\neq \beta'$. Then
$c_0 = c\neren{\alpha}{\alpha'}\neren{\beta}{\beta'} \simeq
c\neren{\beta}{\beta'}\neren{\alpha}{\alpha'} = c_1$.
\end{lemma}

\begin{proof}
We have $c_0 \simeq_{\eqrho}
(\termname{\alpha'}{\termcont{\alpha}{c})\neren{\beta}{\beta'}}
\simeq_{\eqexren} 
\termname{\alpha'}{\termcont{\alpha}{c \neren{\beta}{\beta'}}} \simeq_{\eqrho} 
c_1$.
\end{proof} 
}

We conclude this section by showing some interesting {\it admissible}
$\simeq$-equalities. First, we state a permutation result between substitution
(resp. replacement) contexts and linear term (resp. command) contexts
that will be useful later in the paper:

\begin{toappendix}
\begin{lemma}\mbox{}
\begin{enumerate}
  \item \label{l:subs-out-of-BM} Let $t \in \terms{\LMfull}$. Then
  $\BMC{\ctxtapp{\ctxt{L}}{\ctxtapp{\ctxtLTT}{t}}} \simeq
  \BMC{\ctxtapp{\ctxtLTT}{\ctxtapp{\ctxt{L}}{t}}}$, if
  $\notatall{\bv{\ctxt{L}}}{\ctxtLTT}$ and
  $\freeFor{\ctxt{L}}{\ctxtLTT}$.
  
  \item \label{l:repl-out-of-BM} Let $c \in \commands{\LMfull}$. Then 
  $\BMC{\ctxtapp{\ctxt{R}}{\ctxtapp{\ctxtLCC}{c}}} \simeq
  \BMC{\ctxtapp{\ctxtLCC}{\ctxtapp{\ctxt{R}}{c}}}$, if
  $\notatall{\bn{\ctxt{R}}}{\ctxtLCC}$ and
  $\freeFor{\ctxt{R}}{\ctxtLCC}$.
\end{enumerate}
\label{l:subs-repl-out-of-BM}
\end{lemma}
\end{toappendix}

Some further admissible equations are:

\begin{enumerate}
  \item $\termsubs{y}{v}{\termsubs{x}{u}{t}} \simeq
  \termsubs{x}{u}{\termsubs{y}{v}{t}}$, where $\notatall{x}{v}$, and
  $\notatall{y}{u}$.
  
  \item $c\exrepl{\alpha}{\alpha'}{s}\exrepl{\beta}{\beta'}{s'} \simeq
  c\exrepl{\beta}{\beta'}{s'}\exrepl{\alpha}{\alpha'}{s}$, where $\alpha \neq
  \beta'$, $\beta \neq \alpha'$, $\notatall{\al'}{s'}$ and
  $\notatall{\beta'}{s}$.
  
  \item
  $\termname{\alpha'}{\termcont{\alpha}{\termname{\beta'}{\termcont{\beta}{c}}}}
  \simeq
  \termname{\beta'}{\termcont{\beta}{\termname{\alpha'}{\termcont{\alpha}{c}}}}$.
\end{enumerate}
\ispaper{}{Item (1) holds by $\simeq_\eqrho$ and $\simeq_\eqexren$;
  item (2) from the former.}

Finally, as already mentioned in
Sec.~\ref{sec:refining_replacement}, \textit{linear} $\mu$-steps are captured
by $\lsigma$-equivalence~\cite{Laurent03}. In our setting, they are essentially
captured by the axiom $\simeq_{\eqlinear}$ of the $\simeq$-equivalence.
\ignore{Formally, a \deft{linear $\mu$-step} has the form
$\termapp{(\termcont{\alpha}{\ctxtapp{\ctxt{N}}{\termname{\al}{u}}})}{v}
\Rew{\mu}
\termcont{\alpha}{\ctxtapp{\ctxt{N}}{\termname{\al}{\termapp{u}{v}}}}$, where
$\notatall{\al}{\ctxt{N}}$ and $\notatall{\al}{u}$, for $\ctxt{N}$ the notion
of linear context obtained by restricting the grammar $\ctxtLTC$ to
$\lmu$-objects, as described in Sec.~\ref{sec:refining_replacement}. They too
are captured by $\simeq$-equivalence, as the following equation turns out to be
admissible in $\simeq$.}

% \edu{Otra es citar Thm~\ref{t:equivalencia-sigma-Sigma}, pero requiere
%   hablar de $\simeq_{er}$} 

% \edu{Los terminos de abajo no estan en BMC forma normal. Asi que
%   quizas debamos agregar BMC de los dos lados. No?}

% \delia{de acuerdo, tal cual como esta no va porque no estan en BMC,
%   quizas decir que la BMC de esos terminos son simeq alcanza}

% \delia{En realidad, pensandolo bien, no se si hay que presentarlo asi,
%   tambien piendo que con BMC va a quedar pesado, quizas solo podemos
%   dejar la mencion al hecho de que esta ecuacion que se puede demostrar
%   en Laurent, va a ser capturada de alguna forma por nuestra equivalencia
%   porque justamente es lo que mostramos en el paper en l aparte corresponde
%   (modulo er). }

% \[
%   (\termcont{\alpha}{\ctxtapp{\ctxt{N}}{\termname{\al}{u}}})v \simeq
% \termcont{\alpha}{\ctxtapp{\ctxt{N}}{\termname{\al}{uv}}}
% \]

A last remark of this section concerns preservation of types for our equivalence:

\begin{toappendix}
\begin{lemma}[Preservation of Types for $\simeq$]
\label{l:PTSigma}
Let $o\in \objects{\LMfull} $. If $\pi \dem \Gam \vdash o:\type \mid \Del$ and $o \simeq o'$, then there
exists $\pi' \dem \Gam \vdash o':\type \mid \Del$.
\end{lemma}
\end{toappendix}

As before, when $o \simeq o'$ we will refer to $\pi_o$ and $\pi_{o'}$ as two
\deft{related} typing derivations.

%%%Local Variables:
%%% mode: latex
%%% TeX-master: "main"
%%% End:

\ispaper{
  \input{completeness-paper}
}{
  %%%%%%%%%%%%%%%%%%%%%%%%%%%%%%%%%%%%%%%%%%%%%%%%%%%%%%%%%%%%%%%%%%%%%%%%%%%%%%%
\section{Two Correspondence Results}
\label{s:completeness}
%%%%%%%%%%%%%%%%%%%%%%%%%%%%%%%%%%%%%%%%%%%%%%%%%%%%%%%%%%%%%%%%%%%%%%%%%%%%%%%

% \delia{Theorem~\ref{t:completeness} pinpoints that,
%   $\lsigma$-equivalence is too big, since it corresponds to $\Sigma$
%   plus renamings. This bigger equivalence is also show to be
%   equivalent to PPN equality modulo structural
%   equivalence (Theorem~\ref{t:equivalencia-sigma-Sigma}). 
%   Stated differently, we have managed to isolate the
%   only reshufflings of $\lsigma$ that do not have the same reduction
%   behavior, thus shedding light on the unexpected importance that
%   explicit renaming plays in producing our strong bisimulation result.
%   This can be summarized by the diagram below, where $\eqsigmaer$ is
%   the \deft{renaming equivalence}  generated by our strong
%    bisimulation $\simeq$ plus the following axiom
%  $c\neren{\alpha}{\beta} \simeq_{{\tt ren}}
%  c\ire{\alpha}{\beta}{\emst}$:}
% $$
% o \simeq_{\sig} p
% \mbox{ if and only if }
% \BMC{o} \eqsigmaer \BMC{p}
% $$

This section studies how our $\simeq$-equivalence relates to $\lsigma$ (and
hence, to PPN equality modulo structural equivalence). In particular, we want
to understand whether reshufflings captured by $\lsigma$-equivalence may have
been left out by $\simeq$. One such set of reshufflings are those captured by
the following equation:
\begin{equation*}
\hfill
\termname{\alpha}{\termcont{\beta}{c}} \simeq_{\elsigma_7} c\iren{\beta}{\alpha} 
\hfill
\end{equation*}
As discussed in Sec.~\ref{s:Sigma-equivalence}
(\cf Fig.~\ref{e:rho-breaks-strong-bisimulation:pre}), this equation breaks
strong bisimulation and motivates the introduction of renaming replacements
into $\LMfull$, as well as the inclusion of
equation $\termname{\alpha}{\termcont{\beta}{c}} \simeq_{\eqrho}
\exrepl{\alpha}{\beta}{\termemst}$ into our relation $\simeq$. This gives
us: \[
\hfill
\termname{\alpha}{\termcont{\beta}{c}} \simeq_{\elsigma_7} c\iren{\beta}{\alpha} \mbox{ in } \lmu
\qquad\mbox{vs.}\qquad
\termname{\alpha}{\termcont{\beta}{c}} \simeq_{\eqrho} \termrepl{\alpha}{\beta}{\termemst}{c} \mbox{ in } \LMfull
\hfill
\] The question that arises is whether the reshufflings captured by
$\simeq_{\elsigma_7}$ are the only ones that are an obstacle to obtaining a
strong bisimulation. We prove in this section that this is indeed the case. 
This observation is materialized by the following property (\cf
Thm.~\ref{t:equivalencia-sigma-Sigma}): \[
\hfill
o \simeq_{\sig} p \mbox{ if and only if } \BMC{o} \eqsigmaer \BMC{p}
\hfill
\] where $\eqsigmaer$ is the \deft{renaming equivalence}  generated by our
strong bisimulation $\simeq$ plus the following axiom
$\replace{\alpha}{\beta}{\emst}{c} \simeq_{\eqren} c\neren{\alpha}{\beta}$,
which equates the implicit renaming used in equation $\elsigma_7$ with the
renaming replacement used in equation $\eqrho$.
This sheds light on the unexpected importance that renaming replacement plays in our strong bisimulation result.

% \odelia{In this section we show two major results of our paper.
%   On the one hand, we formally relate $\lsigma$-equivalence on
% $\lmu$-terms (Fig.~\ref{f:sigma-laurent}) to $\Sigma$-equivalence on
% $\LMfull$-terms (Fig.~\ref{f:Sigma-equivalence}), by showing they
% are equivalent.
% On the other hand we show soundness and completeness of
% our $\simeq$ relation, in the sense that two terms are equivalent
% if and only if they have the same proof-net representation.
% }

\ignore{We start by describing how our new equivalence is able to
  capture $\lsigma$-equivalence (the opposite direction is proved at
  the end of this section). Since
$\simeq$ is only defined on $\BMCform$, given
any pair of terms $o$ and $p$ which are $\lsigma$-equivalent, we need
to compute $\BMC{o}$ and $\BMC{p}$ in order to compare them with
$\simeq$. There is a special case however, which is the one of
equation $\simeq_{\lsigma_7}$ recalled below: \[
  \termname{\alpha}{\termcont{\beta}{c}}  \simeq_{\lsigma_7}
  c\iren{\beta}{\alpha}
  \]}
\ignore{ which does not translate to $\simeq_{\eqrho}$ simply by
 $\BMC{\_}$. This is not surprising, since renaming replacements was
 introduced in $\simeq_{\eqrho}$ exactly to repair
 Diag.~\ref{e:rho-breaks-strong-bisimulation}.  At this point, renaming equivalence $\eqsigmaer$ comes
 into play.}

The $(\Rightarrow)$ direction of Thm.~\ref{t:equivalencia-sigma-Sigma}, stated
below, is relatively straightforward to prove:

\begin{toappendix}
\begin{lemma}
\label{l:relation-sigma-Sigma}
Let $o,p\in \objects{\lmu}$. If $o \simeq_{\lsigma} p$, then $\BMC{o} \eqsigmaer \BMC{p}$.
\end{lemma}
\end{toappendix}

\begin{proof}
\ignore{
By induction on $o \simeq_{\lsigma} p$. Cases $\simeq_{\lsigma_1}$ and
$\simeq_{\lsigma_3}$ translate to $\simeq_{\eqexsubs}$, case
$\simeq_{\lsigma_2}$ holds by
Lem.~\ref{l:subs-out-of-BM}.\ref{l:subs-out-of-BM}, case $\simeq_{\lsigma_4}$
translates to $\simeq_{\eqexrepl}$, case $\simeq_{\lsigma_5}$ translates to
$\simeq_{\eqrho}$ and $\simeq_{\eqexren}$, case $\simeq_{\lsigma_6}$ translates
to $\simeq_{\eqpoppop}$ and case $\simeq_{\lsigma_8}$ translates to
$\simeq_{\eqtheta}$. Last, $o = \termname{\alpha}{\termcont{\beta}{c}}
\simeq_{\lsigma_7} \replace{\beta}{\alpha}{\termemst}{c} = p$ implies
$\BMC{\termname{\alpha}{\termcont{\beta}{c}}} =
\termname{\alpha}{\termcont{\beta}{\BMC{c}}} \simeq_{\eqrho}
\BMC{c}\neren{\beta}{\alpha} \simeq_{\exren}
\BMC{\replace{\beta}{\alpha}{\temremst}}$.
}
Uses Lem.~\ref{l:subs-repl-out-of-BM}.\ref{l:subs-out-of-BM}. Full details
\ispaper{in~\cite{TR}.}{in App.~\ref{a:completeness}.}
\end{proof}

In what follows we focus on the $(\Leftarrow)$ direction of
Thm.~\ref{t:equivalencia-sigma-Sigma}. We will structure the
presentation into two correspondence results
(Thm.~\ref{t:completeness} and Thm.~\ref{t:equivalencia-sigma-Sigma}).

\partitle{Relating $\eqsigmaer$ and PPN-Structural Equivalence.}
We first discuss the soundness and completeness properties of the equivalence
relation $\eqsigmaer$. This result is based on Laurent's completeness result
for $\lsigma$-equivalence, and works as described below.

A first translation from typed $\lmu$-objects to polarized proof-nets, written
$\tradlaurentu{\_}$, is defined in~\cite{Laurent03}, together with a second
translation, written $\tradlaurentd{\_}$, which is defined as the
multiplicative normal-form of $\tradlaurentu{\_}$, where multiplicative
normal-forms are obtained by reducing all the
multiplicative cuts.

The $\lsigma$-equivalence relation on $\lmu$-terms has the remarkable
property that $o$ and $p$ are $\lsigma$-equivalent iff their
proof-net (second) translation $\tradlaurentd{\_}$ are structurally
equivalent, \cf $\equiv$-equivalence introduced in
  Sec.~\ref{s:proofnets}.  That is, two $\lsigma$-equivalent objects
have the same {\it structural} proof-net representation modulo
   multiplicative cuts. 

  \begin{theorem}[\cite{Laurent03}]
    \label{t:laurent}
    Let $o$ and $p$ be typed $\lmu$-objects
    such that $o \simeq_{\sig} p$ and
    let $\pi_o, \pi_p$
    be two related corresponding typing derivations. Then
    $o \simeq_{\sig} p$ iff
    $\tradlaurentd{\pi_o} \equiv \tradlaurentd{\pi_p}$.
  \end{theorem}

As mentioned in Sec.~\ref{s:translation}, we have extended the translation $\tradlaurentu{\_}$ to
the new constructors of $\LMfull$,
where  the resulting function is written
$\topn{\_}$.
Since the set $\objects{\LMfull}$ strictly contains the
set $\objects{\lmu}$, it is evident that for every $\lmu$-object $o$ we have
$\topn{\pi_o} \equiv \tradlaurentu{\pi_o}$. 

   \label{l:bm-implies-multiplicative-nf}
   Let $o \in \objects{\LMfull}$ be in $\BMCform$.  Let $\pi_o$ be a
   typing derivation for $o$.  Then $\topn{\pi_o}$ is not necessarily
   in multiplicative normal-form: consider the term
   $ (\termname{\al}{x})\exrepl{\al'}{\al}{y} $, which is in
   $\BMCform$.  In its $\topn{\_}$-translation, the proof-net of
   $\termname{\al}{x}$ is connected with the box containing the
   renaming replacements $\exrepl{\al'}{\al}{y}$ by means of a
   multiplicative cut.  This is also one of the
     reasons why we have extended the translation $\topn{\_}$ to
   a new one, written $\topnd{\_}$, in such a way that $\topnd{\pi}$
   is the multiplicative normal-form of $\topn{\pi}$ (\cf
     Sec.~\ref{s:translation}). Again, since
   $\objects{\lmu} \subset \objects{\LMfull}$, then for every
   $\lmu$-object $o$ we have
   $\topnd{\pi_o} \equiv \tradlaurentd{\pi_o}$.

  %% In this paper, we take a complementary approach, since our notion
  %% of $\eqsigma$-equivalence is only defined on 
  %% $\BMform$-normal forms, which are in some sense, the
  %% terms whose proof-nets are in normal-form w.r.t.
  %% $\nomformpar$-$\nomformten$ and
  %% $\tax$ cuts. We then obtain that
  %% $o \eqsigma p$ if and only if 
  %% $\topn{\pi_o} \equiv  \topn{\pi_p}$. 
\medskip

The following table summarizes all the above mentioned translations
    to PPNs:
\begin{center}
  \begin{tabular}{|c|c|p{6cm}|}
  %\begin{tabular}{|l|c|c|}
    \hline
    \textbf{Name}    &  \textbf{Calculus} & \textbf{Observation} \\\hline
    $\tradlaurentu{\_}$ & $\lmu$ & Not necessarily in multiplicative NF\\\hline
    $\tradlaurentd{\_}$  & $\lmu$ & Multiplicative NF of $\tradlaurentu{\_}$\\\hline
    $\topn{\_}$ & $\LMfull$ & Not necessarily in multiplicative NF\\\hline
    $\topnd{\_}$ & $\LMfull$ & Multiplicative NF of $\topn{\_}$ \\\hline
    \end{tabular}
\end{center}

A first interesting remark concerns the relation between term reduction to
canonical form and proof-net reduction to multiplicative normal form. 

\begin{toappendix}
\begin{lemma}
\label{l:bmc-to-proofnets}
Let $o, p$ be typed $\LMfull$-objects. If $o \Rew{\Can} p$, then
$\topnd{\pi_o} \equiv \topnd{\pi_p}$, where $\pi_o$ and $\pi_p$ are
two corresponding related typing derivations.
\end{lemma}
\end{toappendix}

On the other hand, our relation $\simeq$, together with the equivalence
$\simeq_{\exren}$, which are both defined on our calculus $\LMfull$, represent
also equivalent proof-nets:

\begin{toappendix}
\begin{lemma}[Soundness]
\label{l:soundness}
Let $o,p \in \objects{\LMfull}$ in $\BMCform$. If $o \eqsigmaer p$, then
$\topnd{\pi_{o}} \equiv \topnd{\pi_{p}}$, where $\pi_{o}$ and $\pi_{p}$
are two related corresponding  typing derivations.
\end{lemma}
\end{toappendix}

In order to obtain our  first correspondence result, we relate
$\objects{\LMfull}$ to $\objects{\lmu}$ by an \deft{expansion
  function} $\expansion{\_}$, which eliminates all explicit operators
by letting
$\expansion{t\exsubs{x}{u}} \eqdef (\l x. \expansion{t})\expansion{u}$
and
$\expansion{c\exrepl{\alpha'}{\alpha}{s}} \eqdef
\termname{\alpha'}{(\termcont{\alpha}{\expansion{c}}) \tconc
  \expansion{s}} $. The expansion function behaves as expected on all
the other cases. In the particular case $s = \termemst$, 
$\expansion{c\exrepl{\alpha'}{\alpha}{\emst}}  \eqdef
\termname{\alpha'}{(\termcont{\alpha}{\expansion{c}})}$. 
\ignore{ Indeed, the \deft{expansion function} $\expansion{\_}$  on $\LMfull$-objects is
 defined by induction as follows:
   \[ \begin{array}{lll@{\hspace{.2cm}}lll}
        \expansion{x} & \eqdef  & x \\
        \expansion{tu} & \eqdef & \expansion{t}\expansion{u} \\
        \expansion{\l x. t} & \eqdef & \l x.  \expansion{t} \\
        \expansion{t\exsubs{x}{u}} & \eqdef & (\l x. \expansion{t})\expansion{u}  \\
        \expansion{\termcont{\alpha}{c}} & \eqdef & \termcont{\alpha}{\expansion{c}} \\
                       \expansion{c\exrepl{\alpha'}{\alpha}{s}} & \eqdef & \termname{\alpha'}{(\termcont{\alpha}{\expansion{c}}) \tconc \expansion{s}}  \\
        \expansion{\emst} & \eqdef & \emst\\
        \expansion{\termpush{t}{s}} & \eqdef & \termpush{\expansion{t}}{\expansion{s}}
      \end{array}  \]}
    
Now, given a typed
$\lmu$-object $o$, we observe that $\expansion{\BMC{o}} \neq o$. Indeed, let
$o =\termapp{\termapp{(\termcont{\alpha}{\termname{\al}{x}})}{y}}{z}$. Then
$\BMC{o} =
\termcont{\al'}{\termrepl{\al'}{\al}{\termpush{y}{z}}{(\termname{\al}{x})}}$
and $\expansion{\BMC{o}} =
\termcont{\al'}{\termname{\al'}{\termapp{\termapp{(\termcont{\al}{\termname{\al}{x}})}{y}}{z}}}$.
However, $\expansion{\BMC{o}} \simeq_{\lsigma_8} o$, so that
$\tradlaurentd{\pi_{\expansion{\BMC{o}}}} \equiv \tradlaurentd{\pi_o}$ by
Thm.~\ref{t:laurent}, where  $\pi_{\expansion{\BMC{o}}}$ and $\pi_o$ denote
the respective type derivations.

     Since the expansion steps in the calculus are only translated to
     multiplicative steps in the proof-nets, then we have:
  \begin{lemma}
    \label{l:bm-expansion}
    Let $o \in \objects{\lmu}$ and let $\pi_o, \pi_{\BMC{o}}$ be two
    related corresponding typing derivations for $o$ and
    $\BMC{o}$. Then $\topnd{\pi_o} \equiv \topnd{\pi_{\BMC{o}}}$.  Let
    $o \in \objects{\LMfull}$ and let $\pi_o, \pi_{\expansion{o}}$ be
    two related corresponding typing derivations for $o$ and
    $\expansion{o}$. Then $\topnd{\pi_o} \equiv
    \topnd{\pi_{\expansion{o}}}$.
\end{lemma}

\begin{proof}
The first point is by induction on $\Rew{\Can}$ and the second one by
induction on $\expansion{\_}$.
\end{proof}

As a corollary we obtain: 
\begin{corollary}
\label{c:bm-expansion}
If  $o \in \objects{\lmu}$, $o' \in \objects{\LMfull}$.
Then $\topnd{\pi_{\expansion{\BMC{o}}}} \equiv
\topnd{\pi_o}$ and
$\topnd{\pi_{\BMC{\expansion{o'}}}} \equiv \topnd{\pi_{o'}}$.
\end{corollary}

All these arguments lead us to the following result:

\begin{theorem}[Correspondence I]
\label{t:completeness}
Let $o,p \in \objects{\lmu}$. Then $\BMC{o} \eqsigmaer \BMC{p}$ iff 
$\topnd{\pi_{\BMC{o}}} \equiv \topnd{\pi_{\BMC{p}}}$,
where $\pi_{\BMC{o}}, \pi_{\BMC{p}}$ are
  two related corresponding
typing derivations for $\BMC{o}$ and $\BMC{p}$.
\end{theorem}

\begin{proof} \mbox{}
$(\Rightarrow)$ If $\BMC{o} \eqsigmaer \BMC{p}$, then $\topnd{\pi_{\BMC{o}}}
\equiv \topnd{\pi_{\BMC{p}}}$ by Lem.~\ref{l:soundness}.

$(\Leftarrow)$ Let $\topnd{\pi_{\BMC{o}}} \equiv \topnd{\pi_{\BMC{p}}}$. Then
$\topnd{\pi_{\expansion{\BMC{o}}}} \equiv \topnd{\pi_{\expansion{\BMC{p}}}}$
by Lem.~\ref{l:bm-expansion} and $\topnd{\pi_{o}} \equiv \topnd{\pi_{p}}$
by Cor.~\ref{c:bm-expansion}. As remarked before,
$\topnd{\pi_{o}} \equiv \tradlaurentd{\pi_o}$ and 
$\topnd{\pi_{p}} \equiv\tradlaurentd{\pi_p}$.
Thus we obtain $o \simeq_{\sig} p$
by Thm.~\ref{t:laurent}. 
Hence $\BMC{o} \eqsigmaer \BMC{p}$ by Lem.~\ref{l:relation-sigma-Sigma}.
\end{proof}

\ignore{
On the other hand, we have related $\lsigma$-equivalence in $\lmu$ to
$\Sigma$-equivalence in $\LMfull$ in such a way that $o \simeq_{\sig} p$
implies $\BMC{o} \eqsigmaer \BMC{p}$ (Sec.~\ref{s:sigma-Sigma}), where $\exren$
is the equivalence relation generated by converting the renaming replacements into
an implicit renaming.
}

\partitle{Relating $\simeq_{\exren}$ and $\simeq_\lsigma$.}
Last but not least, $\lsigma$-equivalence is not only captured by $\eqsigmaer$,
but the converse implication also holds. 

\begin{theorem}[Correspondence II]
\label{t:equivalencia-sigma-Sigma}
Let $o, p \in \objects{\lmu}$. Then $o \simeq_{\lsigma} p$ iff $\BMC{o}
\eqsigmaer \BMC{p}$.
\end{theorem}

\begin{proof}
The left-to-right implication holds by Lem.~\ref{l:relation-sigma-Sigma}. For
the right-to-left implication, take $\BMC{o} \eqsigmaer \BMC{p}$. Then
$\topnd{\pi_{\BMC{o}}} \equiv \topnd{\pi_{\BMC{p}}}$ by Lem.~\ref{l:soundness},
and thus $\topnd{\pi_{o}} \equiv \topnd{\pi_{p}}$ by Lem.~\ref{l:bm-expansion}.
As already remarked $\topnd{\pi_{o}} \equiv \tradlaurentd{\pi_o}$ and
$\topnd{\pi_{p}}\equiv \tradlaurentd{\pi_p}$. Then $\tradlaurentd{\pi_o} \equiv
\tradlaurentd{\pi_p}$ implies $o \simeq_{\sig} p$ by Thm.~\ref{t:laurent}.
\end{proof}

The main results of this section can be depicted in the diagram below: \[
  \hfill
  \begin{tikzcd}
o \simeq_{\sig} p
  \arrow[<->,dotted]{d}[anchor=north,left]{Thm.~\ref{t:laurent}}
%  \arrow[<->,dotted]{r}[anchor=north,below]{\BMCform}
  \arrow[<->,dotted]{r}[anchor=north,above]{Thm.~\ref{t:equivalencia-sigma-Sigma}}
  &[27pt] \BMC{o} \eqsigmaer \BMC{p}
  \arrow[<->,dotted]{d}[anchor=north,right]{Thm.~\ref{t:completeness}} \\
\tradlaurentd{\pi_o} \equiv \tradlaurentd{\pi_p} 
% \LRewn{\mbox{multiplicative cuts}}
  &[27pt] \topnd{\pi_{\BMC{o}}} \equiv \topnd{\pi_{\BMC{p}}}
\end{tikzcd} \hfill \]

%%% Local Variables:
%%% mode: latex
%%% TeX-master: "main"
%%% End:

}
%%%%%%%%%%%%%%%%%%%%%%%%%%%%%%%%%%%%%%%%%%%%%%%%%%%%%%%%%%%%%%%%%%%%%%%%%%%%%%%
\section{The Strong Bisimulation Result}
\label{s:bisimulation}
%%%%%%%%%%%%%%%%%%%%%%%%%%%%%%%%%%%%%%%%%%%%%%%%%%%%%%%%%%%%%%%%%%%%%%%%%%%%%%%

As stated in Thm.~\ref{t:laurent}, $\lmu$-objects that map to the
same PPN (modulo structural equivalence) are captured exactly by
Laurent's $\lsigma$-equivalence, which may also been seen as providing
a natural relation of \emph{reshuffling}.  Unfortunately, as explained
in the introduction, $\lsigma$-equivalence is not a
strong bisimulation. We set out to devise a new term calculus in which
reshuffling can be formulated as a strong bisimulation \emph{without
  changing} the PPN semantics. The relation we obtain, $\simeq$, is
indeed a strong bisimulation over canonical forms, \ie
$\simeq$-equivalent canonical terms have exactly the same redexes.
This is the result we present in this section.  It relies crucially on
our decomposition of replacement $\Rew{\rRfull}$
(\cf Sec.~\ref{sec:refining_replacement}) into {\it linear} and {\it
  non-linear} replacements, the former having no computational content
(\ie structurally equivalent PPNs modulo multiplicative cuts),
and thus included in our $\simeq$-equivalence, while
the latter corresponding to exponential cut elimination steps, and
thus considered as part of our meaningful reduction.
%\odelia{Indeed, linear explicit replacement is
%included in our proposed notion of $\simeq$-equivalence, whereas
%non-linear reduction remains as a reduction step, written $\Rew{\rR}$.}

%\delia{aca hay algo que deberiamos explicar mejor, linear replacement es la regla de composition y tambien la ecuacion lineal, mientras non-linear es todo el resto. La forma de linear composition no es la misma que la forma de
%linear replacement. Esto es confuso y esta mal explicado por el momento. }
%\delia{Explicar que una vez que separamos R en
%  lineal y no lineal, la composicion que es lineal
%  no se puede poner como ecuacion y entonces se pone como regla para BMC. }
Before stating the bisimulation result, we mention some
important technical
lemmas: 

\begin{toappendix}
\begin{lemma}
\label{l:app-bm}
Let $o\in \objects{\LMfull}$. If $o \simeq o'$, then $\BMC{\ctxtapp{\ctxt{O}}{o}} \simeq
\BMC{\ctxtapp{\ctxt{O}}{o'}}$.
\end{lemma}
\end{toappendix}

%\begin{proof} By induction on $\simeq$. Details in 
%App.~\ref{a:bisimulation-lemmas}.
%\end{proof}

\begin{toappendix}
\begin{lemma}
\label{l:equiv-for-substitutions-and-replacements} 
Let $u, s, o \in \objects{\LMfull}$ be in $\BMCform$. Assume $p \simeq p'$
and $v \simeq v'$ and $q \simeq q'$. Then, 
\begin{itemize}
\item $\BMC{\substitute{x}{u}{p}} \simeq \BMC{\substitute{x}{u}{p'}}$
  and  $\BMC{\substitute{x}{v}{o}} \simeq \BMC{\substitute{x}{v'}{o}}$.
  \item $\BMC{\replace{\gamma'}{\gamma}{s}{p}} \simeq
    \BMC{\replace{\gamma'}{\gamma}{s}{p'}}$ and 
    $\BMC{\replace{\gamma'}{\gamma}{q}{o}} \simeq
  \BMC{\replace{\gamma'}{\gamma}{q'}{o}}$.
\end{itemize}
\end{lemma}
\end{toappendix}

\begin{proof}
Uses Lem.~\ref{l:app-bm}.
%See Lem.~\ref{l:equiv-substitution},~\ref{l:Sigma_closed_under_substitution:target} in App.~\ref{a:bisimulation-lemmas}.
Details \ispaper{in~\cite{TR}.}{in App.~\ref{a:bisimulation-lemmas}.}
\end{proof}

\ignore{
\begin{lemma}
\label{l:equiv-for-substitutions} \mbox{}
\begin{itemize}
  \item If $w \in \BMCform$ and $o \simeq o'$, then
  $\BMC{\substitute{x}{w}{o}} \simeq \BMC{\substitute{x}{w}{o'}}$. 
  \item If $o \in \BMCform$ and $w \simeq w'$, then
  $\BMC{\substitute{x}{w}{o}} \simeq \BMC{\substitute{x}{w'}{o}}$. 
\end{itemize}
\end{lemma}
   
\begin{proof}
Uses Lem.~\ref{l:app-bm}.
Details in \ispaper{~\cite{TR}.}{App.~\ref{a:bisimulation-lemmas}.}
\end{proof}

\begin{lemma}
  \label{l:equiv-for-replacements} \mbox{}
  \begin{itemize}
\item If $s \in \BMform$ and $o \simeq o'$, then
  $\BMC{\replace{\gamma'}{\gamma}{s}{o}} \simeq
  \BMC{\replace{\gamma'}{\gamma}{s}{o'}}$.
    \item If $o \in \BMform$ and $s \simeq s'$, then
  $\BMC{\replace{\gamma'}{\gamma}{s}{o}} \simeq
  \BMC{\replace{\gamma'}{\gamma}{s'}{o}}$.
\end{itemize}
\end{lemma}

\begin{proof}
Uses Lem.~\ref{l:app-bm}.
%See Lem.~\ref{l:equiv-replacement},~\ref{l:Sigma_closed_under_BM_plus_replacement} in App.~\ref{a:bisimulation-lemmas}.
Details in App.~\ref{a:bisimulation-lemmas}.
\end{proof}
}

We are now able to state the promised result, namely, the fact that $\simeq$ is
a strong $\reduce$-bisimulation. 

%, $$
%\begin{tikzcd}
%o  \arrow[rightsquigarrow]{d}[anchor=north,left]{}
  %&[-25pt] \simeq
  %&[-25pt] p \arrow[rightsquigarrow]{d}[anchor=north,left]{} \\
%o'  
  %&[-25pt] \simeq
  %&[-25pt]  p'
%\end{tikzcd} $$

\begin{toappendix}
\begin{theorem}[Strong Bisimulation]
\label{t:bisimulation}
Let $o\in \objects{\LMfull}$. If $o \simeq p$ and $o \reduce o'$, then $\exists p'$ s.t. $p \reduce p'$ and
$o' \simeq p'$.
\end{theorem}
\end{toappendix}

\begin{proof}
Uses all the previous lemmas of this section. See\ispaper{~\cite{TR}}{~App.~\ref{a:bisimulation-lemmas}} for full
  details.
\end{proof}

%%%Local Variables:
%%% mode: latex
%%% TeX-master: "main"
%%% End:

%%%%%%%%%%%%%%%%%%%%%%%%%%%%%%%%%%%%%%%%
\section{Conclusion}
\label{s:conclusion}
%%%%%%%%%%%%%%%%%%%%%%%%%%%%%%%%%%%%%%%%

\newcommand{\lmex}{\lambda\mu\mathtt{r}}

\ignore{
  \begin{deliae}
  \begin{remark}
$\LM$ is not confluent. Consider
$t=\termrepl{\alpha'}{\alpha}{w}{\termname{\alpha}{(\termsubs{x}{\termname{\alpha}{v}}{u})}}$,
with $\alpha\notin\fn{u,v}$ and $x\notin\fv{u}$. Then
$t \Rew{\rS} \termrepl{\alpha'}{\alpha}{w}{(\termname{\alpha}{u})}
=t_2$ and $t\Rew{\rR}
\termname{\alpha}{(\termapp{\termsubs{x}{\termname{\alpha}{\termapp{v}{w}}}{u}}{w})}=t_3$. However,
$t_2$ and $t_3$ are not joinable.
\delia{esto no es mas de actualidad no? el ejemplo es confluente usando
las ecuaciones para cerrar}
\edu{S'i, es cierto. Puede que con las ecuaciones valga
confluencia modulo. A seguir.}
\end{remark}
\end{deliae}
}

This paper refines the $\lmu$-calculus by splitting its rules into multiplicative and
exponential fragments.  This new presentation of $\lmu$ allows to
reformulate $\lsigma$-equivalence on $\lmu$-terms as a strong
bisimulation relation $\simeq$ on the extended term language
$\LMfull$. In addition, $\simeq$ is conservative
w.r.t. $\lsigma$-equivalence, and $\simeq$-equivalent terms share the
same PPN representation.  \ispaper{}{Moreover, $\Rew{\LMfull}$ is
  shown to be confluent. }

Besides~\cite{Laurent03}, which  inspired this paper
and has been discussed at length, we briefly mention further related work.
In~\cite{Accattoli13}, polarized MELL are represented by proof-nets
without boxes, by using the polarity information to transform explicit
$!$-boxes into more compact structures.
In~\cite{KesnerV17}, the $\lmu$-calculus is
refined to a calculus $\lmex$ with explicit operators, together with a
small-step substitution/replacement operational semantics {\it at a
  distance}. At first sight $\lmex$   seems to be more atomic than
$\LMfull$.
%\odelia{, basically because of the linear reduction steps it implements
%in contrast to our $\rS$ and $\rR$ reductions.}
However,  $\lmex$ forces the explicit replacements to be
evaluated from left to right, as there is no mechanism of
composition, and thus only replacements on named locations can be performed. 
Other refinements of the $\lmu$-calculus
were defined in~\cite{Audebaud94,Polonovski04,vanBakelVigliotti14}.
 A further related reference is~\cite{DBLP:journals/tcs/HondaL10}. A
precise correspondence is established between PPN and
a typed version of the asynchronous $\pi$-calculus. Moreover, they
show that Laurent's $\eqlaurent$ corresponds exactly to structural
equivalence of $\pi$-calculus processes (Prop. 1 in op.cit).
In~\cite{LaurentRegnier03} Laurent and Regnier show that there is a precise correspondence between CPS
translations from classical calculi (such as $\lmu$) into
intuitionistic ones on the one hand, and translations between LLP and
LL on the other. 
% \edu{In~\cite{DBLP:journals/jsyml/DanosJS97}, Danos et al include a
% translation from Parigot's CND into (a fragment of) their
% $\textbf{LK}^{tq}$, the latter of which can be translated to Linear
% Logic. This work does not discuss proof-nets nor $\sigma$-equivalence
% \delia{entonces porque lo citamos? sino deberiamos citar miles de otros
%   papers relacionados con lambda -mu...}.
% Selinger~\cite{DBLP:journals/mscs/Selinger01} introduces a categorical
% semantics for $\lmu$. \delia{mismo comentario}}

Besides confluence, studied in Sec.~\ref{s:lmu-calculus}, 
it would be interesting  to analyse other
rewriting properties of our term language
such as  preservation of
$\lmu$-strong normalization
of the reduction relations
  $\Rew{\LMfull}$ and $\reduce$,
  or confluence of $\reduce$.
  Moreover, a reformulation of $\LMfull$ in terms
    of two different syntactical operators, one for
    renaming replacement, and another one for stack replacements,
    would probably enlighten the intuitions on PPNs that have been
    used in this work.
    We have however chosen a unified syntax for explicit replacements
  in order to shorten the inductive
cases of many of our proofs.

Another further topic would be to explore how our notion of strong
bisimulation behaves on different calculi for Classical Logic, such as
for example $\lambda\mu\widetilde{\mu}$~\cite{CurienH00}. Moreover,
following the computational interpretation of deep inference provided
by the intuitionistic atomic lambda-calculus~\cite{GundersenHP13}, it
would be interesting to investigate a classical extension and its
corresponding notion of strong bisimulation. It is also natural to
wonder what would be an
  ideal syntax for Classical Logic,  that is able to capture strong bisimulation
  by reducing the syntactical axioms to a small and simple set of equations.

  We believe the relation
  $\reduce$ is well-suited for devising a residual
theory for $\lmu$. That is, treating $\reduce$ as an orthogonal system,
from a diagrammatic point of view~\cite{AccattoliBKL14},  in
spite of the critical pairs introduced by $\rho$ and $\theta$. This
could, in turn, shed light on call-by-need for $\lmu$ via
the standard notion of neededness 
defined using residuals.

Finally, our notion of $\simeq$-equivalence could facilitate proofs of
correctness between abstract machines and $\lmu$
(like~\cite{DBLP:conf/icfp/AccattoliBM14} for lambda-calculus) and
help establish whether abstract machines for $\lmu$ are
``reasonable''~\cite{DBLP:conf/icfp/AccattoliBM14}.

%%% Local Variables:
%%% mode: latex
%%% TeX-master: "main"
%%% End:

\section*{Acknowledgment}
\noindent To Olivier Laurent for fruitful discussions. This work was partially
supported by LIA INFINIS, and the ECOS-Sud program PA17C01. 
  
%%% Bibliography
\bibliographystyle{plain}
\ispaper{
  \input{biblio.bbl}
}{
  \bibliography{biblio} 
}

%%Appendix
\ispaper{}{
  \newpage
  \appendix
  %%%%%%%%%%%%%%%%%%%%%%%%%%%%%%%%%%%%%%%%%%%%%%%%%%%%%%%%%%%%%%%%%%%%%%%%%%%%%%%
\section{\texorpdfstring{Appendix: Confluence of $\LMfull$}{Appendix: Confluence of LM}}
\label{a:confluence}
%%%%%%%%%%%%%%%%%%%%%%%%%%%%%%%%%%%%%%%%%%%%%%%%%%%%%%%%%%%%%%%%%%%%%%%%%%%%%%%

The following explains how substitution and replacement
decompose on contexts and terms. 

\begin{lemma}
  \label{l:context-substitution-replacement} \mbox{}
  \begin{itemize}
  \item  \label{l:stable-contexts}
$\substitute{x}{u}{\ctxtapp{\ctxt{O}}{o}} =
\ctxtapp{\substitute{x}{u}{\ctxt{O}}}{\substitute{x}{u}{o}}$.
Moreover, if $x \notin \fv{\ctxt{O}}$ and $\freeFor{u}{\ctxt{O}}$,
then $\substitute{x}{u}{\ctxtapp{\ctxt{O}}{o}} =
  \ctxtapp{\ctxt{O}}{\substitute{x}{u}{o}}$.
\item \label{l:transforming-contexts}
$\replace{\beta}{\alpha}{s}{\ctxtapp{\ctxt{O}}{o}}=
\ctxtapp{\replace{\beta}{\alpha}{s}{\ctxt{O}}}{\replace{\beta}{\alpha}{s}{o}}$.
Moreover, if  $\alpha \notin \fn{\ctxt{O}}$ and $\freeFor{s}{\ctxt{O}}$,
then $\replace{\beta}{\alpha}{s}{\ctxtapp{\ctxt{O}}{o}} =
  \ctxtapp{\ctxt{O}}{\replace{\beta}{\alpha}{s}{o}}$. 
\end{itemize}
\end{lemma}   

{\it E.g.}  let $\ctxt{O}_1=\termapp{\Box}{t_1}$,
$\ctxt{O}_2=\termname{\alpha}{\Box}$
and
$o=t_0$. We have the following equalities:
\[ \begin{array}{l}
  \substitute{x}{u}{\ctxtapp{\ctxt{O}_1}{o}} =
\substitute{x}{u}{(\termapp{t_0}{t_1})} =
\ctxtapp{\substitute{x}{u}{(\termapp{\Box}{t_1})}}{\substitute{x}{u}{t_0}}
= \ctxtapp{\substitute{x}{u}{\ctxt{O}_1}}{\substitute{x}{u}{o}} \\
\replace{\beta}{\alpha}{s}{\ctxtapp{\ctxt{O}_2}{o}} =
\replace{\beta}{\alpha}{s}{(\termname{\alpha}{t_0})} = 
\ctxtapp{\replace{\beta}{\alpha}{s}{(\termname{\alpha}{\Box})}}{\replace{\beta}{\alpha}{s}{t_0}}
=
\ctxtapp{\replace{\beta}{\alpha}{s}{\ctxt{O}_2}}{\replace{\beta}{\alpha}{s}{t_0}}
\end{array} \] 

\ignore{
  \begin{corollary}
\label{c:special-cases-affecting-contexts} \mbox{}
\begin{itemize}
  \item Let $x \notin \fv{\ctxt{O}}$ and $\freeFor{u}{\ctxt{O}}$.
    Then $\substitute{x}{u}{\ctxtapp{\ctxt{O}}{o}} =
    \ctxtapp{\ctxt{O}}{\substitute{x}{u}{o}}$.
  \item Let $\alpha \notin \fn{\ctxt{O}}$ and $\freeFor{s}{\ctxt{O}}$.
    Then $\replace{\beta}{\alpha}{s}{\ctxtapp{\ctxt{O}}{o}} =
    \ctxtapp{\ctxt{O}}{\replace{\beta}{\alpha}{s}{o}}$.
\end{itemize}
\end{corollary}
}

\begin{remark}
\label{rem:projections_are_pure_terms}
$\project{o}$ is a pure term, \ie it has no explicit operators. 
\end{remark}

\begin{lemma}
\label{l:commutation}
Let $o \in \objects{\lmu}$. Then,
\begin{enumerate}
  \item \label{l:commutation-subs-subs}
  $\substitute{x}{u}{\substitute{y}{v}{o}} =
  \substitute{y}{\substitute{x}{u}{v}}{\substitute{x}{u}{o}}$,
  if $y \notin \fv{u}$. 
  \item \label{l:commutation-subs-repl}
  $\substitute{x}{u}{\replace{\beta}{\al}{s}{o}} =
  \replace{\beta}{\al}{\substitute{x}{u}{s}}{\substitute{x}{u}{o}}$,
  if $\al\notin \fn{u}$.
  \item \label{l:commutation-repl-subs}
  $\replace{\beta}{\al}{s}{\substitute{x}{u}{o}} =
  \substitute{x}{\replace{\beta}{\al}{s}{u}}{\replace{\beta}{\al}{s}{o}}$,
  if $x \notin \fv{s}$.
  \item \label{l:commutation-repl-repl-eq}
  $\replace{\beta}{\al}{s}{\replace{\beta'}{\al'}{s'}{o}} =
  \replace{\beta'}{\al'}{\replace{\beta}{\al}{s}{s'}}{\replace{\beta}{\al}{s}{o}}$,
  if $\al \neq \beta'$ and $\al' \notin \fn{s}$.
  \item \label{l:commutation-repl-repl-neq}
  $\replace{\beta}{\al}{s}{\replace{\beta'}{\al'}{s'}{o}} =
  \replace{\beta}{\al'}{\termpush{\replace{\beta}{\al}{s}{s'}}{s}}{\replace{\beta}{\al}{s}{o}}$,
  if $\al = \beta'$ and $\al' \notin \fn{s}$.
\end{enumerate}
\end{lemma}

\begin{proof}
By induction on $o$. 
\end{proof}

\ignore{
\begin{corollary}
\label{l:subst_normal_form_LM}
Let $o,s$ be pure objects.  Let $\gamma_1,\ldots,\gamma_{n-1}$ be distinct
fresh names. Let $\gamma_0 = \alpha$ and $\gamma_n = \alpha'$. Then 
$\replace{\gamma_n}{\gamma_{n-1}}{u_n}{\replace{\gamma_1}{\gamma_0}{u_1}{o}\ldots}
= \replace{\gamma_n}{\gamma_0}{s}{o}$.
\end{corollary}

\begin{proof}
Let us use $\replaceMulti{\gamma_i}{\gamma_{i-1}}{u_i}{o}$ to abbreviate the
left hand side of the equation.
\begin{itemize}
  \item $o = x$. Then $\replaceMulti{\gamma_i}{\gamma_{i-1}}{u_i}{x} = x =
  \replace{\gamma_n}{\gamma_0}{s}{x}$.

  \item $o = \termapp{t}{u}$. Then $$
\begin{array}{lll}
\replaceMulti{\gamma_i}{\gamma_{i-1}}{u_i}{(\termapp{t}{u})}                                           & = & \\
\termapp{\replaceMulti{\gamma_i}{\gamma_{i-1}}{u_i}{t}}{\replaceMulti{\gamma_i}{\gamma_{i-1}}{u_i}{u}} & = & \\
\termapp{\replace{\gamma_n}{\gamma_0}{s}{t}}{\replace{\gamma_n}{\gamma_0}{s}{u}}                       & = & (\ih) \\
\replace{\gamma_n}{\gamma_0}{s}{(\termapp{t}{u})}
\end{array} $$

  \item $o = \termabs{x}{t}$. Then $$
\begin{array}{lll}
\replaceMulti{\gamma_i}{\gamma_{i-1}}{u_i}{(\termabs{x}{t})} & = & \\
\termabs{x}{\replaceMulti{\gamma_i}{\gamma_{i-1}}{u_i}{t}}   & = & \\
\termabs{x}{\replace{\gamma_n}{\gamma_0}{s}{t}}              & = & (\ih) \\
\replace{\gamma_n}{\gamma_0}{s}{(\termabs{x}{t})}
\end{array} $$

  \item $o = \termcont{\beta}{t}$. Then $$
\begin{array}{lll}
\replaceMulti{\gamma_i}{\gamma_{i-1}}{u_i}{(\termcont{\beta}{t})} & = & \\
\termcont{\beta}{\replaceMulti{\gamma_i}{\gamma_{i-1}}{u_i}{t}}   & = & \\
\termcont{\beta}{\replace{\gamma_n}{\gamma_0}{u_i}{t}}            & = & (\ih) \\
\replace{\gamma_n}{\gamma_0}{s}{(\termcont{\beta}{t})}
\end{array} $$

  \item $o = \termname{\gamma}{t}$. We consider two cases:
  \begin{itemize}
    \item $\gamma = \gamma_0$ (other cases are not possible since the
    $\gamma_i$ are fresh for $i = 1 \ldots n-1$). Then $$
\begin{array}{lll}
\replaceMulti{\gamma_i}{\gamma_{i-1}}{u_i}{(\termname{\gamma_0}{t})}                                               & = & \\
\termname{\gamma_n}{(\replaceMulti{\gamma_i}{\gamma_{i-1}}{u_i}{t} \tconc \termpush{u_1}{\termpush{\ldots}{u_n}})} & = & \\
\termname{\gamma_n}{(\replaceMulti{\gamma_i}{\gamma_{i-1}}{u_i}{t} \tconc s)}                                      & = & \\
\termname{\gamma_n}{\replace{\gamma_n}{\gamma_0}{s}{t} \tconc s}                                                   & = & (\ih) \\
\replace{\gamma_n}{\gamma_0}{s}{(\termname{\gamma_{0}}{t})}
\end{array} $$

    \item $\gamma \neq \gamma_0$. Then $$
\begin{array}{lll}
\replaceMulti{\gamma_i}{\gamma_{i-1}}{u_i}{(\termname{\gamma}{t})} & = & \\
\termname{\gamma}{(\replaceMulti{\gamma_i}{\gamma_{i-1}}{u_i}{t})} & = & \\
\termname{\gamma}{\replace{\gamma_n}{\gamma_0}{s}{t}}              & = & (\ih) \\
\replace{\gamma_n}{\gamma_0}{s}{(\termname{\gamma}{t})}
\end{array} $$
  \end{itemize}
\end{itemize}
\end{proof}
}

\begin{lemma}
\label{l:projection_commutes_with_substitution}
Let $o \in \objects{\LMfull}$. Then $\substitute{x}{\project{u}}{\project{o}} =
\project{(\substitute{x}{u}{o})}$.
\end{lemma}

\begin{proof}
By induction on $o$ using Lemma~\ref{l:commutation}. We only show the
interesting cases.
\begin{itemize}
\item $o = \termsubs{y}{v}{t}$. Then $$
\begin{array}{l@{\enspace}l@{\enspace}l@{\enspace}l}
\substitute{x}{\project{u}}{\project{(\termsubs{y}{v}{t})}}                                        & = & 
\substitute{x}{\project{u}}{\substitute{y}{\project{v}}{\project{t}}}                              & =_{(L.\ref{l:commutation}.\ref{l:commutation-subs-subs})} \\
\substitute{y}{\substitute{x}{\project{u}}{\project{v}}}{\substitute{x}{\project{u}}{\project{t}}} & =_{(\ih)} &
\substitute{y}{\project{(\substitute{x}{u}{v})}}{\project{(\substitute{x}{u}{t})}}                 & = \\
\project{(\termsubs{y}{\substitute{x}{u}{v}}{\substitute{x}{u}{t}})}                               & = &
\project{(\substitute{x}{u}{\termsubs{y}{v}{t}})}
\end{array} $$

  \item $o = \termrepl{\beta}{\al}{s}{c}$. Then $$
\begin{array}{l@{\enspace}l@{\enspace}l@{\enspace}l}
\substitute{x}{\project{u}}{\project{(\termrepl{\beta}{\al}{s}{c})}}                                     & = & 
\substitute{x}{\project{u}}{\replace{\al}{\beta}{\project{s}}{\project{c}}}                              & =_{(L.\ref{l:commutation}.\ref{l:commutation-subs-repl})} \\
\replace{\al}{\beta}{\substitute{x}{\project{u}}{\project{s}}}{\substitute{x}{\project{u}}{\project{c}}} & =_{(\ih)} &
\replace{\al}{\beta}{\project{\substitute{x}{u}{s}}}{\project{(\substitute{x}{u}{c})}}                   & = \\
\project{(\termrepl{\beta}{\al}{\substitute{x}{u}{s}}{\substitute{x}{u}{c}})}                            & = &
\project{(\substitute{x}{u}{\termrepl{\beta}{\al}{s}{c}})}
\end{array} $$

\ignore{
  \item $o = \termrepl{\beta}{\alpha}{\termpush{t_1}{\termpush{\ldots}{t_n}}}{c}$ $$
\begin{array}{l@{\enspace}l@{\enspace}l@{\enspace}l}
\project{(\termrepl{\beta}{\alpha}{\termpush{t_1}{\termpush{\ldots}{t_n}}}{c})}\subsdos{x}{\project{u}}                                    & = &
\project{c}\ire{\alpha}{\gamma_1}{\project{t_1}} \ldots \ire{\gamma_{n-1}}{\beta}{\project{t_n}}\subsdos{x}{\project{u}}                   & =_{(L.\ref{l:commutation}.\ref{l:commutation-subs-repl})} \\
\project{c}\subsdos{x}{\project{u}}\ire{\alpha}{\gamma_1}{\project{t_1}\subsdos{x}{\project{u}}} \ldots \ire{\gamma_{n-1}}{\beta}{\project{t_n}\subsdos{x}{\project{u}}}                                                                                                                  & =_{(\ih)} &
\project{c\subsdos{x}{u}}\ire{\alpha}{\gamma_1}{\project{t_1\subsdos{x}{u}}} \ldots \ire{\gamma_{n-1}}{\beta}{\project{t_n\subsdos{x}{u}}} & = \\
\project{(\termrepl{\beta}{\alpha}{\termpush{t_1\subsdos{x}{u}}{\termpush{\ldots}{t_n\subsdos{x}{u}}}}{c\subsdos{x}{u}})}                  & = &
\project{(\termrepl{\beta}{\alpha}{\termpush{t_1}{\termpush{\ldots}{t_n}}}{c}\subsdos{x}{u})}
\end{array} $$
}
\end{itemize}
\end{proof}

\begin{lemma}
\label{l:lmsc_reduction_compatible_with_substitution}
Let $o, p, u \in \objects{\lmu}$. Then $o \Rew{\lmu} o'$ and $u \Rew{\lmu} u'$
implies
\begin{enumerate}
  \item \label{l:lmsc-compatible-subs-o} $\substitute{x}{u}{o} \Rew{\lmu}
  \substitute{x}{u}{o'}$.
  \item \label{l:lmsc-compatible-subs-u} $\substitute{x}{u}{p} \Rewn{\lmu}
  \substitute{x}{u'}{p}$.
\end{enumerate}
\end{lemma}

\begin{proof}
Point~\ref{l:lmsc-compatible-subs-o} by induction on $o \Rew{\lmu} o'$ and
point~\ref{l:lmsc-compatible-subs-u} is by induction on $p$.
\end{proof}

\ignore{
\begin{proof}
The first point is by induction on $p$. The second one is by induction on
$o \Rew{\lmu} o'$. We only show the interesting cases:
\begin{itemize}
  \item $o = \termsubs{y}{v}{t}\subsdos{x}{u} \Rew{\rS}
  t\subsdos{y}{v}\subsdos{x}{u} = o'$. $$
\begin{array}{lll}
\termsubs{y}{v}{t}\subsdos{x}{u}                & = & \\
\termsubs{y}{v\subsdos{x}{u}}{t\subsdos{x}{u}}  & \Rew{\rS} & \\
t\subsdos{x}{u}\subsdos{y}{v\subsdos{x}{u}}     & = & (L.\ref{l:commutation}.\ref{l:commutation-subs-subs}) \\
t\subsdos{y}{v}\subsdos{x}{u}
\end{array} $$

  \item $o = \termrepl{\beta}{\al}{s}{c}\subsdos{x}{u} \Rew{\rRfull}
  c\ire{\al}{\beta}{s}\subsdos{x}{u} = o'$. $$
\begin{array}{lll}
\termrepl{\beta}{\al}{s}{c}\subsdos{x}{u}               & = & \\
\termrepl{\beta}{\al}{s\subsdos{x}{u}}{c\subsdos{x}{u}} & \Rew{\rRfull} & \\
c\subsdos{x}{u}\ire{\al}{\beta}{s\subsdos{x}{u}}        & = & (L.\ref{l:commutation}.\ref{l:commutation-subs-repl}) \\
c\ire{\al}{\beta}{s}\subsdos{x}{u}
\end{array} $$
\end{itemize}
\end{proof}
}

\begin{lemma}
\label{l:lmsc_reduction_compatible_with_replacement}
Let $o, p, s \in \objects{\lmu}$. Then $o \Rew{\lmu} o'$ and $s \Rew{\lmu} s'$
implies
\begin{enumerate}
  \item \label{l:lmsc-compatible-repl-o} $\replace{\alpha}{\alpha'}{s}{o}
  \Rew{\lmu} \replace{\alpha}{\alpha'}{s}{o'}$.
  \item \label{l:lmsc-compatible-repl-s} $\replace{\alpha}{\alpha'}{s}{p}
  \Rewn{\lmu} \replace{\alpha}{\alpha'}{s'}{p}$.
\end{enumerate}
\end{lemma}

\begin{proof}
Point~\ref{l:lmsc-compatible-repl-o} by induction on $o \Rew{\lmu} o'$ and
point~\ref{l:lmsc-compatible-repl-s} is by induction on $p$.
\end{proof}

\begin{lemma}
\label{l:projection_of_replace_commutes_with_replaceplus}
Let $o, s \in \objects{\LMfull}$. Then
$\replace{\alpha'}{\alpha}{\project{s}}{\project{o}} =
\project{(\replace{\alpha'}{\alpha}{s}{o})}$  
\end{lemma}

\begin{proof}
By induction on $o$. We only show the interesting cases. 
\begin{itemize}
  \item $o = \termsubs{y}{v}{t}$ $$
\begin{array}{l@{\enspace}l@{\enspace}l@{\enspace}l}
\project{(\termsubs{y}{v}{t})}\ire{\al}{\al'}{\project{s}}                                  & = &
\project{t}\subsdos{y}{\project{v}}\ire{\al}{\al'}{\project{s}}                             & =_{(L.\ref{l:commutation}.\ref{l:commutation-repl-subs})} \\
\project{t}\ire{\al}{\al'}{\project{s}}\subsdos{y}{\project{v}\ire{\al}{\al'}{\project{s}}} & =_{(\ih)} &
\project{(t\ire{\al}{\al'}{s})}\subsdos{y}{\project{(v\ire{\al}{\al'}{s})}}                 & = \\
\project{(\termsubs{y}{v\ire{\al}{\al'}{s}}{}t\ire{\al}{\al'}{s})}                          & = &
\project{(\termsubs{y}{v}{t}\ire{\al}{\al'}{s})}
\end{array} $$

  %\item $o = \termname{\gamma}{c}$. There are two cases:
  %\begin{enumerate}
    %\item If $\gamma \neq \al$, then $$
%\begin{array}{lll}
%\project{(\termname{\gamma}{c})}\ire{\al}{\al'}{\project{s}} & = & \\ 
%(\termname{\gamma}{\project{c}})\ire{\al}{\al'}{\project{s}} & = & \\
%\termname{\gamma}{\project{c}\ire{\al}{\al'}{\project{s}}}   & = & (\ih) \\
%\termname{\gamma}{\project{c\ire{\al}{\al'}{s}}}             & = & \\
%\project{(\termname{\gamma}{c\ire{\al}{\al'}{s}})}           & = & \\
%\project{((\termname{\gamma}{c})\ire{\al}{\al'}{s})}
%\end{array} $$
%
    %\item If $\gamma = \al$, then $$
%\begin{array}{lll}
%\project{(\termname{\gamma}{c})}\ire{\al}{\al'}{\project{s}}                & = & \\
%(\termname{\gamma}{\project{c}})\ire{\al}{\al'}{\project{s}}                & = & \\
%\termname{\al'}{\project{c}\ire{\al}{\al'}{\project{s}} \tconc \project{s}} & = & (\ih) \\
%\termname{\al'}{\project{c\ire{\al}{\al'}{s}} \tconc \project{s}}           & = & \\
%\project{(\termname{\al'}{c\ire{\al}{\al'}{s} \tconc s})}                   & = & \\
%\project{((\termname{\gamma}{c})\ire{\al}{\al'}{s})}
%\end{array} $$
  %\end{enumerate}

  \item $o = \termrepl{\beta}{\gamma}{\termemst}{c}$. There are two cases:
  \begin{enumerate}
    \item If $\beta = \al$, then $$\kern-2em
\begin{array}{l@{\enspace}l@{\enspace}l@{\enspace}l@{\enspace}}
\project{(\termrepl{\beta}{\gamma}{\termemst}{c})}\ire{\al}{\al'}{\project{s}}                                                                & = \\
\project{c}\ire{\gamma}{\beta}{\termemst}\ire{\al}{\al'}{\project{s}}                                                                         & =_{(L.\ref{l:commutation}.\ref{l:commutation-repl-repl-neq})} \\
\project{c}\ire{\al}{\al'}{\project{s}}\ire{\gamma}{\al'}{\termpush{\termemst\ire{\al}{\al'}{\project{s}}}{\project{s}}}                      & =_{(\ih)} \\
\project{c\ire{\al}{\al'}{s}}\ire{\gamma}{\al'}{\termpush{\termemst\ire{\al}{\al'}{\project{s}}}{\project{s}}}                                & = \\
\project{c\ire{\al}{\al'}{s}}\ire{\gamma}{\al'}{\project{s}}                                                                                  & =_{(\beta' \text{ fresh})} \\
\project{c\ire{\al}{\al'}{s}}\ire{\beta'}{\al'}{\termemst} \ire{\gamma}{\al'}{\termpush{\project{s}\ire{\beta'}{\al'}{\termemst}}{\termemst}} & =_{(L.\ref{l:commutation}.\ref{l:commutation-repl-repl-neq})} \\
\project{c\ire{\al}{\al'}{s}} \ire{\gamma}{\beta'}{\project{s}} \ire{\beta'}{\al'}{\termemst}                                                 & = \\
\project{(\termrepl{\al'}{\beta'}{\termemst}{\termrepl{\beta'}{\gamma}{s}{c\ire{\al}{\al'}{s}}})}                                             & =_{(\beta' \text{ fresh})} \\
\project{(\termrepl{\beta}{\gamma}{\termemst}{c}\ire{\al}{\al'}{s})}
\end{array} $$

    \item If $\beta \neq \al$, then $$
\begin{array}{l@{\enspace}l@{\enspace}l@{\enspace}l@{\enspace}}
\project{(\termrepl{\beta}{\gamma}{\termemst}{c})}\ire{\al}{\al'}{\project{s}}                    & = &
\project{c}\ire{\gamma}{\beta}{\termemst}\ire{\al}{\al'}{\project{s}}                             & =_{(L.\ref{l:commutation}.\ref{l:commutation-repl-repl-eq})} \\
\project{c}\ire{\al}{\al'}{\project{s}}\ire{\gamma}{\beta}{\termemst\ire{\al}{\al'}{\project{s}}} & = &
\project{c}\ire{\al}{\al'}{\project{s}}\ire{\gamma}{\beta}{\termemst}                             & =_{(\ih)} \\
\project{(c\ire{\al}{\al'}{s})} \ire{\gamma}{\beta}{\termemst}                                    & = &
\project{(\termrepl{\beta}{\gamma}{\termemst}{c\ire{\al}{\al'}{s}})}                              & = \\
\project{(\termrepl{\beta}{\gamma}{\termemst}{c}\ire{\al}{\al'}{s})}
\end{array} $$
  \end{enumerate}

  \item $o = \termrepl{\beta}{\gamma}{s_0}{c}$, where $s_0 \neq \termemst$
  \begin{enumerate}
    \item If $\beta = \al$, then $$
\begin{array}{l@{\enspace}l@{\enspace}l@{\enspace}l@{\enspace}}
\project{(\termrepl{\beta}{\gamma}{s_0}{c})}\ire{\al}{\al'}{\project{s}}                                                     & = &
\project{c}\ire{\gamma}{\beta}{\project{s_0}}\ire{\al}{\al'}{\project{s}}                                                    & =_{(L.\ref{l:commutation}.\ref{l:commutation-repl-repl-neq})} \\
\project{c}\ire{\al}{\al'}{\project{s}}\ire{\gamma}{\al'}{\termpush{\project{s_0}\ire{\al}{\al'}{\project{s}}}{\project{s}}} & =_{(\ih)} &
\project{c\ire{\al}{\al'}{s}}\ire{\gamma}{\al'}{\termpush{\project{s_0\ire{\al}{\al'}{s}}}{\project{s}}}                     & = \\
\project{(\termrepl{\al'}{\gamma}{\termpush{s_0\ire{\al}{\al'}{s}}{s}}{c\ire{\al}{\al'}{s}})}                                & = &
\project{(\termrepl{\beta}{\gamma}{s_0}{c}\ire{\al}{\al'}{s})}
\end{array} $$

    \item If $\beta \neq \al$, then $$
\begin{array}{l@{\enspace}l@{\enspace}l@{\enspace}l@{\enspace}}
\project{(\termrepl{\beta}{\gamma}{s_0}{c})}\ire{\al}{\al'}{\project{s}}                              & = &
\project{c}\ire{\gamma}{\beta}{\project{s_0}}\ire{\al}{\al'}{\project{s}}                             & =_{(L.\ref{l:commutation}.\ref{l:commutation-repl-repl-eq})} \\
\project{c}\ire{\al}{\al'}{\project{s}}\ire{\gamma}{\beta}{\project{s_0}\ire{\al}{\al'}{\project{s}}} & =_{(\ih)} &
\project{c\ire{\al}{\al'}{s}} \ire{\gamma}{\beta}{\project{s_0\ire{\al}{\al'}{s}}}                    & = \\
\project{(\termrepl{\beta}{\gamma}{s_0\ire{\al}{\al'}{s}}{c\ire{\al}{\al'}{s}})}                      & = &
\project{(\termrepl{\beta}{\gamma}{s_0}{c}\ire{\al}{\al'}{s})}
\end{array} $$
  \end{enumerate}
\end{itemize}
\end{proof}

\begin{lemma}
\label{l:simulation-projection}\mbox{} 
\begin{enumerate}
  \item Let $o \in \objects{\LMfull}$. Then, $o\Rewn{\LMfull} \project{o}$.
  \item Let $o \in \objects{\lmu}$. Then $o \Rew{\lmu} o'$ implies $o
  \Rewplus{\LMfull} o'$. 
  \item Let $o \in \objects{\LMfull}$. Then, $o \Rew{\LMfull} o'$ implies
  $\project{o} \Rewn{\lmu} \project{o'}$.
\end{enumerate}
\end{lemma}

\begin{proof}
Item 1 is by induction on $o$.
\begin{itemize}
  \item $o = x$. Then $\project{x} = x$ and the result is immediate.
  \item $o = \termapp{t}{u}$. By the \ih $t\Rewn{\LMfull} \project{t}$ and
  $u\Rewn{\LMfull} \project{u}$. Hence $\termapp{t}{u}\Rewn{\LMfull}
  \termapp{\project{t}}{\project{u}} = \project{(\termapp{t}{u})}$.

  \item $o = \termabs{x}{t}$. By the \ih $t\Rewn{\LMfull}\project{t}$.
  Therefore $\termabs{x}{t}\Rewn{\LMfull}\termabs{x}{\project{t}} =
  \project{(\termabs{x}{t})}$.

  \item $o = \termsubs{x}{u}{t}$. By the \ih $t\Rewn{\LMfull}\project{t}$ and
  $u\Rewn{\LMfull}\project{u}$. Thus $\termsubs{x}{u}{t} \Rewn{\LMfull}
  \termsubs{x}{\project{u}}{\project{t}} \Rew{\rS}
  \substitute{x}{\project{u}}{\project{t}} = \project{\termsubs{x}{u}{t}}$. 

  \item $o = \termcont{\alpha}{c}$. By the \ih $c\Rewn{\LMfull}\project{c}$.
  Then $\termcont{\alpha}{c}\Rewn{\LMfull}\termcont{\alpha}{\project{c}} =
  \project{(\termcont{\alpha}{c})}$.

  \item $o = \termname{\alpha}{t}$. By the \ih $t\Rewn{\LMfull}\project{t}$.
  Then $\termname{\alpha}{t}\Rewn{\LMfull} \termname{\alpha}{\project{t}} =
  \project{(\termname{\alpha}{t})}$.

%  \item $o = c\neren{\beta}{\alpha}$. Then by \ih $c\Rewn{\LMfull}\project{c}$.
%  Thus $c\neren{\beta}{\alpha} \Rewn{\rRfull} \project{c}\neren{\beta}{\alpha}
%  \Rew{\LMfull} \project{c}\replaceo{\alpha}{\beta}{\termemst} =
%  \project{(c\neren{\beta}{\alpha})}$.
%  este caso esta comprendido en el de abajo
  
  \item $o = \termrepl{\alpha'}{\alpha}{s}{c}$. By the \ih $s\Rewn{\LMfull}
  \project{s}$ and $c \Rewn{\LMfull} \project{c}$. Then
  $\termrepl{\alpha'}{\alpha}{s}{c} \Rewn{\LMfull}
  \termrepl{\alpha'}{\alpha}{\project{s}}{\project{c}}\Rew{\rRfull}
  \replace{\alpha'}{\alpha}{\project{s}}{\project{c}} =
  \projectSort{c}{(\termrepl{\alpha'}{\alpha}{s}{c})}$.
\end{itemize}

Item 2 is by induction on the context $\ctxt{O}$ where the step $o \Rew{\lmu}
o'$ takes place.
\begin{itemize}
  \item $\ctxt{O} = \Box$. We have two cases:
  \begin{itemize}
    \item $o = \termapp{(\termabs{x}{t})}{u}\rrule{\beta}
    \substitute{x}{u}{t} = o'$. Then $o \rrule{\rB}  \termsubs{x}{u}{t}
    \rrule{\rS} \substitute{x}{u}{t}$ and we conclude.

    \item $o = (\termcont{\alpha_0}{c}) u \rrule{\mu}
    \termcont{\alpha_1}{\replace{\alpha_1}{\alpha_0}{u}{c}} = o'$ and
    $\alpha_1 \notin \fn{c,s,\alpha}$. Then $$(\termcont{\alpha_0}{c}) u \\
    \rrule{\rM} \termcont{\alpha_1}{\termrepl{\alpha_1}{\alpha_0}{u}{c}}
    \rrule{\rRfull} \termcont{\alpha_1}{\replace{\alpha_1}{\alpha_0}{u}{c}}$$
  \end{itemize}

  \item $\ctxt{O} = \termapp{\ctxt{C}}{u}$. Then $o = \termapp{o_1}{u}$,
  $o' = \termapp{o_1'}{u}$ and $o_1 \Rew{\lmu} o_1'$. We conclude from the \ih\ 
 
  \item $\ctxt{O} = \termapp{t}{\ctxt{C}}$, $\ctxt{O} = \termabs{x}{\ctxt{C}}$
  and $\ctxt{O} = \termcont{\alpha}{\ctxt{K}}$, $\ctxt{O} =
  \termname{\alpha}{\ctxt{C}}$ are similar.
\end{itemize}

Item 3 is by induction on the context where the step $o \Rew{\LMfull} o'$
takes place. 
\begin{itemize}
  \item $\ctxt{O} = \Box$. There are three four possible cases:
  \begin{itemize}
    \item $\rB$. $o = \termapp{\ctxtapp{\ctxt{L}}{\termabs{x}{t}}}{u}$ and
    $o' = \ctxtapp{\ctxt{L}}{\termsubs{x}{u}{t}}$ and $\ctxt{L} =
    \termsubs{y_n}{v_n}{\termsubs{y_1}{v_1}{\Box}\ldots}$. Then $$
\begin{array}{l@{\enspace}l@{\enspace}l@{\enspace}l@{\enspace}}
\project{(\termapp{\ctxtapp{\ctxt{L}}{\termabs{x}{t}}}{u})}                                                               & = &
\termapp{\substitute{y_n}{\project{v_n}}{\substitute{y_1}{\project{v_1}}{(\termabs{x}{\project{t}})}\ldots}}{\project{u}} & = \\
\termapp{(\termabs{x}{\substitute{y_n}{\project{v_n}}{\substitute{y_1}{\project{v_1}}{\project{t}}\ldots}})}{\project{u}} & \rrule{\beta} &
\substitute{x}{\project{u}}{\substitute{y_n}{\project{v_n}}{\substitute{y_1}{\project{v_1}}{\project{t}}\ldots}}          & = \\
\substitute{y_n}{\project{v_n}}{\substitute{y_1}{\project{v_1}}{\substitute{x}{\project{u}}{\project{t}}}\ldots}          & = &
\project{\ctxtapp{\ctxt{L}}{\termsubs{x}{u}{t}}}
\end{array} $$

    \item $\rS$. $o = \termsubs{x}{u}{t}$ and $o' = \substitute{x}{u}{t}$. Then
    $\project{(\termsubs{x}{u}{t} )} = \substitute{x}{\project{u}}{\project{t}}
    = \project{(\substitute{x}{u}{t})}$ by
    Lem.~\ref{l:projection_commutes_with_substitution}.

    \item $\rM$. $o = \termapp{\ctxtapp{\ctxt{L}}{\termcont{\alpha}{c}}}{u}$ and
    $o'= \ctxtapp{\ctxt{L}}{\termcont{\alpha'}{\termrepl{\alpha'}{\alpha}{u}{c}}}$
    with $\alpha' \notin \fn{c,u,\alpha}$ and $\ctxt{L} =
    \termsubs{y_n}{v_n}{\termsubs{y_1}{v_1}{\Box}\ldots}$. Then $$
\begin{array}{l@{\enspace}l@{\enspace}l@{\enspace}l@{\enspace}}
\project{(\termapp{\ctxtapp{\ctxt{L}}{\termcont{\alpha}{c}}}{u})}                                                                                 & = &
\termapp{\substitute{y_n}{\project{v_n}}{\substitute{y_1}{\project{v_1}}{(\termcont{\alpha}{\project{c}})}\ldots}}{\project{u}}                   & = \\
\termapp{(\termcont{\alpha}{\substitute{y_n}{\project{v_n}}{\substitute{y_1}{\project{v_1}}{\project{c}}\ldots}})}{\project{u}}                   & \rrule{\mu} &
\termcont{\alpha'}{\replace{\alpha'}{\alpha}{\project{u}}{\substitute{y_n}{\project{v_n}}{\substitute{y_1}{\project{v_1}}{\project{c}}\ldots}}}   & = \\
\substitute{y_n}{\project{v_n}}{\substitute{y_1}{\project{v_1}}{(\termcont{\alpha'}{\replace{\alpha'}{\alpha}{\project{u}}{\project{c}}})}\ldots} & = &
\project{(\ctxtapp{\ctxt{L}}{\termcont{\alpha'}{\termrepl{\alpha'}{\alpha}{u}{c}}})}
\end{array} $$

    \item $\rRfull$. $o = \termrepl{\al'}{\al}{s}{c}$ and $o' =
    c\ire{\al}{\al'}{s}$. Then $\project{\termrepl{\al'}{\al}{s}{c}} =
    \project{c}\ire{\al}{\al'}{\project{s}}
    = \project{c\ire{\al'}{\al}{s}}$ by
    Lem.~\ref{l:projection_of_replace_commutes_with_replaceplus}.
  \end{itemize}

  \item $\ctxt{O} = \termapp{\ctxt{C}}{u}$, $\ctxt{O} = \termapp{t}{\ctxt{C}}$,
  $\ctxt{O} = \termabs{x}{\ctxt{C}}$ and $\ctxt{O} =
  \termcont{\alpha}{\ctxt{K}}$ are immediate from the \ih\ 

  \item $\ctxt{O} = \termsubs{x}{u}{\ctxt{C}}$. Then $o = \termsubs{x}{u}{o_1}$
  and $o'= \termsubs{x}{u}{o_1'}$ and $o_1 \Rew{\LMfull} o_1'$. By the \ih
  $\project{o_1}\Rewn{\lmu}\project{o_1'}$. We conclude that
  $$
\begin{array}{l}
\project{(\termsubs{x}{u}{o_1})}             = 
\substitute{x}{\project{u}}{\project{o_1}}   \Rewn{\lmu} 
\substitute{x}{\project{u}}{\project{o_1'}}  =_{(L.\ref{l:lmsc_reduction_compatible_with_substitution}.\ref{l:lmsc-compatible-subs-o})} 
\project{(\termsubs{x}{u}{o_1'})}
\end{array} $$

  \item $\ctxt{O} = \termsubs{x}{\ctxt{C}}{t}$. Then $o = \termsubs{x}{o_1}{t}$
  and $o'= \termsubs{x}{o_1'}{t}$ and $o_1 \Rew{\LMfull} o_1'$. By the \ih
  $\project{o_1}\Rewn{\lmu}\project{o_1'}$. We conclude that $$
\begin{array}{l}
\project{(\termsubs{x}{o_1}{t})}             = 
\substitute{x}{\project{o_1}}{\project{t}}   \Rewn{\lmu} 
\substitute{x}{\project{o_1'}}{\project{t}}  =_{(L.\ref{l:lmsc_reduction_compatible_with_substitution}.\ref{l:lmsc-compatible-subs-u})} 
\project{(\termsubs{x}{o_1'}{t})}
\end{array} $$

  \item $\ctxt{O} = \boxdot$. Then it is a $\rRfull$ step and $o =
  \termrepl{\alpha'}{\alpha}{s}{c}$ and $o' = \replace{\alpha'}{\alpha}{s}{c}$.
  Then $$
\begin{array}{l}
\project{(\termrepl{\alpha'}{\alpha}{s}{c})}         =  
\replace{\alpha'}{\alpha}{\project{s}}{\project{c}}  =_{(L.\ref{l:projection_of_replace_commutes_with_replaceplus})}
\project{(\replace{\alpha'}{\alpha}{s}{c})}
\end{array} $$

  \item $\ctxt{O} = \termname{\alpha}{\ctxt{C}}$. This case is immediate from
  the \ih\

  \item $\ctxt{O} = \termrepl{\alpha'}{\alpha}{s}{\ctxt{K}}$. Then $o =
  \termrepl{\alpha'}{\alpha}{s}{c}$ and $c\Rew{\LMfull}c'$ and $o' =
  \termrepl{\alpha'}{\alpha}{s}{c'}$. We reason as follows: $$
\begin{array}{l}
\project{(\termrepl{\alpha'}{\alpha}{s}{c})}          = 
\replace{\alpha'}{\alpha}{\project{s}}{\project{c}}   \Rew{\lmu}{}_{(L.\ref{l:lmsc_reduction_compatible_with_replacement}.\ref{l:lmsc-compatible-repl-o})} 
\replace{\alpha'}{\alpha}{\project{s}}{\project{c'}}  =  
\project{(\termrepl{\alpha'}{\alpha}{s}{c'})}
\end{array} $$
 
  \item $\ctxt{O} = \termrepl{\alpha'}{\alpha}{\ctxt{S}}{c}$ with $\alpha'$
  fresh. Then $o = \termrepl{\alpha'}{\alpha}{t}{c}$ and $t \Rew{\LMfull} t'$
  and $o' = \termrepl{\alpha'}{\alpha}{t'}{c}$. We reason as follows: $$
\begin{array}{lll}
\project{(\termrepl{\alpha'}{\alpha}{t}{c})}          =  
\replace{\alpha'}{\alpha}{\project{t}}{\project{c}}   \Rewn{\lmu} {}_{(L.\ref{l:lmsc_reduction_compatible_with_replacement}.\ref{l:lmsc-compatible-repl-s})}
\replace{\alpha'}{\alpha}{\project{t'}}{\project{c}}  = 
\project{(\termrepl{\alpha'}{\alpha}{t'}{c})}
\end{array} $$
\end{itemize}
\end{proof}

%% Concluence Theorem
\gettoappendix{t:confluence_of_LM}

\begin{proof}
The proof can be graphically depicted as follows:
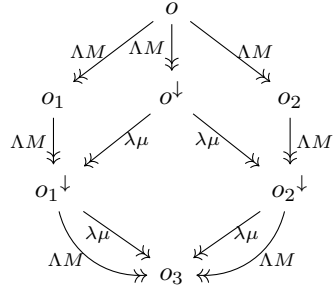
\begin{center}
\begin{tikzcd}
  & o \arrow[->>]{dl}[left]{\LMfull} \arrow[->>]{dr}[right]{\LMfull} \arrow[->>]{d}[left]{\LMfull} \\
o_1 \arrow[->>]{d}[left]{\LMfull} 
  & \project{o} \arrow[->>]{dr}[left]{\lmu} \arrow[->>]{dl}[right]{\lmu}
  & o_2 \arrow[->>]{d}[right]{\LMfull}  \\
\project{o_1} \arrow[->>]{dr}[left]{\lmu} \arrow[->>, bend right=40]{dr}[left]{\LMfull}
  &
  & \project{o_2} \arrow[->>]{dl}[right]{\lmu} \arrow[->>, bend left=40]{dl}[right]{\LMfull} \\
  & o_3
\end{tikzcd}
\end{center}
The arrows $o \Rewn{\LMfull} o_i\ (i=1,2)$ are the hypothesis of the theorem.
The three vertical arrows of the form $p \Rewn{\LMfull} \project{p}$
are given by Lem.~\ref{l:simulation-projection}-(1).
The arrows $\project{o} \Rewn{\LMfull} \project{o_i}\ (i=1,2)$ are give by
Lem.~\ref{l:simulation-projection}-(3).
The arrows $\project{o_i} \Rewn{\lmu} o_3\ (i=1,2)$
are given by the confluence property of $\lmu$~\cite{Parigot93}.
And the arrows $\project{o_i} \Rewn{\LMfull} o_3\ (i=1,2)$
are given by Lem.~\ref{l:simulation-projection}-(2). 
\end{proof}

%%% Local Variables:
%%% mode: latex
%%% TeX-master: "main"
%%% End:

  %%%%%%%%%%%%%%%%%%%%%%%%%%%%%%%%%%%%%%%%%%%%%%%%%%%%%%%%%%%%%%%%%%%%%%%%%%%%%%%
\section{Appendix: Preservation of Types}
\label{a:preservation-types}
%%%%%%%%%%%%%%%%%%%%%%%%%%%%%%%%%%%%%%%%%%%%%%%%%%%%%%%%%%%%%%%%%%%%%%%%%%%%%%%

\gettoappendix{l:relevance}

\begin{proof}
 Straightforward induction on typing derivations.
\end{proof}

\gettoappendix{l:PTLaurent}

\begin{proof}
The proof is by induction on the relation $\eqlaurent$. All the cases are
straightforward by using the variable and names conditions used in the
definition of the relation.
\end{proof}

\begin{lemma}
\label{l:application} \mbox{}
\begin{itemize}
  \item If $\Gam \vdash u:S \tarrow B \mid \Del$ and $\Gam' \vdash s : S \mid
  \Del'$, then $\Gam \cup \Gam' \vdash \termconc{u}{s} : B \mid \Del \cup
  \Del'$.
  \item If $\Gam^* \vdash \termconc{u}{s} : B \mid \Del^*$, then there exist
  $S$, $\Gam, \Gam', \Del, \Del'$ such that $\Gam^* = \Gam \cup \Gam'$, $\Del^*
  = \Del \cup \Del'$, $\Gam \vdash u : S \tarrow B \mid \Del$ and $\Gam' \vdash
  s : S \mid \Del'$. 
\end{itemize}
\end{lemma}

\begin{proof}
Straightforward induction on $s$.
\end{proof}

\begin{lemma}
\label{l:substitution}
Let $o\in\objects{\LMfull}$ and $u\in \terms{\LMfull}$. If $\Gam, (x : B)^{\esta} \vdash o : \type \mid \Del$ and $\Gam' \vdash u : B
\mid \Del'$ then $\Gam^\ast \vdash \substitute{x}{u}{o} : \type \mid \Del^\ast$
where $\Gam^\ast \subseteq \Gam \cup \Gam'$ and $\Del^\ast \subseteq \Del \cup
\Del'$.
\end{lemma}

\begin{proof}
By induction on $o$.
\begin{itemize}
  \item $o = x$. Then $\substitute{x}{u}{x} = u$. The statement holds with
  $\Gam = \emptyset = \Del$, $\Gam^\ast = \Gam'$ and $\Del^\ast = \Del'$.
  
  \item $o = y$ with $y \neq x$. Then $\substitute{x}{u}{y} = y$,
  $(x:B)^{\esta}$ is empty, $\Gam=y:C$ for some $C$, and $\Del=\emptyset$. The
  statement holds with $\Gam^\ast = \Gam$ and $\Del^\ast = \emptyset$.
  
  \item $o = \termapp{t}{r}$. Then $\substitute{x}{u}{(\termapp{t}{r})} =
  \termapp{\substitute{x}{u}{t}}{\substitute{x}{u}{r}}$. This holds by the \ih
  
  \item $o = \termabs{y}{t}$. Then $\substitute{x}{u}{(\termabs{y}{t})} =
  \termabs{y}{\substitute{x}{u}{t}}$ where $x \neq y$ and $y \notin \fv{u}$.
  The typing derivation of $o$ has the following form with $T =
  \typefunc{A'}{B'}$ \[
  \hfill
  \Rule{\Gam, (x:B)^{\esta}, (y:A')^{\esta} \vdash t:B' \mid \Del}
       {\Gam, (x:B)^{\esta} \vdash \termabs{y}{t}:\typefunc{A'}{B'} \mid \Del}
       {}
  \hfill
  \] The \ih\ gives $\Gam^{\ast\ast} \vdash \substitute{x}{u}{t}:B' \mid
  \Del^\ast$ with $\Gam^{\ast\ast} \subseteq \Gam, (y:A')^{\esta} \cup \Gam'$
  and $\Del^\ast \subseteq \Del \cup \Del'$. Moreover, since $x \neq y$ and
  $y \notin \fv{u}$, then $y \in \fv{t}$ iff $y \in \fv{\substitute{x}{u}{t}}$.
  Thus, by Lem.~\ref{l:relevance}, $\Gam^{\ast\ast} = \Gam^\ast,
  (y:A')^{\esta}$ with $\Gam^\ast \subseteq \Gam \cup \Gam'$. Finally, we
  conclude by rule $(\abs)$, $\Gam^\ast \vdash
  \substitute{x}{u}{(\termabs{y}{t})}:T \mid \Del^\ast$.
  
  \item $o = \termcont{\alpha}{c}$. Then
  $\substitute{x}{u}{(\termcont{\alpha}{c})} =
  \termcont{\alpha}{\substitute{x}{u}{c}}$ where $\alpha \notin \fn{u}$. This
  case follows from \ih\ and Lem.~\ref{l:relevance} in a similar way to the
  previous one.
  
  \item $o = \termsubs{y}{r}{t}$. Then $\substitute{x}{u}{(\termsubs{y}{r}{t})}
  = \termsubs{y}{\substitute{x}{u}{r}}{\substitute{x}{u}{t}}$ where $y \neq x$
  and $y \notin \fv{u}$. This case follows from \ih\ and Lem.~\ref{l:relevance}
  in a similar way to the previous one.
  
  \item $o = \termname{\alpha}{t}$. Then
  $\substitute{x}{u}{(\termname{\alpha}{t})} =
  \termname{\alpha}{\substitute{x}{u}{t}}$. This holds by the \ih
  
  \item $o = \termrepl{\alpha'}{\alpha}{s}{c}$. Then
  $\substitute{x}{u}{(\termrepl{\alpha'}{\alpha}{s}{c})} =
  \termrepl{\alpha'}{\alpha}{\substitute{x}{u}{s}}{\substitute{x}{u}{c}}$ where
  $\alpha, \alpha' \notin \fn{u}$. The typing derivation of $o$ has the
  following form \[
  \hfill
  \Rule{\Gam_1, (x:B)^{\esta} \vdash c \mid  \Del_1, (\alpha : S \tarrow A)^{\esta}, (\alpha' : A)^{\esta}
        \quad
        \Gam'_1, (x:B)^{\esta} \vdash s : S  \mid \Del'_1, (\alpha' : A)^{\esta}
       }
       {\Gam_1 \cup \Gam'_1, (x:B)^{\esta} \vdash \termrepl{\alpha'}{\alpha}{s}{c} \mid  \Del_1 \cup \Del'_1, \alpha' : A}
       {} 
  \hfill
  \] where $\Gam = \Gam_1 \cup \Gam'_1$ and $\Del = \Del_1 \cup \Del'_1,
  \alpha' : A$. \\
  The \ih gives $\Gam^{**} \vdash \substitute{x}{u}{c} \mid \Del^{**}$, where
  $\Gam^{**} \subseteq \Gam_1 \cup \Gam'$ and $\Del^{**} \subseteq \Del_1 \cup
  \Del', (\alpha : S \tarrow A)^{\esta}, (\alpha' : A)^{\esta}$, as well as
  $\Gam^{***} \vdash \substitute{x}{u}{s} : S \mid \Del^{***}$, where
  $\Gam^{***} \subseteq \Gam'_1 \cup \Gam'$ and $\Del^{***} \subseteq \Del'_1
  \cup \Del', (\alpha' : A)^{\esta}$. Then rule $(\trepl)$ gives $\Gam^{**}
  \cup \Gam^{***} \vdash
  \termrepl{\alpha'}{\alpha}{\substitute{x}{u}{s}}{\substitute{x}{u}{c}}
  \mid (\Del^{**} \setminus (\alpha : S \tarrow A)) \cup \Del^{***}, \alpha' :
  A$. We conclude with
  \begin{itemize}
    \item $\Gam^{*} = \Gam^{**} \cup \Gam^{***} \subseteq (\Gam_1 \cup \Gam')
    \cup (\Gam'_1 \cup \Gam') = \Gam \cup \Gam'$, and
    \item $\Del^{*} = (\Del^{**} \setminus (\alpha : S \tarrow A)) \cup
    \Del^{***}, \alpha' : A \subseteq (\Del_1 \cup \Del') \cup (\Del'_1 \cup \Del'), \alpha' : A
    = \Del \cup \Del'$.
  \end{itemize}
  
  \item $o = \termemst$. Then $\substitute{x}{u}{\termemst} = \termemst$. The
  statement holds with $\Gam^\ast = \emptyset = \Del^\ast$ and $(x:B)^{\esta}$
  equal to empty.
  
  \item $o = \termpush{t}{s}$. Then $\substitute{x}{u}{(\termpush{t}{s})} =
  \termpush{\substitute{x}{u}{t}}{\substitute{x}{u}{s}}$. This holds by the \ih
\end{itemize}
\end{proof}

%%% Local Variables:
%%% mode: latex
%%% TeX-master: "main"
%%% End:

\begin{lemma}
\label{l:replacement}
Let $o\in \objects{\LMfull}$. If $\Gam \vdash o : \type \mid \Del, (\alpha : S \tarrow B)^{\esta}$ and $\Gam'
\vdash s : S \mid \Del'$ ($\al \notin \Del'$), then $\Gam^\ast \vdash
\replace{\alpha'}{\alpha}{s}{o} : \type \mid \Del^\ast$, where $\Gam^\ast
\subseteq \Gam \cup \Gam'$ and $\Del^\ast \subseteq \Del \cup \Del' \cup
(\alpha' : B)$.
\end{lemma}

\begin{proof} We reason by induction on $o$. 
  \begin{itemize}
    \item $o = x$. Then $\replace{\alpha'}{\alpha}{s}{x} = x$. The statement
    holds for $\Gam^* = \Gam$ and $\Del^*  = \emptyset$ and
    $(\al:S\tarrow B)^{\esta}$ equal to empty.
    
    \item $o = \termapp{t}{u}$. Then
    $\replace{\alpha'}{\alpha}{s}{(\termapp{t}{u})}  =
    \termapp{\replace{\alpha'}{\alpha}{s}{t}}{\replace{\alpha'}{\alpha}{s}{u}}$.
    This holds by the \ih
    
    \item $o= \termabs{x}{t}$. Then
    $\replace{\alpha'}{\alpha}{s}{(\termabs{x}{t})} =
    \termabs{x}{\replace{\alpha'}{\alpha}{s}{t}}$, where $x \notin \fv{s}$.
    This holds by the \ih
    
    \item $o =\termcont{\beta}{c}$. Then
    $\replace{\alpha'}{\alpha}{s}{(\termcont{\beta}{c})} =
    \termcont{\beta}{\replace{\alpha'}{\alpha}{s}{c}}$, where $\beta \notin
    \fn{s, \alpha, \al'}$. This holds by the \ih

    \item $o= \termsubs{x}{u}{t}$. Then
    $\replace{\alpha'}{\alpha}{s}{\termsubs{x}{u}{t}} =
    \termsubs{x}{\replace{\alpha'}{\alpha}{s}{u}}{\replace{\alpha'}{\alpha}{s}{t}}$,
    where $x \notin \fv{s}$. This holds by the \ih
    
    \item $o=\termname{\alpha}{t}$. Then
    $\replace{\alpha'}{\alpha}{s}{(\termname{\alpha}{t})} =
    \termname{\alpha'}{(\termconc{\replace{\alpha'}{\alpha}{s}{t}}{s})}$. The
    typing derivation for $o$ has the following form: \[
    \hfill
    \Rule{\Gam \vdash t:S \tarrow B \mid \Del, (\al: S \tarrow B)^{\leq 1}}
         {\Gam \vdash \termname{\alpha}{t} \mid \Del, \al: S \tarrow B}
         {}
    \hfill
    \] The \ih gives $\Gam^{**} \vdash \replace{\alpha'}{\alpha}{s}{t} : S
    \tarrow B \mid \Del^{**}$, where $\Gam^{**} \subseteq \Gam \cup \Gam'$ and
    $\Del^{**} \subseteq \Del  \cup \Del'\cup (\alpha':B)$.
    Lemma~\ref{l:application} gives $\Gam^{**} \cup \Gam' \vdash
    \termconc{\replace{\alpha'}{\alpha}{s}{t}}{s} :  B \mid \Del^{**} \cup
    \Del'$. Then we get $\Gam^{**} \cup \Gam' \vdash
    \termname{\al'}{\termconc{\replace{\alpha'}{\alpha}{s}{t}}{s}} \mid
    \Del^{**} \cup \Del', \al' : B$ by rule $(\nametr)$. We let $\Gam^* =
    \Gam^{**} \cup \Gam'$ and $\Del^* = \Del^{**} \cup \Del' \cup (\al':B)$.
    We conclude since 
    \begin{itemize}
      \item $\Gam^* = \Gam^{**} \cup \Gam' \subseteq \Gam \cup \Gam'\cup \Gam'
      = \Gam \cup \Gam'$, and
      \item $\Del^* = \Del^{**} \cup \Del' \subseteq \Del \cup \Del' \cup
      (\al':B) \cup \Del' = \Del \cup \Del' \cup (\al':B)$.
    \end{itemize}
    
    \item $o=\termname{\beta}{t}$, where $\alpha \neq \beta$. Then
    $\replace{\alpha'}{\alpha}{s}{(\termname{\beta}{t})} =
    \termname{\beta}{\replace{\alpha'}{\alpha}{s}{t}}$. This holds by the \ih
    
    \item $o=\termrepl{\beta}{\gamma}{s'}{c}$, where $ \alpha \neq \beta$. Then
    $\replace{\alpha'}{\alpha}{s}{\termrepl{\beta}{\gamma}{s'}{c}} =
    \termrepl{\beta}{\gamma}{\replace{\alpha'}{\alpha}{s}{s'}}{\replace{\alpha'}{\alpha}{s}{c}}$.
    The typing derivation for $o$ has the following form: \[
    \hfill
    \Rule{\Gam_1  \vdash c \mid \Del_1, (\gamma : S' \tarrow B')^{\esta}, (\beta : B')^{\esta}, (\alpha : S \tarrow B)^{\esta}
          \quad
          \Gam'_1 \vdash s' \mid \Del'_1, (\beta : B')^{\esta}, (\alpha: S \tarrow B)^{\esta}
         }
         {\Gam_1 \cup \Gam'_1 \vdash \termrepl{\beta}{\gamma}{s'}{c} \mid \Del_1 \cup \Del'_1, \beta : B', (\alpha : S \tarrow B)^{\esta}}
         {}
    \hfill
    \] where $\Gam =\Gam_1 \cup \Gam'_1$ and $\Del = \Del_1 \cup \Del'_1,
    \beta : B', (\alpha : S \tarrow B)^{\esta}$.
    
    The \ih gives $\Gam^{**} \vdash \replace{\alpha'}{\alpha}{s}{c} \mid
    \Del^{**}$, where $\Gam^{**} \subseteq \Gam_1 \cup \Gam'$ and $\Del^{**}
    \subseteq (\Del_1, (\gamma : S' \tarrow B')^{\esta}) \cup \Del' \cup
    (\alpha' : B)$ as well as $\Gam^{***} \vdash
    \replace{\alpha'}{\alpha}{s}{s'} \mid \Del^{***}$, where $\Gam^{***}
    \subseteq \Gam'_1 \cup \Gam'$ and $\Del^{***} \subseteq \Del'_1 \cup \Del'
    \cup (\alpha':B)$. Then rule $(\trepl)$ gives $\Gam^{**} \cup \Gam^{***}
    \vdash \replace{\alpha'}{\alpha}{s}{c}\exrepl{\beta}{\gamma}{\replace{\alpha'}{\alpha}{s}{s'}}
    \mid (\Del^{**} \setminus (\gamma : S' \tarrow B')) \cup \Del^{***}, \beta
    : B'$. We conclude with
    \begin{itemize}
      \item $\Gam^{*} = \Gam^{**} \cup \Gam^{***} \subseteq (\Gam_1 \cup \Gam')
      \cup (\Gam'_1 \cup \Gam') = \Gam \cup \Gam'$, and
      \item $\Del^{*} = (\Del^{**} \setminus (\gamma : S' \tarrow B')) \cup
      \Del^{***} \cup (\beta:B') \subseteq \Del_1 \cup \Del' \cup (\alpha':B)
      \cup \Del'_1 \cup \Del' \cup (\alpha':B) \subseteq \Del \cup \Del' \cup
      (\alpha':B)$.
  \end{itemize}
  
  \item $o=\termrepl{\alpha}{\gamma}{\termemst}{c}$. Then the typing derivation
  for $o$ has the following form: \[
  \hfill
  \Rule{\Gam \vdash c \mid \Del_1, (\gamma : \typeemst \tarrow  B')^{\esta}, (\alpha : S \tarrow B)^{\esta}
        \quad
        \emptyset \vdash \termemst: \typeemst \mid \emptyset
       }
       {\Gam \vdash \termrepl{\al}{\gamma}{\termemst}{c} \mid \Del_1, \al : S \tarrow B}
       {}
  \hfill
  \] where $\Del = \Del_1, \al: S \tarrow B$.

  The \ih gives $\Gam^{**} \vdash \replace{\alpha'}{\alpha}{s}{c} \mid
  \Del^{**}$, where $\Gam^{**} \subseteq \Gam \cup \Gam'$ and $\Del^{**}
  \subseteq (\Del_1, (\gamma : \typeemst \tarrow  B')^{\esta})\cup \Del' \cup
  (\alpha' : B)$.

  There are two cases to consider.
  \begin{enumerate}
    \item If $s \neq \termemst$. Then,
    $\replace{\alpha'}{\alpha}{s}{\termrepl{\alpha}{\gamma}{\termemst}{c}} =
    \termrepl{\alpha'}{\beta}{\termemst}{\termrepl{\beta}{\gamma}{s}{\replace{\alpha'}{\alpha}{s}{c}}}$
    and application of rule $(\trepl)$ gives $\Gam^{**} \cup \Gam' \vdash
    \replace{\alpha'}{\alpha}{s}{c}\exrepl{\beta}{\gamma}{s} \mid (\Del^{**}
    \setminus (\gamma : \typeemst \tarrow  B')) \cup \Del', \beta : B'$.
    Another application of rule $(\trepl)$ gives $\Gam^{**} \cup \Gam' \vdash
    \replace{\alpha'}{\alpha}{s}{c}\exrepl{\beta}{\gamma}{s}
    \exrepl{\al'}{\beta}{\termemst} \mid (((\Del^{**}\setminus (\gamma :
    \typeemst \tarrow  B')) \cup \Del') \setminus (\beta : B')), \al' : B$.
    We conclude with
    \begin{itemize}
      \item $\Gam^* = \Gam^{**} \cup \Gam' \subseteq \Gam \cup \Gam'$, and
      \item $\Del^* = (((\Del^{**} \setminus (\gamma : \typeemst \tarrow  B'))
      \cup \Del') \setminus (\beta:B')) \cup (\al':B) \subseteq \Del \cup \Del'
      \cup (\alpha':B)$.
    \end{itemize}
    
    \item Otherwise (\ie $s = \termemst$),
    $\replace{\alpha'}{\alpha}{s}{\termrepl{\alpha}{\gamma}{\termemst}{c}}
    = \termrepl{\alpha'}{\gamma}{\termemst}{\replace{\alpha'}{\gamma}{s}{c}}$.
    Then application of rule $(\trepl)$ to $\Gam^{**} \vdash
    \replace{\alpha'}{\alpha}{s}{c} \mid \Del^{**}$ gives $\Gam^{**} \vdash
    \termrepl{\alpha'}{\gamma}{\termemst}{\replace{\alpha'}{\alpha}{s}{c}} \mid
    (\Del^{**} \setminus (\gamma : \typeemst \tarrow B')), \alpha':B'$. 
    We
    conclude with
    \begin{itemize}
      \item $\Gam^* = \Gam^{**} \subseteq \Gam \cup \Gam'$, and
      \item $\Del^* = (\Del^{**} \setminus (\gamma : \typeemst \tarrow B'))
      \cup (\alpha' : B') \subseteq (\Del_1 \setminus (\al : S \tarrow B)) \cup
      \Del' \cup (\al' : B) \subseteq \Del \cup \Del' \cup (\alpha' : B)$.
    \end{itemize}
  \end{enumerate}
  
  \item $o = \termrepl{\alpha}{\gamma}{s'}{c}$, where $s' \neq \termemst$.
  Then, $\replace{\alpha'}{\alpha}{s}{\termrepl{\alpha}{\gamma}{s'}{c}} =
  \termrepl{\alpha'}{\gamma}{\termpush{\replace{\alpha'}{\alpha}{s}{s'}}{s}}{\replace{\alpha'}{\alpha}{s}{c}}$.
  Similar to the first case of explicit replacement.
\end{itemize}  
\end{proof}

%%% Local Variables:
%%% mode: latex
%%% TeX-master: "main"
%%% End:

\gettoappendix{l:SR}

\begin{proof}
Let $\pi_o \dem \Gam \vdash o:A \mid \Del$. We reason by induction on $o
\Rew{\LMfull} o'$.

\begin{itemize}
  \item $o = \termapp{\ctxtapp{\ctxt{L}}{\termabs{x}{t}}}{u} \rrule{\rB}
  \ctxtapp{\ctxt{L}}{\termsubs{x}{u}{t}} = o'$. The proof proceeds by induction
  on $\ctxt{L}$. We only show the case $\ctxt{L} = \Box$, as the other ones
  follow directly. The derivation $\pi_o$ has the following form. \[
  \hfill
  \Rule{\Rule{\Gam, (x:B)^{\esta} \vdash t : A \mid \Del}
             {\Gam \vdash \l x. t : B \tarrow A \mid \Del}
             {}
        \quad
        \Gam' \vdash u : B \mid \Del'  
       }
       {\Gam \cup \Gam' \vdash (\l x. t) u : A \mid \Del \cup \Del'}
       {}
  \hfill
  \] We construct the following derivation $\pi_{o'}$: \[
  \hfill
  \Rule{\Gam, (x:B)^{\esta} \vdash t : A \mid \Del
        \quad
        \Gam' \vdash u : B \mid \Del'
       }
       {\Gam \cup \Gam' \vdash \termsubs{x}{u}{t} : A \mid \Del \cup \Del'}
       {}
  \hfill
  \]
  
  \item $o = \termsubs{x}{u}{t} \rrule{\rS} \substitute{x}{u}{t} = o'$. The
  derivation $\pi_o$ has the following form. \[
  \hfill
  \Rule{\Gam, (x:B)^{\esta} \vdash t : A \mid \Del
        \quad
        \Gam' \vdash u : B \mid \Del'
       }
       {\Gam \cup \Gam' \vdash \termsubs{x}{u}{t} : A \mid \Del \cup \Del'}
       {}
  \hfill
  \] We conclude by Lemma~\ref{l:substitution}.
  
  \item $o = \termapp{\ctxtapp{\ctxt{L}}{\termcont{\alpha}{c}}}{u} \rrule{\rM}
  \ctxtapp{\ctxt{L}}{\termcont{\alpha'}{\termrepl{\alpha'}{\alpha}{u}{c}}} =
  o'$. The proof proceeds by induction on $\ctxt{L}$. We only show the case
  $\ctxt{L} = \Box$, as the other ones follow directly. The derivation $\pi_o$
  has the following form. \[
  \hfill
  \Rule{\Rule{\Gam \vdash c \mid \Del, (\alpha:B \tarrow A)^{\esta}}
             {\Gam \vdash \mu \al. c : B \tarrow A \mid \Del}
             {}
        \quad
        \Gam' \vdash u : B \mid \Del'
       }
       {\Gam \cup \Gam' \vdash (\mu \al. c) u : A \mid \Del \cup \Del'}
       {}
  \hfill
  \] We construct the following derivation $\pi_{o'}$ (recall $\al'$ is fresh
  by hypothesis of rule $\rM$): \[
  \hfill
  \Rule{\Rule{\Gam \vdash c \mid \Del, (\al : \typepush{B}{\typeemst} \tarrow A)^{\esta}, (\al' : A)^{0}
              \quad
              \Rule{\Gam' \vdash u : B \mid \Del', (\al' : A)^{0}
                    \quad
                    \emptyset \vdash \termemst: \typeemst \mid \emptyset
                   }
                   {\Gam' \vdash \termpush{u}{\termemst} : \typepush{B}{\typeemst} \mid \Del', (\al' : A)^{0}}
                   {}
             }
             {\Gam \cup \Gam' \vdash \termrepl{\al'}{\al}{u}{c} \mid \Del \cup \Del', \al' : A}
             {}
       }
       {\Gam \cup \Gam' \vdash \termcont{\alpha'}{\termrepl{\alpha'}{\alpha}{u}{c}}:A \mid \Del \cup \Del'}
       {}
  \hfill
  \]
  
  \item $o = \termrepl{\alpha'}{\alpha}{s}{c} \rrule{\rRfull}
  \replace{\alpha'}{\alpha}{s}{c} = o'$. The derivation $\pi_o$ has the
  following form. \[
  \hfill
  \Rule{\Gam \vdash c \mid \Del, (\alpha : S \tarrow A)^{\esta}, (\al':A)^{\esta}
        \quad
        \Gam' \vdash s : S \mid \Del', (\al' : A)^{\esta}
       }
       {\Gam \cup \Gam' \vdash \termrepl{\alpha'}{\alpha}{s}{c} \mid \Del \cup \Del', \al': A}
       {}
  \hfill
  \] We conclude by Lemma~\ref{l:replacement}.
\end{itemize}
\end{proof}

%%% Local Variables:
%%% mode: latex
%%% TeX-master: "main"
%%% End:

%%% Local Variables:
%%% mode: latex
%%% TeX-master: "main"
%%% End:

  %\input{app-translation}  % no se usa mas
  %%%%%%%%%%%%%%%%%%%%%%%%%%%%%%%%%%%%%%%%%%%%%%%%%%%%%%%%%%%%%%%%%%%%%%%%%%%%%%%
\section{Appendix: Structural Equivalence}
\label{a:general-lemmas}
%%%%%%%%%%%%%%%%%%%%%%%%%%%%%%%%%%%%%%%%%%%%%%%%%%%%%%%%%%%%%%%%%%%%%%%%%%%%%%%

Before proceeding with the proof of Lem.~\ref{l:subs-repl-out-of-BM} we
introduce two technical lemmas on $\ctxtLTT$ and $\ctxtLCC$ linear contexts
respectively.

\begin{lemma}
\label{l:tt-bisimulation}
Let $t\in \terms{\LMfull}$. $\{(\ctxtapp{\ctxt{L}}{\ctxtapp{\ctxtLTT}{t}},
\ctxtapp{\ctxtLTT}{\ctxtapp{\ctxt{L}}{t}}) \mid
\notatall{\bv{\ctxt{L}}}{\ctxtLTT} \land
\freeFor{\ctxt{L}}{\ctxtLTT}\}$ is a strong $\Can$-bisimulation.
\end{lemma}
  
\begin{proof} 
Let $\mathcal{R}$ be the relation stated above. We prove that: $$
\begin{tikzcd}
\ctxtapp{\ctxt{L}}{\ctxtapp{\ctxtLTT}{t}} \arrow{d}[left]{\Can} &[-20pt] \mathcal{R} &[-20pt]
\ctxtapp{\ctxtLTT}{\ctxtapp{\ctxt{L}}{t}} \arrow[dotted]{d}[left]{\Can}
\\
\ctxtapp{\ctxt{L'}}{\ctxtapp{\ctxtLTT'}{t'}}                    &[-20pt] \mathcal{R} &[-20pt]
\ctxtapp{\ctxtLTT'}{\ctxtapp{\ctxt{L'}}{t'}}
\end{tikzcd} $$
for some $\ctxt{L'}$, $\ctxtLTT'$ and $t'$. The case $$
\begin{tikzcd}
\ctxtapp{\ctxt{L}}{\ctxtapp{\ctxtLTT}{t}} \arrow[dotted]{d}[left]{\Can} &[-20pt] \mathcal{R} &[-20pt]
\ctxtapp{\ctxtLTT}{\ctxtapp{\ctxt{L}}{t}} \arrow{d}[left]{\Can}
\\
\ctxtapp{\ctxt{L'}}{\ctxtapp{\ctxtLTT'}{t'}}                            &[-20pt] \mathcal{R} &[-20pt]
\ctxtapp{\ctxtLTT'}{\ctxtapp{\ctxt{L'}}{t'}}
\end{tikzcd} $$
is similar and hence omitted. The possible overlapping between the   $\Can$-step and $\ctxtapp{\ctxt{L}}{\ctxtapp{\ctxtLTT}{t}}$ leads to the following
cases:
\begin{itemize}
  \item It overlaps with $\ctxt{L}$. In other words, it occurs entirely inside
  the body of a replacement in $\ctxt{L}$. Then we set $\ctxtLTT' \coloneq
  \ctxtLTT$ and $t' \coloneq t$.
  
  \item It occurs entirely in $t$. Then we set $\ctxt{L'} \coloneq \ctxt{L}$
  and $\ctxtLTT' \coloneq \ctxtLTT$.
  
  \item It occurs entirely in $\ctxtLTT$. In this case it either occurs in the
  body $s$ of an explicit replacement
  $\termrepl{\alpha'}{\alpha}{s}{\ctxtLCT}$, or in the body $u$ of an explicit
  substitution $\termsubs{x}{u}{\ctxtLTT_1}$ or in the argument $u$ of an
  application $\ctxtLTT_1\ u$. In any of these cases, the reduct is again a
  linear TT-context $\ctxtLTT'$. We thus take $\ctxt{L'} \coloneq \ctxt{L}$
  and $t' \coloneq t$.
  
  \item It overlaps with $\ctxtLTT$.
  \begin{itemize}
    \item Suppose the $\Can$-step is a $\rB$-step. There are two further cases:
    \begin{itemize}     
      \item  It does not overlap with $t$. Then $\ctxtLTT =
      \ctxtapp{\ctxtLTT_1}{\termapp{\ctxtapp{\ctxt{L_1}}{\termabs{x}{\ctxtLTT_2}}}{t_2}}$,
      $\ctxt{L'} \coloneq \ctxt{L}$, $\ctxtLTT' \coloneq
      \ctxtapp{\ctxtLTT_1}{\ctxtapp{\ctxt{L_1}}{\termsubs{x}{t_2}{\ctxtLTT_2}}}$ and $t'
      \coloneq t$. Thus,  $$
\begin{tikzcd}
\ctxtapp{\ctxt{L}}{\ctxtapp{\ctxtLTT}{t}} \arrow{d}[left]{\rB}  &[-20pt] \mathcal{R} &[-20pt]
\ctxtapp{\ctxtLTT_1}{\termapp{\ctxtapp{\ctxt{L_1}}{\termabs{x}{\ctxtapp{\ctxtLTT_2}{\ctxtapp{\ctxt{L}}{t}}}}}{t_2}} \arrow[dotted]{d}[left]{\rB}
\\
\ctxtapp{\ctxt{L}'}{\ctxtapp{\ctxtLTT'}{t'}}                    &[-20pt] \mathcal{R} &[-20pt]
\ctxtapp{\ctxtLTT_1}{\ctxtapp{\ctxt{L_1}}{\termsubs{x}{t_2}{\ctxtapp{\ctxtLTT_2}{\ctxtapp{\ctxt{L}'}{t}}}}}
\end{tikzcd} $$

      \item It overlaps with $t$. Then $t =
      \ctxtapp{\ctxt{L_2}}{\termabs{x}{t_1}}$, $\ctxtLTT =
      \ctxtapp{\ctxtLTT_1}{\termapp{\ctxt{L_1}}{t_2}}$ and the LHS of the
      $\rB$-step is
      $\termapp{\ctxtapp{\ctxt{L_1}}{\ctxtapp{\ctxt{L_2}}{\termabs{x}{t_1}}}}{t_2}$. 
      Moreover, $\ctxt{L'} \coloneq \ctxt{L}$, $\ctxtLTT' \coloneq
      \ctxtapp{\ctxtLTT_1}{\ctxt{L_1}}$ and $t' \coloneq
      \ctxtapp{\ctxt{L_2}}{\termsubs{x}{t_2}{t_1}}$. Thus, $$
\begin{tikzcd}
\ctxtapp{\ctxt{L}}{\ctxtapp{\ctxtLTT}{t}} \arrow{d}[left]{\rB}  &[-20pt] \mathcal{R} &[-20pt]
\ctxtapp{\ctxtLTT_1}{\termapp{\ctxtapp{\ctxt{L_1}}{\ctxtapp{\ctxt{L}}{\ctxtapp{\ctxt{L_2}}{\termabs{x}{t_1}}}}}{t_2}} \arrow[dotted]{d}[left]{\rB}
\\
\ctxtapp{\ctxt{L}'}{\ctxtapp{\ctxtLTT'}{t'}}                    &[-20pt] \mathcal{R} &[-20pt]
\ctxtapp{\ctxtLTT_1}{\ctxtapp{\ctxt{L_1}}{\ctxtapp{\ctxt{L}'}{t'}}}
\end{tikzcd} $$
    \end{itemize}

    \item Suppose the $\Can$-step is a $\rM$-step. There are two further cases:
    \begin{itemize}
      \item It does not overlap with $t$. Then $\ctxtLTT =
      \ctxtapp{\ctxtLTT_1}{\termapp{\ctxtapp{\ctxt{L_1}}{\termcont{\alpha}{\ctxtLTC_2}}}{t_2}}$,
      $\ctxt{L'} \coloneq \ctxt{L}$, $\ctxtLTT' \coloneq
      \ctxtapp{\ctxtLTT_1}{\ctxtapp{\ctxt{L_1}}{\termcont{\alpha'}{\termrepl{\alpha'}{\alpha}{t_2}{\ctxtLTC_2}}}}$
      and $t' \coloneq t$. Thus, $$
\begin{tikzcd}
\ctxtapp{\ctxt{L}}{\ctxtapp{\ctxtLTT}{t}} \arrow{d}[left]{\rM}  &[-20pt] \mathcal{R} &[-20pt]
\ctxtapp{\ctxtLTT_1}{\termapp{\ctxtapp{\ctxt{L_1}}{\termcont{\alpha}{\ctxtapp{\ctxtLTC_2}{\ctxtapp{\ctxt{L}}{t}}}}}{t_2}} \arrow[dotted]{d}[left]{\rM}
\\
\ctxtapp{\ctxt{L}'}{\ctxtapp{\ctxtLTT'}{t'}}                    &[-20pt] \mathcal{R} &[-20pt]
\ctxtapp{\ctxtLTT_1}{\ctxtapp{\ctxt{L_1}}{\termcont{\alpha'}{\termrepl{\alpha'}{\alpha}{t_2}{\ctxtapp{\ctxtLTC_2}{\ctxtapp{\ctxt{L}'}{t}}}}}}
\end{tikzcd} $$
      
      \item It overlaps with $t$. Then $t =
      \ctxtapp{\ctxt{L_2}}{\termcont{\alpha}{c}}$, for some $c$, $\ctxtLTT =
      \ctxtapp{\ctxtLTT_1}{\termapp{\ctxt{L_1}}{t_2}}$, $\ctxt{L'} \coloneq
      \ctxt{L}$, $\ctxtLTT' \coloneq \ctxtapp{\ctxtLTT_1}{\ctxt{L_1}}$ and $t'
      \coloneq
      \ctxtapp{\ctxt{L_2}}{\termcont{\alpha'}{\termrepl{\alpha'}{\alpha}{t_2}{c}}}$.
      Thus, $$
\begin{tikzcd}
\ctxtapp{\ctxt{L}}{\ctxtapp{\ctxtLTT}{t}} \arrow{d}[left]{\rM}  &[-20pt] \mathcal{R} &[-20pt]
\ctxtapp{\ctxtLTT_1}{\termapp{\ctxtapp{\ctxt{L_1}}{\ctxtapp{\ctxt{L}}{\ctxtapp{\ctxt{L_2}}{\termcont{\alpha}{c}}}}}{t_2}} \arrow[dotted]{d}[left]{\rM}
\\
\ctxtapp{\ctxt{L}'}{\ctxtapp{\ctxtLTT'}{t'}}                    &[-20pt] \mathcal{R} &[-20pt]
\ctxtapp{\ctxtLTT_1}{\ctxtapp{\ctxt{L_1}}{\ctxtapp{\ctxt{L}'}{\ctxtapp{\ctxt{L_2}}{\termcont{\alpha'}{\termrepl{\alpha'}{\alpha}{t_2}{c}}}}}}
\end{tikzcd} $$
\end{itemize}
      
      \item Suppose the $\Can$-step is a $\rC$-step. Then $\ctxtLTT =
      \termrepl{\beta}{\alpha'}{s'}{\ctxtapp{\ctxtLCC}{\termrepl{\alpha'}{\alpha}{s}{\ctxtLCT}}}$,
      where $s, s' \neq \termemst$. Note that $t$ itself cannot be of the form
      $\ctxtapp{\ctxtLCC_2}{\termrepl{\alpha'}{\alpha}{s}{\ctxtLCT}}$,
      for any $\ctxtLCC_2$, since $t$ is not a command. Moreover, $\ctxt{L}'
      \coloneq \ctxt{L}$, $\ctxtLTT' \coloneq
      \ctxtapp{\ctxtLCC}{\termrepl{\beta}{\alpha}{\termpush{s}{s'}}{\ctxtLCT}}$
      and $t' \coloneq t$. Thus, $$
\begin{tikzcd}
\ctxtapp{\ctxt{L}}{\ctxtapp{\ctxtLTT}{t}} \arrow{d}[left]{\rC}  &[-20pt] \mathcal{R} &[-20pt]
\termrepl{\beta}{\alpha'}{s'}{\ctxtapp{\ctxtLCC}{\termrepl{\alpha'}{\alpha}{s}{\ctxtapp{\ctxtLCT}{\ctxtapp{\ctxt{L}}{t}}}}} \arrow[dotted]{d}[left]{\rC}
\\
\ctxtapp{\ctxt{L}}{\ctxtapp{\ctxtLTT'}{t'}}                     &[-20pt] \mathcal{R} &[-20pt]
\ctxtapp{\ctxtLCC}{\termrepl{\beta}{\alpha}{\termpush{s}{s'}}{\ctxtapp{\ctxtLCT}{\ctxtapp{\ctxt{L}}{t'}}}}
\end{tikzcd} $$
      
      \item Suppose the $\Can$-step is a $\rW$-step. Then $\ctxtLTT =
      \termrepl{\beta}{\alpha'}{s'}{\ctxtapp{\ctxtLCC}{\termrepl{\alpha'}{\alpha}{\termemst}{\ctxtLCT}}}$,
      where in particular $s' \neq \termemst$.  Note that $t$ itself cannot be
      of the form
      $\ctxtapp{\ctxtLCC_2}{\termrepl{\alpha'}{\alpha}{\termemst}{\ctxtLCT}}$,
      for any $\ctxtLCC_2$, since $t$ is not a command. Moreover, $\ctxt{L}'
      \coloneq \ctxt{L}$, $\ctxtLTT' \coloneq
      \ctxtapp{\ctxtLCC}{\termrepl{\beta}{\alpha'}{\termemst}{\termrepl{\alpha'}{\alpha}{s'}{\ctxtLCT}}}$
      and $t' \coloneq t$. Thus, $$
\begin{tikzcd}
\ctxtapp{\ctxt{L}}{\ctxtapp{\ctxtLTT}{t}} \arrow{d}[left]{\rW}  &[-20pt] \mathcal{R} &[-20pt]
\termrepl{\beta}{\alpha'}{s'}{\ctxtapp{\ctxtLCC}{\termrepl{\alpha'}{\alpha}{\termemst}{\ctxtapp{\ctxtLCT}{\ctxtapp{\ctxt{L}}{t}}}}} \arrow[dotted]{d}[left]{\rW}
\\
\ctxtapp{\ctxt{L}'}{\ctxtapp{\ctxtLTT'}{t'}}                    &[-20pt] \mathcal{R} &[-20pt]
\ctxtapp{\ctxtLCC}{\termrepl{\beta}{\alpha'}{\termemst}{\termrepl{\alpha'}{\alpha}{s'}{\ctxtapp{\ctxtLCT}{\ctxtapp{\ctxt{L}'}{t'}}}}}
\end{tikzcd} $$
      
      \item There is no other possible case because of the hypothesis of the
      initial relation.
  \end{itemize}
\end{itemize}
\end{proof}

%%% Local Variables:
%%% mode: latex
%%% TeX-master: "main"
%%% End:

\ignore{
\begin{lemma} \label{l:subs-out-of-BM}
Let $t, u$ be terms. Then \delia{$\BM{\ctxtapp{\ctxt{L}}{\ctxtapp{\ctxtLTT}{t}}
u} \simeq \BMLoapp{\BM{\ctxtapp{\ctxtLTT}{\Ltapp{t}} u}}$.} for $\ctxt{L} =
\Loapp{\ctxLt}$. \delia{esto no se usa mas}
\end{lemma}

\begin{proof}
By  induction on \delia{$\sz{\ctxtapp{\ctxtLTT}{t}}$}. Since
$\sz{\ctxtapp{\ctxtLTT}{t}} = \sz{\BM{\ctxtapp{\ctxtLTT}{t}}}$ by
Lem.~\ref{l:BM-preserves-sz} then w.l.o.g. we can suppose
$\ctxtapp{\ctxtLTT}{t} \in \BMform$ which implies also
$\ctxtapp{\ctxtLTT}{\BMLapp{t}} \in \BMform$.
\begin{itemize}
  \item $\ctxTT = \Box$ and $t = x$. Then \[
  \BM{\ctxtapp{\ctxt{L}}{x} u} = \BMLoapp{\BMLtapp{x}} \BM{u}
  \simeq_{\eqexsubs} \BMLoapp{\BMLtapp{x} \BM{u}} =
  \BMLoapp{\BM{\Ltapp{x}u}} \]
  
  \item $\ctxTT = \Box$ and $t = \termapp{t_0}{t_1}$. Since $t \in \BMform$,
  then $t_0$ cannot be headed by any kind of abstraction, so \[
\begin{array}{ll}
\BM{\termapp{\ctxtapp{\ctxt{L}}{\termapp{t_0}{t_1}}}{u}}    & = \\
\termapp{\BMLoapp{\BM{\Ltapp{\termapp{t_0}{t_1}}}}}{\BM{u}} & \simeq_{\eqexsubs} \\
\BMLoapp{\termapp{\BM{\Ltapp{\termapp{t_0}{t_1}}}}{\BM{u}}} & = \\
\BMLoapp{\BM{\termapp{\Ltapp{\termapp{t_0}{t_1}}}{u}}}
\end{array} \]
  
  \item $\ctxTT = \Box$ and  $t = \termabs{y}{t_0}$. Then let $\ctxt{L'} =
  \ctxtapp{\ctxt{L}}{\termsubs{y}{u}{\Box}}$ and $\ctxt{L'_2} =
  \Ltapp{\termsubs{y}{u}{\Box}}$. We have \[
\begin{array}{ll}
\BM{\termapp{\ctxtapp{\ctxt{L}}{(\termabs{y}{t_0})}}{u}}  & = \\
\BM{\ctxtapp{\ctxt{L}}{\termsubs{y}{u}{t_0}}}             & = \\
\BM{\Lpapp{t_0}}                                          & = \\
\BMLoapp{\BM{\Ltpapp{t_0}}}                               & = \\
\BMLoapp{\BM{\Ltapp{\termsubs{y}{u}{t_0}}}}               & = \\
\BMLoapp{\BM{\termapp{\Ltapp{\termabs{y}{t_0}}}{u}}}
\end{array} \]
  
  \item $\ctxTT = \Box$ and $t = \termsubs{y}{u_0}{t_0}$. Then, let $\ctxt{L'}
  = \ctxtapp{\ctxt{L}}{\termsubs{y}{u_0}{\Box}}$ and $\ctxt{L'_2} =
  \Ltapp{\termsubs{y}{u_0}{\Box}}$. We have \[
\begin{array}{ll}
\BM{\termapp{\ctxtapp{\ctxt{L}}{\termsubs{y}{u_0}{t_0}}}{u}}  & = \\
\BM{\termapp{\Lpapp{t_0}}{u}}                                 & \simeq_{\ih} \\
\BMLoapp{\BM{\termapp{\Ltpapp{t_0}}{u}}}                      & = \\
\BMLoapp{\BM{\termapp{\Ltapp{\termsubs{y}{u_0}{t_0}}}{u}}}
\end{array} \]
  
  \item $\ctxTT = \Box$ and  $t = \termcont{\alpha}{c}$. Then \[
\begin{array}{ll}
\BM{\termapp{\ctxtapp{\ctxt{L}}{\termcont{\alpha}{c}}}{u}}                    & = \\
\BM{\ctxtapp{\ctxt{L}}{\termcont{\alpha}{\termrepl{\alpha'}{\alpha}{u}{c}}}}  & = \\
\BMLoapp{\BM{\Ltapp{\termcont{\alpha}{\termrepl{\alpha'}{\alpha}{u}{c}}}}}    & = \\
\BMLoapp{\BM{\termapp{\Ltapp{\termcont{\alpha}{c}}}{u}}}
\end{array} \]
  
  \item $\ctxTT= \termabs{y}{\ctxTTp}$. Let $\ctxLp =
  \ctxtapp{\ctxt{L}}{\termsubs{y}{u}{\Box}}$ and let $\ctxLop =
  \Loapp{\termsubs{y}{u}{\Box}}$. Then, \[
\begin{array}{ll}
\BM{\termapp{\ctxtapp{\ctxt{L}}{\termabs{y}{\TTpapp{t}}}}{u}} & = \\
\BM{\ctxtapp{\ctxt{L}}{\termsubs{y}{u}{\TTpapp{t}}}}          & = \\
\BM{\Lpapp{\TTpapp{t}}}                                       & \simeq_{\eqexsubs}  \\
\BM{\Lopapp{\Ltapp{\TTpapp{t}}}}                              & = \\
\BM{\Lopapp{\BMLtapp{\TTpapp{t}}}}                            & = \\
\BMLopapp{\BMLtapp{\TTpapp{t}}}                               & \simeq_{\eqexsubs} \\
\BMLopapp{\TTpapp{\BMLtapp{t}}}                               & = \\
\BMLopapp{\BM{\TTpapp{\Ltapp{t}}}}                            & = \\
\BMLoapp{\BM{\termsubs{y}{u}{\TTpapp{\Ltapp{t}}}}}            & = \\
\BMLoapp{\BM{\termapp{(\termabs{y}{\TTpapp{\Ltapp{t}}})}{u}}}
\end{array} \]
  
  \item $\ctxTT= \termapp{\ctxTTp}{v}$. The property is by straightforward
  induction since $\termapp{\ctxTTp}{v} \in \BMform$ implies that $\ctxTTp$
  cannot by any abstraction.
  
  \item $\ctxTT = \termcont{\alpha}{\ctxCT}$. Since $\ctxtapp{\ctxtLTT}{t} \in
  \BMform$, then also $\ctxtapp{\ctxtLCT}{t} \in \BMform$ and
  $\ctxtapp{\ctxtLCT}{\Ltapp{t}} \in \BMform$. Thus we have \[
\begin{array}{ll}
\BM{\ctxtapp{\ctxt{L}}{\termapp{(\termcont{\alpha}{\ctxtapp{\ctxtLCT}{t}}})}{u}}                    & = \\
\BM{\ctxtapp{\ctxt{L}}{\termcont{\alpha'}{\ctxtapp{\ctxtLCT}{t}\exrepl{\alpha'}{\alpha}{u}}}}       & = \\
\BMLoapp{\BMLtapp{\termcont{\alpha'}{\BM{\ctxtapp{\ctxtLCT}{t}}\exrepl{\alpha'}{\alpha}{\BM{u}}}}}  & = \\
\BMLoapp{\BMLtapp{\termcont{\alpha'}{\ctxtapp{\ctxtLCT}{t}\exrepl{\alpha'}{\alpha}{\BM{u}}}}}       & \simeq_{\eqexsubs}  \\
\BMLoapp{\termcont{\alpha'}{\ctxtapp{\ctxtLCT}{\BMLtapp{t}}\exrepl{\alpha'}{\alpha}{\BM{u}}}}       & = \\
\BMLoapp{\termcont{\alpha}{\BM{\ctxtapp{\ctxtLCT}{\Ltapp{t}}}}\exrepl{\alpha'}{\alpha}{\BM{u}}}     & = \\
\BMLoapp{\BM{\termcont{\alpha'}{\ctxtapp{\ctxtLCT}{\Ltapp{t}}}\exrepl{\alpha'}{\alpha}{u}}}         & = \\
\BMLoapp{\BM{\termapp{(\termcont{\alpha}{\ctxtapp{\ctxtLCT}{\Ltapp{t}}})}{u}}}
\end{array} \]
  
  \item $\ctxTT = \termcont{\alpha}{\ctxCT\exrepl{\beta'}{\beta}{s}}$. This
  case is similar to the previous one. 
  
  \item $\ctxTT = \termcont{\alpha}{\ctxCT\neren{\beta}{\gamma}}$. This case is
  trivial since no new $\BMform$-redex can be created. 
\end{itemize}
\end{proof}

%%% Local Variables:
%%% mode: latex
%%% TeX-master: "main"
%%% End:

\delia{
\begin{corollary}
\label{c:substitutions-out-of-BM}
Let $t, u $ be terms. Then
$\BM{\termapp{\ctxtapp{\ctxt{L}}{\ctxtapp{\ctxtLTT}{t}}}{u}}
\simeq \BMLapp{\BM{\termapp{\ctxtapp{\ctxtLTT}{t}}{u}}}$. Thus in particular
$\BM{\termapp{\ctxtapp{\ctxt{L}}{{t}}}{u}} \simeq
\BMLapp{\BM{\termapp{t}{u}}}$. 
\end{corollary}}
}

\begin{lemma}
\label{l:cc-bisimulation}
Let $c\in \commands{\LMfull}$. $\{(\ctxtapp{\ctxt{R}}{\ctxtapp{\ctxtLCC}{c}},
\ctxtapp{\ctxtLCC}{\ctxtapp{\ctxt{R}}{c}}) \mid
\notatall{\bn{\ctxt{R}}}{\ctxtLCC} \land
\freeFor{\ctxt{R}}{\ctxtLCC}\}$ is a strong $\Can$-bisimulation.
\end{lemma}

\begin{proof}
Let $\mathcal{R}$ be the relation stated above. We prove that: $$
\begin{tikzcd}
\ctxtapp{\ctxt{R}}{\ctxtapp{\ctxtLCC}{c}} \arrow{d}[left]{\Can} &[-20pt] \mathcal{R} &[-20pt]
\ctxtapp{\ctxtLCC}{\ctxtapp{\ctxt{R}}{c}} \arrow[dotted]{d}[left]{\Can}
\\
\ctxtapp{\ctxt{R'}}{\ctxtapp{\ctxtLCC'}{c'}}                    &[-20pt] \mathcal{R} &[-20pt]
\ctxtapp{\ctxtLCC'}{\ctxtapp{\ctxt{R'}}{c'}}
\end{tikzcd} $$
for some $\ctxt{R'}$, $\ctxtLCC'$ and $c'$. The case $$
\begin{tikzcd}
\ctxtapp{\ctxt{R}}{\ctxtapp{\ctxtLCC}{c}} \arrow[dotted]{d}[left]{\Can} &[-20pt] \mathcal{R} &[-20pt]
\ctxtapp{\ctxtLCC}{\ctxtapp{\ctxt{R}}{c}} \arrow{d}[left]{\Can}
\\
\ctxtapp{\ctxt{R'}}{\ctxtapp{\ctxtLCC'}{c'}}                            &[-20pt] \mathcal{R} &[-20pt]
\ctxtapp{\ctxtLCC'}{\ctxtapp{\ctxt{R'}}{c'}}
\end{tikzcd} $$
is similar and hence omitted. The possible overlapping between the $\Can$-step
and $\ctxtapp{\ctxt{R}}{\ctxtapp{\ctxtLCC}{c}}$ leads to the following cases:
\begin{itemize}
  \item It overlaps with $\ctxt{R}$. In other words, it occurs entirely inside
  the body of a replacement in $\ctxt{R}$. Then we set $\ctxtLCC' \coloneq
  \ctxtLCC$ and $c' \coloneq c$.
  
  \item It occurs entirely in $c$. Then we set $\ctxt{R'} \coloneq \ctxt{R}$
  and $\ctxtLCC' \coloneq \ctxtLCC$.
  
  \item It occurs entirely in $\ctxtLCC$. In this case it either occurs in the
  body $s$ of an explicit replacement
  $\termrepl{\alpha'}{\alpha}{s}{\ctxtLCC_1}$, or in the body $u$ of an
  explicit substitution $\termsubs{x}{u}{\ctxtLTC}$ or in the argument $u$ of
  an application $\termapp{\ctxtLTC}{u}$. In any of these cases, the reduct is
  a linear CC-context $\ctxtLCC'$.
  
  \item It overlaps with $\ctxtLCC$.
  \begin{itemize}
    \item Suppose the $\Can$-step is a $\rB$-step. Then $\ctxtLCC =
    \ctxtapp{\ctxtLCC_1}{\termname{\alpha}{\ctxtapp{\ctxtLTC_1}{\termapp{\ctxtapp{\ctxt{L}}{\termabs{x}{\ctxtLTC_2}}}{t_2}}}}$,
    $\ctxt{R'} \coloneq \ctxt{R}$, $\ctxtLCC' \coloneq
    \ctxtapp{\ctxtLCC_1}{\termname{\alpha}{\ctxtapp{\ctxtLTC_1}{\ctxtapp{\ctxt{L}}{\termsubs{x}{t_2}{\ctxtLTC_2}}}}}$
    and $c' \coloneq c$. Thus, $$
\begin{tikzcd}
\ctxtapp{\ctxt{R}}{\ctxtapp{\ctxtLCC}{c}} \arrow{d}[left]{\rB}  &[-20pt] \mathcal{R} &[-20pt]
\ctxtapp{\ctxtLCC_1}{\termname{\alpha}{\ctxtapp{\ctxtLTC_1}{\termapp{\ctxtapp{\ctxt{L}}{\termabs{x}{\ctxtapp{\ctxtLTC_2}{\ctxtapp{\ctxt{R}}{c}}}}}{t_2}}}} \arrow[dotted]{d}[left]{\rB}
\\
\ctxtapp{\ctxt{R}'}{\ctxtapp{\ctxtLCC'}{c'}}                    &[-20pt] \mathcal{R} &[-20pt]
\ctxtapp{\ctxtLCC_1}{\termname{\alpha}{\ctxtapp{\ctxtLTC_1}{\ctxtapp{\ctxt{L}}{\termsubs{x}{t_2}{\ctxtapp{\ctxtLTC_2}{\ctxtapp{\ctxt{R}'}{c}}}}}}}
\end{tikzcd} $$

    \item Suppose the $\Can$-step is a $\rM$-step. Then $\ctxtLCC =
    \ctxtapp{\ctxtLCC_1}{\termname{\alpha}{\ctxtapp{\ctxtLTC_1}{\termapp{\ctxtapp{\ctxt{L}}{\termcont{\alpha}{\ctxtLTC_2}}}{t_2}}}}$,
    $\ctxt{R'} \coloneq \ctxt{R}$, $\ctxtLCC' \coloneq
    \ctxtapp{\ctxtLCC_1}{\termname{\alpha}{\ctxtapp{\ctxtLTC_1}{\ctxtapp{\ctxt{L}}{\termcont{\alpha'}{\termrepl{\alpha'}{\alpha}{t_2}{\ctxtLTC_2}}}}}}$
    and $c' \coloneq c$. Thus, $$\kern-2em
\begin{tikzcd}
\ctxtapp{\ctxt{R}}{\ctxtapp{\ctxtLCC}{c}} \arrow{d}[left]{\rM}  &[-25pt] \mathcal{R} &[-25pt]
\ctxtapp{\ctxtLCC_1}{\termname{\alpha}{\ctxtapp{\ctxtLTC_1}{\termapp{\ctxtapp{\ctxt{L}}{(\termcont{\alpha}{\ctxtapp{\ctxtLTC_2}{\ctxtapp{\ctxt{R}}{c}}})}}{t_2}}}} \arrow[dotted]{d}[left]{\rM}
\\
\ctxtapp{\ctxt{R}'}{\ctxtapp{\ctxtLCC'}{c'}}                    &[-25pt] \mathcal{R} &[-25pt]
\ctxtapp{\ctxtLCC_1}{\termname{\alpha}{\ctxtapp{\ctxtLTC_1}{\ctxtapp{\ctxt{L}}{\termcont{\alpha'}{\termrepl{\alpha'}{\alpha}{t_2}{\ctxtapp{\ctxtLTC_2}{\ctxtapp{\ctxt{R}'}{c}}}}}}}}
\end{tikzcd} $$

    \item Suppose the $\Can$-step is a $\rC$-step. There are two further cases:
    \begin{itemize}
      \item  It does not overlap with $c$. Then $\ctxtLCC =
      \termrepl{\beta}{\alpha'}{s'}{\ctxtapp{\ctxtLCC_1}{\termrepl{\alpha'}{\alpha}{s}{\ctxtLCT}}}$,
      where in particular $s, s' \neq \termemst$. Moreover, $\ctxt{L'} \coloneq
      \ctxt{L}$, $\ctxtLCC' \coloneq
      \ctxtapp{\ctxtLCC_1}{\termrepl{\beta}{\alpha}{\termpush{s}{s'}}{\ctxtLCT}}$
      and $c' \coloneq c$. Thus, $$
\begin{tikzcd}
\ctxtapp{\ctxt{R}}{\ctxtapp{\ctxtLCC}{c}} \arrow{d}[left]{\rC}  &[-20pt] \mathcal{R} &[-20pt]
\termrepl{\beta}{\alpha'}{s'}{\ctxtapp{\ctxtLCC_1}{\termrepl{\alpha'}{\alpha}{s}{\ctxtapp{\ctxtLCT}{\ctxtapp{\ctxt{R}}{c}}}}} \arrow[dotted]{d}[left]{\rC}
\\
\ctxtapp{\ctxt{R}'}{\ctxtapp{\ctxtLCC'}{c'}}                    &[-20pt] \mathcal{R} &[-20pt]
\ctxtapp{\ctxtLCC_1}{\termrepl{\beta}{\alpha}{\termpush{s}{s'}}{\ctxtapp{\ctxtLCT}{\ctxtapp{\ctxt{R}'}{c'}}}}
\end{tikzcd} $$
     
      \item It overlaps with $c$. Then $c =
      \ctxtapp{\ctxtLCC_3}{\termrepl{\alpha'}{\alpha}{s}{c_1}}$, $\ctxtLCC =
      \ctxtapp{\ctxt{LCC}_1}{\termrepl{\beta}{\alpha'}{s'}{\ctxtLCC_2}}$, where
      $s, s' \neq \termemst$. Moreover, $c' \coloneq
      \ctxtapp{\ctxt{LCC}_3}{\termrepl{\beta}{\alpha}{\termpush{s}{s'}}{c_1}}$,
      $\ctxt{R'} \coloneq \ctxt{R}$, $\ctxtLCC' \coloneq
      \ctxtapp{\ctxtLCC_1}{\ctxt{LCC}_2}$. Thus, $$
\begin{tikzcd}
\ctxtapp{\ctxt{R}}{\ctxtapp{\ctxtLCC}{c}} \arrow{d}[left]{\rC}  &[-20pt] \mathcal{R} &[-20pt]
\ctxtapp{\ctxtLCC_1}{\termrepl{\beta}{\alpha'}{s'}{\ctxtapp{\ctxt{LCC_2}}{\ctxtapp{\ctxt{R}}{\ctxtapp{\ctxtLCC_3}{\termrepl{\alpha'}{\alpha}{s}{c_1}}}}}} \arrow[dotted]{d}[left]{\rC}
\\
\ctxtapp{\ctxt{R}'}{\ctxtapp{\ctxtLCC'}{c'}}                    &[-20pt] \mathcal{R} &[-20pt]
\ctxtapp{\ctxtLCC_1}{\ctxtapp{\ctxt{LCC}_2}{\ctxtapp{\ctxt{R}'}{\ctxtapp{\ctxt{LCC}_3}{\termrepl{\beta}{\alpha}{\termpush{s}{s'}}{c_1}}}}}
\end{tikzcd} $$
    \end{itemize}
      
    \item Suppose the $\Can$-step is a $\rW$-step. There are two further cases:
    \begin{itemize}
      \item It does not overlap with $c$. Then $\ctxtLCC =
      \termrepl{\alpha'}{\alpha}{s}{\ctxtapp{\ctxtLCC_1}{\termrepl{\alpha}{\beta}{\termemst}{\ctxtLCT}}}$,
      where $s \neq \termemst$. Moreover, $\ctxt{R'} \coloneq \ctxt{R}$,
      $\ctxtLCC' \coloneq
      \ctxtapp{\ctxtLCC_1}{\termrepl{\alpha'}{\alpha}{\termemst}{\termrepl{\alpha}{\beta}{s}{\ctxtLCT}}}$
      and $c' \coloneq c$. Thus, $$
\begin{tikzcd}
\ctxtapp{\ctxt{R}}{\ctxtapp{\ctxtLCC}{c}} \arrow{d}[left]{\rW}  &[-20pt] \mathcal{R} &[-20pt]
\termrepl{\alpha'}{\alpha}{s}{\ctxtapp{\ctxtLCC_1}{\termrepl{\alpha}{\beta}{\termemst}{\ctxtapp{\ctxtLCT}{\ctxtapp{\ctxt{R}}{c}}}}} \arrow[dotted]{d}[left]{\rW}
 \\
\ctxtapp{\ctxt{R}'}{\ctxtapp{\ctxtLCC'}{c'}}                    &[-20pt] \mathcal{R} &[-20pt]
\ctxtapp{\ctxtLCC_1}{\termrepl{\alpha'}{\alpha}{\termemst}{\termrepl{\alpha}{\beta}{s}{\ctxtapp{\ctxtLCT}{\ctxtapp{\ctxt{R'}}{c'}}}}}
\end{tikzcd} $$

    \item It overlaps with $c$. Then $c =
    \ctxtapp{\ctxtLCC_3}{\termrepl{\alpha}{\beta}{\termemst}{c_1}}$, $\ctxtLCC
    = \ctxtapp{\ctxtLCC_1}{\termrepl{\alpha'}{\alpha}{s}{\ctxtLCC_2}}$, where
    $s \neq \termemst$. Moreover, $\ctxt{R'} \coloneq \ctxt{R}$, $\ctxtLCC'
    \coloneq \ctxtapp{\ctxtLCC_1}{\ctxtLCC_2}$ and $c' \coloneq
    \ctxtapp{\ctxtLCC_3}{\termrepl{\alpha'}{\alpha}{\termemst}{\termrepl{\alpha}{\beta}{s}{c_1}}}$.
    Thus, $$
\begin{tikzcd}
\ctxtapp{\ctxt{R}}{\ctxtapp{\ctxtLCC}{c}} \arrow{d}[left]{\rW}  &[-20pt] \mathcal{R} &[-20pt]
\ctxtapp{\ctxtLCC_1}{\termrepl{\alpha'}{\alpha}{s}{\ctxtapp{\ctxtLCC_2}{\ctxtapp{\ctxt{R}}{\ctxtapp{\ctxtLCC_3}{\termrepl{\alpha}{\beta}{\termemst}{c_1}}}}}} \arrow[dotted]{d}[left]{\rW}
\\
\ctxtapp{\ctxt{R}'}{\ctxtapp{\ctxtLCC'}{c'}}                    &[-20pt] \mathcal{R} &[-20pt]
\ctxtapp{\ctxtLCC_1}{\ctxtapp{\ctxtLCC_2}{\ctxtapp{\ctxt{R}'}{\ctxtapp{\ctxtLCC_3}{\termrepl{\alpha'}{\alpha}{\termemst}{\termrepl{\alpha}{\beta}{s}{c_1}}}}}} 
\end{tikzcd} $$
\end{itemize}

    \item There is no other possible case because of the hypothesis of the
    initial relation.
  \end{itemize}
\end{itemize}
\end{proof}

%%% Local Variables:
%%% mode: latex
%%% TeX-master: "main"
%%% End:

%% Substitution and Replacement out ot BM
\gettoappendix{l:subs-repl-out-of-BM}

\begin{proof}
\begin{enumerate}
  \item Consider the term $\ctxtapp{\ctxt{L}}{\ctxtapp{\ctxtLTT}{t}}$. Clearly
  $(\ctxtapp{\ctxt{L}}{\ctxtapp{\ctxtLTT}{t}},
  \ctxtapp{\ctxtLTT}{\ctxtapp{\ctxt{L}}{t}}) \in \mathcal{R}$ where $\mathcal{R}
  \eqdef \{(\ctxtapp{\ctxt{L}}{\ctxtapp{\ctxtLTT}{t}},
  \ctxtapp{\ctxtLTT}{\ctxtapp{\ctxt{L}}{t}}) \mid
  \notatall{\bv{\ctxt{L}}}{\ctxtLTT} \land \freeFor{\ctxt{L}}{\ctxtLTT}\}$.
  Moreover, since $\mathcal{R}$ is a strong $\Can$-bisimulation
  (Lem.~\ref{l:tt-bisimulation}), the
  $\BMCform$ of $\ctxtapp{\ctxt{L}}{\ctxtapp{\ctxtLTT}{t}}$ must be of the form
  $\ctxtapp{\ctxt{L}'}{\ctxtapp{\ctxtLTT'}{t'}}$. Moreover, for the same
  reason, the $\BMCform$ of $\ctxtapp{\ctxtLTT}{\ctxtapp{\ctxt{L}}{t}}$ is
  $\ctxtapp{\ctxtLTT'}{\ctxtapp{\ctxt{L}'}{t'}}$. By definition of the
  equivalence, using $\eqexsubs$ multiple times
  $\ctxtapp{\ctxt{L}'}{\ctxtapp{\ctxtLTT'}{t'}} \simeq
  \ctxtapp{\ctxtLTT'}{\ctxtapp{\ctxt{L}'}{t'}}$.

  \item Consider the term $\ctxtapp{\ctxt{R}}{\ctxtapp{\ctxtLCC}{c}}$. Clearly
  $(\ctxtapp{\ctxt{R}}{\ctxtapp{\ctxtLCC}{c}},
  \ctxtapp{\ctxtLCC}{\ctxtapp{\ctxt{R}}{c}}) \in \mathcal{R} \eqdef
  \{(\ctxtapp{\ctxt{R}}{\ctxtapp{\ctxtLCC}{c}},
  \ctxtapp{\ctxtLCC}{\ctxtapp{\ctxt{R}}{c}}) \mid
  \notatall{\bn{\ctxt{R}}}{\ctxtLCC} \land
  \freeFor{\ctxt{R}}{\ctxtLCC}\}$. Moreover, since $\mathcal{R}$ is a strong
  $\Can$-bisimulation (Lem.~\ref{l:cc-bisimulation}), the $\BMCform$ of
  $\ctxtapp{\ctxt{R}}{\ctxtapp{\ctxtLCC}{c}}$ must be of the form
  $\ctxtapp{\ctxt{R}'}{\ctxtapp{\ctxtLCC'}{c'}}$. Moreover, for the same
  reason, the $\BMCform$ of $\ctxtapp{\ctxtLCC}{\ctxtapp{\ctxt{R}}{c}}$ is
  $\ctxtapp{\ctxtLCC'}{\ctxtapp{\ctxt{R}'}{c'}}$. By definition of the
  equivalence, using $\eqexrepl$ multiple times
  $\ctxtapp{\ctxt{R}'}{\ctxtapp{\ctxtLCC'}{c'}} \simeq
  \ctxtapp{\ctxtLCC'}{\ctxtapp{\ctxt{R}'}{c'}}$.
\end{enumerate}
\end{proof}

%%% Local Variables:
%%% mode: latex
%%% TeX-master: "main"
%%% End:

\begin{lemma}
    \label{l:concatenation-stacks}
    If $\Gam_s \vdash s: S \mid \Del_s$ and
    $\Gam_{s'} \vdash s':S' \mid \Del_{s'}$, then
    $\Gam_s \cup \Gam_{s'} \vdash \termpush{s}{s'}: \typepush{S}{S'} \mid \Del_s \cup \Del_{s'}$.
  \end{lemma}

\begin{lemma}[Subject Reduction/Expansion for $\BMCName$]
    \label{l:BMC}
    Let $o\in \objects{\LMfull} $. Let $\Gam \vdash o:\type \mid \Del$.
    If $o \Rew{\BMCName} o'$ (resp. $o' \Rew{\BMCName} o$),
    then $\Gam \vdash o':\type \mid \Del$.
  \end{lemma}

\begin{proof}
The cases $o \Rew{\mathsf{BM}} o'$ were already proved in Lem.~\ref{l:SR}. The
cases $o' \Rew{\mathsf{BM}} o$ are similar.

Let us consider $o =
\termrepl{\beta'}{\beta}{s'}{\ctxtapp{\ctxtLCC}{\termrepl{\beta}{\al}{s}{c}}}
\Rew{\rC} \ctxtapp{\ctxtLCC}{\termrepl{\beta'}{\al}{\termpush{s}{s'}}{c}} =
o'$, where $s, s' \neq \termemst$, and  $\beta \notin \fn{c,\ctxtLCC,s}$ and
$\freeFor{(s',\beta')}{\ctxtLCC}$ (the case $o' \Rew{\rC} o$ is similar). Also,
we can assume $\beta \notin \fn{s'}$ by $\alpha$-conversion and
Lem.~\ref{l:relevance}. The proof is by induction on $\ctxtLCC$. We only show
here the case of the empty context, as the other ones are straightforward. The
starting typing derivation has the following form: 
{\scriptsize \[
\hfill
\Rule{\Rule{\Gam_c \vdash c \mid \Del_c, (\alpha : T_\alpha)^{\esta}, (\beta : T_\beta)^{0}, (\beta' : B)^{\esta}
            \quad
            \Gam_s \vdash s : S \mid \Del_s, (\beta : T_\beta)^{0}, (\beta' : B)^{\esta}
           }
           {\Gam_s \cup \Gam_c \vdash \termrepl{\beta}{\al}{s}{c} \mid \Del_s \cup \Del_c, \beta : T_\beta, (\beta' : B)^{\esta}}
           {}
      \quad
      \Gam_{s'} \vdash s' : S' \mid \Del_{s'}, (\beta' : B)^{\esta}
     }
     {\Gam_s \cup \Gam_c \cup \Gam_{s'} \vdash \termrepl{\beta'}{\beta}{s'}{\termrepl{\beta}{\al}{s}{c}} \mid \Del_s \cup \Del_c \cup \Del_{s'}, \beta' : B}
     {}
\hfill
\]}
where $T_\alpha = S \tarrow S' \tarrow B$, $T_\beta =  S' \tarrow B$, and, by
definition, $S \tarrow S' \tarrow B$ is equal to $\typepush{S}{S'} \tarrow B$.
Also, we can assume $\alpha \notin \dom{\Del_s}$ by $\alpha$-conversion and
Lem.~\ref{l:relevance}. The resulting type derivation can be constructed as
follows, using in particular Lem.~\ref{l:concatenation-stacks} to
derive the right-hand side of the derivation:
{\scriptsize \[
\hfill
\Rule{\Gam_c \vdash c \mid \Del_c, (\alpha : T_\alpha)^{\esta}, (\beta : T_\beta)^{0}, (\beta' : B)^{\esta}
      \quad
      \Rule{\Gam_s \vdash s : S \mid \Del_s, (\beta : T_\beta)^{0}, (\beta' : B)^{\esta}
            \quad
            \Gam_{s'} \vdash s' : S' \mid \Del_{s'}, (\beta' : B)^{\esta}
           }
           {\Gam_s \cup \Gam_{s'} \vdash \termpush{s}{s'} : \termpush{S}{S'} \mid \Del_s \cup \Del_{s'}, (\beta : T_\beta)^{0}, (\beta' : B)^{\esta}}
           {}
     }
     {\Gam_c \cup \Gam_s \cup \Gam_{s'} \vdash \termrepl{\beta'}{\al}{\termpush{s}{s'}}{c} \mid \Del_c \cup \Del_s \cup \Del_{s'}, (\beta : T_\beta)^{0}, \beta' : B}
     {}
\hfill
\]}
Let us consider $o =
\termrepl{\beta'}{\beta}{s'}{\ctxtapp{\ctxtLCC}{\termrepl{\beta}{\al}{\termemst}{c}}}
\Rew{\rW}
\ctxtapp{\ctxtLCC}{\termrepl{\beta'}{\beta}{\termemst}{\termrepl{\beta}{\al}{s'}{c}}}
= o'$, where $s' \neq \termemst$, and $\beta \notin \fn{c, \ctxtLCC}$
and $\freeFor{(s',\beta')}{\ctxtLCC}$ (the case $o' \Rew{\rW} o$ is similar).
Also, we can assume $\beta \notin \fn{s'}$ by $\alpha$-conversion and
Lem.~\ref{l:relevance}. The proof is by induction on $\ctxtLCC$. We only show
here the case of the empty context, as the other ones are straightforward.
The starting typing derivation has the following form:
{\scriptsize \[
\hfill
\Rule{\Rule{\Gam_c \vdash c \mid \Del_c, (\alpha : T_\alpha)^{\esta}, (\beta : T_\beta)^{0}, (\beta' : B)^{\esta}
            \quad
            \emptyset \vdash \termemst : \typeemst \mid \emptyset
           }
           {\Gam_c \vdash \termrepl{\beta}{\al}{\termemst}{c} \mid \Del_c, \beta : T_\beta, (\beta' : B)^{\esta}}
           {}
      \quad
      \Gam_{s'} \vdash s' : S' \mid \Del_{s'}, (\beta' : B)^{\esta}
     }
     {\Gam_c \cup \Gam_{s'} \vdash \termrepl{\beta'}{\beta}{s'}{\termrepl{\beta}{\al}{\termemst}{c}} \mid \Del_c \cup \Del_{s'}, \beta' : B}
     {}
\hfill
\]}
where $T_\alpha = S \tarrow S' \tarrow B$, $T_\beta = S' \tarrow B$, and, by
definition, $S \tarrow S' \tarrow B$ is equal to $\typepush{S}{S'} \tarrow B$.
Also, we can assume $\alpha \notin \dom{\Del_s}$ by $\alpha$-conversion and
Lem.~\ref{l:relevance}. The resulting type derivation can be constructed as
follows, using in particular Lem.~\ref{l:concatenation-stacks}:
{\scriptsize \[
\hfill
\Rule{\Rule{\Gam_c \vdash c \mid \Del_c, (\alpha : T_\alpha)^{\esta}, (\beta : T_\beta)^{0}, (\beta' : B)^{\esta}
            \quad
            \Gam_{s'} \vdash s' : S' \mid \Del_{s'}, (\beta' : B)^{\esta}
           }
           {\Gam_c \cup \Gam_{s'} \vdash \termrepl{\beta}{\al}{s'}{c} \mid \Delta_c \cup \Delta_{s'}, \beta : T_\beta, (\beta' : B)^{\esta}}
           {}
      \quad
      \emptyset \vdash \termemst : \typeemst \mid \emptyset
     }
     {\Gam_c \cup \Gam_{s'} \vdash \termrepl{\beta'}{\beta}{\termemst}{\termrepl{\beta}{\al}{s'}{c}} \mid \Del_c \cup \Del_{s'}, \beta' : B}
     {}
\hfill
\]}
\end{proof}

%%%Local Variables:
%%% mode: latex
%%% TeX-master: "main"
%%% End:

%% Preservation of Types for simeq
\gettoappendix{l:PTSigma}

\begin{proof} 
By induction on the relation $\simeq$.
\begin{itemize}
  \item $o = \termsubs{x}{u}{\ctxtapp{\ctxtLTT}{v}} \simeq_{\eqexsubs}
  \ctxtapp{\ctxtLTT}{\termsubs{x}{u}{v}} = o'$, where $\notatall{x}{\ctxtLTT}$
  and $\freeFor{u}{\ctxtLTT}$. The proof is by induction on $\ctxtLTT$. The
  base case $\ctxtLTT = \Box$ is trivial, and the inductive cases are
  straightforward.
  
  \item $o = \termrepl{\alpha'}{\alpha}{s}{\ctxtapp{\ctxtLCC}{c}}
  \simeq_{\eqexrepl} \ctxtapp{\ctxtLCC}{\termrepl{\alpha'}{\alpha}{s}{c}} =
  o'$, where $\notatall{\alpha}{\ctxtLCC}$ and
  $\freeFor{(s,\alpha')}{\ctxtLCC}$. The proof is by induction on $\ctxtLCC$.
  The base case $\ctxtLCC = \boxdot$ is trivial and the inductive cases are
  straightforward.
  
  \item $o = \termrepl{\alpha'}{\alpha}{s}{(\termname{\alpha}{u})}
  \simeq_{\eqlinear} \termname{\alpha'}{\BMC{\termconc{u}{s}}} = o'$, where
  $\al \notin \fn{u}$, and $s \neq \termemst$. We can also assume by
  $\alpha$-conversion that $\alpha \notin \fn{s}$.
  
  We first show the left-to-right lemma. Let consider a typing derivation for
  $o$, which is necessarily of the following form: \[
  \hfill
  \Rule{\Rule{\Gam_u \vdash u : S \tarrow B \mid \Del_u, (\alpha : S \tarrow B)^{0}, (\alpha' : B)^{\esta}}
             {\Gam_u \vdash \termname{\alpha}{u} \mid \Del_u, \alpha : S \tarrow B, (\alpha' : B)^{\esta}}
             {}
        \quad
        \Gam_s \vdash s: S \mid \Del_s, (\alpha':B)^{\esta}
       }
       {\Gam_u \cup \Gam_s \vdash \termrepl{\alpha'}{\alpha}{s}{(\termname{\alpha}{u})} \mid \Del_u \cup \Del_s, \alpha' : B}
       {}
  \hfill
  \] Now, to construct the resulting derivation we proceed as follows. First of
  all we know by $\alpha$-conversion that $\alpha \notin\fn{s}$. From $\Gam_u
  \vdash u : S \tarrow B \mid \Del_u, (\alpha' : B)^{\esta}$ and $\Gam_s \vdash
  s : S \mid \Del_s, (\alpha' : B)^{\esta}$ we obtain $\Gam_u \cup \Gam_s
  \vdash \termconc{u}{s} : B \mid \Del_u \cup \Del_s, (\alpha' : B)^{\esta}$ by
  Lem.~\ref{l:application}. Then Lem.~\ref{l:BMC} (reduction) gives $\Gam_u
  \cup \Gam_s \vdash \BMC{\termconc{u}{s}} : B \mid \Del_u \cup \Del_s,
  (\alpha' : B)^{\esta}$ because variable and name contexts do not change by
  $\BMCName$-reduction. Then rule $(\nametr)$ gives $\Gam_u \cup \Gam_s \vdash
  o' \mid \Del_u \cup \Del_s, \alpha':B$, as expected.
  
  We now show the right-to-left lemma. Let consider a typing derivation for
  $o'$, whose typing judgement is necessarily of the following form $\Gam \vdash
  \termname{\alpha'}{\BMC{\termconc{u}{s}}} \mid \Del^*, \alpha' : B$, for some
  $B$. This gives $\Gam \vdash \BMC{\termconc{u}{s}} : B \mid \Del^*, (\alpha'
  : B)^{\esta}$ by rule $(\nametr)$. Then Lem.~\ref{l:BMC} (expansion) gives
  $\Gam \vdash \termconc{u}{s} : B \mid \Del^*, (\alpha' : B)^{\esta}$. Now,
  Lem.~\ref{l:application} provides $\Gam = \Gam_u \cup \Gam_s$, $\Del^* =
  \Del_u \cup \Del_s$, $\Gam_u \vdash u : S' \tarrow B \mid \Del_u, (\alpha' :
  B)^{\esta}$ and $\Gam_s \vdash s : S' \mid \Del_s, (\alpha' : B)^{\esta}$.
  Then we conclude $\Gam \vdash
  \termrepl{\alpha'}{\alpha}{s}{(\termname{\alpha}{u})} \mid \Del^*, \alpha' :
  B$ by rules $(\nametr)$ and $(\stackd)$ as expected.
  
  \item $o =
  \termname{\alpha'}{\termabs{x}{\termcont{\alpha}{\termname{\beta'}{\termabs{y}{\termcont{\beta}{u}}}}}}
  \simeq_{\eqpoppop}
  \termname{\beta'}{\termabs{y}{\termcont{\beta}{\termname{\alpha'}{\termabs{x}{\termcont{\alpha}{u}}}}}}$,
  if $\alpha\neq\beta'$ and $\alpha'\neq\beta$. This case is straightforward.
  
  \item $o = \termname{\alpha}{\termcont{\beta}{c}} \simeq_{\eqrho}
  c\exrepl{\alpha}{\beta}{\termemst} = o'$. This case is straightforward and
  uses the fact that $\typeemst \tarrow A = A$ for all type $A$.
  
  \item $o = \termcont{\alpha}{\termname{\alpha}{t}} \simeq_{\eqtheta} t$, if
  $\notfreen{\alpha}{t}$. This is straightforward.
\end{itemize}
\end{proof}

%%% Local Variables:
%%% mode: latex
%%% TeX-master: "main"
%%% End:

%%% Local Variables:
%%% mode: latex
%%% TeX-master: "main"
%%% End:

  %%%%%%%%%%%%%%%%%%%%%%%%%%%%%%%%%%%%%%%%%%%%%%%%%%%%%%%%%%%%%%%%%%%%%%%%%%%%%%%
\section{Appendix: Correspondence Results}
\label{a:completeness}
%%%%%%%%%%%%%%%%%%%%%%%%%%%%%%%%%%%%%%%%%%%%%%%%%%%%%%%%%%%%%%%%%%%%%%%%%%%%%%%

%% Relation between sigma and Sigma
\gettoappendix{l:relation-sigma-Sigma}

\begin{proof}
By induction on $o \simeq_{\sig} p$. We only show the base cases. 
\begin{itemize}
  \item If $o = \termapp{(\termabs{x}{\termabs{y}{u}})}{v} \simeq_{\sigma_1}
  \termabs{y}{\termapp{(\termabs{x}{u})}{v}} = p$, where $x \neq y,
  \notatall{y}{v}$. Then $$
\begin{array}{l@{\enspace}l@{\enspace}l@{\enspace}l@{\enspace}l@{\enspace}l@{\enspace}l}
\BMC{\termapp{(\termabs{x}{\termabs{y}{u}})}{v}} & = &
\BMC{\termsubs{x}{v}{(\termabs{y}{u})}}          & \simeq_{(L.\ref{l:subs-repl-out-of-BM}.\ref{l:subs-out-of-BM})} &
\BMC{\termabs{y}{\termsubs{x}{v}{u}}}            & = &
\BMC{\termabs{y}{\termapp{(\termabs{x}{u})}{v}}}
\end{array} $$

  \item If $o = \termapp{(\termapp{(\termabs{x}{u})}{v})}{w} \simeq_{\sigma_2}
  \termapp{(\termabs{x}{\termapp{u}{w}})}{v} = p$, where $\notatall{x}{w}$. Then $$
\begin{array}{l@{\enspace}l@{\enspace}l@{\enspace}l@{\enspace}l@{\enspace}l@{\enspace}l}
\BMC{\termapp{(\termapp{(\termabs{x}{u})}{v})}{w}} & = &
\BMC{\termapp{\termsubs{x}{v}{u}}{w}}              & \simeq_{(L.\ref{l:subs-repl-out-of-BM}.\ref{l:subs-out-of-BM})} &
\BMC{\termsubs{x}{v}{(\termapp{u}{w})}}            & = &
\BMC{\termapp{(\termabs{x}{\termapp{u}{w}})}{v}}
\end{array} $$

  \item If $o =
  \termapp{(\termabs{x}{\termcont{\alpha}{\termname{\beta}{u}}})}{v}
  \simeq_{\sigma_3}
  \termcont{\alpha}{\termname{\beta}{\termapp{(\termabs{x}{u})}{v}}} = p$,
  where $\notatall{\alpha}{v}$. Then $$\kern-2em
\begin{array}{l@{\enspace}l@{\enspace}l@{\enspace}l@{\enspace}l@{\enspace}l@{\enspace}l}
\BMC{\termapp{(\termabs{x}{\termcont{\alpha}{\termname{\beta}{u}}})}{v}} & = &
\BMC{\termsubs{x}{v}{(\termcont{\alpha}{\termname{\beta}{u}})}}          & \simeq_{(L.\ref{l:subs-repl-out-of-BM}.\ref{l:subs-out-of-BM})} &
\BMC{\termcont{\alpha}{\termname{\beta}{\termsubs{x}{v}{u}}}}            & = &
\BMC{\termcont{\alpha}{\termname{\beta}{\termapp{(\termabs{x}{u})}{v}}}}
\end{array} $$

  \item If $o =
  \termname{\alpha'}{\termapp{(\termcont{\alpha}{\termname{\beta'}{\termapp{(\termcont{\beta}{c})}{v}}})}{w}}
  \simeq_{\sigma_4}
  \termname{\beta'}{\termapp{(\termcont{\beta}{\termname{\alpha'}{\termapp{(\termcont{\alpha}{c})}{w}}})}{v}}
  = p$, where $\notatall{\alpha}{v}, \notatall{\beta}{w}$. Let $\alpha'', \beta''$ be
  two distinct fresh names. Then $$\kern-1em
\begin{array}{l@{\enspace}l@{\enspace}l@{\enspace}l}
\BMC{\termname{\alpha'}{(\termcont{\alpha}{\termname{\beta'}{(\termcont{\beta}{c})v}})w}}                                                               & = & %(\alpha'', \beta'' \text{ dist. and fresh}) \\
\termname{\alpha'}{\termcont{\alpha''}{\BMC{\termrepl{\alpha''}{\alpha}{w}{(\termname{\beta'}{\termcont{\beta''}{\termrepl{\beta''}{\beta}{v}{c}}})}}}} & \simeq_{(L.\ref{l:subs-repl-out-of-BM}.\ref{l:repl-out-of-BM})} \\
\termname{\alpha'}{\termcont{\alpha''}{\BMC{\termrepl{\alpha''}{\alpha}{w}{\termname{\beta'}{\termcont{\beta''}{\termrepl{\beta''}{\beta}{v}{c}}}}}}}   & = &
\termname{\alpha'}{\termcont{\alpha''}{\termname{\beta'}{\termcont{\beta''}{\BMC{c\exrepl{\beta''}{\beta}{v}\exrepl{\al''}{\al}{w}}}}}}                 & \simeq_{(L.\ref{l:subs-repl-out-of-BM}.\ref{l:repl-out-of-BM})} \\
\termname{\alpha'}{\termcont{\alpha''}{\termname{\beta'}{\termcont{\beta''}{\BMC{c\exrepl{\al''}{\al}{w}\exrepl{\beta''}{\beta}{v}}}}}}                 & \simeq &
\termname{\beta'}{\termcont{\beta''}{\termname{\alpha'}{\termcont{\alpha''}{\BMC{c\exrepl{\al''}{\al}{w}\exrepl{\beta''}{\beta}{v}}}}}}                 & = \\
\BMC{\termname{\beta'}{\termapp{(\termcont{\beta}{\termname{\alpha'}{\termapp{(\termcont{\alpha}{c})}{w}}})}{v}}}
\end{array} $$

  \item If $o =
  \termname{\alpha'}{\termapp{(\termcont{\alpha}{\termname{\beta'}{\termabs{x}{\termcont{\beta}{c}}}})}{v}}
  \simeq_{\sigma_5}    
  \termname{\beta'}{\termabs{x}{\termcont{\beta}{\termname{\alpha'}{\termapp{(\termcont{\alpha}{c})}{v}}}}}
  = p$, where $\notatall{x}{v}, \notatall{\beta}{v}$. Let $\alpha''$ be a fresh name.
  Then $$%\kern-1em
\begin{array}{l@{\enspace}l@{\enspace}l@{\enspace}l}
\BMC{\termname{\alpha'}{\termapp{(\termcont{\alpha}{\termname{\beta'}{\termabs{x}{\termcont{\beta}{c}}}})}{v}}}                    & = & %(\alpha'' \text{ fresh}) \\
\termname{\alpha'}{\termcont{\alpha''}{\termname{\beta'}{\termabs{x}{\termcont{\beta}{\BMC{\termrepl{\alpha''}{\alpha}{v}{c}}}}}}} & \simeq_{\eqrho} \\
(\termname{\beta'}{\termabs{x}{\termcont{\beta}{\BMC{\termrepl{\alpha''}{\alpha}{v}{c}}}}})\neren{\alpha''}{\alpha'}               & \simeq_{(L.\ref{l:subs-repl-out-of-BM}.\ref{l:repl-out-of-BM})} &
\termname{\beta'}{\termabs{x}{\termcont{\beta}{\BMC{\termrepl{\alpha''}{\alpha}{v}{c}}\neren{\alpha''}{\alpha'}}}}                 & \simeq_{\eqrho} \\
\termname{\beta'}{\termabs{x}{\termcont{\beta}{\termname{\alpha'}{\termcont{\alpha''}{\BMC{\termrepl{\alpha''}{\alpha}{v}{c}}}}}}} & = &
\termname{\beta'}{\termabs{x}{\termcont{\beta}{\termname{\alpha'}{\BMC{\termapp{(\termcont{\alpha}{c})}{v}}}}}}                    & = \\ 
\BMC{\termname{\beta'}{\termabs{x}{\termcont{\beta}{\termname{\alpha'}{\termapp{(\termcont{\alpha}{c})}{v}}}}}}
\end{array} $$

  \item If $o =
  \termname{\alpha'}{\termabs{x}{\termcont{\alpha}{\termname{\beta'}{\termabs{y}{\termcont{\beta}{c}}}}}}
  \simeq_{\sigma_6}
  \termname{\beta'}{\termabs{y}{\termcont{\beta}{\termname{\alpha'}{\termabs{x}{\termcont{\alpha}{c}}}}}}
  = p$, where $x \neq y$. Then $$
\begin{array}{l@{\enspace}l@{\enspace}l@{\enspace}l}
\BMC{\termname{\alpha'}{\termabs{x}{\termcont{\alpha}{\termname{\beta'}{\termabs{y}{\termcont{\beta}{c}}}}}}} & = &
\termname{\alpha'}{\termabs{x}{\termcont{\alpha}{\termname{\beta'}{\termabs{y}{\termcont{\beta}{\BMC{c}}}}}}} & \simeq_{\eqpoppop}  \\
\termname{\beta'}{\termabs{y}{\termcont{\beta}{\termname{\alpha'}{\termabs{x}{\termcont{\alpha}{\BMC{c}}}}}}} & = &
\BMC{\termname{\beta'}{\termabs{y}{\termcont{\beta}{\termname{\alpha'}{\termabs{x}{\termcont{\alpha}{c}}}}}}}
\end{array} $$

  \item If $o = \termname{\alpha}{\termcont{\beta}{c}} \simeq_{\sigma_7}
  \replace{\beta}{\alpha}{\termemst}{c} = p$. Then $$\kern-1em
\begin{array}{l@{\enspace}l@{\enspace}l@{\enspace}l@{\enspace}l@{\enspace}l@{\enspace}l@{\enspace}l@{\enspace}l}
\BMC{\termname{\alpha}{\termcont{\beta}{c}}} & = &
\termname{\alpha}{\termcont{\beta}{\BMC{c}}} & \simeq_{\eqrho} &
\BMC{c}\neren{\beta}{\alpha}                 & \simeq_{\exren} &
\replace{\beta}{\alpha}{\termemst}{\BMC{c}}  & = &
\BMC{\replace{\beta}{\alpha}{\termemst}{c}}
\end{array} $$

  \item If $o = \termcont{\alpha}{\termname{\alpha}{t}} \simeq_{\sigma_8} t =
  p$, where $\notatall{\alpha}{t}$. Then $$
\begin{array}{l@{\enspace}l@{\enspace}l@{\enspace}l@{\enspace}l}
\BMC{\termcont{\alpha}{\termname{\alpha}{t}} } & = &
\termcont{\alpha}{\termname{\alpha}{\BMC{t}}}  & \simeq_{\eqtheta} &
\BMC{t}
\end{array} $$
\end{itemize}
\end{proof}

%%% Local Variables:
%%% mode: latex
%%% TeX-master: "main"
%%% End:

In Sec.~\ref{s:translation}, we have introduced the notation $\topnd{\pi}$ to
denote the multiplicative normal form of the proof-net associated to the
typying derivation $\pi$. By abuse of notation we use now the same notation
$\topnd{\_}$ directly on proof-nets to refer to its multiplicative normal forms
as well.

%% BMC to proof-nets
\gettoappendix{l:bmc-to-proofnets}

\begin{proof}
By induction on $o \Rew{\Can} p$. We only show the base cases.
\begin{itemize}
  \item $\rB$: then $o = \termapp{\ctxtapp{\ctxt{L}}{\termabs{x}{t}}}{u}$ and
  $p = \ctxtapp{\ctxt{L}}{\termsubs{x}{u}{t}}$ with $\freeFor{u}{\ctxt{L}}$.
  The proof follows by induction on $\ctxt{L}$. We only illustrate here the
  base case ($\ctxt{L} = \Box$) where $x \in \fv{t}$,
  the case $x \notin \fv{t}$ being similar.
  The  inductive case  follows immediately from the \ih
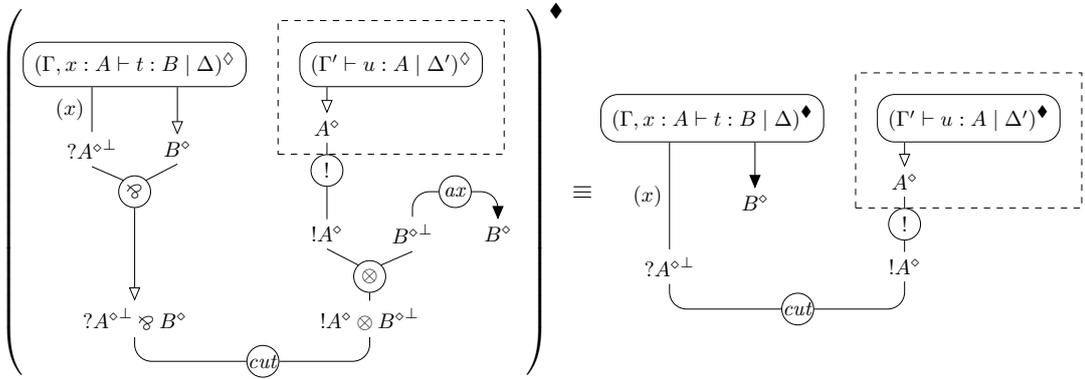
  $$\kern-2em \mcnf{\scalebox{.8}{\input{proofnets/bmc/ruleB2}}} \equiv \scalebox{.8}{\input{proofnets/bmc/ruleB3}}$$
  where each side of the structural equivalence is the expansion of %the definition of
  $\topnd{(\Gam, \Gam' \vdash \termapp{(\termabs{x}{t})}{u}:B \mid \Del, \Del')}$ and
  $\topnd{(\Gam, \Gam'\vdash \termsubs{x}{u}{t}:B \mid \Del, \Del')}$ respectively.
  
  \item $\rM$: then $o = \termapp{\ctxtapp{\ctxt{L}}{\termcont{\alpha}{c}}}{u}$
  and $p =
  \ctxtapp{\ctxt{L}}{\termcont{\alpha'}{\termrepl{\alpha'}{\alpha}{u}{c}}}$
  with $\freeFor{u}{\ctxt{L}}$ and $\alpha' \notin \fn{c,u,\alpha,\ctxt{L}}$.
  This case also follows by induction on $\ctxt{L}$. As before, we only
  illustrate here the base case ($\ctxt{L} = \Box$) for $\alpha \in \fn{c}$,
  the case $\alpha \notin \fn{c}$ being similar.
  The inductive case follows immediately from the \ih\ In this case
  $\topnd{(\Gam, \Gam' \vdash \termapp{(\termcont{\alpha}{c})}{u}:B \mid \Del, \Del')}$
  and
  $\topnd{(\Gam, \Gam' \vdash \termcont{\alpha'}{\termrepl{\alpha'}{\alpha}{u}{c}} \mid \Del, \Del')}$
  are both structurally equivalent, by definition, to
  $$\mcnf{\scalebox{.8}{\input{proofnets/bmc/ruleM2}}}$$
  
  \item $\rC$: then $o =
  \termrepl{\al'}{\al}{s}{\ctxtapp{\ctxtLCC}{\termrepl{\al}{\beta}{s'}{c}}}$
  and $p = \ctxtapp{\ctxtLCC}{\termrepl{\al'}{\beta}{\termpush{s'}{s}}{c}}$
  with $\al \notin (c,\ctxtLCC,s')$, $\freeFor{(s,\al')}{\ctxtLCC}$
  and $s, s' \neq \termemst$. The proof follows by induction on $\ctxtLCC$, the
  most interesting case being the base case $(\ctxtLCC = \boxdot)$ which we
  illustrate next. We consider the case $\beta \in \fn{c}$ and $\alpha' \notin
  \fn{s}$, the other one being similar.
  $$\kern-7em \mcnf{\scalebox{.8}{\input{proofnets/bmc/ruleC2}}} \equiv \mcnf{\scalebox{.8}{\input{proofnets/bmc/ruleC3}}}$$
  where each proof-net is the unfolding of %the definition of
  $\topnd{(\Gam, \Gam', \Gam'' \vdash
  \termrepl{\al'}{\al}{s}{\termrepl{\al}{\beta}{s'}{c}} : B \mid \Del, \Del',
  \Del'')}$ and $\topnd{(\Gam, \Gam', \Gam'' \vdash
  \termrepl{\al'}{\beta}{\termpush{s'}{s}}{c} : B \mid \Del, \Del', \Del'')}$
  respectively. It is worth noticing that the cut between the wire $\alpha$ in
  $\topn{(\Gam' \vdash s':S' \mid \Del')_{\typefunc{S}{B}}}$ and the
  distinguished output of $\topn{(\Gam'' \vdash s : S \mid \Del'')_{B}}$ is
  multiplicative (recall $s, s' \neq \termemst$ and stacks are translated into
  $\otimes$-trees, whose base case is an axiom). The multiplicative normal
  form contracts this multiplicative cut, thus resulting  in a new
  $\otimes$-tree that turns out to be $\topn{(\Gam',\Gam'' \vdash
  \termpush{s'}{s} : \typepush{S'}{S} \mid \Del', \Del'')_{B}}$.
  
  \item $\rW$: then $o =
  \termrepl{\al'}{\al}{s}{\ctxtapp{\ctxtLCC}{\termrepl{\al}{\beta}{\termemst}{c}}}$
  and $p = 
  \ctxtapp{\ctxtLCC}{\termrepl{\al'}{\al}{\termemst}{\termrepl{\al}{\beta}{s}{c}}}$
  with $\notatall{\al}{(c,\ctxtLCC)}$, $\freeFor{(s,\al')}{\ctxtLCC}$ and $s \neq
  \termemst$. The proof follows by induction on $\ctxtLCC$, the most
  interesting case being the base case $(\ctxtLCC = \boxdot)$ which we
  illustrate next. We consider the case $\beta \in \fn{c}$ and $\alpha' \notin
  \fn{s}$, the other ones being similar.
  $$\kern-7em \mcnf{\scalebox{.8}{\input{proofnets/bmc/ruleW2}}} \equiv \mcnf{\scalebox{.8}{\input{proofnets/bmc/ruleW3}}}$$
  where each proof-net is the expansion of the
  definition of $\topnd{(\Gam, \Gam' \vdash
  \termrepl{\al'}{\al}{s}{\termrepl{\al}{\beta}{\termemst}{c}} : B \mid \Del,
  \Del')}$ and $\topnd{(\Gam, \Gam', \vdash
  \termrepl{\al'}{\al}{\termemst}{\termrepl{\al}{\beta}{s}{c}} : B \mid \Del,
  \Del')}$ respectively. The equivalence follows form the fact that
  $\topn{(\emptyset \vdash \termemst:\typeemst \mid \emptyset)_{C}}$ is an
  axiom node for any given type $C$. Thus, the internal cut on left-hand
  side proof-net is multiplicative, as well as the external cut on the
  right-hand side one. The multiplicative normal form contracts these
  multiplicative cuts, thus resulting in equicalente proof-nets.
\end{itemize}
\end{proof}

%%% Local Variables:
%%% mode: latex
%%% TeX-master: "main"
%%% End:

The following lemma introduces an alternative top down definition of linear
contexts instead of the original more natural bottom up definition in
Sec.~\ref{s:Sigma-equivalence}. This will prove convenient when analyzing
Lem.~\ref{l:soundness}.

\begin{lemma}[Alternative Definition of Linear Contexts]
\label{l:contexts}
Linear contexts can be alternatively be defined by the following grammars: $$
\begin{array}{r@{\enspace}r@{\enspace}c@{\enspace}l}
{\textbf{(Linear TT Contexts)}} & 
\ctxtLTT & \Coloneq & \Box \mid \termapp{\Box}{t} \mid \termabs{x}{\Box} \mid \termcont{\alpha}{\ctxtLCT} \mid \termsubs{x}{t}{\Box} \\
{\textbf{(Linear TC Contexts)}} &
\ctxtLTC & \Coloneq & \termapp{\ctxtLTC}{t} \mid \termabs{x}{\ctxtLTC} \mid \termcont{\alpha}{\boxdot} \mid \termsubs{x}{t}{\ctxtLTC} \\
{\textbf{(Linear CC Contexts)}} & 
\ctxtLCC & \Coloneq & \boxdot \mid \termname{\alpha}{\ctxtLTC} \mid \termrepl{\alpha'}{\alpha}{s}{\boxdot} \\
{\textbf{(Linear CT Contexts)}} & 
\ctxtLCT & \Coloneq & \termname{\alpha}{\Box} \mid \termrepl{\alpha'}{\alpha}{s}{\ctxtLCT}
\end{array} $$
\end{lemma}

\begin{proof}
By induction on the grammar of contexts.
\end{proof}

%% Soundness
\gettoappendix{l:soundness}

\begin{proof}
By induction on the derivation of $o \eqsigmaer p$. We only show the base
cases.
\begin{itemize}
  \item $\eqexsubs$: then $o = \termsubs{x}{u}{\ctxtapp{\ctxtLTT}{v}}$ and
  $p = \ctxtapp{\ctxtLTT}{\termsubs{x}{u}{v}}$ with $\notatall{x}{\ctxtLTT}$
  and $\freeFor{u}{\ctxtLTT}$. We analyze the shape of $\ctxtLTT$ resorting
  to Lem.~\ref{l:contexts}.
  \begin{itemize}
    \item $\ctxtLTT = \termapp{\Box}{r}$ with $x \notin \fv{r}$. We only
    illustrate the case where $x \in \fv{v}$. All the other cases are similar.
    %$$\kern-4em
    %\begin{array}{c}
    %\scaleinput{.8}{proofnets/soundness/ttapp1}         \enskip=
    %\mcnf{\scaleinput{.8}{proofnets/soundness/ttapp2}}  \equiv\quad
    %\scaleinput{.8}{proofnets/soundness/ttapp3}
    %\end{array} $$
    In this case
    $\topnd{(\Gam, \Gam', \Gam'' \vdash \termsubs{x}{u}{(\termapp{v}{r})} : B \mid \Del, \Del', \Del'')}$
    and
    $\topnd{(\Gam, \Gam', \Gam'' \vdash \termapp{\termsubs{x}{u}{v}}{r} : B \mid \Del, \Del', \Del'')}$
    are both structurally equivalent, by definition, to
    $$\mcnf{\scalebox{.8}{\input{proofnets/soundness/ttapp2}}}$$
    Note that there is an equivalence $\equiv$ between the second and third
    proof-net, and not an equality, because of the associativity of
    contractions (first equation of the structural equivalence definition).

    \item $\ctxtLTT = \termabs{y}{\Box}$ with $y \notin \fv{u}$. We only
    illustrate the case where $x, y \in \fv{v}$. All the other cases are
    similar.
    %$$
    %\begin{array}{c}
    %\scaleinput{.8}{proofnets/soundness/ttabs1}  \enskip=
    %\scaleinput{.8}{proofnets/soundness/ttabs2}  =\enskip
    %\scaleinput{.8}{proofnets/soundness/ttabs3}
    %\end{array} $$
    In this case
    $\topnd{(\Gam, \Gam' \vdash \termsubs{x}{u}{(\termabs{y}{v})} : \typefunc{A_y}{B} \mid \Del, \Del')}$
    and
    $\topnd{(\Gam, \Gam' \vdash \termabs{y}{\termsubs{x}{u}{v}} : \typefunc{A_y}{B} \mid \Del, \Del')}$
    are both equal, by definition, to
    $$\scalebox{.8}{\input{proofnets/soundness/ttabs2}}$$
    
    \item $\ctxtLTT = \termcont{\alpha}{\ctxtLCT}$. We proceed by induction
    on $\ctxtLCT$:
    \begin{itemize}
      \item $\ctxtLCT = \termname{\beta}{\boxdot}$ with $\alpha, \beta \notin
      \fn{u}$. We only illustrate the case where $x \in \fv{v}$, $\alpha \in
      \fn{v}$ and $\beta \notin \fn{v}$. All the other cases are similar.
      %$$
      %\begin{array}{c}
      %\scaleinput{.8}{proofnets/soundness/ttname1}  \enskip=
      %\scaleinput{.8}{proofnets/soundness/ttname2}  =\enskip
      %\scaleinput{.8}{proofnets/soundness/ttname3}
      %\end{array} $$
      By definition of the translation to proof-nets
      $\topnd{(\Gam, \Gam' \vdash \termsubs{x}{u}{(\termcont{\alpha}{\termname{\beta}{v}})} : A \mid \Del, \Del')}$
      and
      $\topnd{(\Gam, \Gam' \vdash \termcont{\alpha}{\termname{\beta}{\termsubs{x}{u}{v}}} : A \mid \Del, \Del')}$
      are both equal to
      $$\scalebox{.8}{\input{proofnets/soundness/ttname2}}$$

      \item $\ctxtLCT = \termrepl{\beta'}{\beta}{s}{\ctxtLCT'}$ with
      $\notatall{x}{(\ctxtLCT',s)}$, $\alpha, \beta \notin \fn{u}$ and
      $\freeFor{u}{\ctxtLCT'}$. We only illustrate the case where $\beta'$ is
      fresh. All the other cases are similar.
      %$$%\kern-1em
      %\begin{array}{c}
      %\scaleinput{.8}{proofnets/soundness/ttrepl1}         \enskip=
      %\mcnf{\scaleinput{.8}{proofnets/soundness/ttrepl2}}  \equiv \\
      %\\
      %\mcnf{\scaleinput{.8}{proofnets/soundness/ttrepl3}}  \equiv_{(\ih)}
      %\mcnf{\scaleinput{.8}{proofnets/soundness/ttrepl4}}  = \\
      %\\
      %\mcnf{\scaleinput{.8}{proofnets/soundness/ttrepl5}}  =\enskip
      %\scaleinput{.8}{proofnets/soundness/ttrepl6}
      %\end{array} $$
      We start by expanding the definition of
      $\topnd{(\Gam, \Gam', \Gam'' \vdash \termsubs{x}{u}{(\termcont{\alpha}{\termrepl{\beta'}{\beta}{s}{\ctxtapp{\ctxtLCT'}{v}}})} : A \mid \Del, \Del', \Del'', \beta' : B)}$
      to obtain $$
      \begin{array}{c}
      \mcnf{\scalebox{.8}{\input{proofnets/soundness/ttrepl2}}}  \equiv
      \end{array} $$ $$\kern-2em
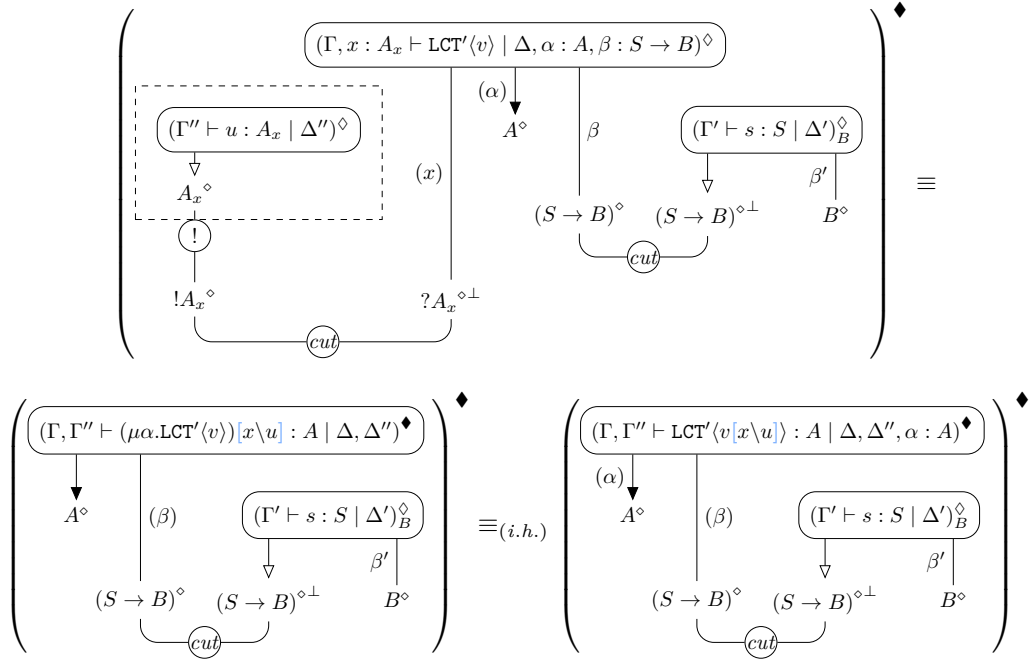
      \begin{array}{c}
      \mcnf{\scalebox{.8}{\input{proofnets/soundness/ttrepl3}}}  \equiv_{(\ih)}
      \mcnf{\scalebox{.8}{\input{proofnets/soundness/ttrepl5}}}
      \end{array} $$
      where the last proof-net is
      $\topnd{(\Gam, \Gam', \Gam'' \vdash \termcont{\alpha}{\termrepl{\beta'}{\beta}{s}{\ctxtapp{\ctxtLCT'}{\termsubs{x}{u}{v}}}} : A \mid \Del, \Del', \Del'', \beta' : B)}$.
      Thus, we conclude.
    \end{itemize}
    
    \item $\ctxtLTT = \termsubs{y}{r}{\Box}$ with $x \notin \fv{r}$ and
    $y \notin \fv{y}$. We only illustrate the case where $x, y \in \fv{v}$. All
    the other cases are similar.
    %$$\kern-4em
    %\begin{array}{c}
    %\scaleinput{.8}{proofnets/soundness/ttsub1}  \enskip=
    %\scaleinput{.8}{proofnets/soundness/ttsub2}  \equiv\quad
    %\scaleinput{.8}{proofnets/soundness/ttsub3}
    %\end{array} $$
    Here
    $\topnd{(\Gam, \Gam', \Gam'' \vdash \termsubs{x}{u}{\termsubs{y}{r}{v}} : B \mid \Del, \Del', \Del'')}$
    and
    $\topnd{(\Gam, \Gam', \Gam'' \vdash \termsubs{y}{r}{\termsubs{x}{u}{v}} : B \mid \Del, \Del', \Del'')}$
    are both structurally equivalent, by definition, to
    $$\scalebox{.8}{\input{proofnets/soundness/ttsub2}}$$
    As before, structural equivalence is necessary to consider associativity of
    contractions.
  \end{itemize}

  \item $\eqexrepl$: then $o =
  \termrepl{\alpha'}{\alpha}{s}{\ctxtapp{\ctxtLCC}{c}}$ and $p =
  \ctxtapp{\ctxtLCC}{\termrepl{\alpha'}{\alpha}{s}{c}}$ with
  $\notatall{\alpha}{\ctxtLCC}$ and $\freeFor{(s,\alpha')}{\ctxtLCC}$. We 
  analyze the shape of $\ctxtLCC$ resorting to Lem.~\ref{l:contexts}.
  \begin{itemize}
    \item $\ctxtLCC = \termrepl{\beta'}{\beta}{s'}{\boxdot}$. We only
    illustrate the case where $\alpha, \beta \in \fn{c}$ and $\alpha', \beta'$
    are fresh. All the other cases are similar. Note that by hypothesis
    $\alpha \notin \fn{s'}$, $\alpha \neq \beta$, $\alpha \neq \beta'$, $\beta
    \notin \fn{s}$ and $\beta \neq \alpha'$. Moreover, by $\alpha$-conversion
    we can assume $\alpha \neq \alpha'$ and $\beta \neq \beta'$ too. By
    definition of translation
    $\topnd{(\Gam, \Gam', \Gam'' \vdash \termrepl{\alpha'}{\alpha}{s}{\termrepl{\beta'}{\beta}{s'}{c}} \mid \Del, \Del', \Del'', \alpha' : A, \beta' : B)}$
    and
    $\topnd{(\Gam, \Gam', \Gam'' \vdash \termrepl{\beta'}{\beta}{s'}{\termrepl{\alpha'}{\alpha}{s}{c}} \mid \Del, \Del', \Del'', \alpha' : A, \beta' : B)}$
    are structurally equivalent to
    $$\mcnf{\scalebox{.8}{\input{proofnets/soundness/ccrepl2}}}$$
    As before, structural equivalence is necessary to consider associativity of
    contractions.
    
    \item $\ctxtLCC = \termname{\beta}{\ctxtLTC}$. We proceed by induction
    on $\ctxtLTC$:
    \begin{itemize}
      \item $\ctxtLTC = \termcont{\gamma}{\boxdot}$. We only illustrate the
      case where $\alpha, \beta, \gamma \in \fn{c}$, $\alpha' \notin \fn{c}$
      and $\beta \neq \gamma$. All the other cases are similar. By hypothesis
      we also have $\alpha \neq \beta$, $\alpha \neq \gamma$,
      % \oandres{$\alpha' \neq\beta$,}
    and $\alpha' \neq \gamma$. In this case
      $\topnd{(\Gam, \Gam' \vdash \termrepl{\alpha'}{\alpha}{s}{(\termname{\beta}{\termcont{\gamma}{c}})} \mid \Del, \Del', \alpha':A, \beta:B)}$
      and
      $\topnd{(\Gam, \Gam' \vdash \termname{\beta}{\termcont{\gamma}{\termrepl{\alpha'}{\alpha}{s}{c}}} \mid \Del, \Del', \alpha':A, \beta:B)}$
      are both equal, by definition, to
      $$\mcnf{\scalebox{.8}{\input{proofnets/soundness/ccmu2}}}$$

      \item $\ctxtLTC = \termapp{\ctxtLTC'}{u}$. We only illustrate the case
      where $\alpha \in \fn{\ctxtapp{\ctxtLTC'}{c}}$, $\alpha' \notin
      \fn{\ctxtapp{\ctxtLTC'}{c}}$. All the other cases are similar. Let
      $\gamma$ be a fresh name. Arbitrary addition/removing of final weakening
      (by means of $\equiv$) allows as to use $\gamma$ to properly construct
      the proof-net of an smaller command for which the \ih applies.
      %$$%\kern-1em
      %\begin{array}{c}
      %\scaleinput{.8}{proofnets/soundness/ccapp1}         \equiv
      %\mcnf{\scaleinput{.8}{proofnets/soundness/ccapp2}}  \equiv \\
      %\\
      %\mcnf{\scaleinput{.8}{proofnets/soundness/ccapp3}}  \equiv_{(\ih)}
      %\mcnf{\scaleinput{.8}{proofnets/soundness/ccapp4}}  \equiv \\
      %\\
      %\scaleinput{.8}{proofnets/soundness/ccapp5}
      %\end{array} $$
      We start by expanding the definition of
      $\topnd{(\Gam, \Gam', \Gam'' \vdash \termrepl{\alpha'}{\alpha}{s}{(\termname{\beta}{\termapp{\ctxtapp{\ctxtLTC'}{c}}{u}})} \mid \Del, \Del', \Del'', \alpha':A, \beta:B)}$
      to obtain the structurally equivalent proof-nets $$%\kern-4em
      \begin{array}{cl}
      \mcnf{\scalebox{.8}{\input{proofnets/soundness/ccapp2}}} & \equiv \\
      \\
      \mcnf{\scalebox{.8}{\input{proofnets/soundness/ccapp3}}} & \equiv_{(\ih)}\\
      \\
      \mcnf{\scalebox{.8}{\input{proofnets/soundness/ccapp4}}}
      \end{array} $$
      By definition of the translation, the last proof-net above turns out to
      be structurally equivalent to
      $\topnd{(\Gam, \Gam', \Gam'' \vdash \termname{\beta}{\termapp{\ctxtapp{\ctxtLTC'}{\termrepl{\alpha'}{\alpha}{s}{c}}}{u}} \mid \Del, \Del', \Del'', \alpha':A, \beta:B)}$.
      Thus, we conclude.
      
      \item $\ctxtLTC = \termabs{x}{\ctxtLTC'}$. We only illustrate the case
      where $x \in \fv{\ctxtapp{\ctxtLTC'}{c}}$, $\alpha \in
      \fn{\ctxtapp{\ctxtLTC'}{c}}$, $\alpha' \notin
      \fn{\ctxtapp{\ctxtLTC'}{c}}$. All the other cases are similar. As before
      we resort to a fresh name $\gamma$ to properly apply the \ih
      %$$%\kern-1em
      %\begin{array}{c}
      %\scaleinput{.8}{proofnets/soundness/ccabs1}         \equiv
      %\mcnf{\scaleinput{.8}{proofnets/soundness/ccabs2}}  \equiv \\
      %\\
      %\scaleinput{.8}{proofnets/soundness/ccabs3}         \equiv_{(\ih)}
      %\scaleinput{.8}{proofnets/soundness/ccabs4}         \equiv \\
      %\\
      %\scaleinput{.8}{proofnets/soundness/ccabs5}
      %\end{array} $$
      Expanding the definition of the translation for
      $\topnd{(\Gam, \Gam' \vdash \termrepl{\alpha'}{\alpha}{s}{(\termname{\beta}{\termabs{x}{\ctxtapp{\ctxtLTC'}{c}}})} \mid \Del, \Del', \alpha':A, \beta:\typefunc{A_x}{B})}$
      to obtain the structurally equivalent proof-nets $$\kern-1em
      \begin{array}{cl}
      \mcnf{\scalebox{.8}{\input{proofnets/soundness/ccabs2}}}  \equiv
      \end{array} $$ $$\kern-5em
      \begin{array}{cl}
      \scalebox{.8}{\input{proofnets/soundness/ccabs3}}         \equiv_{(\ih)}
      \scalebox{.8}{\input{proofnets/soundness/ccabs4}}
      \end{array} $$
      As in the previous case, the last proof-net above is in turn
      structurally equivalent to
      $\topnd{(\Gam, \Gam' \vdash \termname{\beta}{\termabs{x}{\ctxtapp{\ctxtLTC'}{\termrepl{\alpha'}{\alpha}{s}{c}}}} \mid \Del, \Del', \alpha':A, \beta:\typefunc{A_x}{B})}$
      by definition of the translation. Thus, we conclude.

      \item $\ctxtLTC = \termsubs{x}{u}{\ctxtLTC'}$ We only illustrate the case
      where $x \in \fv{\ctxtapp{\ctxtLTC'}{c}}$, $\alpha \in
      \fn{\ctxtapp{\ctxtLTC'}{c}}$, $\alpha' \notin
      \fn{\ctxtapp{\ctxtLTC'}{c}}$. All the other cases are similar. Once again
      we resort to a fresh name $\gamma$ to properly apply the \ih
      %$$%\kern-1em
      %\begin{array}{c}
      %\scaleinput{.8}{proofnets/soundness/ccsub1}         \equiv
      %\mcnf{\scaleinput{.8}{proofnets/soundness/ccsub2}}  \equiv \\
      %\\
      %\scaleinput{.8}{proofnets/soundness/ccsub3}         \equiv_{(\ih)}
      %\scaleinput{.8}{proofnets/soundness/ccsub4}         \equiv \\
      %\\
      %\scaleinput{.8}{proofnets/soundness/ccsub5}
      %\end{array} $$
      We start by expanding the definition of
      $\topnd{(\Gam, \Gam', \Gam'' \vdash \termrepl{\alpha'}{\alpha}{s}{(\termname{\beta}{\termsubs{x}{u}{\ctxtapp{\ctxtLTC'}{c}}})} \mid \Del, \Del', \Del'', \alpha':A, \beta:B)}$
      to obtain the structurally equivalent proof-nets $$\kern-1em
      \begin{array}{cl}
      \mcnf{\scalebox{.8}{\input{proofnets/soundness/ccsub2}}}  \equiv 
      \end{array} $$ $$\kern-7em
      \begin{array}{cl}
      \scalebox{.8}{\input{proofnets/soundness/ccsub3}}         \equiv_{(\ih)}
      \scalebox{.8}{\input{proofnets/soundness/ccsub4}}
      \end{array} $$
      Once again, by definition of the translation, the last proof-net above
      turns out to be structurally equivalent to
      $\topnd{(\Gam, \Gam', \Gam'' \vdash \termname{\beta}{\termsubs{x}{u}{\ctxtapp{\ctxtLTC'}{\termrepl{\alpha'}{\alpha}{s}{c}}}} \mid \Del, \Del', \Del'', \alpha':A, \beta:B)}$.
      Thus, we conclude.
    \end{itemize}
  \end{itemize}

  \item $\eqlinear$: then $o =
  \termrepl{\alpha'}{\alpha}{s}{(\termname{\alpha}{u})}$ and $p =
  \termname{\alpha'}{\BMC{\termconc{u}{s}}}$ with $\alpha \notin \fn{u}$ and $s
  = \termpush{t_1}{\termpush{t_2}{\termpush{\ldots}{\termpush{t_n}{\termemst}}}}$
  $(n > 0)$. This case is illustrated on Fig.~\ref{f:linear}, where we assume
  $\alpha'$ fresh for simplicity. We start from
  $\topnd{(\Gam, \Gam_1, \ldots, \Gam_n \vdash \termrepl{\alpha'}{\alpha}{s}{(\termname{\alpha}{u})} \mid \Del, \Del_1, \ldots, \Del_n, \alpha':B)}$
  which expands, by definition, to the first proof-net in the figure,  while
  the last one illustrates the unfolding of
  $\topnd{(\Gam, \Gam_1, \ldots, \Gam_n \vdash \termname{\alpha'}{\BMC{u \tconc s}} \mid \Del, \Del_1, \ldots, \Del_n, \alpha':B)}$.
\begin{figure} $$
\begin{array}{cl}
\mcnf{\scalebox{.8}{\input{proofnets/soundness/lin2}}} & = \\
\\
\\
\mcnf{\scalebox{.8}{\input{proofnets/soundness/lin3}}} & \equiv \\
\\
\\
\mcnf{\scalebox{.8}{\input{proofnets/soundness/lin4}}} \\
\\
\\
\end{array} $$
\caption{Case $\eqlinear$ of Lemma~\ref{l:soundness} proof.}
\label{f:linear}
\end{figure}

  \item $\eqpoppop$: then $o =
  \termname{\alpha'}{\termabs{x}{\termcont{\alpha}{\termname{\beta'}{\termabs{y}{\termcont{\beta}{c}}}}}}$
  and $p =
  \termname{\beta'}{\termabs{y}{\termcont{\beta}{\termname{\alpha'}{\termabs{x}{\termcont{\alpha}{c}}}}}}$
  with $\alpha \neq \beta'$ and $\beta \neq \alpha'$. There are many different
  possibilities depending on whether each variable/name appears or not free in
  $c$. We only illustrate two cases where $x \notin \fv{c}$, $y \in \fv{c}$,
  $\alpha \in \fn{c}$ and $\beta \notin \fn{c}$. The remaining ones are
  similar. Notice that the branch involving $x$, $\alpha$ and $\alpha'$ is
  parallel to the one involving $y$, $\beta$ and $\beta'$.
  \begin{enumerate}
    \item $\alpha' = \beta' \notin \fn{c}$.
    %$$%\kern-1em
    %\begin{array}{c}
    %\scaleinput{.8}{proofnets/soundness/popeq1}         \enskip=
    %\mcnf{\scaleinput{.8}{proofnets/soundness/popeq2}}  =\enskip
    %\scaleinput{.8}{proofnets/soundness/popeq3}
    %\end{array} $$
    In this case
    $\topnd{(\Gam \vdash \termname{\alpha'}{\termabs{x}{\termcont{\alpha}{\termname{\beta'}{\termabs{y}{\termcont{\beta}{c}}}}}} \mid \Del, \alpha':\typefunc{A}{B})}$
    and
    $\topnd{(\Gam \vdash \termname{\beta'}{\termabs{y}{\termcont{\beta}{\termname{\alpha'}{\termabs{x}{\termcont{\alpha}{c}}}}}} \mid \Del, \alpha':\typefunc{A}{B})}$
    are both equal, by definition, to
    $$\scalebox{.8}{\input{proofnets/soundness/popeq2}}$$

    \item $\alpha' \neq \beta'$ assuming $\alpha' \notin \fn{c}$ and $\beta'
    \in \fn{c}$.
    %$$%\kern-1em
    %\begin{array}{c}
    %\scaleinput{.8}{proofnets/soundness/popneq1}         \enskip=
    %\mcnf{\scaleinput{.8}{proofnets/soundness/popneq2}}  =\enskip
    %\scaleinput{.8}{proofnets/soundness/popneq3}
    %\end{array} $$
    In this case, by definition of the translation to proof-nets,
    $\topnd{(\Gam \vdash \termname{\alpha'}{\termabs{x}{\termcont{\alpha}{\termname{\beta'}{\termabs{y}{\termcont{\beta}{c}}}}}} \mid \Del, \alpha':\typefunc{A_x}{A}, \beta':\typefunc{B_y}{B})}$
    and
    $\topnd{(\Gam \vdash \termname{\beta'}{\termabs{y}{\termcont{\beta}{\termname{\alpha'}{\termabs{x}{\termcont{\alpha}{c}}}}}} \mid \Del, \alpha':\typefunc{A_x}{A}, \beta':\typefunc{B_y}{B})}$
    are both equal to
    $$\scalebox{.8}{\input{proofnets/soundness/popneq2}}$$
  \end{enumerate}

	\item $\eqrho$: then $\pi_{o} \dem \Gam \vdash
  \termname{\alpha}{\termcont{\beta}{c}} \mid \Del, \alpha: A$ and $\pi_{p}
  \dem \Gam \vdash c \exrepl{\alpha}{\beta}{\emst} \mid \Del, \alpha: A$. We
  illustrate only the cases where $\alpha \notin \fn{c}$, the remaining ones
  are similar adding contractions between the already exiting $\alpha$ edges
  and the newly added one. We distinguish two cases:
  \begin{enumerate}
    \item $\beta \in \fn{c}$.
    %$$%\kern-1em
    %\begin{array}{c}
    %\scaleinput{.8}{proofnets/soundness/rhoin1}         \enskip=\enskip
    %\scaleinput{.8}{proofnets/soundness/rhoin2}         \enskip=\enskip \\
    %\mcnf{\scaleinput{.8}{proofnets/soundness/rhoin3}}  \enskip=\enskip 
    %\mcnf{\scaleinput{.8}{proofnets/soundness/rhoin4}}  \enskip=\enskip \\
    %\scaleinput{.8}{proofnets/soundness/rhoin5}
    %\end{array} $$
    We start by expanding the definition of
    $\topnd{(\Gam \vdash \termname{\alpha}{\termcont{\beta}{c}} \mid \Del, \alpha: A)}$
    to obtain $$
    \begin{array}{c}
    \scalebox{.8}{\begin{tikzpicture}[baseline=(current bounding box.center),node distance=\PNdist]
\node[PNsub] (TC) [minimum width=\PNdist*4] {$\topnd{(\Gam \vdash c \mid \Del, \alpha: A)}$};
\node (N1) [below of=TC, yshift=-\PNdist] {$\toform{A}$};

\draw[PNarrow] (TC.south) -- node[left]{$(\beta)$} node[right]{$\alpha$} (N1.north);
\end{tikzpicture}
}         \enskip=\enskip
    \mcnf{\scalebox{.8}{\input{proofnets/soundness/rhoin3}}}  \enskip=\enskip 
    \mcnf{\scalebox{.8}{\begin{tikzpicture}[baseline=(current bounding box.center),node distance=\PNdist]
\node[PNsub] (TC) [minimum width=\PNdist*4] {$\topn{(\Gam \vdash c\exrepl{\alpha}{\beta}{\emst} \mid \Del, \alpha: A)}$};
\node (N1) [below of=TC, yshift=-\PNdist] {$\toform{A}$};

\draw[PNarrow] (TC.south) -- node[left]{$\alpha$} (N1.north);
\end{tikzpicture}
}}
    \end{array} $$
    where the last proof-net is
    $\topnd{(\Gam \vdash c\exrepl{\alpha}{\beta}{\emst} \mid \Del, \alpha: A)}$.
    Thus, we conclude.

    \item $\beta \notin \fn{c}$.
    %$$%\kern-1em
    %\begin{array}{c}
    %\scaleinput{.8}{proofnets/soundness/rhonotin1}         \enskip=\enskip
    %\scaleinput{.8}{proofnets/soundness/rhonotin2}         \enskip=\enskip \\
    %\mcnf{\scaleinput{.8}{proofnets/soundness/rhonotin3}}  \enskip=\enskip 
    %\mcnf{\scaleinput{.8}{proofnets/soundness/rhonotin4}}  \enskip=\enskip \\
    %\scaleinput{.8}{proofnets/soundness/rhonotin5}
    %\end{array} $$
    We start by expanding the definition of
    $\topnd{(\Gam \vdash \termname{\alpha}{\termcont{\beta}{c}} \mid \Del, \alpha: A)}$
    to obtain $$
    \begin{array}{c}
    \scalebox{.8}{\begin{tikzpicture}[baseline=(current bounding box.center),node distance=\PNdist]
\PNnode{W1}{\textit{w}};
\node[PNsub] (TC) [right of=W1, xshift=\PNdist*2] {$\topnd{(\Gam \vdash c \mid \Del)}$};
\node (N1) [below of=W1, yshift=-\PNdist*0.5] {$\toform{A}$};

\draw[PNarrow] (W1.south) -- node[left]{$(\beta)$} node[right]{$\alpha$} (N1.north);
\end{tikzpicture}
}         \equiv
    \mcnf{\scalebox{.8}{\input{proofnets/soundness/rhonotin3}}}  \equiv 
    \mcnf{\scalebox{.8}{\begin{tikzpicture}[baseline=(current bounding box.center),node distance=\PNdist]
\node[PNsub] (TC) [minimum width=\PNdist*4] {$\topn{(\Gam \vdash c\exrepl{\alpha}{\beta}{\emst} \mid \Del, \alpha: A)}$};
\node (N1) [below of=TC, yshift=-\PNdist] {$\toform{A}$};

\draw[PNarrow] (TC.south) -- node[left]{$\alpha$} (N1.north);
\end{tikzpicture}
}}
    \end{array} $$
    where the last proof-net is
    $\topnd{(\Gam \vdash c\exrepl{\alpha}{\beta}{\emst} \mid \Del, \alpha: A)}$.
    Thus, we conclude.
  \end{enumerate}

  \item $\eqtheta$: then $\pi_{o} \dem \Gam \vdash
  \termcont{\alpha}{\termname{\alpha}{t}} \mid \Del$ and $\pi_{p} \dem \Gam
  \vdash t \mid \Del$ with $\alpha \notin \fn{t}$. $$
  \begin{array}{c}
  \scalebox{.8}{\begin{tikzpicture}[baseline=(current bounding box.center),node distance=\PNdist]
\node[PNsub] (TC) [minimum width=\PNdist*4] {$\topnd{(\Gam \vdash \termcont{\alpha}{\termname{\alpha}{t}} \mid \Del)}$};
\node (N1) [below of=TC, yshift=-\PNdist] {$\toform{A}$};

\draw[PNarout] (TC.south) -- node{} (N1.north);
\end{tikzpicture}
}  \enskip=\enskip
  \scalebox{.8}{\begin{tikzpicture}[baseline=(current bounding box.center),node distance=\PNdist]
\node[PNsub] (TC) [minimum width=\PNdist*4] {$\topnd{(\Gam \vdash t \mid \Del)}$};
\node (N1) [below of=TC, yshift=-\PNdist] {$\toform{A}$};

\draw[PNarout] (TC.south) -- node[left]{$(\alpha)$} (N1.north);
\end{tikzpicture}
}  \enskip=\enskip 
  \scalebox{.8}{\begin{tikzpicture}[baseline=(current bounding box.center),node distance=\PNdist]
\node[PNsub] (TC) [minimum width=\PNdist*4] {$\topnd{(\Gam \vdash t \mid \Del)}$};
\node (N1) [below of=TC, yshift=-\PNdist] {$\toform{A}$};

\draw[PNarout] (TC.south) -- node{} (N1.north);
\end{tikzpicture}
}
  \end{array} $$

  \item $\eqren$: then $\pi_{o} \dem \Gam \vdash c\ire{\alpha}{\beta}{\emst}
  \mid \Del, \beta: A$ and $\pi_{p} \dem \Gam \vdash
  c\exrepl{\beta}{\alpha}{\emst} \mid \Del, \beta: A$. The interesting case
  is when $\beta \in \fn{c}$. We start by expanding the definition of
  $\topnd{(\Gam \vdash c\exrepl{\beta}{\alpha}{\emst} \mid \Del, \beta: A)}$
  to obtain $$
  \begin{array}{c}
  \mcnf{\scalebox{.8}{\input{proofnets/soundness/ren1}}}  \equiv
  \mcnf{\scalebox{.8}{\input{proofnets/soundness/ren2}}}
  \end{array} $$
  and resort to Lem. 11.6 in~\cite{Laurent02} to get rid of the contraction
  node and conclude with $\topnd{(\Gam \vdash c\ire{\alpha}{\beta}{\emst} \mid
  \Del, \beta: A)}$.
\end{itemize}
\end{proof}

%%% Local Variables:
%%% mode: latex
%%% TeX-master: "main"
%%% End:

%%% Local Variables:
%%% mode: latex
%%% TeX-master: "main"
%%% End:

  %%%%%%%%%%%%%%%%%%%%%%%%%%%%%%%%%%%%%%%%%%%%%%%%%%%%%%%%%%%%%%%%%%%%%%%%%%%%%%%
\section{Appendix: Strong Bisimulation}
\label{a:bisimulation-lemmas}
%%%%%%%%%%%%%%%%%%%%%%%%%%%%%%%%%%%%%%%%%%%%%%%%%%%%%%%%%%%%%%%%%%%%%%%%%%%%%%%

We start by definition the size of terms as follows:
\[ \hfill
\begin{array}{lll}
\sz{x}                                & \coloneq & 1 \\
\sz{\termapp{t}{u}}                   & \coloneq & \sz{t} + \sz{u} + 1 \\
\sz{\termabs{x}{t}}                   & \coloneq & \sz{t} + 1 \\
\sz{\termcont{\alpha}{c}}             & \coloneq & \sz{c} + 1 \\
\sz{\termsubs{x}{t}{u}}               & \coloneq & \sz{t} + \sz{u} + 1 \\
\sz{\termname{\alpha}{t}}             & \coloneq & \sz{t} + 1 \\
\sz{\termrepl{\alpha'}{\alpha}{s}{c}} & \coloneq & \sz{c} + \sz{s} + 1 \\
\sz{\termemst}                        & \coloneq & 0 \\
\sz{\termpush{t}{s}}                  & \coloneq & \sz{t} + \sz{s} + 1
\end{array} \hfill 
\] The size of contexts is defined accordingly, where  in particular $\sz{\Box}
= \sz{\boxdot} = 0$. A straightforward induction allows us to show: 

\begin{lemma}
Let $o\in \objects{\LMfull}$. If $o \Rew{\rB\rM\rC\rW} o'$, then $\sz{o} \geq \sz{o'}$.
\end{lemma}

%% Application of BMC
\gettoappendix{l:app-bm}

%\delia{LEER BIEN ESTA PRUEBA QUE SE SIMPLIFICO MUCHO}
\begin{proof}
We prove the following statements to conclude with the general (last) statement
of the lemma.
\begin{enumerate}
  \item If $t \simeq t'$, then $\BMC{\termapp{t}{u_0}} \simeq
  \BMC{\termapp{t'}{u_0}}$.
  \item If $c \simeq c'$, then $\BMC{\termrepl{\al'}{\al}{s}{c}} \simeq
  \BMC{\termrepl{\al'}{\al}{s}{c'}}$.
  \item If $o \simeq o'$, then $\BMC{\ctxtapp{\ctxt{O}}{o}} \simeq
    \BMC{\ctxtapp{\ctxt{O}}{o'}}$.
\end{enumerate}

\begin{enumerate}
  \item By induction on the equations. Note that equations
    $\eqexrepl,\eqlinear,\eqpoppop$ and $\eqrho$ are not applicable
    since these relate commands.  This leaves the following two base cases:
    
  \begin{itemize}
    \item $t = \termsubs{x}{u}{\ctxtapp{\ctxtLTT}{v}} \simeq_{\eqexsubs}
      \ctxtapp{\ctxtLTT}{\termsubs{x}{u}{v}} = t'$. By
      applying twice 
      Lem.~\ref{l:subs-repl-out-of-BM}.\ref{l:subs-out-of-BM} we obtain
      $$\BMC{\termapp{\termsubs{x}{u}{\ctxtapp{\ctxtLTT}{v}}}{u_0}} \simeq
\BMC{\termsubs{x}{u}{(\termapp{\ctxtapp{\ctxtLTT}{v}}{u_0})}}
\simeq
    \BMC{\termapp{\ctxtapp{\ctxtLTT}{\termsubs{x}{u}{v}}}{u_0}}$$

  \item $t = \termcont{\alpha}{\termname{\alpha}{v}} \simeq_{\eqtheta} v = t'$,
    where $\alpha \notin \fn{v}$. $$
\begin{array}{l@{\enspace}l@{\enspace}l@{\enspace}l}
\BMC{\termapp{(\termcont{\alpha}{\termname{\alpha}{v}})}{u_0}}                          & = &
\BMC{\termcont{\alpha'}{\termrepl{\alpha'}{\alpha}{u_0}{(\termname{\alpha}{v})}}}       & = \\
\termcont{\alpha'}{\BMC{\termrepl{\alpha'}{\alpha}{u_0}{(\termname{\alpha}{v})}}}       & =_{(\ast)} &
\termcont{\alpha'}{\termrepl{\alpha'}{\alpha}{\BMC{u_0}}{(\termname{\alpha}{\BMC{v}})}} & \simeq_{\eqlinear} \\
\termcont{\alpha'}{\termname{\alpha'}{\BMC{\termapp{\BMC{v}}{\BMC{u_0}}}}}              & = &
\termcont{\alpha'}{\termname{\alpha'}{\BMC{\termapp{v}{u_0}}}}                          & \simeq_{\eqtheta} \\
\BMC{\termapp{v}{u_0}}
\end{array} $$
    The equality $(\ast)$ is justified by the fact that $\alpha \notin \fn{v}$
    hence $\termrepl{\alpha'}{\alpha}{u_0}{(\termname{\alpha}{v})}$ cannot be
    a $\rC$-redex nor a $\rW$-redex.

  \item The inductive cases are easy.

  \end{itemize}

  \item By induction on the equations. Note that applicable equations
    are only those relating commands.  This leaves the following four base cases:
  \begin{itemize}
    \item $c = \termrepl{\beta'}{\beta}{s_0}{\ctxtapp{\ctxtLCC}{c_0}}
    \simeq_{\eqexrepl} \ctxtapp{\ctxtLCC}{\termrepl{\beta'}{\beta}{s_0}{c_0}}=c'$,
    where $\notatall{\beta}{\ctxtLCC,c_0}$ and
    $\freeFor{(s_0,\beta')}{\ctxtLCC}$. By $\alpha$-conversion we can also
    assume $\freeFor{(s,\alpha')}{\ctxtLCC}$.

    If $\alpha = \beta'$, $\notatall{\beta'}{(c_0,\ctxtLCC,s_0)}$ and $s_0, s
    \neq \termemst$, $$
\begin{array}{l@{\enspace}l@{\enspace}l@{\enspace}l}
\BMC{\termrepl{\alpha'}{\alpha}{s}{\termrepl{\beta'}{\beta}{s_0}{\ctxtapp{\ctxtLCC}{c_0}}}} & = &
\BMC{\termrepl{\alpha'}{\beta}{\termpush{s_0}{s}}{\ctxtapp{\ctxtLCC}{c_0}}}                 & =_{L.\ref{l:subs-repl-out-of-BM}.\ref{l:repl-out-of-BM}} \\
\BMC{\ctxtapp{\ctxtLCC}{\termrepl{\alpha'}{\beta}{\termpush{s_0}{s}}{c_0}}}                 & = & 
\BMC{\termrepl{\alpha'}{\alpha}{s}{\ctxtapp{\ctxtLCC}{\termrepl{\beta'}{\beta}{s_0}{c_0}}}} 
\end{array} $$

    If $\alpha = \beta'$, $\notatall{\beta'}{(c_0,\ctxtLCC)}$ and $s_0 =
    \termemst$ and $s \neq \termemst$, $$
\begin{array}{l@{\enspace}l@{\enspace}l@{\enspace}l}
\BMC{\termrepl{\alpha'}{\alpha}{s}{\termrepl{\beta'}{\beta}{\termemst}{\ctxtapp{\ctxtLCC}{c_0}}}}  & = &
\BMC{\termrepl{\alpha'}{\alpha}{\termemst}{\termrepl{\beta'}{\beta}{s}{\ctxtapp{\ctxtLCC}{c_0}}}}  & =_{L.\ref{l:subs-repl-out-of-BM}.\ref{l:repl-out-of-BM}} \\
\BMC{\ctxtapp{\ctxtLCC}{\termrepl{\alpha'}{\alpha}{\termemst}{\termrepl{\beta'}{\beta}{s'}{c_0}}}} & = & 
\BMC{\termrepl{\alpha'}{\alpha}{s}{\ctxtapp{\ctxtLCC}{\termrepl{\beta'}{\beta}{\termemst}{c_0}}}} 
\end{array} $$

    Otherwise, $\BMC{\termrepl{\alpha'}{\alpha}{s}{c}} =
    \termrepl{\alpha'}{\alpha}{\BMC{s}}{c} \simeq
    \termrepl{\alpha'}{\alpha}{\BMC{s}}{c'} =
    \BMC{\termrepl{\alpha'}{\alpha}{s}{c'}}$.

    \item $c = \termrepl{\beta'}{\beta}{s_0}{(\termname{\beta}{u})}
    \simeq_{\eqlinear} \termname{\beta'}{\BMC{\termconc{u}{s_0}}} = c'$, where
    $\notfreen{\beta,\beta'}{u}$ and $s_0 \neq \termemst$.

    If $\alpha = \beta'$, $\notatall{\beta'}{s_0}$ and $s \neq \termemst$, $$
\begin{array}{l@{\enspace}l@{\enspace}l@{\enspace}l}
\BMC{\termrepl{\alpha'}{\alpha}{s}{\termrepl{\beta'}{\beta}{s_0}{(\termname{\beta}{u})}}} & = &
\BMC{\termrepl{\al'}{\beta}{\termpush{s_0}{s}}{(\termname{\beta}{u})}}                    & = \\
\termrepl{\al'}{\beta}{\termpush{s_0}{\BMC{s}}}{(\termname{\beta}{u})}                    & \simeq_{\eqlinear} &
\termname{\al'}{\BMC{\termconc{u}{(\termpush{s_0}{\BMC{s}})}}}                            & = \\
\termname{\al'}{\BMC{\termconc{u}{(\termpush{s_0}{s})}}}                                  & = &
\termname{\al'}{\BMC{\termconc{\BMC{\termconc{u}{s_0}}}{\BMC{s}}}}                        & \simeq_{\eqlinear} \\
\termrepl{\alpha'}{\alpha}{\BMC{s}}{(\termname{\alpha}{\BMC{\termconc{u}{s_0}}})}         & = &
\BMC{\termrepl{\alpha'}{\alpha}{s}{(\termname{\alpha}{\BMC{\termconc{u}{s_0}}})}}         & = \\
\BMC{\termrepl{\alpha'}{\alpha}{s}{c'}}   
\end{array} $$

    Otherwise, $\BMC{\termrepl{\alpha'}{\alpha}{s}{c}} = 
    \termrepl{\alpha'}{\alpha}{\BMC{s}}{c} \simeq_{\eqlinear} 
    \termrepl{\alpha'}{\alpha}{\BMC{s}}{c'} = 
    \BMC{\termrepl{\alpha'}{\alpha}{s}{c'}}$. Remark that
    $\exrepl{\alpha'}{\alpha}{s}$ cannot fire a $\rW$-redex since
    $s_0 \neq \termemst$ by hypothesis.
    
%\begin{array}{l@{\enspace}l@{\enspace}l@{\enspace}l@{\enspace}l@{\enspace}l@{\enspace}l}
%\BMC{\termrepl{\alpha'}{\alpha}{s}{c}}  & = &
%\termrepl{\alpha'}{\alpha}{\BMC{s}}{c}  & \simeq_{\eqlinear} &
%\termrepl{\alpha'}{\alpha}{\BMC{s}}{c'} & = &
%\BMC{\termrepl{\alpha'}{\alpha}{s}{c'}}
%\end{array} $$

    \item $c =
    \termname{\gamma'}{\termabs{x}{\termcont{\gamma}{\termname{\beta'}{\termabs{y}{\termcont{\beta}{u}}}}}}
    \simeq_{\eqpoppop}
    \termname{\beta'}{\termabs{y}{\termcont{\beta}{\termname{\gamma'}{\termabs{x}{\termcont{\gamma}{u}}}}}}
    = c'$, where $\gamma \neq \beta'$ and $\gamma' \neq \beta$.

    If $\BMC{c} =
    \termname{\gamma'}{\termabs{x}{\termcont{\gamma}{\termname{\beta'}{\termabs{y}{\termcont{\beta}{\ctxtapp{\ctxtLCC}{\termrepl{\alpha}{\delta}{s'}{c_1}}}}}}}}$
    verifying the conditions for rule $\rC$, $$
\begin{array}{l@{\enspace}l}
\BMC{\termrepl{\alpha'}{\alpha}{s}{c}} & = \\
\BMC{\termname{\gamma'}{\termabs{x}{\termcont{\gamma}{\termname{\beta'}{\termabs{y}{\termcont{\beta}{\ctxtapp{\ctxtLCC}{\termrepl{\alpha'}{\delta}{\termpush{s'}{s}}{c_1}}}}}}}}} & = \\
\termname{\gamma'}{\termabs{x}{\termcont{\gamma}{\termname{\beta'}{\termabs{y}{\termcont{\beta}{\BMC{\ctxtapp{\ctxtLCC}{\termrepl{\alpha'}{\delta}{\termpush{s'}{s}}{c_1}}}}}}}}} & \simeq_{\eqpoppop} \\
\termname{\beta'}{\termabs{y}{\termcont{\beta}{\termname{\gamma'}{\termabs{x}{\termcont{\gamma}{\BMC{\ctxtapp{\ctxtLCC}{\termrepl{\alpha'}{\delta}{\termpush{s'}{s}}{c_1}}}}}}}}} & = \\
\BMC{\termrepl{\alpha'}{\alpha}{s}{c'}}
\end{array} $$

 If $\BMC{c} =
    \termname{\gamma'}{\termabs{x}{\termcont{\gamma}{\termname{\beta'}{\termabs{y}{\termcont{\beta}{\ctxtapp{\ctxtLCC}{\termrepl{\alpha}{\delta}{s'}{c_1}}}}}}}}$
    verifying the conditions for rule $\rW$, then the analysis is similar to the previous case. 
    
    Otherwise, $\BMC{\termrepl{\alpha'}{\alpha}{s}{c}} = 
    \termrepl{\alpha'}{\alpha}{\BMC{s}}{c} \simeq_{\eqrho}
    \termrepl{\alpha'}{\alpha}{\BMC{s}}{c'} =
    \BMC{\termrepl{\alpha'}{\alpha}{s}{c'}}$.

    \item $c = \termname{\gamma'}{\termcont{\beta}{c_0}} \simeq_{\eqrho}
      \termrepl{\gamma'}{\beta}{\termemst}{c_0} = c'$.

      If $\BMC{c} =
    \termname{\gamma'}{\termcont{\beta}{\ctxtapp{\ctxtLCC}{\termrepl{\alpha}{\delta}{s'}{c_1}}}}$
    verifying the conditions for rule $\rC$ $$
\begin{array}{l@{\enspace}l@{\enspace}l@{\enspace}l}
\BMC{\termrepl{\alpha'}{\alpha}{s}{c}}                                                                                               & = &
\BMC{\termrepl{\alpha'}{\alpha}{s}{(\termname{\gamma'}{\termcont{\beta}{\ctxtapp{\ctxtLCC}{\termrepl{\alpha}{\delta}{s'}{c_1}}}})}} & = \\
\BMC{(\termname{\gamma'}{\termcont{\beta}{\ctxtapp{\ctxtLCC}{\termrepl{\alpha'}{\delta}{\termpush{s'}{s}}{c_1}}}})}                 & = &
\termname{\gamma'}{\termcont{\beta}{\BMC{\ctxtapp{\ctxtLCC}{\termrepl{\alpha'}{\delta}{\termpush{s'}{s}}{c_1}}}}}                   & \simeq_{\eqrho} \\
\termrepl{\gamma'}{\beta}{\termemst}{\BMC{\ctxtapp{\ctxtLCC}{\termrepl{\alpha'}{\delta}{\termpush{s'}{s}}{c_1}}}}                   & =_{(\ast)} &
\BMC{\termrepl{\gamma'}{\beta}{\termemst}{\ctxtapp{\ctxtLCC}{\termrepl{\alpha'}{\delta}{\termpush{s'}{s}}{c_1}}}}                   & =_{(\star)} \\
\BMC{\termrepl{\alpha'}{\alpha}{s}{\termrepl{\gamma'}{\beta}{\termemst}{\ctxtapp{\ctxtLCC}{\termrepl{\alpha}{\delta}{s'}{c_1}}}}}
\end{array} $$
The equality $(\ast)$ is justified by the fact that $\beta \neq \alpha'$
and $\termrepl{\gamma'}{\beta}{\termemst}{\BMC{c}}
  =\BMC{\termrepl{\gamma'}{\beta}{\termemst}{c}}$,
    while $(\star)$ relies on $\alpha \neq \beta,\gamma'$.
    
If $\BMC{c} =
    \termname{\gamma'}{\termcont{\beta}{\ctxtapp{\ctxtLCC}{\termrepl{\alpha}{\delta}{\termemst}{c_1}}}}$
    verifying the conditions for rule $\rW$ $$
\begin{array}{l@{\enspace}l@{\enspace}l@{\enspace}l}
\BMC{\termrepl{\alpha'}{\alpha}{s}{c}}                                                                                               & = &
\BMC{\termrepl{\alpha'}{\alpha}{s}{(\termname{\gamma'}{\termcont{\beta}{\ctxtapp{\ctxtLCC}{\termrepl{\alpha}{\delta}{\termemst}{c_1}}}})}} & = \\
\BMC{\termname{\gamma'}{\termcont{\beta}{\ctxtapp{\ctxtLCC}{\termrepl{\alpha'}{\alpha}{\termemst}{\termrepl{\alpha}{\delta}{s}{c_1}}}}}} & = &
\termname{\gamma'}{\termcont{\beta}{\BMC{\ctxtapp{\ctxtLCC}{\termrepl{\alpha'}{\alpha}{\termemst}{\termrepl{\alpha}{\delta}{s}{c_1}}}}}}        & \simeq_{\eqrho} \\
\termrepl{\gamma'}{\beta}{\termemst}{\BMC{\ctxtapp{\ctxtLCC}{\termrepl{\alpha'}{\alpha}{\termemst}{\termrepl{\alpha}{\delta}{s}{c_1}}}}}            & =_{(\ast)} &
\BMC{ \termrepl{\gamma'}{\beta}{\termemst}{\ctxtapp{\ctxtLCC}{\termrepl{\alpha'}{\alpha}{\termemst}{\termrepl{\alpha}{\delta}{s}{c_1}}}}}              & =_{(\star)} \\
\BMC{\termrepl{\alpha'}{\alpha}{s}{\termrepl{\gamma'}{\beta}{\termemst}{\ctxtapp{\ctxtLCC}{\termrepl{\alpha}{\delta}{\termemst}{c_1}}}}}
\end{array} $$
    The equalities  $(\ast)$ and $(\star)$ are  justified as before.

    Otherwise, $\BMC{\termrepl{\alpha'}{\alpha}{s}{c}} = 
    \termrepl{\alpha'}{\alpha}{\BMC{s}}{c}  \simeq_{\eqrho} 
    \termrepl{\alpha'}{\alpha}{\BMC{s}}{c'} = 
    \BMC{\termrepl{\alpha'}{\alpha}{s}{c'}}$.
  \item  The inductive cases are easy. 

  \end{itemize}

  \item By induction on $\sz{\ctxt{O}}$ using the previous points for the
    key cases:

    \begin{itemize}
      
  \item $\ctxt{O} = \Box$ or $\ctxt{O} = \boxdot$. The result is
    immediate from the hypothesis.

  \item $\ctxt{O} = \termapp{\ctxt{T}}{t}$. Then
    $\BMC{\ctxtapp{\ctxt{T}}{o}} \simeq_{\ih} \BMC{\ctxtapp{\ctxt{T}}{o'}}$,
    $$ \begin{array}{lllll}
       \BMC{\termapp{\ctxtapp{\ctxt{T}}{o}}{t}} & = & 
       \BMC{\termapp{\BMC{\ctxtapp{\ctxt{T}}{o}}}{\BMC{t}}} & \simeq_{\mbox{item 1}} \\
       \BMC{\termapp{\BMC{\ctxtapp{\ctxt{T}}{o'}}}{\BMC{t}}} & = & 
       \BMC{\termapp{\ctxtapp{\ctxt{T}}{o'}}{t}} 
      \end{array} $$

  \item $\ctxt{O} = \termapp{t}{\ctxt{T}}$.  We reason by cases analysis. 

%%%%%%%%%%%%%%%%%%%%%%%%
  \begin{itemize}
  \item If $\BMC{t} = \ctxtapp{\ctxt{L}}{\termabs{x}{s}}$, then
    $\sz{ \termapp{t}{\ctxt{T}}} =
    \sz{t} + \sz{\ctxt{T}} + 1 \geq
    \sz{\ctxtapp{\ctxt{L}}{\termabs{x}{s}}} + \sz{\ctxt{T}} + 1 =
    \sz{\ctxt{L}} + 1 + \sz{s} + \sz{\ctxt{T}} + 1 >
    \sz{\ctxt{L}} + \sz{s} + \sz{\ctxt{T}} + 1 =
    \sz{\ctxtapp{\ctxt{L}}{\termsubs{x}{\ctxt{T}}{s}}}$.

    We then conclude as follows   $$
\begin{array}{l@{\enspace}l@{\enspace}l@{\enspace}l@{\enspace}l@{\enspace}l}
\BMC{\termapp{t}{\ctxtapp{\ctxt{T}}{o}}}                                   & = &
\BMC{\termapp{\ctxtapp{\ctxt{L}}{\termabs{x}{s}}}{\ctxtapp{\ctxt{T}}{o}}}    & = &
\BMC{\ctxtapp{\ctxt{L}}{\termsubs{x}{\ctxtapp{\ctxt{T}}{o}}{s}}}             & \simeq_{\ih} \\ 
%\ctxtapp{\ctxt{\BMC{L}}}{\termsubs{x}{\BMC{\ctxtapp{\ctxt{T}}{o}}}{\BMC{s}}} & =_{\ih} &
%\delia{\ctxtapp{\ctxt{\BMC{L}}}{\termsubs{x}{\BMC{\ctxtapp{\ctxt{T}}{o'}}}{\BMC{s}}}} & = &
\BMC{\ctxtapp{\ctxt{L}}{\termsubs{x}{\ctxtapp{\ctxt{T}}{o'}}{s}}}    & = &
 \BMC{\termapp{t}{\ctxtapp{\ctxt{T}}{o'}}}
\end{array} $$

  \item If $\BMC{t} = \ctxtapp{\ctxt{L}}{\termcont{\alpha}{c}}$,
    then $\sz{ \termapp{t}{\ctxt{T}}} =
    \sz{t} + \sz{\ctxt{T}} + 1 \geq
    \sz{\ctxtapp{\ctxt{L}}{\termcont{\alpha}{c}}} + \sz{\ctxt{T}} + 1 =
    \sz{\ctxt{L}} + 1 + \sz{c} + \sz{\ctxt{T}} + 1 >
    \sz{c} + \sz{\ctxt{T}} + 1 =
    \sz{\termrepl{\alpha'}{\alpha}{\ctxt{T}}{c}}$.

    $$
\begin{array}{l@{\enspace}l@{\enspace}l@{\enspace}l}
\BMC{\termapp{t}{\ctxtapp{\ctxt{T}}{o}}}  & = &
\BMC{\termapp{\ctxtapp{\ctxt{L}}{\termcont{\alpha}{c}}}{\ctxtapp{\ctxt{T}}{o}}}    & = \\
\BMC{\ctxtapp{\ctxt{L}}{\termcont{\alpha'}{\termrepl{\alpha'}{\alpha}{\ctxtapp{\ctxt{T}}{o}}{c}}}}                                      & = &
                                                                                                              \ctxtapp{\ctxt{L}}{\termcont{\alpha'}{\BMC{\termrepl{\alpha'}{\alpha}{\ctxtapp{\ctxt{T}}{o}}{c}}}}                                      & \simeq_{\ih} \\
\ctxtapp{\ctxt{L}}{\termcont{\alpha'}{\BMC{\termrepl{\alpha'}{\alpha}{\ctxtapp{\ctxt{T}}{o'}}{c}}}}                                      & =  & 
\BMC{\termapp{t}{\ctxtapp{\ctxt{T}}{o'}}}
\end{array} $$

\item Otherwise, since $\sz{\termapp{t}{\ctxt{T}}} > \sz{\ctxt{T}}$,  we have
  $$\BMC{\termapp{t}{\ctxtapp{\ctxt{T}}{o}}} = \termapp{\BMC{t}}{\BMC{\ctxtapp{\ctxt{T}}{o}}} \simeq_{\ih}
    \termapp{\BMC{t}}{\BMC{\ctxtapp{\ctxt{T}}{o'}}} = \BMC{\termapp{t}{\ctxtapp{\ctxt{T}}{o'}}}$$
  \end{itemize}

%%%%%%%%%%%%%%%%%%%%%%%%%

  \item The cases $\ctxt{O} = \termabs{x}{\ctxt{T}}$,
     $\ctxt{O} =\termcont{\alpha}{\ctxt{C}}$,
    $\ctxt{O} = \termsubs{x}{t}{\ctxt{T}}$,
$\ctxt{O} = \termsubs{x}{\ctxt{T}}{t}$, 
$\ctxt{O} = \termname{\alpha}{\ctxt{T}}$,
$\ctxt{O} = \termpush{\ctxt{T}}{s}$, and
$\ctxt{O} = \termpush{t}{\ctxt{S}}$
are all straightforward. 

    \ignore{
  \item $\ctxt{O} = \termabs{x}{\ctxt{T}}$.
    \[\begin{array}{rll}
        & \BMC{\ctxtapp{(\termabs{x}{\ctxt{T}})}{o}} \\
        \simeq & \BMC{\termabs{x}{\ctxtapp{\ctxt{T}}{o}}} \\
        \simeq & \termabs{x}{\BMC{\ctxtapp{\ctxt{T}}{o}}} \\
        \simeq & \termabs{x}{\BMC{\ctxtapp{\ctxt{T}}{o'}}} & (\ih) \\
        \simeq & \BMC{\termabs{x}{\ctxtapp{\ctxt{T}}{o'}}} \\
        \simeq & \BMC{\ctxtapp{(\termabs{x}{\ctxt{T}})}{o'}}
                 \end{array}\]

             \item $\ctxt{O} = \termcont{\alpha}{\ctxt{C}}$.
               \[\begin{array}{rll}
        & \BMC{\ctxtapp{(\termcont{\alpha}{\ctxt{T}})}{o}} \\
        \simeq & \BMC{\termcont{\alpha}{\ctxtapp{\ctxt{T}}{o}}} \\
        \simeq & \termcont{\alpha}{\BMC{\ctxtapp{\ctxt{T}}{o}}} \\
        \simeq & \termcont{\alpha}{\BMC{\ctxtapp{\ctxt{T}}{o'}}} & (\ih) \\
        \simeq & \BMC{\termcont{\alpha}{\ctxtapp{\ctxt{T}}{o'}}} \\
        \simeq & \BMC{\ctxtapp{(\termcont{\alpha}{\ctxt{T}})}{o'}}
                 \end{array}\]

             \item $\ctxt{O} = \termsubs{x}{t}{\ctxt{T}}$.
                      \[\begin{array}{rll}
        & \BMC{\ctxtapp{(\termsubs{x}{t}{\ctxt{T}})}{o}} \\
        \simeq & \BMC{\termsubs{x}{t}{\ctxtapp{\ctxt{T}}{o}}} \\
        \simeq & \termsubs{x}{\BMC{t}}{\BMC{\ctxtapp{\ctxt{T}}{o}}} \\
        \simeq  & \termsubs{x}{\BMC{t}}{\BMC{\ctxtapp{\ctxt{T}}{o'}}} & (\ih) \\
        \simeq & \BMC{\termsubs{x}{t}{\ctxtapp{\ctxt{T}}{o'}}} \\
        \simeq & \BMC{\ctxtapp{(\termsubs{x}{t}{\ctxt{T}})}{o'}}
                 \end{array}\]

             \item $\ctxt{O} = \termsubs{x}{\ctxt{T}}{t}$.
               \[\begin{array}{rll}
        & \BMC{\ctxtapp{(\termsubs{x}{\ctxt{T}}{t})}{o}} \\
        \simeq & \BMC{\termsubs{x}{\ctxtapp{\ctxt{T}}{o}}{t}} \\
        \simeq & \termsubs{x}{\BMC{\ctxtapp{\ctxt{T}}{o}}}{\BMC{t}} \\
        \simeq  & \termsubs{x}{\BMC{\ctxtapp{\ctxt{T}}{o'}}}{\BMC{t}} & (\ih) \\
        \simeq & \BMC{\termsubs{x}{\ctxtapp{\ctxt{T}}{o'}}{t}} \\
        \simeq & \BMC{\ctxtapp{(\termsubs{x}{\ctxt{T}}{t})}{o'}}
                 \end{array}\]
                                                 
             \item $\ctxt{O} = \termname{\alpha}{\ctxt{T}}$.
                 \[\begin{array}{rll}
        & \BMC{\ctxtapp{(\termname{\alpha}{\ctxt{T}})}{o}} \\
        \simeq & \BMC{\termname{\alpha}{\ctxtapp{\ctxt{T}}{o}}} \\
        \simeq & \termname{\alpha}{\BMC{\ctxtapp{\ctxt{T}}{o}}} \\
        \simeq  & \termname{\alpha}{\BMC{\ctxtapp{\ctxt{T}}{o'}}} & (\ih) \\
        \simeq & \BMC{\termname{\alpha}{\ctxtapp{\ctxt{T}}{o'}}} \\
        \simeq & \BMC{\ctxtapp{(\termname{\alpha}{\ctxt{T}})}{o'}}
                 \end{array}\]
             }
        
             \item $\ctxt{O} = \termrepl{\alpha'}{\alpha}{\ctxt{S}}{c}$.
               There are two cases. Before analysing them we remark that
  $o = \termemst$ iff $o' = \termemst$.

               %%%%%%%%%%%%%% 

  \begin{itemize}
  \item $\BMC{c} = \ctxtapp{\ctxtLCC}{\termrepl{\alpha}{\beta}{s_0}{c'}}$,
    where $\alpha \notin \fn{c',\ctxtLCC,s_0}$ and $s_0,
    \ctxtapp{\ctxt{S}}{o} \neq \termemst$ (thus also  $\ctxtapp{\ctxt{S}}{o'}
    \neq \termemst$). We have $\sz{\termrepl{\alpha'}{\alpha}{\ctxt{S}}{c}} =
    \sz{c} + \sz{\ctxt{S}} + 1 \geq \sz{\ctxtLCC} + \sz{c'} + \sz{s_0} + 1  +
    \sz{\ctxt{S}} +  1 > \sz{s_0} + \sz{\ctxt{S}} + 1 =
    \sz{\termpush{s_0}{\ctxt{S}}}$. Then \[
\begin{array}{l@{\enspace}l@{\enspace}l@{\enspace}l@{\enspace}l@{\enspace}l}
\BMC{\termrepl{\alpha'}{\alpha}{\ctxtapp{\ctxt{S}}{o}}{c}}                                       & = &
\BMC{\ctxtapp{\ctxtLCC}{\termrepl{\alpha'}{\beta}{\termpush{s_0}{\ctxtapp{\ctxt{S}}{o}}}{c'}}}  & = \\
\ctxtapp{\ctxtLCC}{\termrepl{\alpha'}{\beta}{\BMC{\termpush{s_0}{\ctxtapp{\ctxt{S}}{o}}}}{c'}}   & \simeq_{\ih}  & 
\ctxtapp{\ctxtLCC}{\termrepl{\alpha'}{\beta}{\BMC{\termpush{s_0}{\ctxtapp{\ctxt{S}}{o'}}}}{c'}}       & = \\
\BMC{\termrepl{\alpha'}{\alpha}{\ctxtapp{\ctxt{S}}{o'}}{c}}
\end{array} \]

  \item $\BMC{c} = \ctxtapp{\ctxtLCC}{\termrepl{\alpha}{\beta}{\termemst}{c'}}$,
    where $\alpha \notin \fn{c',\ctxtLCC}$ and $\ctxtapp{\ctxt{S}}{o} \neq
    \termemst$ (thus also  $\ctxtapp{\ctxt{S}}{o'} \neq \termemst$). We have
    $\sz{\termrepl{\alpha'}{\alpha}{\ctxt{S}}{c}} = \sz{c} + \sz{\ctxt{S}} + 1
    \geq \sz{\ctxtLCC} + \sz{c'} +  1  +\sz{\ctxt{S}} +  1 > \sz{c'} +
    \sz{\ctxt{S}} + 1 = \sz{\termrepl{\alpha}{\beta}{\ctxt{S}}{c'}}$. Then, \[
\begin{array}{l@{\enspace}l@{\enspace}l@{\enspace}l@{\enspace}l@{\enspace}l}
\BMC{\termrepl{\alpha'}{\alpha}{\ctxtapp{\ctxt{S}}{o}}{c}}                                       & = &
\BMC{\ctxtapp{\ctxtLCC}{\termrepl{\alpha'}{\alpha}{\termemst}{\termrepl{\alpha}{\beta}{\ctxtapp{\ctxt{S}}{o}}{c'}}}}  & = \\
\ctxtapp{\ctxtLCC}{\termrepl{\alpha'}{\alpha}{\termemst}{\BMC{\termrepl{\alpha}{\beta}{\ctxtapp{\ctxt{S}}{o}}{c'}}}}   & \simeq_{\ih}  & 
\ctxtapp{\ctxtLCC}{\termrepl{\alpha'}{\alpha}{\termemst}{\BMC{\termrepl{\alpha}{\beta}{\ctxtapp{\ctxt{S}}{o'}}{c'}}}}   & =  \\ 
\BMC{\termrepl{\alpha'}{\alpha}{\ctxtapp{\ctxt{S}}{o'}}{c}}
\end{array} \]

\item Otherwise, since $\sz{\termrepl{\alpha'}{\alpha}{\ctxt{S}}{c}} > \sz{\ctxt{S}}$, we have
  $$\begin{array}{ll}
      \BMC{\termrepl{\alpha'}{\alpha}{\ctxtapp{\ctxt{S}}{o}}{c}} &
      = \\
    \termrepl{\alpha'}{\alpha}{\BMC{\ctxtapp{\ctxt{S}}{o}}}{\BMC{c}}
      & \simeq_{\ih} \\
    \termrepl{\alpha'}{\alpha}{\BMC{\ctxtapp{\ctxt{S}}{o'}}}{\BMC{c}}
      & =  \\
          \BMC{\termrepl{\alpha'}{\alpha}{\ctxtapp{\ctxt{S}}{o'}}{c}}
          \end{array}$$
  \end{itemize}

               %%%%%%%%%%%%%%%%%
     
             \item $\ctxt{O} = \termrepl{\alpha'}{\alpha}{s}{\ctxt{C}}$.
       \[\begin{array}{llll}
           \BMC{\termrepl{\alpha'}{\alpha}{s}{\ctxtapp{\ctxt{C}}{o}}} & = &
          \BMC{\termrepl{\alpha'}{\alpha}{\BMC{s}}{\BMC{\ctxtapp{\ctxt{C}}{o}}}} & \simeq_{\mbox{item 2}} \\
     \BMC{\termrepl{\alpha'}{\alpha}{\BMC{s}}{\BMC{\ctxtapp{\ctxt{C}}{o'}}}}  & = & 
           \BMC{\termrepl{\alpha'}{\alpha}{s}{\ctxtapp{\ctxt{C}}{o'}}} 
           \end{array} \]

     \end{itemize}

\end{enumerate}
\end{proof}

%%% Local Variables:
%%% mode: latex
%%% TeX-master: "main"
%%% End:

\begin{lemma}
\label{l:equiv-substitution}
Let $o\in \objects{\LMfull}$ and $w\in \terms{\LMfull}$. If $w \in \BMCform$ and $o \simeq o'$, then $\BMC{\substitute{x}{w}{o}} \simeq
\BMC{\substitute{x}{w}{o'}}$. 
\end{lemma}

\begin{proof}
%\delia{Este lemma ya esta chequeado para la nueva version de BMC}
By induction on $o \simeq o'$. 
\begin{itemize}
  \item If $o = \termsubs{y}{u}{\ctxtapp{\ctxtLTT}{t}} \simeq_{\eqexsubs}
  \ctxtapp{\ctxtLTT}{\termsubs{y}{u}{t}} = o'$, where
  $\notatall{x}{\ctxtLTT}$ and $\freeFor{u}{\ctxtLTT}$. Let us write
  $t' = \substitute{x}{w}{t}$, $\ctxtLTT' = \substitute{x}{w}{\ctxtLTT}$,
  and $u' = \substitute{x}{w}{u}$. Then, $$
\begin{array}{l@{\enspace}l@{\enspace}l@{\enspace}l}
\BMC{\substitute{x}{w}{\termsubs{y}{u}{\ctxtapp{\ctxtLTT}{t}}}} & = &
\BMC{\termsubs{y}{u'}{\ctxtapp{\ctxtLTT'}{t'}}}                 & \simeq_{(L.\ref{l:subs-repl-out-of-BM}.\ref{l:subs-out-of-BM})} \\
\BMC{\ctxtapp{\ctxtLTT'}{\termsubs{y}{u'}{t'}}}                 & = &
\BMC{\substitute{x}{w}{\ctxtapp{\ctxtLTT}{\termsubs{y}{u}{t}}}} 
\end{array} $$

  \item The case $\simeq_{\eqexrepl}$ is similar and uses
  Lem.~\ref{l:subs-repl-out-of-BM}.\ref{l:repl-out-of-BM}. 

  \item $o = \termrepl{\alpha'}{\alpha}{s}{(\termname{\alpha}{u})}
  \simeq_{\eqlinear} \termname{\alpha'}{\BMC{\termconc{u}{s}}} = o'$, where
  $\alpha \notin \fn{u}$ and $s \neq \termemst$. Let us write $s' = \substitute{x}{w}{s}$, and $u' =
  \substitute{x}{w}{u}$. By $\alpha$-conversion we can also assume $\alpha
  \notin \fn{w}$. Then we have $$
\begin{array}{l@{\enspace}l@{\enspace}l@{\enspace}l}
\BMC{\substitute{x}{w}{\termrepl{\alpha'}{\alpha}{s}{(\termname{\alpha}{u})}}} & = &
\BMC{\termrepl{\alpha'}{\alpha}{s'}{(\termname{\alpha}{u'})}}                  & = \\
\termrepl{\alpha'}{\alpha}{\BMC{s'}}{(\termname{\alpha}{\BMC{u'}})}            & \simeq_{(\ast)} &
\termname{\alpha'}{\BMC{\termconc{\BMC{u'}}{\BMC{s'}}}}                        & \simeq \\
\termname{\alpha'}{\BMC{\termconc{u'}{s'}}}                                    & = &
\termname{\alpha'}{\BMC{\substitute{x}{w}{\BMC{\termconc{u}{s}}}}}             & = \\
\BMC{\substitute{x}{w}{\termname{\alpha'}{\BMC{\termconc{u}{s}}}}}
\end{array} $$
  The equality $(\ast)$ is justified by the fact that $\alpha \notin \fn{u,w}$.

  \item All the other cases are straightforward.
\end{itemize}
\end{proof}

\begin{lemma}
\label{l:Sigma_closed_under_substitution:target}
Let $o\in \objects{\LMfull} $ and $w\in \terms{\LMfull} $. If $o \in \BMCform$ and $w \simeq w'$, then $\BMC{\substitute{x}{w}{o}} \simeq
\BMC{\substitute{x}{w'}{o}}$. 
\end{lemma}

\begin{proof} %\delia{Este lemma ya esta chequeado para la nueva version de BMC}
We first remark that $o \in \Cform$ implies $\substitute{x}{w}{o} \in \Cform$,
since $w \in \Cform$. We proceed by induction on $o$, by only showing the
interesting cases.  
\begin{itemize}
  \item $o = y$. If $x = y$, then $\BMC{\substitute{x}{w}{o}} = \BMC{w} = w
  \simeq w'= \BMC{w'} = \BMC{\substitute{x}{w'}{o}}$. If $x \neq y$, then
  $\BMC{\substitute{x}{w}{o}} = \BMC{y} = y \simeq y = \BMC{y} =
  \BMC{\substitute{x}{w'}{o}}$.

%%   \item $o = \termabs{y}{v}$. We have $$
%% \begin{array}{lll}
%% \BMC{\substitute{x}{w}{(\termabs{y}{v})}}   = 
%% \termabs{y}{\BMC{\substitute{x}{w}{v}}}     \simeq_{(\ih)}
%% \termabs{y}{\BMC{\substitute{x}{w'}{v}}}    = 
%% \BMC{\substitute{x}{w'}{(\termabs{y}{v})}}
%% \end{array} $$

  \item $o = \termapp{v}{u}$. Then we have $$
\begin{array}{l@{\enspace}l@{\enspace}l@{\enspace}l}
\BMC{\substitute{x}{w}{(\termapp{v}{u})}}                          & = &
\BMC{\termapp{\substitute{x}{w}{v}}{\substitute{x}{w}{u}}}         & = \\
\BMC{\termapp{\BMC{\substitute{x}{w}{v}}}{\substitute{x}{w}{u}}}   & \simeq_{(\ih + L.\ref{l:app-bm})} &
\BMC{\termapp{\BMC{\substitute{x}{w'}{v}}}{\substitute{x}{w}{u}}}  & = \\
\BMC{\termapp{\substitute{x}{w'}{v}}{\BMC{\substitute{x}{w}{u}}}}  & \simeq_{(\ih + L.\ref{l:app-bm})} &
\BMC{\termapp{\substitute{x}{w'}{v}}{\BMC{\substitute{x}{w'}{u}}}} & = \\
\BMC{\substitute{x}{w'}{(\termapp{v}{u})}}
\end{array} $$

%%   \item $o = \termname{\alpha}{t}$. Then $$
%% \begin{array}{lll}
%% \BMC{\substitute{x}{w}{(\termname{\alpha}{t})}}  & = & \\
%% \BMC{\termname{\alpha}{\substitute{x}{w}{t}}}    & = & \\
%% \termname{\alpha}{\BMC{\substitute{x}{w}{t}}}    & \simeq & (\ih) \\
%% \termname{\alpha}{\BMC{\substitute{x}{w'}{t}}}   & = & \\
%% \BMC{\substitute{x}{w'}{\termname{\alpha}{t}}}   & = & \\
%% \BMC{\substitute{x}{w'}{(\termname{\alpha}{t})}}
%% \end{array} $$

%%   \item $o = \termsubs{y}{u}{v}$. Then $$
%% \begin{array}{lll}
%% \BMC{\substitute{x}{w}{\termsubs{y}{u}{v}}}                            & = & \\
%% \BMC{\termsubs{y}{\substitute{x}{w}{u}}{\substitute{x}{w}{v}}}         & = & \\
%% \termsubs{y}{\BMC{\substitute{x}{w}{u}}}{\BMC{\substitute{x}{w}{v}}}   & \simeq & (\ih) \\
%% \termsubs{y}{\BMC{\substitute{x}{w'}{u}}}{\BMC{\substitute{x}{w'}{v}}} & = & \\
%% \BMC{\termsubs{y}{\substitute{x}{w'}{u}}{\substitute{x}{w'}{v}}}       & = & \\
%% \BMC{\substitute{x}{w'}{\termsubs{y}{u}{v}}}
%% \end{array} $$

%%   \item $o = \termcont{\alpha}{c}$. Then $$
%% \begin{array}{lll}
%% \BMC{\substitute{x}{w}{(\termcont{\alpha}{c})}}  & = & \\
%% \BMC{\termcont{\alpha}{\substitute{x}{w}{c} }}   & = & \\
%% \termcont{\alpha}{\BMC{\substitute{x}{w}{c} }}   & \simeq & (\ih) \\
%% \termcont{\alpha}{\BMC{\substitute{x}{w'}{c} }}  & = & \\
%% \BMC{\termcont{\alpha}{\substitute{x}{w'}{c} }}  & = & \\
%% \BMC{\substitute{x}{w'}{(\termcont{\alpha}{c})}}
%% \end{array} $$

  \item $o = \termrepl{\alpha'}{\alpha}{s}{c}$. Then $$%\kern-4em
\begin{array}{l@{\enspace}l@{\enspace}l@{\enspace}l}
\BMC{\substitute{x}{w}{\termrepl{\alpha'}{\alpha}{s}{c}}}                                                                                        & = &
\BMC{\termrepl{\alpha'}{\alpha}{\substitute{x}{w}{s}}{\substitute{x}{w}{c}}}                                                                     & =_{(\ast)} \\
\termrepl{\alpha'}{\alpha}{\BMC{\substitute{x}{w}{s}}}{\BMC{\substitute{x}{w}{c}}}                                                               & =_{(\ih)} &
\termrepl{\alpha'}{\alpha}{\BMC{\substitute{x}{w'}{s}}}{\BMC{\substitute{x}{w'}{c}}}                                                             & =_{(\ast)} \\
\BMC{\termrepl{\alpha'}{\alpha}{\substitute{x}{w'}{s}}{\substitute{x}{w'}{c}}}                                                                   & = &
\BMC{\substitute{x}{w'}{\termrepl{\alpha'}{\alpha}{s}{c}}}
\end{array} $$
  The equalities $(\ast)$ are justified by the fact that $o \in \Cform$.

\end{itemize}
\end{proof}

\begin{corollary}
Let $o\in \objects{\LMfull} $ and $w\in \terms{\LMfull} $. Let $\ctxt{O}$ be a context s.t. $x \notin \fv{\ctxt{O}}$. Then if $o \in \BMCform$
and $w \simeq w'$, then $\BMC{\substitute{x}{w}{\ctxtapp{\ctxt{O}}{o}}} \simeq
\BMC{\ctxtapp{\ctxt{O}}{\substitute{x}{w'}{o}}}$. 
\end{corollary}

\begin{proof}
By Lem.~\ref{l:context-substitution-replacement} we have
$$\BMC{\substitute{x}{w}{\ctxtapp{\ctxt{O}}{o}}} =
\BMC{\ctxtapp{\substitute{x}{w}{\ctxt{O}}}{\substitute{x}{w}{o}}} = 
\BMC{\ctxtapp{\ctxt{O}}{\BMC{\substitute{x}{w}{o}}}}$$ and
$$\BMC{\substitute{x}{w'}{\ctxtapp{\ctxt{O}}{o}}} =
\BMC{\ctxtapp{\substitute{x}{w'}{\ctxt{O}}}{\substitute{x}{w}{o}}} =
\BMC{\ctxtapp{\ctxt{O}}{\BMC{\substitute{x}{w'}{o}}}}$$
Lem.~\ref{l:Sigma_closed_under_substitution:target} gives
$\BMC{\substitute{x}{w}{o}} \simeq \BMC{\substitute{x}{w'}{o}}$. Then
Lem.~\ref{l:app-bm} allows to conclude.
\end{proof}

\begin{lemma}
\label{l:Sigma_closed_under_BM_plus_replacement}
Let $o\in \objects{\LMfull} $. If $s \in \BMCform$ and $o \simeq o'$, then
$\BMC{\replace{\gamma'}{\gamma}{s}{o}} \simeq
\BMC{\replace{\gamma'}{\gamma}{s}{o'}}$.
\end{lemma}

\begin{proof} %\delia{ESTE LEMMA YA ESTA VERIFICADO PARA $\rW$}
By induction on $o \simeq o'$. We only show the interesting cases.
\begin{itemize}
  \item If $o = \termsubs{y}{u}{\ctxtapp{\ctxtLTT}{t}} \simeq_{\eqexsubs}
  \ctxtapp{\ctxtLTT}{\termsubs{y}{u}{t}} = o'$, where $\notatall{y}{\ctxtLTT}$ and
  $\freeFor{u}{\ctxtLTT}$. Then let us write $t' =
  \replace{\gamma'}{\gamma}{s}{t}$, $\ctxtLTT' =
  \replace{\gamma'}{\gamma}{s}{\ctxtLTT}$, and $u' =
  \replace{\gamma'}{\gamma}{s}{u}$. Then, $$
\begin{array}{l@{\enspace}l@{\enspace}l@{\enspace}l}
\BMC{\replace{\gamma'}{\gamma}{s}{\termsubs{y}{u}{\ctxtapp{\ctxtLTT}{t}}}} & = &
\BMC{\termsubs{y}{u'}{\ctxtapp{\ctxtLTT'}{t'}}}                            & \simeq_{(L.\ref{l:subs-repl-out-of-BM}.\ref{l:subs-out-of-BM})} \\
\BMC{\ctxtapp{\ctxtLTT'}{\termsubs{y}{u'}{t'}}}                            & = &
\BMC{\replace{\gamma'}{\gamma}{s}{\ctxtapp{\ctxtLTT}{\termsubs{y}{u}{t}}}}
\end{array} $$

  \item If $o = \termrepl{\alpha'}{\alpha}{s_0}{\ctxtapp{\ctxtLCC}{c}}
  \simeq_{\eqexrepl} \ctxtapp{\ctxtLCC}{\termrepl{\alpha'}{\alpha}{s_0}{c}} =
  o'$ and $\ctxtLCC$ is $\alpha$-free. Let $\ctxtLCC' =
  \replace{\gamma'}{\gamma}{s}{\ctxtLCC}$, $s'_0 =
  \replace{\gamma'}{\gamma}{s}{s_0}$ and $c' =
  \replace{\gamma'}{\gamma}{s}{c}$.
  There are three cases:
  \begin{itemize}
    \item If $\gamma = \alpha'$, $s_0 = \termemst$ and $s \neq \termemst$, then $$
\begin{array}{l@{\enspace}l@{\enspace}l@{\enspace}l}
\BMC{\replace{\gamma'}{\gamma}{s}{\termrepl{\alpha'}{\alpha}{\termemst}{\ctxtapp{\ctxtLCC}{c}}}}    & = &  
\BMC{\termrepl{\gamma'}{\gamma''}{\termemst}{\termrepl{\gamma''}{\al}{s}{\ctxtapp{\ctxtLCC'}{c'}}}} & \simeq_{(L.\ref{l:subs-repl-out-of-BM}.\ref{l:repl-out-of-BM})} \\
\BMC{\ctxtapp{\ctxtLCC'}{\termrepl{\gamma'}{\gamma''}{\termemst}{\termrepl{\gamma''}{\al}{s}{c'}}}} & = & 
\BMC{\replace{\gamma'}{\gamma}{s}{\ctxtapp{\ctxtLCC}{\termrepl{\alpha'}{\alpha}{\termemst}{c}}}}
\end{array} $$

\item If $\gamma = \alpha'$ and $s_0 = s = \termemst$ or
  $s_0, s \neq \termemst$, %\delia{revisar condicion},
  then $$
\begin{array}{l@{\enspace}l@{\enspace}l@{\enspace}l}
\BMC{\replace{\gamma'}{\gamma}{s}{\termrepl{\alpha'}{\alpha}{s_0}{\ctxtapp{\ctxtLCC}{c}}}} & = & 
\BMC{\termrepl{\gamma'}{\alpha}{\termpush{s_0}{s}}{\ctxtapp{\ctxtLCC'}{c'}}}               & \simeq_{(L.\ref{l:subs-repl-out-of-BM}.\ref{l:repl-out-of-BM})} \\
\BMC{\ctxtapp{\ctxtLCC'}{\termrepl{\gamma'}{\alpha}{\termpush{s_0}{s}}{c'}}}               & = &  
\BMC{\replace{\gamma'}{\gamma}{s}{\ctxtapp{\ctxtLCC}{\termrepl{\alpha'}{\alpha}{s_0}{c}}}}
\end{array} $$

    \item Otherwise,  $$
\begin{array}{l@{\enspace}l@{\enspace}l@{\enspace}l}
\BMC{\replace{\gamma'}{\gamma}{s}{\termrepl{\alpha'}{\alpha}{s_0}{\ctxtapp{\ctxtLCC}{c}}}} & = & 
\BMC{\termrepl{\alpha'}{\alpha}{s'_0}{\ctxtapp{\ctxtLCC'}{c'}}}                            & \simeq_{(L.\ref{l:subs-repl-out-of-BM}.\ref{l:repl-out-of-BM})} \\
\BMC{\ctxtapp{\ctxtLCC'}{\termrepl{\alpha'}{\alpha}{s'_0}{c'}}}                            & = & 
\BMC{\replace{\gamma'}{\gamma}{s}{\ctxtapp{\ctxtLCC}{\termrepl{\alpha'}{\alpha}{s_0}{c}}}}
\end{array} $$
  \end{itemize}

  \item If $o = \termrepl{\alpha'}{\alpha}{s_0}{(\termname{\alpha}{u})}
  \simeq_{\eqlinear} \termname{\alpha'}{\BMC{\termconc{u}{s_0}}} = o'$, where
  $\alpha, \alpha' \notin \fn{u}$ and $s_0 \neq \termemst$. Let us write $u' =
  \replace{\gamma'}{\gamma}{s}{u}$ and $s'_0 =
  \replace{\gamma'}{\gamma}{s}{s_0}$. By $\alpha$-conversion we can assume
  $\alpha \neq \gamma'$ and $\alpha \notin \fn{s}$. There are two cases:
  \begin{itemize}
   \item If $\gamma = \alpha'$  then $\alpha'$ fresh in $u$ implies in
    particular $\gamma$ fresh in $u$. Also $\al \neq \al'$ implies $\al \neq
    \gamma$. Then, $$
\begin{array}{l@{\enspace}l@{\enspace}l@{\enspace}l}
\BMC{\replace{\gamma'}{\gamma}{s}{\termrepl{\alpha'}{\alpha}{s_0}{(\termname{\alpha}{u})}}}   & = &
\BMC{\termrepl{\gamma'}{\alpha}{\termpush{s_0}{s}}{(\termname{\alpha}{u})}}                   & = \\
\termrepl{\gamma'}{\alpha}{\BMC{\termpush{s_0}{s}}}{(\termname{\alpha}{\BMC{u}})}             & \simeq_{\eqlinear} & 
\termname{\gamma'}{\BMC{\termconc{\BMC{u}}{\BMC{\termpush{s_0}{s}}}}}                         & = \\
\termname{\gamma'}{\BMC{\termconc{u}{\termpush{s_0}{s}}}}                                     & = & 
\termname{\gamma'}{\BMC{\termconc{\BMC{\termconc{u}{s_0}}}{s}}}                               & = \\
\BMC{\termname{\gamma'}{\termconc{\replace{\gamma'}{\gamma}{s}{\BMC{\termconc{u}{s_0}}}}{s}}} & = & 
\BMC{\replace{\gamma'}{\gamma}{s}{(\termname{\alpha'}{\BMC{\termconc{u}{s_0}})}}}
\end{array} $$
   \item Otherwise,   $$
\begin{array}{l@{\enspace}l@{\enspace}l@{\enspace}l}
\BMC{\replace{\gamma'}{\gamma}{s}{\termrepl{\alpha'}{\alpha}{s_0}{(\termname{\alpha}{u})}}} & = & 
\BMC{\termrepl{\alpha'}{\alpha}{s'_0}{(\termname{\alpha}{u'})}}                             & = \\
\termrepl{\alpha'}{\alpha}{\BMC{s'_0}}{(\termname{\alpha}{\BMC{u'}})}                       & \simeq_{\eqlinear} & 
\termname{\alpha'}{\BMC{\termconc{\BMC{u'}}{\BMC{s'_0}}}}                                   & = \\
\termname{\alpha'}{\BMC{\termconc{u'}{s'_0}}}                                               & = & 
\termname{\alpha'}{\BMC{\replace{\gamma'}{\gamma}{s}{(\termconc{u}{s_0})}}}                 & = \\
\termname{\alpha'}{\BMC{\replace{\gamma'}{\gamma}{s}{\BMC{\termconc{u}{s_0}}}}}             & = & 
\BMC{\replace{\gamma'}{\gamma}{s}{(\termname{\alpha'}{\BMC{\termconc{u}{s_0}})}}}
\end{array} $$
    The use of $\eqlinear$ is justified by the fact that $\alpha \notin \fn{u,s}$.

    \end{itemize}

  \item If $o =
  \termname{\alpha'}{\termcont{\alpha}{\termname{\beta'}{\termcont{\beta}{u}}}}
  \simeq_{\eqpoppop} 
  \termname{\beta'}{\termcont{\beta}{\termname{\alpha'}{\termcont{\alpha}{u}}}}
  = o'$. Let us write $u' = \replace{\gamma'}{\gamma}{s}{u}$. By
  $\alpha$-conversion we can assume $\gamma \neq \alpha, \beta$. There are four
  cases:
  \begin{itemize}
    \item If $\gamma \neq \alpha'$ and $\gamma \neq \beta'$, then $$
\begin{array}{l@{\enspace}l@{\enspace}l@{\enspace}l}
\BMC{\replace{\gamma'}{\gamma}{s}{(\termname{\alpha'}{\termcont{\alpha}{\termname{\beta'}{\termcont{\beta}{u}}}})}} & = & 
\BMC{\termname{\alpha'}{\termcont{\alpha}{\termname{\beta'}{\termcont{\beta}{u'}}}}}                                & = \\
\termname{\alpha'}{\termcont{\alpha}{\termname{\beta'}{\termcont{\beta}{\BMC{u'}}}}}                                & \simeq_{\eqpoppop} & 
\termname{\beta'}{\termcont{\beta}{\termname{\alpha'}{\termcont{\alpha}{\BMC{u'}}}}}                                & = \\
\BMC{\replace{\gamma'}{\gamma}{s}{(\termname{\beta'}{\termcont{\beta}{\termname{\alpha'}{\termcont{\alpha}{u}}}})}}
\end{array} $$

    \item If $\gamma = \alpha'$ and $\gamma \neq \beta'$, then let $\alpha_1
    \eqdef \alpha$, and consider fresh names $\al_2, \ldots, \al_{k+1}$. Let us
    write $s = \termpush{u_1}{\termpush{\ldots}{u_k}})\ (k \geq 0)$. Then $$
\begin{array}{l@{\enspace}l@{\enspace}l@{\enspace}l}
\BMC{\replace{\gamma'}{\gamma}{s}{(\termname{\alpha'}{\termcont{\alpha}{\termname{\beta'}{\termcont{\beta}{u}}}})}} & = &
\BMC{\termname{\gamma'}{\termconc{(\termcont{\alpha}{\termname{\beta'}{\termcont{\beta}{u'}}})}{s}}}                & = \\
\BMC{\termname{\gamma'}{\termcont{\alpha_{k+1}}(\termname{\beta'}{\termcont{\beta}{u'})}
  \lvec{\exrepl{\alpha_{i+1}}{\alpha_i}{u_i}}
}}    & = &
\termname{\gamma'}{\termcont{\alpha_{k+1}}{\BMC{(\termname{\beta'}{\termcont{\beta}{u'}})
  \lvec{\exrepl{\alpha_{i+1}}{\alpha_i}{u_i}}
}}}   & \simeq_{(L.\ref{l:subs-repl-out-of-BM}.\ref{l:repl-out-of-BM})} \\
\termname{\gamma'}{\termcont{\alpha_{k+1}}{\BMC{\termname{\beta'}{\termcont{\beta}{u'}
  \lvec{\exrepl{\alpha_{i+1}}{\alpha_i}{u_i}}
}}}}  & = &
\termname{\gamma'}{\termcont{\alpha_{k+1}}{\termname{\beta'}{\termcont{\beta}{\BMC{u'
  \lvec{\exrepl{\alpha_{i+1}}{\alpha_i}{u_i}}
}}}}} & \simeq_{\eqpoppop} \\
\termname{\beta'}{\termcont{\beta}{\termname{\gamma'}{\termcont{\alpha_{k+1}}{\BMC{u'
  \lvec{\exrepl{\alpha_{i+1}}{\alpha_i}{u_i}}
}}}}} & = &
\BMC{\replace{\gamma'}{\gamma}{s}{(\termname{\beta'}{\termcont{\beta}{\termname{\alpha'}{\termcont{\alpha}{u}}}})}}
\end{array} $$

    \item The case $\gamma \neq \alpha'$ and $\gamma = \beta'$ is similar to
    the previous one.

    \item If $\gamma = \alpha' = \beta'$, then let $\al_1 = \alpha$, $\beta_1 =
    \beta$ and consider fresh names $\al_2, \ldots, \al_{k}, \beta_2, \ldots,
    \beta_k$. Let us write $s = \termpush{u_1}{\termpush{\ldots}{u_k}}\ (k \geq 0)$. Then $$\kern-3.5em
\begin{array}{l@{\enspace}l}
\BMC{\replace{\gamma'}{\gamma}{s}{(\termname{\alpha'}{\termcont{\alpha}{\termname{\beta'}{\termcont{\beta}{u}}}})}}   & = \\
\BMC{\termname{\gamma'}{\termconc{(\termcont{\alpha}{\termname{\gamma'}{\termconc{(\termcont{\beta}{u'})}{s}}})}{s}}} & = \\
\BMC{\termname{\gamma'}{\termcont{\alpha'_k}{(\termname{\gamma'}{\termcont{\beta'_k}{u'
  \lvec{\exrepl{\beta'_i}{\beta_i}{u_i}}
}})}
  \lvec{\exrepl{\alpha'_i}{\alpha_i}{u_i}}
}}    & = \\
\termname{\gamma'}{\termcont{\alpha'_k}{\BMC{(\termname{\gamma'}{\termcont{\beta'_k}{u'
  \lvec{\exrepl{\beta'_i}{\beta_i}{u_i}}
}})
  \lvec{\exrepl{\alpha'_i}{\alpha_i}{u_i}}
}}}   & \simeq_{(L.\ref{l:subs-repl-out-of-BM}.\ref{l:repl-out-of-BM})} \\
\termname{\gamma'}{\termcont{\alpha'_k}{\BMC{\termname{\gamma'}{\termcont{\beta'_k}{u'
  \lvec{\exrepl{\beta'_i}{\beta_i}{u_i}}
  \lvec{\exrepl{\alpha'_i}{\alpha_i}{u_i}}
}}}}} & = \\
\termname{\gamma'}{\termcont{\alpha'_k}{\termname{\gamma'}{\termcont{\beta'_k}{\BMC{u'
  \lvec{\exrepl{\beta'_i}{\beta_i}{u_i}}
  \lvec{\exrepl{\alpha'_i}{\alpha_i}{u_i}}
}}}}} & \simeq_{\eqpoppop} \\
\termname{\gamma'}{\termcont{\beta'_k}{\termname{\gamma'}{\termcont{\alpha'_k}{\BMC{u'
  \lvec{\exrepl{\beta'_i}{\beta_i}{u_i}}
  \lvec{\exrepl{\alpha'_i}{\alpha_i}{u_i}}
}}}}} & = \\
\termname{\beta'}{\termcont{\beta'_k}{\termname{\alpha'}{\termcont{\alpha'_k}{\BMC{u'
  \lvec{\exrepl{\beta'_i}{\beta_i}{u_i}}
  \lvec{\exrepl{\alpha'_i}{\alpha_i}{u_i}}
}}}}} & = \\
\BMC{\replace{\gamma'}{\gamma}{s}{(\termname{\beta'}{\termcont{\beta}{\termname{\alpha'}{\termcont{\alpha}{u}}}})}}
\end{array} $$
  \end{itemize}

  \item If $o = \termname{\alpha}{\termcont{\beta}{c}} \simeq_{\eqrho}
  c\neren{\beta}{\alpha} = o'$. Let us write $c' =
  \replace{\gamma'}{\gamma}{s}{c}$. There are two cases:
  \begin{itemize}
    \item If $\gamma \neq  \alpha$, then $$
\begin{array}{l@{\enspace}l@{\enspace}l@{\enspace}l@{\enspace}l@{\enspace}l}
\BMC{\replace{\gamma'}{\gamma}{s}{(\termname{\alpha}{\termcont{\beta}{c}})}} & = &
\termname{\alpha}{\termcont{\beta}{\BMC{c'}}}                                & \simeq_{\eqrho} &
\BMC{c'}\neren{\beta}{\alpha}                                                & = \\
\BMC{c'\neren{\beta}{\alpha}}                                                & = &
\BMC{\replace{\gamma'}{\gamma}{s}{(c\neren{\beta}{\alpha})}}
\end{array} $$

    \item If $\gamma = \alpha$.

    \begin{itemize}
      \item If $s = \termemst$, then $$
\begin{array}{l@{\enspace}l@{\enspace}l@{\enspace}l@{\enspace}l}
\BMC{\replace{\gamma'}{\alpha}{\termemst}{(\termname{\alpha}{\termcont{\beta}{c}})}} & = &
\BMC{\termname{\gamma'}{\termcont{\beta}{c'}}}                                       & = \\
\termname{\gamma'}{\termcont{\beta}{\BMC{c'}}}                                       & \simeq_{\eqrho} &
\BMC{c'}\neren{\beta}{\gamma'}                                                       & = \\
\BMC{c'\exrepl{\gamma'}{\beta}{\termemst}}                                           & = &
\BMC{\replace{\gamma'}{\alpha}{\termemst}{c\neren{\beta}{\alpha}}}
\end{array} $$

      \item If $s \neq  \termemst$, then suppose $s =
      \termpush{u_1}{\termpush{\ldots}{u_n}}$ ($n \geq 1$) and let $\beta_1 =
      \beta$. We consider two further cases below:
      \begin{itemize}
        \item $\BMC{c'} =
        \ctxtapp{\ctxtLCC}{\termrepl{\beta_1}{\delta}{s''}{c''}}$,
        where $\beta_1 \notin \fn{c'', \ctxtLCC}$ and $s'' \neq \termemst$ . $$
\begin{array}{l@{\enspace}l@{\enspace}l}
\BMC{\replace{\gamma'}{\gamma}{s}{(\termname{\alpha}{\termcont{\beta}{c}})}}                                                                   & = & \\
\BMC{\termname{\gamma'}{\termconc{(\termcont{\beta}{c'})}{s}}}                                                                                 & = & \\
\BMC{\termname{\gamma'}{\termcont{\beta_{n+1}}{c'
  \lvec{\exrepl{\beta_{i+1}}{\beta_i}{u_i}}
}}} & = & \\
\termname{\gamma'}{\termcont{\beta_{n+1}}{\BMC{c'
  \lvec{\exrepl{\beta_{i+1}}{\beta_i}{u_i}}
}}} & = & \\
\termname{\gamma'}{\termcont{\beta_{n+1}}{\BMC{\termrepl{\beta_{n+1}}{\beta_1}{s}{c'}}}}                                                       & = & \\
\termname{\gamma'}{\termcont{\beta_{n+1}}{\BMC{\termrepl{\beta_{n+1}}{\beta_1}{s}{\BMC{c'}}}}}                                                 & = & \\
\termname{\gamma'}{\termcont{\beta_{n+1}}{\BMC{\termrepl{\beta_{n+1}}{\beta_1}{s}{\ctxtapp{\ctxtLCC}{\termrepl{\beta_1}{\delta}{s''}{c''}}}}}} & = & \\
\termname{\gamma'}{\termcont{\beta_{n+1}}{\BMC{\ctxtapp{\ctxtLCC}{\termrepl{\beta_{n+1}}{\delta}{\termpush{s''}{s}}{c''}}}}}                   & \simeq_{\eqrho} & \\
\BMC{\ctxtapp{\ctxtLCC}{\termrepl{\beta_{n+1}}{\delta}{\termpush{s''}{s}}{c''}}}\neren{\beta_{n+1}}{\gamma'}                                   & = & \\
\BMC{\termrepl{\beta_{n+1}}{\beta_1}{s}{\ctxtapp{\ctxtLCC}{\termrepl{\beta_1}{\delta}{s''}{c''}}}}\neren{\beta_{n+1}}{\gamma'}                 & = & \\
\BMC{\termrepl{\beta_{n+1}}{\beta_1}{s}{\BMC{c'}}}\neren{\beta_{n+1}}{\gamma'}                                                                 & = & \\
\BMC{\termrepl{\beta_{n+1}}{\beta_1}{s}{c'}}\neren{\beta_{n+1}}{\gamma'}                                                                       & = & \\
\BMC{\termrepl{\beta_{n+1}}{\beta_1}{s}{\replace{\gamma'}{\gamma}{s}{c}}}\neren{\beta_{n+1}}{\gamma'}                                          & = & \\
\BMC{\replace{\gamma'}{\gamma}{s}{c\neren{\beta}{\alpha}}}
\end{array} $$
        \item Otherwise,  $$
\begin{array}{l@{\enspace}l@{\enspace}l@{\enspace}l}
\BMC{\replace{\gamma'}{\gamma}{s}{(\termname{\alpha}{\termcont{\beta}{c}})}}                          & = &
\BMC{\termname{\gamma'}{\termconc{(\termcont{\beta}{c'})}{s}}}                                        & = \\
\BMC{\termname{\gamma'}{\termcont{\beta_{n+1}}{c'
  \lvec{\exrepl{\beta_{i+1}}{\beta_i}{u_i}}
}}} & = &
\termname{\gamma'}{\termcont{\beta_{n+1}}{\BMC{c'
  \lvec{\exrepl{\beta_{i+1}}{\beta_i}{u_i}}
}}} & = \\
\termname{\gamma'}{\termcont{\beta_{n+1}}{\BMC{\termrepl{\beta_{n+1}}{\beta_1}{s}{c'}}}}              & \simeq_{\eqrho} &
\BMC{\termrepl{\beta_{n+1}}{\beta_1}{s}{c'}}\neren{\beta_{n+1}}{\gamma'}                              & = \\
\BMC{\termrepl{\beta_{n+1}}{\beta_1}{s}{\replace{\gamma'}{\gamma}{s}{c}}}\neren{\beta_{n+1}}{\gamma'} & = &
\BMC{\replace{\gamma'}{\gamma}{s}{c\neren{\beta}{\alpha}}}
\end{array} $$
      \end{itemize}
    \end{itemize}
  \end{itemize}

  \item If $o = \termcont{\alpha}{\termname{\alpha}{t}} \simeq_{\eqtheta} t =
  o'$, where $\alpha \notin \fn{t}$. Then, $$
\begin{array}{lllll}
\BMC{\replace{\gamma'}{\gamma}{s}{(\termcont{\alpha}{\termname{\alpha}{t}})}}  & = & 
\BMC{\termcont{\alpha}{\termname{\alpha}{\replace{\gamma'}{\gamma}{s}{t}}}}    & = \\
\termcont{\alpha}{\termname{\alpha}{\BMC{\replace{\gamma'}{\gamma}{s}{t}}}}    & \simeq_{\eqtheta} & 
\BMC{\replace{\gamma'}{\gamma}{s}{t}}
\end{array} $$
\end{itemize}
\end{proof}

\begin{lemma}
\label{l:equiv-replacement}
Let $o\in \objects{\LMfull} $. If $o \in \BMCform$ and $s \simeq s'$, then
$\BMC{\replace{\gamma'}{\gamma}{s}{o}} \simeq
\BMC{\replace{\gamma'}{\gamma}{s'}{o}}$. 
\end{lemma}

\begin{proof} %\delia{ESTE LEMMA YA ESTA VERIFICADO PARA $\rW$}
  By induction on $o$. There are two interesting cases: 
\begin{itemize}
  \item If $o = \termname{\beta}{u}$, then
  $\BMC{\replace{\gamma'}{\gamma}{s}{u}} \simeq
  \BMC{\replace{\gamma'}{\gamma}{s'}{u}}$ by the \ih
  There are two cases:
  \begin{itemize}
    \item If $\beta = \gamma$, then $$
\begin{array}{l@{\enspace}l@{\enspace}l@{\enspace}l}
\BMC{\replace{\gamma'}{\gamma}{s}{(\termname{\gamma}{u})}}                     & = &
\BMC{\termname{\gamma'}{\termconc{\replace{\gamma'}{\gamma}{s}{u}}{s}}}        & = \\
\BMC{\termname{\gamma'}{\BMC{\termconc{\replace{\gamma'}{\gamma}{s}{u}}{s}}}}  & = &
\termname{\gamma'}{\BMC{\termconc{\BMC{\replace{\gamma'}{\gamma}{s}{u}}}{s}}}  & \simeq_{(\ih + L.\ref{l:app-bm})} \\
\termname{\gamma'}{\BMC{\termconc{\BMC{\replace{\gamma'}{\gamma}{s'}{u}}}{s}}} & = &
\termname{\gamma'}{\BMC{\termconc{\replace{\gamma'}{\gamma}{s'}{u}}{s}}}       & \simeq_{(L.\ref{l:app-bm})} \\
\termname{\gamma'}{\BMC{\termconc{\replace{\gamma'}{\gamma}{s'}{u}}{s'}}}      & = &
\BMC{\replace{\gamma'}{\gamma}{s'}{(\termname{\gamma}{u})}}
\end{array} $$

    \item If $\beta \neq \gamma$, then $$
\begin{array}{l@{\enspace}l@{\enspace}l@{\enspace}l@{\enspace}l@{\enspace}l}
\BMC{\replace{\gamma'}{\gamma}{s}{(\termname{\beta}{u})}}   & = & 
\BMC{\termname{\beta}{\replace{\gamma'}{\gamma}{s}{u}}}     & = &
\termname{\beta}{\BMC{\replace{\gamma'}{\gamma}{s}{u}}}     & \simeq_{(\ih)} \\
\termname{\beta}{\BMC{\replace{\gamma'}{\gamma}{s'}{u}}}    & = &
\BMC{\replace{\gamma'} {\gamma}{s'}{(\termname{\beta}{u})}}
\end{array} $$
  \end{itemize}

  \item If $o = \termrepl{\beta'}{\beta}{s_0}{c}$, then w.l.o.g. we assume
  $\beta \notin \fn{s_0}$. Let $s'_0 = \replace{\gamma'}{\gamma}{s}{s_0}$,
  $s''_0  = \replace{\gamma'}{\gamma}{s'}{s_0}$, $c' =
  \replace{\gamma'}{\gamma}{s}{c}$, $c''= \replace{\gamma'}{\gamma}{s'}{c}$. By
  the \ih\ we have $\BMC{s'_0} \simeq \BMC{s''_0}$ and $\BMC{c'} \simeq
  \BMC{c''}$. We have three cases:
  \begin{itemize}
    \item If $\gamma \neq \beta'$, then $$
\begin{array}{l@{\enspace}l@{\enspace}l@{\enspace}l}
\BMC{\replace{\gamma'}{\gamma}{s}{\termrepl{\beta'}{\beta}{s_0}{c}}}  & = & 
\BMC{\termrepl{\beta'}{\beta}{s'_0}{c'}}                              & = \\
\BMC{\termrepl{\beta'}{\beta}{s'_0}{\BMC{c'}}}                        & \simeq_{(\ih + L.\ref{l:app-bm})} & 
\BMC{\termrepl{\beta'}{\beta}{s'_0}{\BMC{c''}}}                       & = \\
\BMC{\termrepl{\beta'}{\beta}{\BMC{s'_0}}{c''}}                       & \simeq_{(\ih + L.\ref{l:app-bm})} & 
\BMC{\termrepl{\beta'}{\beta}{\BMC{s''_0}}{c''}}                      & = \\
\BMC{\termrepl{\beta'}{\beta}{s''_0}{c''}}                            & = & 
\BMC{\replace{\gamma'}{\gamma}{s'}{\termrepl{\beta'}{\beta}{s_0}{c}}}
\end{array} $$

\item If $\gamma = \beta'$, then there are two cases:

  \begin{itemize}
    \item If $s_0 \neq \termemst$ or $s_0 = s = \termemst$, then $$
\begin{array}{l@{\enspace}l@{\enspace}l@{\enspace}l}
\BMC{\replace{\gamma'}{\gamma}{s}{\termrepl{\beta'}{\beta}{s_0}{c}}}  & = & 
\BMC{\termrepl{\gamma'}{\beta}{\termpush{s_0}{s}}{c'}}               & = \\
\BMC{\termrepl{\gamma'}{\beta}{\termpush{s_0}{s}}{\BMC{c'}}}         & \simeq_{(\ih + L.\ref{l:app-bm})} & 
\BMC{\termrepl{\gamma'}{\beta}{\termpush{s_0}{s}}{\BMC{c''}}}        & = \\
\BMC{\termrepl{\gamma'}{\beta}{\termpush{\BMC{s_0}}{s}}{c''}}        & \simeq_{(L.\ref{l:app-bm})} & 
\BMC{\termrepl{\gamma'}{\beta}{\termpush{\BMC{s_0}}{s'}}{c''}}       & = \\
\BMC{\termrepl{\gamma'}{\beta}{\termpush{s_0}{s'}}{c''}}             & = & 
\BMC{\replace{\gamma'}{\gamma}{s'}{\termrepl{\beta'}{\beta}{s_0}{c}}}
\end{array} $$

\item  Otherwise,  $$
\begin{array}{l@{\enspace}l@{\enspace}l@{\enspace}l}
\BMC{\replace{\gamma'}{\gamma}{s}{\termrepl{\beta'}{\beta}{\termemst}{c}}}              & = & 
\BMC{\termrepl{\gamma'}{\gamma''}{\termemst}{\termrepl{\gamma''}{\beta}{s}{c'}}}        & = \\
\BMC{\termrepl{\gamma'}{\gamma''}{\termemst}{\termrepl{\gamma''}{\beta}{s}{\BMC{c'}}}}  & \simeq_{(\ih + L.\ref{l:app-bm})} & 
\BMC{\termrepl{\gamma'}{\gamma''}{\termemst}{\termrepl{\gamma''}{\beta}{s'}{\BMC{c''}}}} & = \\
\BMC{\replace{\gamma'}{\gamma}{s'}{\termrepl{\beta'}{\beta}{\termemst}{c}}}
\end{array} $$

\end{itemize}
\end{itemize}
\end{itemize}
\end{proof}

%% Equivalence for Substitution and Replacement
\gettoappendix{l:equiv-for-substitutions-and-replacements}

\begin{proof}
By
Lem.~\ref{l:equiv-substitution},~\ref{l:Sigma_closed_under_substitution:target},~\ref{l:Sigma_closed_under_BM_plus_replacement}
and~\ref{l:equiv-replacement}.
\end{proof}

%%%%%%%%%%%%%%%%%%%%%%%%%%%%%%%%%%%%%%%%%%%%%%%%%%%%%%%%%%%%%%%%%%%%%%%%%%%%%%%
\subsection{Appendix: Bisimulation Proof}
\label{a:bisimulation-proof}
%%%%%%%%%%%%%%%%%%%%%%%%%%%%%%%%%%%%%%%%%%%%%%%%%%%%%%%%%%%%%%%%%%%%%%%%%%%%%%%

%% Bisumulation
\gettoappendix{t:bisimulation}

\begin{proof}
Let $o = \ctxtapp{\ctxt{O}}{l}$ and $p = \ctxtapp{\ctxt{O}}{r}$, for $(l,r)$ an
axiom in $\simeq$ and $\ctxt{O}$ a context, and let $o = \ctxtapp{\ctxt{P}}{g}$
and $p = \ctxtapp{\ctxt{P}}{d}$, for $(g,d)$ either $\rR$ or $\rS$ and
$\ctxt{P}$ a context. We will consider all possible forms for $\ctxt{O}$ and
$\ctxt{P}$. We begin with the case in which $\ctxt{O}$ is just a hole: $\Box$
and $\boxdot$.
\begin{itemize}
  \item $\ctxt{O} = \Box$. We have two cases to consider, $\eqexsubs$ and
  $\eqtheta$, since these are the only equations that relate terms:
  \begin{description}
    \item[Case \eqexsubs] $o = \termsubs{x}{u}{\ctxtapp{\ctxtLTT}{v}}$ and
    $p = \ctxtapp{\ctxtLTT}{\termsubs{x}{u}{v}}$ and $\ctxtLTT$ is $x$-free
    and $\freeFor{u}{\ctxtLTT}$.
    \begin{enumerate}
      \item $\rS$ redex at the root. $$
\begin{tikzcd}
o = \termsubs{x}{u}{\ctxtapp{\ctxtLTT}{v}} \arrow[rightsquigarrow]{d}[anchor=north,left]{\rS}
  &[-30pt] \simeq_{\eqexsubs}
  &[-30pt] \ctxtapp{\ctxtLTT}{\termsubs{x}{u}{v}} = p \arrow[rightsquigarrow]{d}[anchor=north,left]{\rS} \\[-5pt]
o' = \BMC{\substitute{x}{u}{\ctxtapp{\ctxtLTT}{v}}}
  &[-30pt] =_{(L.\ref{l:context-substitution-replacement})}
  &[-30pt] \BMC{\ctxtapp{\ctxtLTT}{\substitute{x}{u}{v}}} = p'
\end{tikzcd} $$

      \item $\rS$ redex overlaps with $\ctxtLTT$. $$
\begin{tikzcd}
o = \termsubs{x}{u}{\ctxtapp{\ctxtLTT_1}{\termsubs{y}{w}{\ctxtapp{\ctxtLTT_2}{v}}}} \arrow[rightsquigarrow]{d}[anchor=north,left]{\rS}
  &[-30pt] \simeq_{\eqexsubs}
  &[-30pt] \ctxtapp{\ctxtLTT_1}{\termsubs{y}{w}{\ctxtapp{\ctxtLTT_2}{\termsubs{x}{u}{ v}}}} = p \arrow[rightsquigarrow]{d}[anchor=north,left]{\rS} \\[-5pt]
o' = \BMC{\termsubs{x}{u}{\ctxtapp{\ctxtLTT_1}{\substitute{y}{w}{\ctxtapp{\ctxtLTT_2}{v}}}}}
  &[-30pt] \simeq
  &[-30pt] \BMC{\ctxtapp{\ctxtLTT_1}{\substitute{y}{w}{\ctxtapp{\ctxtLTT_2}{\termsubs{x}{u}{ v}}}}} = p'
\end{tikzcd} $$

% The bottom line of the diagram is justified as follows:
% \[\begin{array}{lll}
%         & \termsubs{x}{u}{\ctxtapp{\ctxtLTT_1}{\substitute{y}{w}{\ctxtapp{\ctxtLTT_2}{v}}}} \\
%    =     & \termsubs{x}{u}{\ctxtapp{\ctxtLTT_1}{\ctxtapp{\ctxtLTT'_2}{\substitute{y}{w}{ v}}}} \\
%   \cdelia{  =}{\simeq} &
%         \ctxtapp{\ctxtLTT_1}{\ctxtapp{\ctxtLTT'_2}{\termsubs{x}{u}{\substitute{y}{w}{ v}}}}
%     & (\eqexsubs) \\
%    = &
%     \ctxtapp{\ctxtLTT_1}{\substitute{y}{w}{\ctxtapp{\ctxtLTT_2}{\termsubs{x}{u}{
%        v}}}} & 
%   \end{array}\]

%\delia{esto no es correcto del todo: de la linea 3 a la cuatro se pasa en ppio por
%  $\simeq$ y no por igualdad, pero al mismo tiempo solo se puede pasar por $\simeq$ si el termino esta en $\BMCform$, y eso no es seguro porque se aplica una sustitucion} \\

      The bottom line of the diagram is justified as follows: Let
      $$o_0 = \termsubs{x}{u}{\ctxtapp{\ctxtLTT_1}{\substitute{y}{w}{\ctxtapp{\ctxtLTT_2}{v}}}} =
      \termsubs{x}{u}{\ctxtapp{\ctxtLTT_1}{\ctxtapp{\ctxtLTT'_2}{v'}}}$$ and
      $$p_0 = \ctxtapp{\ctxtLTT_1}{\substitute{y}{w}{\ctxtapp{\ctxtLTT_2}{\termsubs{x}{u}{v}}}} =
      \ctxtapp{\ctxtLTT_1}{\ctxtapp{\ctxtLTT'_2}{\termsubs{x}{u}{v'}}}$$
      Moreover, let $\ctxtLTT_3 = \ctxtapp{\ctxtLTT_1}{\ctxtLTT'_2}$. Then
      $o_0 = \termsubs{x}{u}{\ctxtapp{\ctxtLTT_3}{v'}}$ and $p_0 =
      \ctxtapp{\ctxtLTT_3}{ \termsubs{x}{u}{v'}}$. Note that $\ctxtLTT_3$ is
      $x$-free since $x \notin \fv{w}$ by the hypothesis that $\ctxtLTT$ is
      $x$-free. Furthermore, clearly $\freeFor{u}{\ctxtLTT_3}$ since
      $\freeFor{u}{\ctxtLTT}$. Hence we can apply
      Lem.~\ref{l:subs-repl-out-of-BM}.\ref{l:subs-out-of-BM} and deduce
      $\BMC{o_0} \simeq \BMC{p_0}$ which concludes the proof of this case. 

      \item $\rR$ redex overlaps with $\ctxtLTT$. $$
\begin{tikzcd}
o = \termsubs{x}{u}{\ctxtapp{\ctxtLTC_1}{\termrepl{\beta}{\alpha}{s}{\ctxtapp{\ctxtLCT_2}{v}}}} \arrow[rightsquigarrow]{d}[anchor=north,left]{\rR}
  &[-30pt] \simeq_{\eqexsubs}
  &[-30pt] \ctxtapp{\ctxtLTC_1}{\termrepl{\beta}{\alpha}{s}{\ctxtapp{\ctxtLCT_2}{\termsubs{x}{u}{ v}}}} = p \arrow[rightsquigarrow]{d}[anchor=north,left]{\rR} \\[-5pt]
o' = \BMC{\termsubs{x}{u}{\ctxtapp{\ctxtLTC_1}{\replace{\beta}{\alpha}{s}{\ctxtapp{\ctxtLCT_2}{v}}}}}
  &[-30pt] \simeq
  &[-30pt] \BMC{\ctxtapp{\ctxtLTC_1}{\replace{\beta}{\alpha}{s}{\ctxtapp{\ctxtLCT_2}{\termsubs{x}{u}{ v}}}}} = p'
\end{tikzcd} $$

% The bottom line of the diagram is justified as follows:
% \[\begin{array}{lll}
%         & \termsubs{x}{u}{\ctxtapp{\ctxtLTC_1}{\replace{\beta}{\alpha}{s}{\ctxtapp{\ctxtLCT_2}{v}}}} \\
%      =   & \termsubs{x}{u}{\ctxtapp{\ctxtLTC_1}{\ctxtapp{\ctxtLCT'_2}{\replace{\beta}{\alpha}{s}{ v}}}} \\
%     = &
%         \ctxtapp{\ctxtLTC_1}{\ctxtapp{\ctxtLCT'_2}{\termsubs{x}{u}{\replace{\beta}{\alpha}{s}{ v}}}}
%     & (\eqexsubs) \\
%    = &
%     \ctxtapp{\ctxtLTC_1}{\replace{\beta}{\alpha}{s}{\ctxtapp{\ctxtLCT_2}{\termsubs{x}{u}{
%        v}}}} & \\
% \end{array}\]
      The bottom line of the diagram is justified as follows:
      $$o_0 = \termsubs{x}{u}{\ctxtapp{\ctxtLTC_1}{\replace{\beta}{\alpha}{s}{\ctxtapp{\ctxtLCT_2}{v}}}} =
      \termsubs{x}{u}{\ctxtapp{\ctxtLTC_1}{\ctxtapp{\ctxtLCT'_2}{v'}}}$$ and
      $$p_0 = \ctxtapp{\ctxtLTC_1}{\replace{\beta}{\alpha}{s}{\ctxtapp{\ctxtLCT_2}{\termsubs{x}{u}{v}}}} =
      \ctxtapp{\ctxtLTC_1}{{\ctxtapp{\ctxtLCT'_2}{\termsubs{x}{u'}{v'}}}}$$
      Let $\ctxtLTT_3 = \ctxtapp{\ctxtLTC_1}{\ctxtLCT'_2}$. Then $o_0 =
      \ctxtapp{\ctxtLTT_3}{v'}\exsubs{x}{u}$ and $o_0 =
      \ctxtapp{\ctxtLTT_3}{v'\exsubs{x}{u}}$. Note that $\ctxtLTT_3$ is
      $\alpha$-free since $\alpha \notin \fn{u}$ by the hypothesis that
      $\ctxtLTT$ is $\alpha$-free. Furthermore, clearly
      $\freeFor{s}{\ctxtLTT_3}$ since $\freeFor{s}{\ctxtLTT}$. Hence we can
      apply Lem.~\ref{l:subs-repl-out-of-BM}.\ref{l:subs-out-of-BM} and
      deduce $\BMC{o_0} \simeq \BMC{p_0}$ which concludes the proof of this
      case.
    \end{enumerate}

    \item [Case $\eqtheta$.] $o = \termcont{\alpha}{\termname{\alpha}{t}}$ and
    $p = t$ with $\alpha \notin \fn{t}$. Trivial since all reductions are
    internal to $t$.
  \end{description}

  \item $\ctxt{O}=\boxdot$. We consider cases for $\eqexrepl, \eqexren,
  \eqlinear, \eqpoppop$ and $\eqtheta$, all of which relate commands.  
    \begin{description}
    \item[Case \eqexrepl] $o =
    \termrepl{\alpha'}{\alpha}{s}{\ctxtapp{\ctxtLCC}{c}}$ and
    $p = \ctxtapp{\ctxtLCC}{\termrepl{\alpha'}{\alpha}{s}{c}}$ and
    $\ctxtLCC$ is $\alpha$-free and $\freeFor{s}{\ctxtLCC}$.
    \begin{enumerate}
      \item $\rR$ redex at the root (so that $s \neq \emst$) $$
\begin{tikzcd}
o = \termrepl{\alpha'}{\alpha}{s}{\ctxtapp{\ctxtLCC}{c}} \arrow[rightsquigarrow]{d}[anchor=north,left]{\rR}
  &[-30pt] \simeq_{\eqexrepl}
  &[-30pt]\ctxtapp{\ctxtLCC}{\termrepl{\alpha'}{\alpha}{s}{c}} = p \arrow[rightsquigarrow]{d}[anchor=north,left]{\rR} \\[-5pt]
o' = \BMC{\replace{\alpha'}{\alpha}{s}{\ctxtapp{\ctxtLCC}{c}}}
  &[-30pt] =_{(L.\ref{l:context-substitution-replacement})}
  &[-30pt] \BMC{\ctxtapp{\ctxtLCC}{\replace{\alpha'}{\alpha}{s}{c}}} = p'
\end{tikzcd}$$
     
      \item $\rS$ redex overlaps with $\ctxtLCC$. $$
\begin{tikzcd}
o = \termrepl{\alpha'}{\alpha}{s}{\ctxtapp{\ctxtLCT_1}{\termsubs{y}{v}{\ctxtapp{\ctxtLTC_2}{c}}}} \arrow[rightsquigarrow]{d}[anchor=north,left]{\rS}
  &[-30pt] \simeq_{\eqexrepl}
  &[-30pt] \ctxtapp{\ctxtLCT_1}{\termsubs{y}{v}{\ctxtapp{\ctxtLTC_2}{\termrepl{\alpha'}{\alpha}{s}{ c}}}} = p \arrow[rightsquigarrow]{d}[anchor=north,left]{\rS} \\[-5pt]
o' = \BMC{\termrepl{\alpha'}{\alpha}{s}{\ctxtapp{\ctxtLCT_1}{\substitute{y}{v}{\ctxtapp{\ctxtLTC_2}{c}}}}}
  &[-30pt] \simeq
  &[-30pt] \BMC{\ctxtapp{\ctxtLCT_1}{\substitute{y}{v}{\ctxtapp{\ctxtLTC_2}{\termrepl{\alpha'}{\alpha}{s}{ c}}}}} = p'
\end{tikzcd}$$

% The bottom line of the diagram is justified as follows:
% \[\begin{array}{lll}
%         & \termrepl{\alpha'}{\alpha}{s}{\ctxtapp{\ctxtLCT_1}{\substitute{y}{v}{\ctxtapp{\ctxtLTC_2}{c}}}} \\
%       = & \termrepl{\alpha'}{\alpha}{s}{\ctxtapp{\ctxtLCT_1}{\ctxtapp{\ctxtLTC'_2}{\substitute{y}{v}{ c}}}} \\
%     = &
%        \ctxtapp{\ctxtLCT_1}{\ctxtapp{\ctxtLTC'_2}{ \termrepl{\alpha'}{\alpha}{s}{\substitute{y}{v}{ c}}}}
%     & (\eqexrepl) \\
%    = &
%     \ctxtapp{\ctxtLCT_1}{\substitute{y}{v}{\ctxtapp{\ctxtLTC_2}{\termrepl{\alpha'}{\alpha}{s}{
%        c}}}}& 
% \end{array}\]

      The bottom line of the diagram is justified as follows: $$o_0 =
      \termrepl{\alpha'}{\alpha}{s}{\ctxtapp{\ctxtLCT_1}{\substitute{y}{v}{\ctxtapp{\ctxtLTC_2}{c}}}} =
      \termrepl{\alpha'}{\alpha}{s}{\ctxtapp{\ctxtLCT_1}{\ctxtapp{\ctxtLTC'_2}{c'}}}$$
      and $$p_0 =
      \ctxtapp{\ctxtLCT_1}{\substitute{y}{v}{\ctxtapp{\ctxtLTC_2}{\termrepl{\alpha'}{\alpha}{s}{c}}}} =
      \ctxtapp{\ctxtLCT_1}{\ctxtapp{\ctxtLTC'_2}{\termrepl{\alpha'}{\alpha}{s'}{c'}}}$$
      Let $\ctxtLCC_3 = \ctxtapp{\ctxtLCT_1}{\ctxtLTC'_2}$. Then $o_0 =
      \ctxtapp{\ctxtLCC_3}{c'}\exrepl{\alpha'}{\alpha}{s}$ and $p_0 =
      \ctxtapp{\ctxtLCC_3}{c'\exrepl{\alpha'}{\alpha}{s}}$. Note that
      $\ctxtLCC_3$ is $y$-free from the hypothesis that $\ctxtLCC$ is
      $y$-free. Furthermore, clearly $\freeFor{v}{\ctxtLCC_3}$ since
      $\freeFor{v}{\ctxtLCC}$. Hence we can apply
      Lem.~\ref{l:subs-repl-out-of-BM}.\ref{l:repl-out-of-BM} and deduce
      $\BMC{o_0} \simeq \BMC{p_0}$ which concludes the proof of this case.  

      \item $\rR$ redex overlaps with $\ctxtLCC$. $$
\begin{tikzcd}
o = \termrepl{\alpha'}{\alpha}{s}{\ctxtapp{\ctxtLCC_1}{\termrepl{\beta'}{\beta}{s_0}{\ctxtapp{\ctxtLCC_2}{c}}}} \arrow[rightsquigarrow]{d}[anchor=north,left]{\rR}
  &[-30pt] \simeq_{\eqexrepl}
  &[-30pt] \ctxtapp{\ctxtLCC_1}{\termrepl{\beta'}{\beta}{s_0}{\ctxtapp{\ctxtLCC_2}{\termrepl{\alpha'}{\alpha}{s}{ c}}}} = p \arrow[rightsquigarrow]{d}[anchor=north,left]{\rR} \\[-5pt]
o' = \BMC{\termrepl{\alpha'}{\alpha}{s}{\ctxtapp{\ctxtLCC_1}{\replace{\beta'}{\beta}{s_0}{\ctxtapp{\ctxtLCC_2}{c}}}}}
  &[-30pt] \simeq
  &[-30pt] \BMC{\ctxtapp{\ctxtLCC_1}{\replace{\beta'}{\beta}{s_0}{\ctxtapp{\ctxtLCC_2}{\termrepl{\alpha'}{\alpha}{s}{ c}}}}} = p'
\end{tikzcd}$$

% The bottom line of the diagram is justified as follows:
% \[\begin{array}{lll}
%         & \termrepl{\alpha'}{\alpha}{s}{\ctxtapp{\ctxtLCC_1}{\replace{\alpha''}{\alpha'}{s_0}{\ctxtapp{\ctxtLCC_2}{c}}}} \\
%      = & \termrepl{\alpha'}{\alpha}{s}{\ctxtapp{\ctxtLCC_1}{\ctxtapp{\ctxtLCC'_2}{\replace{\alpha''}{\alpha'}{s_0}{ c}}}} \\
%     = &
%         \ctxtapp{\ctxtLCC_1}{\ctxtapp{\ctxtLCC'_2}{\termrepl{\alpha'}{\alpha}{s}{\replace{\alpha''}{\alpha'}{s_0}{ c}}}}
%     & (\eqexrepl) \\
%    = &
%     \ctxtapp{\ctxtLCC_1}{\replace{\alpha''}{\alpha'}{s_0}{\ctxtapp{\ctxtLCC_2}{\termrepl{\alpha'}{\alpha}{s}{
%        c}}}} & 
% \end{array}\]

      The bottom line of the diagram is justified as follows: $$o_0 =
      \termrepl{\alpha'}{\alpha}{s}{\ctxtapp{\ctxtLCC_1}{\replace{\beta'}{\beta}{s_0}{\ctxtapp{\ctxtLCC_2}{c}}}} =
      \termrepl{\alpha'}{\alpha}{s}{\ctxtapp{\ctxtLCC_1}{\ctxtapp{\ctxtLCC'_2}{c'}}}$$
      and $$p_0 =
      \ctxtapp{\ctxtLCC_1}{\replace{\beta'}{\beta}{s}{\ctxtapp{\ctxtLCC_2}{\termrepl{\alpha'}{\alpha}{s}{c}}}} =
      \ctxtapp{\ctxtLCC_1}{\ctxtapp{\ctxtLCC'_2}{\termrepl{\alpha'}{\alpha}{s_0}{c'}}}$$
      Let $\ctxtLCC_3 = \ctxtapp{\ctxtLCC_1}{\ctxtLCC'_2}$. Then $o_0 =
      \ctxtapp{\ctxtLCC_3}{c'}\exrepl{\alpha'}{\alpha}{s}$ and $o_0 =
      \ctxtapp{\ctxtLCC_3}{c'\exrepl{\alpha'}{\alpha}{s}}$. Note that
      $\ctxtLCC_3$ is $\beta$-free since $\beta\notin\fn{u}$ by the hypothesis
      that $\ctxtLCC$ is$\beta$-free. Furthermore, clearly
      $\freeFor{s}{\ctxtLCC_3}$ since $\freeFor{s}{\ctxtLCC}$. Hence we can
      apply Lem.~\ref{l:subs-repl-out-of-BM}.\ref{l:repl-out-of-BM} and deduce
      $\BMC{o_0} \simeq \BMC{p_0}$ which concludes the proof of this case.
    \end{enumerate}

    \item[Case \eqlinear] $o =
    \termrepl{\alpha'}{\alpha}{s}{\ctxtapp{\ctxtLCC}{\termname{\alpha}{u}}}$
    and $p = \ctxtapp{\ctxtLCC}{\termname{\alpha'}{\BMC{u \tconc s}}}$ and
    $\alpha\notin\fn{u}$ and $\ctxtLCC$ is $\alpha$-free. There are only two
    further cases to consider (note that $\rR$ cannot occur at the root since
    $\rR$ only contracts non-linear $\rRfull$-redexes):
    \begin{enumerate}
      \item $\rS$ redex overlaps with $\ctxtLTT$. $$
\begin{tikzcd}
o = \termrepl{\alpha'}{\alpha}{s}{\ctxtapp{\ctxtLCT_1}{\termsubs{y}{w}{\ctxtapp{\ctxtLTC_2}{\termname{\alpha}{u}}}}} \arrow[rightsquigarrow]{d}[anchor=north,left]{\rS}
  &[-35pt] \simeq_{\eqlinear}
  &[-35pt] \ctxtapp{\ctxtLCT_1}{\termsubs{y}{w}{\ctxtapp{\ctxtLTC_2}{ \termname{\alpha'}{\BMC{u \tconc s}}}}} = p \arrow[rightsquigarrow]{d}[anchor=north,left]{\rS} \\[-5pt]
  o'=
\BMC{\termrepl{\alpha'}{\alpha}{s}{\ctxtapp{\ctxtLCT_1}{\substitute{y}{w}{\ctxtapp{\ctxtLTC_2}{\termname{\alpha}{u}}}}}}
  &[-35pt] \simeq
  &[-35pt] \BMC{\ctxtapp{\ctxtLCT_1}{\substitute{y}{w}{\ctxtapp{\ctxtLTC_2}{ \termname{\alpha'}{\BMC{u \tconc s}}}}}} = p'
\end{tikzcd} $$

% The bottom line of the diagram is justified as follows:
% \[\begin{array}{lll}
%         & \termrepl{\alpha'}{\alpha}{s}{\ctxtapp{\ctxtLCC_1}{\substitute{y}{w}{\ctxtapp{\ctxtLTC_2}{\termname{\alpha}{u}}}}} \\
%     = & \termrepl{\alpha'}{\alpha}{s}{\ctxtapp{\ctxtLCC_1}{\ctxtapp{\ctxtLTC'_2}{\termname{\alpha}{\substitute{y}{w}{ u}}}}}
%    & \\
%    = &
%     \ctxtapp{\ctxtLCC_1}{\ctxtapp{\ctxtLTC'_2}{
%        \termname{\alpha'}{\BMC{\substitute{y}{w}{u} \tconc s}}}} & (\eqlinear) \\
%    = &
%     \ctxtapp{\ctxtLCC_1}{\substitute{y}{w}{\ctxtapp{\ctxtLTC_2}{
%        \termname{\alpha'}{\BMC{u \tconc s}}}}} & 
% \end{array}\]
     
      Let $\ctxtLCC_3 = \ctxtapp{\ctxtLCT_1}{\ctxtLTC'_2}$ and
      $u' = \substitute{y}{w}{u}$. Then we have $$
\begin{array}{r@{\enspace}c@{\enspace}l}
\BMC{o'}
  & =                  & \ctxtapp{\BMC{\ctxtLCC_3}}{\termname{\alpha}{\BMC{u'}}}\exrepl{\alpha'}{\alpha}{\BMC{s}} \\
  & \simeq_{\eqlinear} & \ctxtapp{\BMC{\ctxtLCC_3}}{\termname{\alpha'}{\BMC{\BMC{u'} \tconc \BMC{s}}}} \\
  & =                  & \ctxtapp{\BMC{\ctxtLCC_3}}{\termname{\alpha'}{\BMC{u' \tconc s}}}
\end{array} $$
    We conclude since $$
\begin{array}{r@{\enspace}c@{\enspace}l}
\BMC{p'}
  & = & \ctxtapp{\BMC{\ctxtLCC_3}}{\termname{\alpha'}{\BMC{\substitute{y}{w}{\BMC{u \tconc s}}}}} \\
  & = & \ctxtapp{\BMC{\ctxtLCC_3}}{\termname{\alpha'}{\BMC{\substitute{y}{w}{(u \tconc s)}}}} \\
  & = & \ctxtapp{\BMC{\ctxtLCC_3}}{\termname{\alpha'}{\BMC{u'\tconc s}}}
\end{array} $$

      \item $\rR$ redex overlaps with $\ctxtLCC$. $$
\begin{tikzcd}
o = \termrepl{\alpha'}{\alpha}{s}{\ctxtapp{\ctxtLCC_1}{\termrepl{\beta'}{\beta}{s'}{\ctxtapp{\ctxtLCC_2}{\termname{\alpha}{u}}}}} \arrow[rightsquigarrow]{d}[anchor=north,left]{\rR}
  &[-35pt] \simeq_{\eqlinear}
  &[-35pt] \ctxtapp{\ctxtLCC_1}{\termrepl{\beta'}{\beta}{s'}{\ctxtapp{\ctxtLCC_2}{ \termname{\alpha'}{\BMC{u \tconc s}}}}} = p \arrow[rightsquigarrow]{d}[anchor=north,left]{\rR} \\[-5pt]
  o'=
\BMC{\termrepl{\alpha'}{\alpha}{s}{\ctxtapp{\ctxtLCC_1}{\replace{\beta'}{\beta}{s'}{\ctxtapp{\ctxtLCC_2}{\termname{\alpha}{u}}}}}}
  &[-35pt] \simeq
  &[-35pt] \BMC{\ctxtapp{\ctxtLCC_1}{\replace{\beta'}{\beta}{s'}{\ctxtapp{\ctxtLCC_2}{ \termname{\alpha'}{\BMC{u \tconc s}}}}}} = p'
\end{tikzcd} $$

% The bottom line of the diagram is justified as follows:
% \[\begin{array}{lll}
%         & \BMC{\termrepl{\alpha'}{\alpha}{s}{\ctxtapp{\ctxtLCC_1}{\replace{\beta'}{\beta}{s'}{\ctxtapp{\ctxtLTC_2}{\termname{\alpha}{u}}}}}} \\
%     = &
%         \BMC{\termrepl{\alpha'}{\alpha}{s}{\ctxtapp{\ctxtLCC_1}{\ctxtapp{\ctxtLTC'_2}{\termname{\alpha}{\replace{\beta'}{\beta}{s'}{
%         u}}}}}} & \alpha\neq \beta\\
%    = &
% \BMC{\ctxtapp{\ctxtLCC_1}{\ctxtapp{\ctxtLTC'_2}{
%        \termname{\alpha'}{\BMC{\replace{\beta'}{\beta}{s'}{ u} \tconc
%        s}}}}} & (\eqlinear) \\
%    = &
% \BMC{\ctxtapp{\ctxtLCC_1}{\replace{\beta'}{\beta}{s'}{\ctxtapp{\ctxtLTC_2}{ \termname{\alpha'}{\BMC{u \tconc s}}}}}} & 
% \end{array}\]

      Let $\ctxtLCC_3 = \ctxtapp{\ctxtLCC_1}{\ctxtLCC'_2}$ and
      $u' = \replace{\beta'}{\beta}{s'}{u}$. Then we have $$
\begin{array}{r@{\enspace}c@{\enspace}l}
\BMC{o'}
  & =                  & \ctxtapp{\BMC{\ctxtLCC_3}}{\termname{\alpha}{\BMC{u'}}}\exrepl{\alpha'}{\alpha}{\BMC{s}} \\
  & \simeq_{\eqlinear} & \ctxtapp{\BMC{\ctxtLCC_3}}{\termname{\alpha'}{\BMC{\BMC{u'} \tconc \BMC{s}}}} \\
  & =                  & \ctxtapp{\BMC{\ctxtLCC_3}}{\termname{\alpha'}{\BMC{u' \tconc s}}}
\end{array} $$
      We conclude since $$
\begin{array}{r@{\enspace}c@{\enspace}l}
\BMC{p'}
  & = & \ctxtapp{\BMC{\ctxtLCC_3}}{\termname{\alpha'}{\BMC{\replace{\beta'}{\beta}{s'}{\BMC{u\tconc s}}}}} \\
  & = & \ctxtapp{\BMC{\ctxtLCC_3}}{\termname{\alpha'}{\BMC{\replace{\beta'}{\beta}{s'}{(u\tconc s)}}}} \\
  & = & \ctxtapp{\BMC{\ctxtLCC_3}}{\termname{\alpha'}{\BMC{u'\tconc s}}}
\end{array} $$
    \end{enumerate}
  
    \item[Case \eqpoppop] $o =
    \termname{\alpha'}{\termabs{x}{\termcont{\alpha}{\termname{\beta'}{\termabs{y}{\termcont{\beta}{u}}}}}}$
    and $p = 
    \termname{\beta'}{\termabs{y}{\termcont{\beta}{\termname{\alpha'}{\termabs{x}{\termcont{\alpha}{u}}}}}}$.
    Trivial since all reductions are internal to $u$.

    \item[Case $\eqrho$] $o = \termname{\alpha}{\termcont{\beta}{c}}$ and
    $p = c \neren{\beta}{\alpha}$. Trivial since all reductions are internal to $c$.
  \end{description}
\end{itemize}

\bigskip
We now address the inductive cases.
\begin{itemize}
  \item $\ctxt{O}=\ctxt{C}$.

  \begin{itemize}
  \item $\ctxt{C} = \termname{\gamma}{\ctxt{T}}$. Then
    $o = \termname{\gamma}{\ctxtapp{\ctxt{T}}{l}}$, $p=\termname{\gamma}{\ctxtapp{\ctxt{T}}{r}}$ and the reduction
    step must take place inside $\ctxtapp{\ctxt{T}}{l}$. In this
      case, (a) below must hold by the
    \ih, allowing us to conclude with (b): $$
  \begin{array}{c@{\qquad}c}
\begin{tikzcd}
\ctxtapp{\ctxt{T}}{l}\arrow[rightsquigarrow]{d}[anchor=north,left]{}
  &[-25pt] \simeq
  &[-25pt] \ctxtapp{\ctxt{T}}{r} \arrow[rightsquigarrow]{d}[anchor=north,left]{} \\[-5pt]
\BMC{t}
  &[-25pt] \simeq
  &[-25pt] \BMC{t'}
\end{tikzcd}
    &
\begin{tikzcd}
o = \termname{\gamma}{\ctxtapp{\ctxt{T}}{l}}\arrow[rightsquigarrow]{d}[anchor=north,left]{}
  &[-25pt] \simeq
  &[-25pt] \termname{\gamma}{\ctxtapp{\ctxt{T}}{r}}= p \arrow[rightsquigarrow]{d}[anchor=north,left]{} \\[-5pt]
o' = \BMC{\termname{\gamma}{t}}
  &[-25pt] \simeq
  &[-25pt] \BMC{\termname{\gamma}{t'}} = p'
\end{tikzcd}
    \\
    \mbox{(a)} & \mbox{(b)}
\end{array} $$
The bottom of (b) follows from Lem.~\ref{l:app-bm} with
$\ctxt{O} = \termname{\gamma}{\Box}$ and $\BMC{t} \simeq \BMC{t'}$.

  \item $\ctxt{C} =
    \termrepl{\alpha'}{\alpha}{s}{\ctxt{C_1}}$. 
    Then
    $o = \termrepl{\alpha'}{\alpha}{s}{\ctxtapp{\ctxt{C_1}}{l}}$,
    $p=\termrepl{\alpha'}{\alpha}{s}{\ctxtapp{\ctxt{C_1}}{r}}$. We
    have three cases:

    \begin{itemize}
    \item The reduction step is in $\ctxtapp{\ctxt{C_1}}{l}$. In this
      case, (a) below must hold by the
    \ih, allowing us to conclude with (b): $$
\begin{array}{c@{\qquad}c}
\begin{tikzcd}
  \ctxtapp{\ctxt{C_1}}{l}\arrow[rightsquigarrow]{d}[anchor=north,left]{}
  &[-25pt] \simeq
  &[-25pt] \ctxtapp{\ctxt{C_1}}{r} \arrow[rightsquigarrow]{d}[anchor=north,left]{} \\[-5pt]
  \BMC{c} &[-25pt] \simeq &[-25pt] \BMC{c'}
\end{tikzcd}
&
\begin{tikzcd}
  \termrepl{\alpha'}{\alpha}{s}{\ctxtapp{\ctxt{C_1}}{l}}\arrow[rightsquigarrow]{d}[anchor=north,left]{}
  &[-25pt] \simeq
  &[-25pt] \termrepl{\alpha'}{\alpha}{s}{\ctxtapp{\ctxt{C_1}}{r}} \arrow[rightsquigarrow]{d}[anchor=north,left]{} \\[-5pt]
  \BMC{\termrepl{\alpha'}{\alpha}{s}{c}} &[-25pt] \simeq &[-25pt] \BMC{\termrepl{\alpha'}{\alpha}{s}{c'}}
\end{tikzcd} \\
\mbox{(a)} & \mbox{(b)}
\end{array} $$
The bottom of (b) follows from Lem.~\ref{l:app-bm}
with $\ctxt{O} = \termrepl{\alpha'}{\alpha}{s}{\Box}$
and $\BMC{c} \simeq \BMC{c'}$.

\item The reduction step is at the root (so that $s \neq \emst$). $$
\begin{tikzcd}
  \termrepl{\alpha'}{\alpha}{s}{\ctxtapp{\ctxt{C_1}}{l}}\arrow[rightsquigarrow]{d}[anchor=north,left]{\rR}
  &[-25pt] \simeq
  &[-25pt] \termrepl{\alpha'}{\alpha}{s}{\ctxtapp{\ctxt{C_1}}{r}} \arrow[rightsquigarrow]{d}[anchor=north,left]{\rR} \\[-5pt]
  \BMC{\replace{\alpha'}{\alpha}{s}{\ctxtapp{\ctxt{C_1}}{l}}} &[-25pt] \simeq &[-25pt] \BMC{\replace{\alpha'}{\alpha}{s}{\ctxtapp{\ctxt{C_1}}{r}}}
\end{tikzcd}$$
The bottom side of the diagram follows from
Lem.~\ref{l:Sigma_closed_under_BM_plus_replacement} with
$s\in\BMCform$ and $\ctxtapp{\ctxt{C_1}}{l}\simeq \ctxtapp{\ctxt{C_1}}{r}$.

\item The reduction step is in $s$. $$
\begin{tikzcd}
  \termrepl{\alpha'}{\alpha}{s}{\ctxtapp{\ctxt{C_1}}{l}}\arrow[rightsquigarrow]{d}[anchor=north,left]{}
  &[-25pt] \simeq
  &[-25pt] \termrepl{\alpha'}{\alpha}{s}{\ctxtapp{\ctxt{C_1}}{r}} \arrow[rightsquigarrow]{d}[anchor=north,left]{} \\[-5pt]
  \BMC{\termrepl{\alpha'}{\alpha}{s'}{\ctxtapp{\ctxt{C_1}}{l}}} &[-25pt] \simeq &[-25pt] \BMC{\termrepl{\alpha'}{\alpha}{s'}{\ctxtapp{\ctxt{C_1}}{r}}}
\end{tikzcd} $$
The bottom of the diagram follows from Lem.~\ref{l:app-bm}
with $\ctxt{O} = \termrepl{\alpha'}{\alpha}{s'}{\Box}$
and $\ctxtapp{\ctxt{C_1}}{l}\simeq \ctxtapp{\ctxt{C_1}}{r}$.
\end{itemize}

  \item $\ctxt{C} =
    \termrepl{\alpha'}{\alpha}{\ctxt{S}}{c}$.  Then
    $o = \termrepl{\alpha'}{\alpha}{\ctxtapp{\ctxt{S}}{l}}{c}$,
    $p=\termrepl{\alpha'}{\alpha}{\ctxtapp{\ctxt{S}}{r}}{c}$. We
    have three cases:
    \begin{itemize}
\item The reduction step is in $c$. $$
\begin{tikzcd}
  \termrepl{\alpha'}{\alpha}{\ctxtapp{\ctxt{S}}{l}}{c}\arrow[rightsquigarrow]{d}[anchor=north,left]{}
  &[-25pt] \simeq
  &[-25pt] \termrepl{\alpha'}{\alpha}{\ctxtapp{\ctxt{S}}{r}}{c} \arrow[rightsquigarrow]{d}[anchor=north,left]{} \\[-5pt]
  \BMC{\termrepl{\alpha'}{\alpha}{\ctxtapp{\ctxt{S}}{l}}{c'}} &[-25pt] \simeq &[-25pt] \BMC{\termrepl{\alpha'}{\alpha}{\ctxtapp{\ctxt{S}}{r}}{c'}}
\end{tikzcd}$$
  
    \item The reduction step is in $\ctxtapp{\ctxt{S}}{l}$. In this
    case, (a) below must hold by the \ih, allowing us to conclude with (b): $$
\begin{array}{c@{\qquad}c}
\begin{tikzcd}
  \ctxtapp{\ctxt{S}}{l}\arrow[rightsquigarrow]{d}[anchor=north,left]{}
  &[-25pt] \simeq
  &[-25pt] \ctxtapp{\ctxt{S}}{r} \arrow[rightsquigarrow]{d}[anchor=north,left]{} \\[-5pt]
  \BMC{s} &[-25pt] \simeq &[-25pt] \BMC{s'}
\end{tikzcd}
&
\begin{tikzcd}
  \termrepl{\alpha'}{\alpha}{\ctxtapp{\ctxt{S}}{l}}{c}\arrow[rightsquigarrow]{d}[anchor=north,left]{}
  &[-25pt] \simeq
  &[-25pt] \termrepl{\alpha'}{\alpha}{\ctxtapp{\ctxt{S}}{r}}{c} \arrow[rightsquigarrow]{d}[anchor=north,left]{} \\[-5pt]
  \BMC{\termrepl{\alpha'}{\alpha}{s}{c}} &[-25pt] \simeq &[-25pt] \BMC{\termrepl{\alpha'}{\alpha}{s'}{c}}
\end{tikzcd} \\
\mbox{(a)} & \mbox{(b)}
\end{array} $$
The bottom of (b) follows from Lem.~\ref{l:app-bm}
with $\ctxt{O} = \termrepl{\alpha'}{\alpha}{\Box}{c}$
and $\BMC{s}\simeq \BMC{s'}$.

\item The reduction step is at the root (so that $\ctxtapp{\ctxt{S}}{l} \neq \emst$). $$
\begin{tikzcd}
  \termrepl{\alpha'}{\alpha}{\ctxtapp{\ctxt{S}}{l}}{c}\arrow[rightsquigarrow]{d}[anchor=north,left]{\rR}
  &[-25pt] \simeq
  &[-25pt] \termrepl{\alpha'}{\alpha}{\ctxtapp{\ctxt{S}}{r}}{c} \arrow[rightsquigarrow]{d}[anchor=north,left]{\rR} \\[-5pt]
  \BMC{\replace{\alpha'}{\alpha}{\ctxtapp{\ctxt{S}}{l}}{c}} &[-25pt] \simeq &[-25pt] \BMC{\replace{\alpha'}{\alpha}{\ctxtapp{\ctxt{S}}{r}}{c}}
\end{tikzcd} $$
The bottom of the diagram follows from Lem.~\ref{l:equiv-replacement}
with $c \in \BMCform$ and $\ctxtapp{\ctxt{S}}{l} \simeq \ctxtapp{\ctxt{S}}{r}$.
\end{itemize}
\end{itemize}

  \item $\ctxt{O}=\ctxt{S}$.
  \begin{itemize}
    \item $\ctxt{S} = \ctxt{T}$.  We use the \ih w.r.t. $\ctxt{T}$.

    \item $\ctxt{S} = \ctxt{T}\cdot s$. Depending on whether the
      reduction is in $\ctxt{T}$ or in $s$ the result either holds
      from the \ih\ (in the former case) or directly (in the latter).

    \item $\ctxt{S} = t \cdot \ctxt{S}$. Depending on whether the
      reduction is in $t$ or in $\ctxt{S}$ the result either holds
      from the \ih\ (in the latter case) or directly (in the former).
  \end{itemize}
  
  \item $\ctxt{O}=\ctxt{T}$.
  \begin{itemize}
    \item $\ctxt{T} = \termapp{\ctxt{T_1}}{u}$. 
    Then
    $o = \termapp{\ctxtapp{\ctxt{T_1}}{l}}{u}$,
    $p=\termapp{\ctxtapp{\ctxt{T_1}}{r}}{u}$. We
    have two cases:

    \begin{itemize}
      \item The reduction step is in $\ctxtapp{\ctxt{T_1}}{l}$. In this
      case, (a) below must hold by the \ih, allowing us to conclude with (b): $$
\begin{array}{c@{\qquad}c}
\begin{tikzcd}
  \ctxtapp{\ctxt{T_1}}{l}\arrow[rightsquigarrow]{d}[anchor=north,left]{}
  &[-25pt] \simeq
  &[-25pt] \ctxtapp{\ctxt{T_1}}{r} \arrow[rightsquigarrow]{d}[anchor=north,left]{} \\[-5pt]
  \BMC{t} &[-25pt] \simeq &[-25pt] \BMC{t'}
\end{tikzcd}
&
\begin{tikzcd}
  \termapp{\ctxtapp{\ctxt{T_1}}{l}}{u}\arrow[rightsquigarrow]{d}[anchor=north,left]{}
  &[-25pt] \simeq
  &[-25pt] \termapp{\ctxtapp{\ctxt{T_1}}{r}}{u}\arrow[rightsquigarrow]{d}[anchor=north,left]{} \\[-5pt]
  \BMC{\termapp{t}{u}} &[-25pt] \simeq &[-25pt] \BMC{\termapp{t'}{u}}
\end{tikzcd} \\
\mbox{(a)} & \mbox{(b)}
\end{array} $$
The bottom of (b) follows from Lem.~\ref{l:app-bm}
with $\ctxt{O} = \Box u$ and $\BMC{t}\simeq \BMC{t'}$.

      \item The reduction step is in $u$. $$
\begin{tikzcd}
  \termapp{\ctxtapp{\ctxt{T_1}}{l}}{u}\arrow[rightsquigarrow]{d}[anchor=north,left]{}
  &[-25pt] \simeq
  &[-25pt] \termapp{\ctxtapp{\ctxt{T_1}}{r}}{u}\arrow[rightsquigarrow]{d}[anchor=north,left]{} \\[-5pt]
  \BMC{\termapp{\ctxtapp{\ctxt{T_1}}{l}}{u'}} &[-25pt] \simeq &[-25pt] \BMC{\termapp{\ctxtapp{\ctxt{T_1}}{r}}{u'}}
\end{tikzcd} $$
The bottom of the diagram follows from Lem.~\ref{l:app-bm}
with $\ctxt{O} = \Box u'$ and $\ctxtapp{\ctxt{T_1}}{l}\simeq \ctxtapp{\ctxt{T_1}}{r}$.
\end{itemize}

  \item $\ctxt{T} = \termapp{u}{\ctxt{T_1}}$.     Then
    $o = \termapp{u}{\ctxtapp{\ctxt{T_1}}{l}}$,
    $p=\termapp{u}{\ctxtapp{\ctxt{T_1}}{r}}$. We
    have two cases:

    \begin{itemize}
    \item The reduction step is in $\ctxtapp{\ctxt{T_1}}{l}$. In this
      case, (a) below must hold by the
    \ih, allowing us to conclude with (b): $$
\begin{array}{c@{\qquad}c}
\begin{tikzcd}
  \ctxtapp{\ctxt{T_1}}{l}\arrow[rightsquigarrow]{d}[anchor=north,left]{}
  &[-25pt] \simeq
  &[-25pt] \ctxtapp{\ctxt{T_1}}{r} \arrow[rightsquigarrow]{d}[anchor=north,left]{} \\[-5pt]
  \BMC{t} &[-25pt] \simeq &[-25pt] \BMC{t'}
\end{tikzcd}
&
\begin{tikzcd}
  \termapp{u}{\ctxtapp{\ctxt{T_1}}{l}}\arrow[rightsquigarrow]{d}[anchor=north,left]{}
  &[-25pt] \simeq
  &[-25pt] \termapp{u}{\ctxtapp{\ctxt{T_1}}{r}}\arrow[rightsquigarrow]{d}[anchor=north,left]{} \\[-5pt]
  \BMC{\termapp{u}{t}} &[-25pt] \simeq &[-25pt] \BMC{\termapp{u}{t'}}
\end{tikzcd} \\
\mbox{(a)} & \mbox{(b)}
\end{array} $$
The bottom of (b) follows from Lem.~\ref{l:app-bm}
with $\ctxt{O} = u\Box$ and $\BMC{t}\simeq \BMC{t'}$.

    \item The reduction step is in $u$. $$
\begin{tikzcd}
  \termapp{u}{\ctxtapp{\ctxt{T_1}}{l}}\arrow[rightsquigarrow]{d}[anchor=north,left]{}
  &[-25pt] \simeq
  &[-25pt] \termapp{u}{\ctxtapp{\ctxt{T_1}}{r}}\arrow[rightsquigarrow]{d}[anchor=north,left]{} \\[-5pt]
  \BMC{\termapp{u'}{\ctxtapp{\ctxt{T_1}}{l}}} &[-25pt] \simeq &[-25pt] \BMC{\termapp{u'}{\ctxtapp{\ctxt{T_1}}{r}}}
\end{tikzcd}$$
The bottom of the diagram follows from Lem.~\ref{l:app-bm}
with $\ctxt{O} = u' \Box$
  and $\ctxtapp{\ctxt{T_1}}{l}\simeq \ctxtapp{\ctxt{T_1}}{r}$.
\end{itemize}

  \item $\ctxt{T} = \termabs{z}{\ctxt{T_1}}$. The result follows from the \ih
    
  \item $\ctxt{T} = \termcont{\gamma}{\ctxt{C}}$. The result follows from the \ih
    
  \item $\ctxt{T} = \termsubs{z}{u}{\ctxt{T_1}}$. Then
    $o=\termsubs{z}{u}{\ctxtapp{\ctxt{T_1}}{l}}$ and
    $p=\termsubs{z}{u}{\ctxtapp{\ctxt{T_1}}{r}}$. 
    There are three further cases:

\begin{itemize}
  \item The reduction step takes place in $\ctxtapp{\ctxt{T_1}}{l}$
  In this case, (a) below must hold by the
  \ih, allowing us to conclude with (b): $$
\begin{array}{c@{\qquad}c}
\begin{tikzcd}
  \ctxtapp{\ctxt{T_1}}{l}\arrow[rightsquigarrow]{d}[anchor=north,left]{}
  &[-25pt] \simeq
  &[-25pt] \ctxtapp{\ctxt{T_1}}{r} \arrow[rightsquigarrow]{d}[anchor=north,left]{} \\[-5pt]
  \BMC{t} &[-25pt] \simeq &[-25pt] \BMC{t'}
\end{tikzcd}
&
\begin{tikzcd}
  \termsubs{z}{u}{\ctxtapp{\ctxt{T_1}}{l}}\arrow[rightsquigarrow]{d}[anchor=north,left]{}
  &[-25pt] \simeq
  &[-25pt] \termsubs{z}{u}{\ctxtapp{\ctxt{T_1}}{r}}\arrow[rightsquigarrow]{d}[anchor=north,left]{} \\[-5pt]
  \BMC{\termsubs{z}{u}{t}} &[-25pt] \simeq &[-25pt] \BMC{\termsubs{z}{u}{t'}}
\end{tikzcd} \\
\mbox{(a)} & \mbox{(b)}
\end{array}$$
  The bottom of (b) follows from Lem.~\ref{l:app-bm}
  with $\ctxt{O} = \termsubs{z}{u}{\Box}$
  and $\BMC{t}\simeq \BMC{t'}$.

\item The reduction step takes place at the root. $$
\begin{tikzcd}
  \termsubs{z}{u}{\ctxtapp{\ctxt{T_1}}{l}}\arrow[rightsquigarrow]{d}[anchor=north,left]{\rS}
  &[-25pt] \simeq
  &[-25pt] \termsubs{z}{u}{\ctxtapp{\ctxt{T_1}}{r}}\arrow[rightsquigarrow]{d}[anchor=north,left]{\rS} \\[-5pt]
  \BMC{\substitute{z}{u}{\ctxtapp{\ctxt{T_1}}{l}}} &[-25pt] \simeq &[-25pt] \BMC{\substitute{z}{u}{\ctxtapp{\ctxt{T_1}}{r}}}
\end{tikzcd}$$
  The bottom of the diagram follows from
  Lem.~\ref{l:equiv-substitution}
  with $u \in \BMCform$ and $\ctxtapp{\ctxt{T_1}}{l}\simeq \ctxtapp{\ctxt{T_1}}{r}$.

  \item The reduction step takes place in $u$.$$
\begin{tikzcd}
  \termsubs{z}{u}{\ctxtapp{\ctxt{T_1}}{l}}\arrow[rightsquigarrow]{d}[anchor=north,left]{}
  &[-25pt] \simeq
  &[-25pt] \termsubs{z}{u}{\ctxtapp{\ctxt{T_1}}{r}}\arrow[rightsquigarrow]{d}[anchor=north,left]{} \\[-5pt]
  \BMC{\termsubs{z}{u'}{\ctxtapp{\ctxt{T_1}}{l}}} &[-25pt] \simeq &[-25pt] \BMC{\termsubs{z}{u'}{\ctxtapp{\ctxt{T_1}}{r}}}
\end{tikzcd}$$
  The bottom of the diagram follows from Lem.~\ref{l:app-bm}
  with $\ctxt{O} = \termsubs{z}{u'}{\Box}$
  and $\ctxtapp{\ctxt{T_1}}{l}\simeq \ctxtapp{\ctxt{T_1}}{r}$.
  \end{itemize}

  \item $\ctxt{T} = \termsubs{z}{\ctxt{T_1}}{u}$. Then
  $o = \termsubs{z}{\ctxtapp{\ctxt{T_1}}{l}}{u}$ and
  $p = \termsubs{z}{\ctxtapp{\ctxt{T_1}}{r}}{u}$.
  There are three further cases:
  \begin{itemize}
    \item The reduction step takes place in $u$. $$
\begin{tikzcd}
  \termsubs{z}{\ctxtapp{\ctxt{T_1}}{l}}{u}\arrow[rightsquigarrow]{d}[anchor=north,left]{}
  &[-25pt] \simeq
  &[-25pt] \termsubs{z}{\ctxtapp{\ctxt{T_1}}{r}}{u}\arrow[rightsquigarrow]{d}[anchor=north,left]{} \\[-5pt]
  \BMC{\termsubs{z}{\ctxtapp{\ctxt{T_1}}{l}}{u'}} &[-25pt] \simeq &[-25pt] \BMC{\termsubs{z}{\ctxtapp{\ctxt{T_1}}{r}}{u'}}
\end{tikzcd}$$
    The bottom of the diagram follows from Lem.~\ref{l:app-bm} with
    $\ctxt{O} = \termsubs{z}{\Box}{u'}$ and
    $\ctxtapp{\ctxt{T_1}}{l} \simeq \ctxtapp{\ctxt{T_1}}{r}$.

    \item The reduction step takes place in $\ctxtapp{\ctxt{T_1}}{l}$
    In this case, (a) below must hold by the \ih, allowing us to conclude
    with (b): $$
\begin{array}{c@{\qquad}c}
\begin{tikzcd}
  \ctxtapp{\ctxt{T_1}}{l}\arrow[rightsquigarrow]{d}[anchor=north,left]{}
  &[-25pt] \simeq
  &[-25pt] \ctxtapp{\ctxt{T_1}}{r} \arrow[rightsquigarrow]{d}[anchor=north,left]{} \\[-5pt]
  \BMC{t} &[-25pt] \simeq &[-25pt] \BMC{t'}
\end{tikzcd}
&
\begin{tikzcd}
  \termsubs{z}{\ctxtapp{\ctxt{T_1}}{l}}{u}\arrow[rightsquigarrow]{d}[anchor=north,left]{}
  &[-25pt] \simeq
  &[-25pt] \termsubs{z}{\ctxtapp{\ctxt{T_1}}{r}}{u}\arrow[rightsquigarrow]{d}[anchor=north,left]{} \\[-5pt]
  \BMC{\termsubs{z}{t}{u}} &[-25pt] \simeq &[-25pt] \BMC{\termsubs{z}{t'}{u}}
\end{tikzcd} \\
\mbox{(a)} & \mbox{(b)}
\end{array}$$
    The bottom of (b) follows from Lem.~\ref{l:app-bm} with $\ctxt{O} =
    \termsubs{z}{\Box}{u}$ and $\BMC{t}\simeq \BMC{t'}$.

    \item The reduction step takes place at the root. $$
\begin{tikzcd}
  \termsubs{z}{\ctxtapp{\ctxt{T_1}}{l}}{u}\arrow[rightsquigarrow]{d}[anchor=north,left]{\rS}
  &[-25pt] \simeq
  &[-25pt] \termsubs{z}{\ctxtapp{\ctxt{T_1}}{r}}{u}\arrow[rightsquigarrow]{d}[anchor=north,left]{\rS} \\[-5pt]
  \BMC{\substitute{z}{\ctxtapp{\ctxt{T_1}}{l}}{u}} &[-25pt] \simeq &[-25pt] \BMC{\substitute{z}{\ctxtapp{\ctxt{T_1}}{r}}{u}}
\end{tikzcd}$$
    The bottom of the diagram follows from
    Lem.~\ref{l:Sigma_closed_under_substitution:target}
    with $u \in \BMCform$ and $\ctxtapp{\ctxt{T_1}}{l} \simeq
    \ctxtapp{\ctxt{T_1}}{r}$.
    \end{itemize}
  \end{itemize} 
\end{itemize}
\end{proof}

%%% Local Variables:
%%% mode: latex
%%% TeX-master: "main"
%%% End:

%%%Local Variables:
%%% mode: latex
%%% TeX-master: "main"
%%% End:

}

\end{document}